\documentclass[rmp,aps,lengthcheck]{revtex4-1}
\usepackage{longtable}
\usepackage{supertabular}

\usepackage{dcolumn}
\usepackage{epsf}
\usepackage{bm}
\usepackage{amsmath}
\usepackage{graphicx}
\usepackage{xr}
\usepackage{afterpage}
\usepackage{color}
\usepackage{inputenc}
\usepackage{fontenc}

\mathchardef\mhyphen="2D %
\def\shyphen{\!-\!}

\begin{document}
\def\elabel#1{\label{#1}}
\def\plabel#1{\label{#1}}
\def\slabel#1{\label{#1}}
\def\tlabel#1{\label{#1}}
\def\flabel#1{\label{#1}}
\def\numcite#1{\cite{#1}}
\def\numtextcite#1{\textcite{#1}}
\def\lref#1{\ref{#1}}
\def\comment#1{}
\def\tfrac#1#2{\displaystyle \frac{#1}{#2}}

\def\b{\hbox to 12pt{}}
\def\s#1{\hbox to #1pt{}}

\renewcommand{\thefootnote}{\arabic{footnote}}
\newcommand{\greeksym}[1]{{\usefont{U}{psy}{m}{n}#1}}
\newcommand{\rmssmu}{\mbox{\scriptsize{{\greeksym{m}}}}}
\newcommand{\rmsstau}{\mbox{\scriptsize{\greeksym{t}}}}
\newcommand{\rmssgamma}{\mbox{\scriptsize{\greeksym{g}}}}
\newcommand{\rmmu}{\mbox{\greeksym{m}}}
\newcommand{\rmtau}{\mbox{\greeksym{t}}}
\newcommand{\rmalpha}{\mbox{\greeksym{a}}}
\newcommand{\rmssalpha}{\mbox{\scriptsize{\greeksym{a}}}}
\newcommand{\rmpi}{\mbox{\greeksym{p}}}
\newcommand{\rmsspi}{\mbox{\scriptsize{\greeksym{p}}}}
\newcommand{\rmphi}{\mbox{\greeksym{f}}}
\newcommand\T{\rule{0pt}{2.4ex}}
\newcommand\B{\rule[-0.8ex]{0pt}{0pt}}
\def\lbar{\lambda\hskip-4.5pt\vrule height4.6pt depth-4.3pt width4pt}
\def\moles{$^{\rm o}\hskip-4.85pt\vrule height4.98pt depth-4.89pt width4.4pt\hskip4.85pt$}
\def\fr#1#2{{\textstyle{#1\over#2}}}
\def\nothing#1{\phantom{#1}}
\def\iG{\it \Gamma}

\title{
       CODATA Recommended Values of the Fundamental Physical
       Constants: 2022}

\thanks{
  This report was prepared by the authors under the auspices of 
  the CODATA Task Group on Fundamental Constants.
  The members of the task group are: \\
  F. Bielsa, Bureau International des Poids et Mesures\\
  K. Fujii, National Metrology Institute of Japan, Japan \\
  S. G. Karshenboim, Pulkovo Observatory, Russian Federation and
Max-Planck-Institut f\"ur Quantenoptik, Germany\\
  H. Margolis, National Physical Laboratory, United Kingdom \\
  P. J. Mohr, National Institute of Standards and Technology, United States of America \\
  D. B. Newell, National Institute of Standards and Technology, United States of America \\
  F. Nez, Laboratoire Kastler-Brossel, France \\
  R. Pohl, Johannes Gutenberg-Universit\"at Mainz, Germany\\
  K. Pachucki, University of Warsaw, Poland \\
  J. Qu, National Institute of Metrology of China, China\\
  T. Quinn (emeritus), Bureau International des Poids et Mesures\\
  A. Surzhykov, Physikalisch-Technische Bundesanstalt, Germany \\
  B. N. Taylor (emeritus), National Institute of Standards and Technology, United States of America\\
  E. Tiesinga, National Institute of Standards and Technology, United States of America\\
  M. Wang, Institute of Modern Physics, Chinese Academy of Sciences, China \\
  B. M. Wood, National Research Council, Canada 
  }
  
\author{Peter J. Mohr$^1$}
\thanks{Electronic address: mohr@nist.gov}
\author{David B. Newell$^1$}
\thanks{Electronic address: dnewell@nist.gov}
\author{Barry N. Taylor$^1$}
\thanks{Electronic address: barry.taylor@nist.gov}
\author{Eite Tiesinga$^{1,2}$}
\thanks{Electronic address: eite.tiesinga@nist.gov} 
\affiliation{
   $^1$National Institute of Standards and Technology,
    Gaithersburg, Maryland 20899, USA}
\affiliation{$^2$Joint
Quantum Institute and Joint Center for Quantum Information and Computer
Science, College Park, Maryland 20742, USA}

\date{\today}

\begin{abstract}

We report the 2022 self-consistent values of constants and conversion
factors of physics and chemistry recommended by the Committee on Data of
the International Science Council (CODATA).  The recommended values can
also be found at physics.nist.gov/constants.  The values are based on a
least-squares adjustment that takes into account all theoretical and
experimental data available through 31 December 2022.  A discussion of
the major improvements as well as inconsistencies within the data is
given.

\end{abstract}

\maketitle %

\tableofcontents

\makeatletter
\let\toc@pre\relax
\let\toc@post\relax
\makeatother

\section{Introduction}
\label{sec:intro}

\subsection{Background}
\label{subsec:bkgnd}

The 2022 CODATA least-squares adjustment (LSA) of the fundamental
physical constants by the Task Group on Fundamental Constants (TGFC) is
the most recent in a series of compilations of recommended values that
arguably began over 90 years ago by \citet{1929002}. The TGFC was
established in 1969 by the Committee on Data for Science and Technology,
now called the Committee on Data of the International Science Council
(ISC); it is given the responsibility of periodically providing the
scientific and technological communities with an internationally
accepted, self-consistent set of values of fundamental physical
constants and related conversion factors. In the same year, researchers
at RCA Laboratories and the University of Pennsylvania published a
comprehensive paper documenting an adjustment they carried out as an
outgrowth of their measurement of the Josephson constant $K_{\rm J}$
that included a complete set of recommended values~\cite{1969012}.
Although well received, it did not have a formal association with an
internationally recognized scientific body.

The first three recommended sets of fundamental constants provided by
the TGFC were from the 1973, 1986 and 1998 adjustments~\cite{1973003,
1987004, 2000035}. Since the 1998 adjustment the TGFC has carried out an
adjustment every four years and the 2022 adjustment is the seventh in
the series.  In addition, there was a special CODATA adjustment
completed by the TGFC in the summer of 2017 to determine the exact
values of the Planck constant $h$, elementary charge $e$, Boltzmann
constant $k$, and Avogadro constant $N_{\rm A}$ as the basis for the
revised International System of Units (SI) that went into effect on 20
May 2019 \cite{2018015, 2018018}.

The 2018 CODATA adjustment \cite{2021035, 2021036} was the first based
on the revised SI, which had a profound effect and will continue to
influence all subsequent adjustments. A comparison of the 2018 with the
2014 adjustment shows that the many data related to the determination of
$h$, $e$, $k$, and $N_{\rm A}$ no longer need to be considered.
Consequently, the relative role of quantum physics in the adjustments
has significantly increased.

\subsection{Overview of the 2022 adjustment}
\label{ssec:ova}

Here we identify the new experimental and theoretical data for possible
inclusion in the 2022 adjustment by considering all data available up
until the closing date of midnight, 31 December 2022.  It was not
necessary for papers reporting new results to have been published by
this date; however, they needed to at least be available as a preprint.
It can therefore be assumed that any cited paper with a 2023 or later
publication date was available before the 31 December 2022, closing
date. For conciseness, no references for the new data are included in
this summary since they are given in the sections of the paper in which
the data are discussed, and those sections are duly noted in the
summary. Although some topics, for example, muonium, the theoretical
values of bound-particle to free-particle ratios such as $g({\rm
H})/g_{\rm e}$, the proton magnetic moment in nuclear magnetons
$\mu_{\rm p}/\mu_{\rm N}$, lattice spacings of silicon crystals, and the
Newtonian constant of gravitation $G$ are reviewed in the main text,
they are not discussed here because nothing of significance relevant to
them has occurred since the 2018 adjustment.

Further, electron-proton and electron-deuteron scattering experiments
are not discussed in the main text. Although values of the
root-mean-square (rms) charge radius of the proton $r_{\rm p}$ and of
the deuteron $r_{\rm d}$ obtained from such experiments are included in
the 2018 adjustment, after due consideration the Task Group decided that
the scattering values of $r_{\rm p}$ and $r_{\rm d}$ should not be
included in the 2022 adjustment. This is because there is a lack of
consensus on how the experimental data should be analyzed to obtain
values of the radii and different methods can yield significantly
different values. Moreover, the two most recent data sets for the proton
can yield conflicting values depending on how the earlier set is
analyzed. Finally, the uncertainties of the scattering values are well
over an order of magnitude larger than those resulting from the
measurement of the Lamb shift in muonic hydrogen and muonic deuterium;
see, for example, the papers by \citet{2022006, 2021059, 2020022,
2019103}.

This summary, which generally follows the order of topics in the paper
as given in the table of contents, concludes with a brief discussion of
the treatment of all the available data to obtain the 2022 set of
recommended values.  Considerable material of a review nature may also
be found at the end of the paper in its final two sections,
Sec.~\ref{sec:2022crv} and Sec.~\ref{sec:sumconcl}. We start here with
relative atomic masses.

\subsubsection{Relative atomic masses}
\label{sssec:ram}

The required relative atomic masses $A_{\rm r}(^{28}{\rm Si})$, $A_{\rm
r}(^{87}{\rm Rb})$, and $A_{\rm r}(^{133}{\rm Cs})$ and their
correlation coefficients are taken from the 2020 Atomic Mass Evaluation
(AME) of the Atomic Mass Data Center (AMDC). The AME value of $A_{\rm
r}({\rm n})$ is not used as an input datum as in the past but is based
on the original capture gamma-ray measurement reported in 1999
[Sec.~\ref{ssec:amdci}; Table~\ref{tab:rmass20}; Sec.~\ref{sec:nmnc};
Tables~\ref{tab:pdata} and \ref{tab:pobseqsb}, D5, D6, D11, D14].

Experimental measurements of transition frequencies between
rovibrational states in HD$^+$ and the theory of the transitions have
achieved an uncertainty that allows the measured frequencies to be
included as input data and contribute to the determination of $A_{\rm
r}({\rm e})$, $A_{\rm r}({\rm p})$, and especially $A_{\rm r}({\rm d})$
[Sec~\ref{sec:HDplus}; Tables~\ref{tab:HDptheory-input} and
\ref{tab:betacoeff}; Tables~\ref{tab:pdata} and \ref{tab:pobseqsb},
D27-D32].

Four new experimentally determined cyclotron frequency ratios that also
contribute to the determination of $A_{\rm r}({\rm e})$, $A_{\rm r}({\rm
p})$, and $A_{\rm r}({\rm d})$ have become available for inclusion as
input data: $\omega_{\rm c}(^{12}{\rm C}^{6+})/\omega_{\rm c}({\rm p})$,
which is a 2017 result included in CODATA 2018 but revised in 2019;
$\omega_{\rm c}(^{12}{\rm C}^{6+})/\omega_{\rm c}({\rm d})$;
$\omega_{\rm c}({\rm H}_2^+)/\omega_{\rm c}({\rm d})$; and $\omega_{\rm
c}(^{12}{\rm C}^{4+})/\omega_{\rm c}({\rm HD}^+)$ [Sec.~\ref{ssec:frmd};
Tables~\ref{tab:pdata} and \ref{tab:pobseqsb}, D15-D18].

\subsubsection{Ionization and binding energies} The ionization energies
of $^1$H, $^3$H, $^3$He, $^4$He, $^{12}$C, and $^{28}$Si are from the
2022 National Institute of Standards and Technology (NIST) Atomic
Spectra Database (ASD). The required ionization energy for $^3$He$^+$
and the binding energies for $^{12}{\rm C}^{4+}$, $^{12}{\rm C}^{5+}$,
$^{12}{\rm C}^{6+}$, and $^{28}{\rm Si}^{13+}$ are derived from these as
appropriate and are input data for the 2022 adjustment
[Sec.~\ref{ssec:amdci}; Table~\ref{tab:be}; Table~\ref{tab:pdata}, D8,
D12, D22-D24]. (The binding energies of the molecular ions H$_2^+$ and
HD$^+$ in Eqs.~(\ref{eq:eih2plus}) and (\ref{eq:eihdplus}) of
Sec.~\ref{sec:ram}, and also D25 and D26 in Table~\ref{tab:pdata}, are
input data too but were available for the 2018 adjustment.)

\subsubsection{Hydrogen and deuterium transition energies}

A new value of the $1{\rm S}_{1/2}-3{\rm S}_{1/2}$ transition frequency
with $u_{\rm r} = 2.5\times 10^{-13}$ and of the $2{\rm S}_{1/2}-8{\rm
}D_{5/2}$ transition frequency with $u_{\rm r} = 2.6\times10^{-12}$,
both for hydrogen, have become available for the 2022 adjustment.
Reported in 2020 and 2022, respectively, the first is from the
Max-Planck-Institut fur Quantenoptik, Garching, Germany (MPQ) and the
second from the Colorado State University in Fort Collins, Colorado
(CSU). The MPQ result agrees well with the previous MPQ value for the
same transition reported in 2016 with $u_{\rm r} = 5.8\times10^{-12}$
and which is an input datum in the 2018 adjustment. However, it is
superseded by the new result since its uncertainty is 23 times smaller.
As will be seen, there are inconsistencies among the 29 H and D
transition frequencies that are input data in the 2022 adjustment; they
are addressed by the application of the same expansion factor to the
uncertainty of each of them.

The theoretical expressions for the transition frequencies are carefully
reviewed and updated with new results as appropriate.  However, the
uncertainties of the theoretical expressions for the experimentally
determined transition frequencies have not changed significantly from
those used in the 2018 adjustment. The H and D transition frequencies
and associated theory play an important role in CODATA adjustments
because they not only determine the Rydberg constant $R_\infty$ but
contribute to the determination of the proton and deuteron radii
discussed in the following paragraph [Sec.~\ref{sec:ryd};
Table~\ref{tab:rydfreq}; Table~\ref{tab:deltaRyd}].

\subsubsection{Muonic atoms and ions, radius of proton, deuteron, and
$\rmalpha$ particle}

Included as input data in the 2022 adjustment are the experimentally
determined values of the Lamb-shift transitions in muonic hydrogen,
$\rmmu$H, reported in 2013, in muonic deuterium, $\rmmu$D, reported in
2016, and in the muonic helium ion, $\rmmu^4{\rm He}^+$, reported in
2021 (in a muonic atom or hydrogenic ion the electron is replaced by a
negative muon).  Although the Lamb shifts in $\rmmu$H and $\rmmu$D were
available for use in previous adjustments, this is the first adjustment
for which $\rmmu^4{\rm He}^+$ is available. Together with the theory of
these Lamb shifts, the measurements contribute to the determination of
the rms charge radius of the proton $r_{\rm p}$, deuteron $r_{\rm d}$,
and alpha particle $r_{\rmssalpha}$ , respectively. The relative
uncertainties $u_{\rm r}$ of the measured Lamb shifts in $\rmmu$H,
$\rmmu$D, and $\rmmu^4{\rm He}^+$ are $1.1\times10^{-5}$,
$1.7\times10^{-5}$, and $3.5\times10^{-5}$, respectively. Based on the
recently published theory discussed in Sec.~\ref{sec:mulsrprd} of this
report, the respective relative uncertainties of the theoretical values
of the three Lamb shifts are now $1.2\times10^{-5}$, $1.0\times10^{-4}$,
and $3.1\times10^{-4}$. The end result is that $u_{\rm r}$ of the 2022
recommended values of the three radii are $7.6\times10^{-4}$,
$1.3\times10^{-4}$, and $12\times10^{-4}$, respectively.  The reduction
of the uncertainties of $r_{\rm p}$ and $r_{\rm d}$ compared to the
uncertainties of these radii in the 2018 adjustment has contributed to
the reduction in the uncertainty of the 2022 recommended value of
$R_\infty$ [Sec.~\ref{sec:mulsrprd}; Tables~\ref{tab:muhdata} and
\ref{tab:muonicatoms}].

Although these improvements are significant, as discussed in
Sec.~\ref{sec:mulsrprd}, a problem remains that future experiments and
theory may resolve. If the final adjustment used to obtain the 2022
recommended values is rerun without the muonic Lamb-shift data in
Tables~\ref{tab:muhdata} and \ref{tab:muonicatoms}, the resulting values
of $r_{\rm p}/$fm and $r_{\rm d}/$fm are $0.8529(13)~[51\times10^{-4}]$
and $2.1326(17)~[8.1\times10^{-4}]$ and arise from the electronic H and
D transition frequency data alone. When compared with the values $0.840
60(66)~[7.8\times10^{-4}]$ and $2.12643(133) ~[6.2\times10^{-4}]$ that
result from the muonic Lamb-shift data, the electronic H and D alone
values for both $r_{\rm p}$ and $r_{\rm p}$ exceed their muonic
Lamb-shift values by $2.8\,\sigma$ (as usual, $\sigma$ is the
root-sum-square uncertainty).  The proton radius ``puzzle'' is not yet
over.

\subsubsection{Electron magnetic-moment anomaly}

Included as an input datum in the 2022 adjustment is a new experimental
value of $a_{\rm e}$ with a relative uncertainty $u_{\rm r}$ of
$1.1\times10^{-10}$, which is 2.2 times smaller than that of the value
reported in 2008 and used as an input datum in the 2018 adjustment.
Both determinations were carried out under the supervision of
Prof.~G.~Gabrielse, but the earlier one at Harvard University and the
later one at Northwestern University. The experimenters view the new
result as superseding the earlier result.  There has not been a
comparable advance in the theory of $a_{\rm e}$; $u_{\rm r}$ of the
theoretical expression in the current adjustment (not including the
uncertainty of the fine-structure constant $\alpha$) is
$1.4\times10^{-11}$, not very different from the value
$1.5\times10^{-11}$ used in the 2018 adjustment. The only change of any
significance is that the value of the coefficient $A_1^{(10)}$ in the
2022 theoretical expression for $a_{\rm e}$ is 6.08(16) whereas it is
6.675(192) in the 2018 expression. The electron anomaly is of great
importance because experiment and theory together provide one of the
three most accurate determinations of the fine-structure constant
$\alpha$ in the 2022 adjustment [Sec.~\ref{sec:elmagmom}].

\subsubsection{Atom recoil}

The two input data $h/m(^{87}{\rm Rb})$ and $h/m(^{133}{\rm Cs})$ in the
2022 adjustment obtained using atom interferometry are also of great
importance because they provide the two other accurate values of
$\alpha$ through the comparatively simple observational equation $h/m(X)
= [A_{\rm r}({\rm e})/A_{\rm r}(X)][c\alpha^2/2R_\infty]$. The
$h/m(^{133}{\rm Cs})$ result with $u_{\rm r} = 4.0\times10^{-10}$ was
reported in 2018 and also used in that adjustment and the $h/m(^{87}{\rm
Rb})$ result with $u_{\rm r} = 1.4\times10^{-10}$ was reported in 2020.
A value for this quotient with $u_{\rm r} = 1.2\times10^{-9}$ obtained
by the same group in an earlier version of the experiment was reported
in 2011 and an input datum in the 2018 adjustment. However, the new
experiment uncovered previously unrecognized systematic effects in the
earlier experiment and as a consequence the 2020 value is fractionally
smaller by about 3 parts in $10^9$ than the previous value. Because that
value could not be corrected retroactively and because its $u_{\rm r}$
is over eight times larger, the new value is viewed as superseding it
[Sec.~\ref{sec:atomrecoil}].

\subsubsection{Muon magnetic-moment anomaly}

There are two experimental input data that determine the recommended
value of $a_{\rmssmu}$. These are values of the quantity
$R^{\,\prime}_{\rmssmu} = \omega_{\rm a}/\omega_{\rm p}^\prime$, where
$\omega_{\rm a}$ is the difference frequency between the spin flip (or
precession) frequency and cyclotron frequency of a muon in an applied
magnetic flux density $B$ and $\omega_{\rm p}^\prime$ is the precession
frequency of a proton in a spherical H$_2$O sample at 25$\,^\circ$C
inserted in $B$. The two are in good agreement: the first is the value
obtained at Brookhaven National Laboratory (BNL), Brookhaven, New York,
and reported in 2006 with $u_{\rm r} = 5.4\times10^{-7}$; and the second
is the value obtained at the Fermi National Accelerator Laboratory
(FNAL) in Batavia, Illinois, and reported in 2021 with $u_{\rm r} =
4.6\times10^{-7}$. The 7-meter diameter, 1.45 T muon storage ring magnet
used in the BNL measurement was moved to FNAL and used in the experiment
there. The BNL result is an input datum in the past four adjustments.
However, in the 2022 adjustment, it is treated same way as the FNAL
result.  There is no impact on its value and uncertainty.  The theory of
$a_{\rmssmu}$ has been thoroughly reviewed and updated by the Particle
Data Group (PDG), with special emphasis on hadronic contributions including
results from lattice quantum chromodynamics (QCD), with the uncertainty
of the value estimated to be $u_{\rm r} = 3.8 \times10^{-7}$.  The Task
Group has omitted it from the 2022 adjustment because of possible
contributions from physics beyond the standard model and because of the
long standing and significant disagreement between the experimental and
theoretical values.  Based on the results of the 2022 adjustment and the
theoretical value, there is a 4.2\,$\sigma$ discrepancy between experiment
and theory [Sec.~\ref{sec:mmma}; Tables~\ref{tab:pdata} and
\ref{tab:pobseqsb}, D33, D34].

\subsubsection{Electron $g$-factors in hydrogenic carbon-12 and
silicon-28}

The experimental values of the spin precession to cyclotron frequency
ratios $\omega_{\rm s}(^{12}{\rm C}^{5+})/\omega_{\rm c}(^{12}{\rm
C}^{5+})$ and $\omega_{\rm s}(^{28}{\rm Si}^{13+})/\omega_{\rm
c}(^{28}{\rm Si}^{13+})$, which determine $A_{\rm r}({ \rm e})$, are
input data in the 2018 adjustment and the same values are input data in
the 2022 adjustment. However, their observational equations, both of
which contain $A_{\rm r}({\rm e})$ as an adjusted constant, also contain
as an adjusted constant the $g$-factors $g_{\rm e}(^{12}{\rm C}^{5+})$
and $g_{\rm e}(^{28}{\rm Si}^{13+})$, respectively. These are calculated
from theory and improvements in the theory have reduced their
uncertainties, thereby leading to a recommended value of $A_{\rm r}({\rm
e})$ with a reduced uncertainty [Secs.~\ref{ssec:thbegf} and
\ref{ssec:bsgfexps}; Tables~\ref{tab:gfactthc} and \ref{tab:gfactthsi};
Tables~\ref{tab:pdata} and \ref{tab:pobseqsb}, D7, D9, D10, D13].

\subsubsection{Helion $g$-factor and magnetic shielding corrections}

The new direct measurement of the $g$-factor of the helion bound in the
$^3{\rm He}^+$ ion, $g_{\rm h}(^3{\rm He^+})$, provides the important
new input datum $\mu_{\rm h}(^3{\rm He}^+)/\mu_{\rm N}$ since $\mu_{\rm
h}(^3{\rm He}^+)/\mu_{\rm N} = g_{\rm h}(^3{\rm He}^+)$/2. Together with
new values of the bound helion magnetic shielding corrections
$\sigma_{\rm h}(^3{\rm He})$ and $\sigma_{\rm h}(^3{\rm He^+})$ with
uncertainties sufficiently small that the ratio $[1 - \sigma_{\rm
h}(^3{\rm He^+})]/[1 - \sigma_{\rm h}(^3{\rm He})]$ can be taken as
exact in the observational equation for this input datum, it leads to an
improved value of the adjusted constant $\mu_{\rm h}(^3{\rm
He})/\mu_{\rm p}^\prime$.  The latter in turn yields improved values of
both $\mu_{\rm h}(^3{\rm He}^+)$ and $\sigma_{\rm p}^\prime$, the proton
magnetic shielding correction in a spherical H$_2$O sample at
25\,$^\circ$C [Sec.~\ref{ssec:dmhe3p}; Sec.~\ref{ssec:thbfrats};
Tables~\ref{tab:pdata} and \ref{tab:pobseqsb}, D45].

\subsubsection{Electroweak quantities}

The recommended values for the mass of the tau lepton $m_{\rmsstau}$,
Fermi coupling constant $G_{\rm F}$, and sine squared of the weak mixing
angle $\sin^2{\theta_{\rm W}}$ are from the 2022 report of the PDG
[Sec.~\ref{sec:xeq}].

\subsubsection{Newtonian constant of gravitation}

We note that the 16 values of the Newtonian constant of gravitation $G$
in Table~\ref{tab:bg} are unrelated to any other data and are treated in
a separate calculation.  Since there is no new value, the same 3.9
expansion factor applied to their uncertainties in 2018 to reduce their
inconsistencies to an acceptable level is also used in 2022; the 2022
and 2018 recommended values of $G$ are therefore identical [See
Sec.~\ref{sec:ncg}].

\subsubsection{Least-squares adjustment (LSA)}

The 2022 CODATA set of recommended values of the constants are based on
a least-squares adjustment of 133 input data and 79 adjusted constants,
thus its degrees of freedom is $\nu = N – M = 54$. The $\chi^2$ of the
initial adjustment with no expansion factors applied to the
uncertainties of any data is 109.6; for 54 degrees of freedom, this
value of $\chi^2$ has only a $0.001\,\%$ probability of occurring by
chance.  Moreover, eight input data account for about $70\,\%$ of it. To
reduce $\chi^2$ to an acceptable level, an expansion factor of 1.7 is
applied to the uncertainties of all the data in
Tables~\ref{tab:rydfreq}, \ref{tab:deltaRyd}, and \ref{tab:muhdata} and
2.5 to those of data items D1 through D6 in Table~\ref{tab:pdata}.
Thus, for the final adjustment on which the 2022 recommended values are
based, $\chi^2$ is 44.2, which for $N = 54$ has a probability of
occurring by chance of 83\,\%.  See Appendices E and F in
\citet{2000035} for details of the adjustment [Sec.~\ref{sec:2022crv}].

\section{Relative atomic masses of light nuclei, neutral
silicon, rubidium, and cesium}
\label{sec:ram}

For the 2022 CODATA adjustment, we determine the relative atomic masses
$A_{\rm r}(X)$ of the neutron n, proton p, deuteron d, triton t, helion
h, the $\rmalpha$ particle, as well as the heavier neutral atoms
$^{28}$Si, $^{87}$Rb and $^{133}$Cs. The relative atomic masses of eight
of these nine particles are adjusted constants. The ninth adjusted
constant is the relative atomic mass of the hydrogenic $^{28}{\rm
Si^{13+}}$ ion rather than that of neutral $^{28}$Si. 

\subsection{Atomic Mass Data Center input}
\label{ssec:amdci}

The input data for the relative atomic masses of $^{28}$Si, $^{87}$Rb
and $^{133}$Cs are taken from the Atomic Mass Data Center (AMDC)
\cite{2021032,2021033}.  Their values with uncertainties and correlation
coefficients are listed in Table~\ref{tab:rmass20}.  The values can also
be found in Table~\ref{tab:pdata} as items D5, D6, and D11.  Of
course, the carbon-12 relative atomic mass is by definition simply the
number 12.  The observational equations for  $^{87}$Rb and $^{133}$C are
simply $A_{\rm r}(X) \doteq A_{\rm r}(X)$.  This  equation and all other
observational equations given in this section can also be found in
Table~\ref{tab:pobseqsb}.

The observational equation relating the relative atomic masses of
$^{12}$C with $^{12}$C$^{4+}$ and $^{12}$C$^{6+}$ as well as $^{28}$Si
with that of $^{28}{\rm Si}^{13+}$ follows from the  general observation
that the mass of any neutral atom  is the sum of its nuclear mass and
the masses of its electrons minus the mass equivalent of the binding
energy of the electrons. In other words, the observational equation for
the relative atomic mass of neutral atom $X$ in terms of that of ion
$X^{n+}$ in charge state $n=1,2,\cdots$ is
\begin{eqnarray}
  A_{\rm \rm r} (X) &\doteq&A_{\rm r}(X^{n+}) + n A_{\rm r} ({\rm e})
      -\frac{\Delta E_{\rm B}(X^{n+})}{{m_{\rm u} c^2}}\, ,
\label{eq:araxn}
\end{eqnarray}
where $A_{\rm r} ({\rm e})$ is the relative atomic mass of the electron
and ${\Delta}E_{\rm B}(X^{n+})>0$ is the binding or removal energy
needed to remove $n$ electrons from the neutral atom.  This binding
energy is the sum of the electron ionization energies $E_{\rm
I}(X^{i+})$ of ion $X^{i+}$. That is,
\begin{eqnarray}
  \Delta E_{\rm B}(X^{n+}) &=& \sum_{i = 0}^{n-1} E_{\rm I}(X^{i+}).
\label{eq:bind}
\end{eqnarray}
For a bare nucleus $n=Z$, while for a neutral atom $n=0$ and $\Delta
E_{\rm B}(X^ {0+}) = 0$.  The quantities $A_{\rm r} ({\rm e})$ and
${\Delta}E_{\rm B}(X^{n+})$ are adjusted constants.  The observational
equations for binding energies are simply
\begin{equation}
 {\Delta}E_{\rm B}(X^{n+})/hc \doteq  {\Delta}E_{\rm B}(X^{n+})/hc \,.
\end{equation}

Ionization energies for all relevant atoms and ions can be found in
Table~\ref{tab:be}. These data were taken from the 2022 NIST Atomic
Spectra Database (ASD) at \url{https://doi.org/10.18434/T4W30F}.  The
four binding energies relevant to the 2022 Adjustment are listed in
Table~\ref{tab:pdata} as items D8, D12, D23, and D24. The uncertainties
of the ionization energies are sufficiently small that correlations
among them or with any other data used in the 2022 adjustment are
inconsequential.  Nevertheless, the binding or new term energies of
$^{12}{\rm C}^{4+}$, $^{12}{\rm C}^{5+}$, and $^{12}{\rm C}^{6+}$ are
highly correlated with correlation coefficients 
\begin{eqnarray}
   r(^{12}{\rm C}^{4+},^{12}{\rm C}^{5+}) 
   &=& 0.996\,820\,, \nonumber\\
   r(^{12}{\rm C}^{4+},^{12}{\rm C}^{6+}) 
   &=& 0.996\,820\,, \label{eq:ccremoval}\\
   r(^{12}{\rm C}^{5+},^{12}{\rm C}^{6+}) 
   &=& 0.999\,999\,743\,, \nonumber
\end{eqnarray}
due to the uncertainties in the common ionization energies at lower
stages of ionization.

Binding energies are tabulated as wave number equivalents $\Delta E_{\rm
B}(X^{n+})/hc$, but are needed in terms of their relative atomic mass
unit equivalents $\Delta E_{\rm B}(X^{n+})/m_{\rm u}c^2$.  Given that
the Rydberg energy $hcR_\infty = \alpha^2m_{\rm e}c^2/2$, the last term
in Eq.~(\ref{eq:araxn}) is then rewritten as
\begin{eqnarray}
\frac{\Delta E_{\rm B}(X^{n+})}{{m_{\rm u} c^2}} &=& 
\frac{\alpha^2A_{\rm r}({\rm e})}{2R_\infty}\,
\frac{\Delta E_{\rm B}(X^{n+})}{hc} \, ,
\label{eq:ebxn}
\end{eqnarray}
where $\alpha$ and $R_\infty$ are also adjusted constants.

\begin{table}
 \caption[Input data for relative atomic masses from AME 2020]{Relative
 atomic masses used as input data in the 2022 CODATA adjustment and
 taken from the 2020 Atomic-Mass-Data-Center (AMDC) mass evaluation
 \cite{2021032,2021033}. 
 }
\label{tab:rmass20}
\begin{tabular}{cD{.}{.}{8.20}c}
 \hline\hline
 Atom &  \multicolumn{1}{c}{Relative atomic}
      & Relative standard \\
      & \multicolumn{1}{c}{\text{mass\footnotemark[1] $A_{\rm r}({X})$}}  
      & \hskip 8pt uncertainty $u_{\rm r}$  \\
   \hline
  $^{12}$C  & 12 & exact \vbox to 10 pt {}\\
  $^{28}$Si   &    27.976\,926\,534\,42(55)  &  $ 2.0\times 10^{-11}$  \\
  $^{87}$Rb   &    86.909\,180\,5291(65)  &  $ 7.5\times 10^{-11}$  \\
 $^{133}$Cs   &    132.905\,451\,9585(86) &  $ 6.5\times 10^{-11}$ \\[1 pt]
 \hline\hline
\end{tabular} 
\begin{tabular}{@{\hbox to 70 pt {}}rcl@{\hbox to 70  pt{}}}

\multicolumn{3}{c}{Correlation coefficients}  \\
 \hline
$r(^{28}$Si,$^{87}$Rb)&$=$&$ 0.0678$ \vbox to 10 pt {}\\
$r(^{28}$Si,$^{133}$Cs)&=&$ 0.0630$ \vbox to 10 pt {}\\
$r(^{87}$Rb,$^{133}$Cs)&=&$ 0.1032$ \vbox to 10 pt {}\\[1 pt]
 \hline\hline
\end{tabular}

{\small \footnotemark[1]The relative atomic mass $A_{\rm
r}(X)$ of particle $X$ with mass $m(X)$ is defined by $A_{\rm r}(X) =
m(X)/m_{\rm u}$, where $m_{\rm u} = m(^{12}{\rm C})/12$ is the atomic
mass constant.  }
\end{table}

\subsection{Neutron mass from the neutron capture gamma-ray measurement}
\label{sec:nmnc}

The mass of the neutron is derived from measurements of the binding
energy of the proton and neutron in a deuteron, $E_{\rm B}({\rm d})$,
and the definition that $m_{\rm d}=m_{\rm n}+m_{\rm p}-E_{\rm B}({\rm
d})/c^2$.  The binding energy is most accurately determined from the
measurement of the wavelength of the gamma-ray photon, $\lambda_{\rm
d}$, emitted from the capture of a neutron by a proton, both nearly at
rest, accounting for the recoil of the deuteron. That is, relativistic
energy and momentum conservation give $m_{\rm n} c^2+m_{\rm p}
c^2=\sqrt{(m_{\rm d}c^2)^2+p_{\rm d}^2c^2}+ hc/\lambda_{\rm d}$ and
$p_{\rm d}=h/\lambda_{\rm d}$, respectively, where $p_{\rm d}$ is the
(absolute value of the) momentum of the deuteron.  

\citet{1999052} measured the wavelength of the gamma rays by Bragg
diffraction of this light from the 220 plane of the natural-silicon
single crystal labeled ${\rm \scriptstyle ILL}$ in Sec.~\ref{sec:msc}.
Their  estimated value of the wavelength is $\lambda_{\rm d}= \eta_{\rm
d} \times d_{220}({\rm {\scriptstyle ILL}})$ per cycle, where
dimensionless measured input datum $\eta_{\rm d}=  2.904\,302\,45(49)\times 10^{-3}$ and
adjusted constant $d_{220}({\rm {\scriptstyle ILL}})$ is the relevant
lattice constant of crystal  ${\rm \scriptstyle ILL}$, constrained by
lattice constant data in Sec.~\ref{sec:msc}.  Over the past twenty
years, the accuracy of $d_{220}({\rm {\scriptstyle ILL}})$ has improved
since the 1999 measurement of $\eta_{\rm d}$. In fact, \citet{2006033}
gave an updated value for $\lambda_{\rm d}$ based on the then
recommended value for  $d_{220}({\rm {\scriptstyle ILL}})$.

The two relativistic conservation  laws for the capture of the neutron
by a proton can be solved for $\lambda_{\rm d}$ and thus input datum
$\eta_{\rm d}$. In fact,  expressed  in terms of  CODATA adjusted
constants, we have
\begin{equation}
 \eta_{\rm d} \doteq \frac{\alpha^2}{R_\infty}
 \,\frac{1}{d_{220}({\rm {\scriptstyle ILL}})} 
 \,\frac{A_{\rm r}({\rm e}) \, [A_{\rm r}({\rm n})+A_{\rm r}({\rm p})]}
 {[A_{\rm r}({\rm n})+A_{\rm r}({\rm p})]^2-A_{\rm r}({\rm d})^2}\,,
\label{eq:etad}
\end{equation}
as the observational equation for the least-squares adjustment to
determine the relative atomic mass of the neutron.  See also Eq.~(50) of
\citet{2000035}.  The approximate expression for the deuteron binding
energy 
\begin{equation}
 E_{\rm B}({\rm d})    \approx\frac{hc}{\lambda_{\rm d}}
 \left\{ 1 +  \frac{1}{2m_{\rm d}c^2}\frac{hc}{\lambda_{\rm d}}\right\} 
 \label{eq:ebd}
\end{equation}
is sufficiently accurate at the level of the current relative
uncertainty of $\eta_{\rm d}$.  We find $E_{\rm B}({\rm
d})= 2224.566\,40(38)$ keV with a relative uncertainty of $ 1.7\times 10^{-7}$. 

\subsection{Frequency ratio mass determinations}
\label{ssec:frmd}

Relative atomic masses of  p, d,  t,  h, and the $\rmalpha$ particle may
be derived from measurements of seven cyclotron frequency ratios of
pairs of the p, d, t, H$_2^+$, HD$^+$, $^3$He$^+$, $^4$He$^{2+}$ ions as
well as the $^{12}{\rm C}^{4+}$ and $^{12}{\rm C}^{6+}$ charge states of
carbon-12. Although several of these frequency measurements have been
used in the 2020 AMDC mass evaluation as well as in the previous CODATA
adjustment, here, we briefly describe these input data.  Since 2020,
ratios of the proton, deuteron, and electron masses are also constrained
by measurements of rotational and vibrational transition frequencies of
HD$^+$. These data and the relevant theory for these transition
frequencies are described in Sec.~\ref{sec:HDplus}.

The cyclotron frequency measurements rely on the fact that atomic or
molecular ions $X^{n+}$ with charge $ne$ in a homogeneous flux density
or magnetic field of strength $B$ undergo circular motion with cyclotron
frequency $\omega_{\rm c}(X^{n+})=ne\hbar B/m(X^{n+})$ that can be
accurately measured.  With the careful experimental design, ratios of
cyclotron frequencies for ions $X^{n+}$ and $Y^{p+}$ in the same
magnetic field environment then satisfy
\begin{equation}
\frac{ \omega_{\rm c}(X^{n+}) }{ \omega_{\rm c}(Y^{p+})} =
\frac{ n A_{\rm r}(Y^{p+})}{p A_{\rm r}(X^{n+})}
\end{equation}
independent of field strength.  For frequency ratios that depend on the
relative atomic masses of $^3$He$^+$, H$_2^+$, or HD$^+$ we use
\begin{eqnarray}
A_{\rm \rm r} (^3{\rm He}^+) &=& A_{\rm r}({\rm h}) + A_{\rm r} ({\rm e})
 -\frac{E_{\rm I}(^3{\rm He}^{+})}{{m_{\rm u} c^2}}\, ,
 \label{eq:arahe3p} \\
 A_{\rm \rm r} ({\rm H}_2^+) &=&2 A_{\rm r}({\rm p}) + A_{\rm r} ({\rm e})
 -\frac{E_{\rm I}({\rm H}_2^{+})}{{m_{\rm u} c^2}}\, ,
\label{eq:arah2}
\end{eqnarray}
or
\begin{eqnarray}
 A_{\rm \rm r} ({\rm HD}^+) &=&A_{\rm r}({\rm p}) 
 + A_{\rm r}({\rm d}) + A_{\rm r} ({\rm e})
 -\frac{ E_{\rm I}({\rm HD}^{+})}{{m_{\rm u} c^2}}\, ,
\label{eq:arahd}
\end{eqnarray}
respectively, with the $^3{\rm He}^+$ ionization energy from
Table~\ref{tab:be} and molecular  ionization energies \cite{2017094}

\begin{eqnarray}   
 E_{\rm I}({\rm H}_2^+)/hc  &=&    1.310\,581\,219\,937(6)  \times 10^7\ {\rm m}^{-1}\,,
 \label{eq:eih2plus}\\
 E_{\rm I}({\rm HD}^+)/hc &=&  1.312\,246\,841\,650(6) \times 10^7\ {\rm m}^{-1} \,.
 \label{eq:eihdplus}
 \end{eqnarray}
These data are used without being updated for current values of the
relevant constants, because their uncertainties and thus any changes
enter at the $10^{-15}$ level, which is completely negligible.

For ease of reference, the seven measured cyclotron frequency ratios are
summarized in Table~\ref{tab:pdata}.  Observational equations are given
in Table~\ref{tab:pobseqsb}. The first of these measurements is relevant
for the determination of the relative atomic mass of the proton. In
2017, the ratio of cyclotron frequencies of the $^{12}$C$^{6+}$ ion and
the proton, $\omega_{\rm c}(^{12}{\rm C}^{6+})/\omega_{\rm c}({\rm p})$,
was measured at a Max Planck Institute in Heidelberg, Germany (MPIK)
\cite{2017056}.  The researchers reanalyzed their experiment in
\citet{2019074} and published a corrected value that supersedes their
earlier value.  The corrected value has shifted by a small fraction of
the standard uncertainty and has the same uncertainty.

\citet{2020060} measured the cyclotron frequency ratio of d and
$^{12}$C$^{6+}$  and the cyclotron frequency ratio of HD$^{+}$ and
$^{12}$C$^{4+}$ to mainly constrain the relative atomic mass of the
deuteron. These two input data were not available for the AME 2020
adjustment and are also new for this 2022 CODATA adjustment. This
experiment as well the experiments by \citet{2017056,2019074} were done
in a cryogenic 4.2 K Penning trap mass spectrometer with multiple
trapping regions for the ions.  The cryogenic temperatures imply a near
perfect vacuum avoiding ion ejection and frequency shifts due
to collisions with molecules in the environment. The main systematic
limitation listed in \citet{2017056} was a residual quadratic magnetic
inhomogeneity.  In 2020, the researchers reduced this effect with a
chargeable superconducting coil placed around the trap chamber but
inside their main magnet and   reduced the  magnetic-field inhomogeneity
by a factor of 100 compared to that reported by \citet{2017056}. The
leading systemic effects are now due to image charges of the ion on the
trap surfaces and limits on the analysis of the lineshape of the
detected axial oscillation frequency of the ions.  The statistical
uncertainty is slightly smaller than the combined systematic
uncertainty.  Finally, the derived deuteron relative atomic mass
differed by $4.8\,\sigma$ from the recommended value of our 2018 CODATA
adjustment.

In 2021, \citet{2021042} at Florida State University, USA measured the
cyclotron frequency ratio of the homonuclear molecular ion H$_2^+$ and
the deuteron. This value supersedes the value published by the same
authors in \citet{2020003} and is new for our 2022 CODATA adjustment. In
their latest experiment, the authors used a cryogenic Penning trap with
one ion trapping region but with the twist that both ions are
simultaneously confined.  The ions have coupled magnetron orbits, such
that the ions orbit the center of the trap in the plane perpendicular to
the magnetic field direction, $180^\circ$ apart, and at a separation of
$\approx 1$ mm.  Simultaneous measurements of the coupled or shifted
cyclotron and axial frequencies then suppressed the role of temporal
variations in the magnetic field by three orders of magnitude compared
to that of sequential measurements.  The four frequencies are  combined
to arrive at the required (uncoupled) cyclotron frequency ratio.  In
fact, the authors could also assign the rovibrational state of the
H$_2^+$ ion from their signals.  The largest systematic uncertainties
are now due to the effects of special relativity on the fast cyclotron
motion and in uncertainties in the cyclotron radius when driving the
cyclotron mode.  The statistical uncertainty in this experiment is
slightly smaller than the combined systematic uncertainty.

Two cyclotron-frequency-ratio measurements determine the triton and
helion relative atomic masses, $A_{\rm r}({\rm t})$ and $A_{\rm r}({\rm
h})$, respectively.  These masses are primarily determined by the ratios
$\omega_{\rm c}(\rm t)/\omega_{\rm c}(^{3}{\rm He}^{+})$ and
$\omega_{\rm c}({\rm HD}^{+})/\omega_{\rm c}(^{3}{\rm He}^{+})$, both of
which were measured at Florida State University. The ratios have been
reported by \citet{2015002} and \citet{2017099}, respectively.  See also
the recent review by \citet{2019054}.  Both data have already been used
in the 2018 CODATA adjustment.

Finally, we use as an input datum the cyclotron frequency ratio
$\omega_{\rm c}(^{4}{\rm He}^{2+})/\omega_{\rm c}(^{12}{\rm C}^{6+})$ as
measured by \cite{2006036} at the University of Washington, USA. We
follow the 2020 AMDC recommendation to expand the published uncertainty
by a factor 2.5 to account for inconsistencies between data from the
University of Washington, the Max Planck Institute in Heidelberg, and
Florida State University on related cyclotron frequency ratios.

A post deadline publication reports a measurement of the $^{4}{\rm
He}^{2+}$ mass by \citet{2023022} made with a high-precision
Penning-trap mass spectrometer (LIONTRAP).  Their result differs from
the CODATA 2022 value by $3\,\sigma$.

\subsection{Mass ratios from frequency measurements in HD$^+$}
\label{sec:HDplus}

Measurements  as well as theoretical determinations of transition
frequencies between rovibrational states in the molecular ion HD$^+$  in
its electronic X$^2\Sigma^+$ ground state have become sufficiently
accurate that their comparison can be used to constrain the
electron-to-proton and electron-to-deuteron mass ratios. This 2022
CODATA adjustment is the first time that these types of data are used.
We follow the analysis of the experiments on and modeling of  the
three-particle system HD$^+$  by \citet{2023021}. Theoretical results
without nuclear hyperfine interactions have been derived by
\citet{2021039}.

Three independent experimental data sets exist. They correspond to
measurements of hyperfine-resolved rovibrational transition
frequencies $f^{\rm exp}(vLQ\to v'L'Q')=(E_{v'L'Q'}-E_{vLQ})/h$ between
pairs of states with energies $E_{vLQ}$ labeled by vibrational level
$v$, quantum number $L$ of the rotational or orbital angular momentum
${\bf L}$ of the three particles  as well as the collective
``hyperfine'' label $Q$ representing spin states produced by fine- and
hyperfine-couplings among ${\bf L}$,  electron spin ${\bf s}_{\rm e}$,
proton spin ${\bf i}_{\rm p}$, and deuteron spin ${\bf i}_{\rm d}$.
(Fine-structure splittings due to operator ${\bf L}\cdot {\bf s}_{\rm
e}$ do exist in HD$^+$, but are small compared to those due to hyperfine
operators like ${\bf i}_{\rm p}\cdot {\bf s}_{\rm e}$. For simplicity,
we use the aggregate term hyperfine to describe all these interactions.)
A limited number of transitions between hyperfine components of state
$vL$ and those of state $v'L'$  with $vL\neq v'L'$ have been measured.
\citet{2020029} at the Heinrich-Heine-Universit\"at in D\"usseldorf,
Germany measured six hyperfine-resolved transition frequencies for
$vL=0,0\to v'L'=0,1$.  \citet{2021049}  in the same laboratory measured
two hyperfine-resolved transition frequencies for $vL=0,0\to v'L'=1,1$,
while \citet{2020061} in the LaserLab at the Vrije Universiteit
Amsterdam, The Netherlands measured two hyperfine-resolved transition
frequencies for the $vL=0,3\to v'L'=9,3$ overtone.

The computation of the theoretical transition frequencies summarized by
\citet{2023021} has multiple steps. The first is a numerical evaluation
of the relevant non-relativistic three-body eigenvalues and
eigenfunctions of the electron, proton, and deuteron system in its
center of mass frame and in atomic units with energies expressed in the
Hartree energy $E_{\rm h}$ and lengths in the Bohr radius $a_0$.  In
these units, the non-relativistic three-body Hamiltonian  has mass
ratios $m_{\rm e}/m_{\rm p}$ and $m_{\rm e}/m_{\rm d}$ as the only
``free'' parameters.  The relevant non-relativistic states have energies
between $-0.6E_{\rm h}$ and $-0.5E_{\rm h}$ and were computed with
standard uncertainties better than $10^{-20}E_{\rm h}$ in the early
2000s.

The non-relativistic three-body Hamiltonian commutes with $\bf L$.
Hence, each eigenstate can be labeled by a unique angular momentum
quantum number ${L=0},\,1,\,\cdots$.  In addition, a Hund's case (b)
$^{2}\Lambda^\pm$ electronic state and a vibrational quantum number $v$
can be assigned to each eigenstate based on the adiabatic,
Born-Oppenheimer approximation for the system.  Here, first the proton
and deuteron are frozen in space,  the energetically lowest eigenvalue
$V_X(R)$ of the remaining electronic Hamiltonian is found as function of
proton-deuteron separation $R$, and, finally, the vibrational states for
the relative motion of the proton and deuteron in potential
$V_X(R)+\hbar^2L(L+1)/2\mu_{pd} R^2$ are computed using the reduced
nuclear mass $\mu_{\rm pd}=m_{\rm p}m_{\rm d}/(m_{\rm p}+m_{\rm d})$ in
the radial kinetic energy operator.  The experimentally observed
rovibrational states all belong to the ground ${\rm X}^2\Sigma^+$
electronic state with $v=0$, 1, and 9.

Relativistic, relativistic-recoil, quantum-electrodynamic (QED), and
hyperfine corrections as well as corrections due to nuclear charge
distributions are computed using first- and second-order perturbation
theory starting from the numerical non-relativistic three-body
eigenstates.  For example, the corrections correspond to the Breit-Pauli
Hamiltonian, three-dimensional delta function potentials modeling the
proton and deuteron nuclear charge distributions, as well as effective
Hamiltonians for the self-energy and vacuum-polarization  of the
electron.  Hyperfine interactions, those that couple the nuclear spins
of the proton and  deuteron to the electron spin and  angular momentum,
form another class of corrections.  Some of the smallest contributions
have not been computed using the non-relativistic three-body eigenstates
but rather using wavefunctions obtained within the Born-Oppenheimer
approximation or  with even simpler variational Ansatzes. Estimates of
the size of missing corrections and of uncertainties due to the
Born-Oppenheimer or variational approximations determine the
uncertainties of the theoretical energy levels and transition
frequencies.

For the 2022 CODATA adjustment, we follow \citet{2023021} and use as
input data the three spin-averaged (SA) transition frequencies $f^{\rm
exp}_{\rm SA} (vL\to v'L')$ between rovibrational states $(v,L)$ and
$(v'L')$ derived from the measured hyperfine resolved $f^{\rm
exp}(vLQ\to v'L'Q')$ for each $vL\to v'L'$ where the effects of the
fine- and hyperfine-structure have been removed. As explained in
\citet{2023021}, this removal is not without problems as the
experimentally observed hyperfine splittings were inconsistent with
theoretical predictions.  An expansion factor had to be introduced.  The
SA transition frequencies and  correlation coefficients among these
frequencies can be found in Tables \ref{tab:pdata} and
\ref{tab:ccpdata}.

The corresponding theoretical spin-averaged transition frequencies
$f^{\rm th}_{\rm SA} (vL\to v'L')$ are functions of the Rydberg
constant $R_\infty$, the ratio $\lambda_{\rm pd}=\mu_{\rm pd}/m_{\rm
e}$, as well $r^2_{\rm p}$ and $r^2_{\rm d}$,  the squares of the
nuclear charge radii of the proton and deuteron, respectively.  To a
lesser extent the spin-averaged transition frequencies also depend on
the deuteron-to-electron mass ratio $\lambda_{\rm d}=m_{\rm d}/m_{\rm
e}$.  In fact, \citet{2023021} showed that the expansion
\begin{eqnarray} 
\lefteqn{   f^{\rm th}_{\rm SA} (vL\to v'L') =
f_{\rm ref}  (vL\to v'L') } \label{eq:HDptheory}   \\ && \quad  +
\beta_{\lambda,{\rm pd}}(vL\to v'L') \left[  \lambda_{\rm pd}-
\lambda_{\rm pd,ref}\right] \nonumber \\ && \quad  + \beta_{\lambda,{\rm
d}}(vL\to v'L') \left[  \lambda_{\rm d}-    \lambda_{\rm d,ref}\right]
\nonumber \\ &&\qquad 
+ \beta_{\rm \infty}(vL\to v'L') \left[
cR_{\infty}-    cR_{\infty,\rm ref}\right] \quad    
\nonumber\\ &&
\qquad\quad+ \beta_{\rm r,p}(vL\to v'L') \left[  r^2_{\rm p}-(r_{\rm
p, ref})^2\right]
\nonumber\\ && \qquad\quad+ \beta_{\rm r,d}(vL\to v'L')
\left[  r^2_{\rm d}-    (r_{\rm d, ref})^2\right]\,,
\nonumber
\end{eqnarray}
around reference values $\lambda_{\rm pd, ref}$, $\lambda_{\rm d, ref}$,
$cR_{\infty,\rm ref}$, $r_{\rm p,ref}$, and $r_{\rm d,ref}$ for the five
constants is sufficient to accurately describe  theoretical transition
frequencies and their dependence on $\lambda_{\rm pd}$, $\lambda_{\rm
d}$, $cR_\infty$, $r^2_{\rm p}$, and $r^2_{\rm d}$.  The reference
values are derived from the recommended values for $A_{\rm r}({\rm e})$,
$A_{\rm r}({\rm p})$, $A_{\rm r}({\rm d})$, $cR_\infty$, $r_{\rm p}$,
and $r_{\rm d}$ from the 2022 CODATA adjustment and can be found in
Table \ref{tab:HDptheory-input}.  The reference transition frequencies
$f_{\rm ref}  ({vL\to v'L'})$ and coefficients $\beta_i({vL\to v'L'})$
are found from  calculations by \citet{2023021} at the reference values
for $\lambda_{\rm pd}$, $\lambda_{\rm d}$, $cR_\infty$, $r_{\rm p}$, and
$r_{\rm d}$. Values for $f_{\rm ref}  ({vL\to v'L'})$ and
$\beta_i({vL\to v'L'})$ can be found in Table \ref{tab:betacoeff}.  The
2018 recommended value for the fine-structure constant was also used in
the theoretical simulations but its uncertainty  does not affect $f^{\rm
th}_{\rm SA} (vL\to v'L')$ at current levels of uncertainty in the
theory.

The observational equations are
\begin{eqnarray}
   f^{\rm exp}_{\rm SA}(vL\to v'L') &\doteq& 
   f^{\rm th}_{\rm SA}(vL\to v'L')   \label{eq:HDpobseqn}  \\
   && \quad +\,\delta_{\rm HD^+}^{\rm th}(vL\to v'L') \nonumber 
\end{eqnarray}
and
\begin{equation}
  \delta_{\rm HD^+}(vL\to v'L') 
  \doteq \delta_{\rm HD^+}^{\rm th}(vL\to v'L') \,,
\end{equation}
where frequencies $\delta_{\rm HD^+}^{\rm th}(vL\to v'L')$ are additive
adjusted constants accounting for the uncomputed terms in the
theoretical expression of Eq.~(\ref{eq:HDptheory}). The values,
uncertainties, and correlation coefficients for  input data $
\delta_{\rm HD^+}(vL\to v'L') $ can be found in Tables
\ref{tab:pdata} and \ref{tab:ccpdata}. We also use
\begin{equation}
  \lambda_{\rm pd} = \frac{A_{\rm r}({\rm p})A_{\rm r}({\rm d})}{A_{\rm r}({\rm p})+A_{\rm r}({\rm d})} \frac{1}{A_{\rm r}({\rm e})}
\end{equation}
and
\begin{equation}
  \lambda_{\rm d} =A_{\rm r}({\rm d})/A_{\rm r}({\rm e})
\end{equation}
in terms of adjusted constants $A_{\rm r}({\rm e})$, $A_{\rm r}({\rm
p})$, and $A_{\rm r}({\rm d})$ in order to evaluate $f^{\rm th}_{\rm
SA}(vL\to v'L')$ using Eq.~(\ref{eq:HDptheory}). The HD$^+$ input data
and observational equations can also be found in Table~\ref{tab:pdata}
as items D27-D32 and in Table~\ref{tab:pobseqsb}, respectively.

\begin{table}
\caption{Values for the five reference constants in the theoretical
expression for the HD$^+$ transition frequencies in
Eq.~(\ref{eq:HDptheory}).  } \label{tab:HDptheory-input}
\begin{tabular}{c|@{\ \ }l} 
\hline\hline
 $\lambda_{\rm pd,ref}$  & $ 1\,223.899\,228\,723\,2$      \\ 
 $\lambda_{\rm d,ref}$   & $ 3\,670.482\,967\,881\,4$      \\   
 $cR_{\infty,\rm ref}$   & $ 3\,289\,841\,960\,250.8$\ kHz \\
 $r_{\rm p,ref}$         & $ 0.8414$\ fm    \\
 $r_{\rm d,ref}$         & $ 2.127\,99$\ fm    \\
\hline\hline
\end{tabular}
\end{table}

\begin{table*}
\caption{Values for  theoretical coefficients appearing  in
Eq.~(\ref{eq:HDptheory}) for  three  $vL \to v'L'$ rovibrational
transitions in HD$^+$.} \label{tab:betacoeff}
\begin{tabular}{c|c@{\hspace{1mm}}c@{\hspace{1mm}}
c@{\hspace{1mm}}c@{\hspace{1mm}}c@{\hspace{1mm}}c} 
\hline\hline
Transition & $f_{\rm ref}$ & $\beta_{\lambda,\rm pd}$ 
& $\beta_{\lambda,\rm d}$ &  $\beta_{\infty}$ 
& $\beta_{\rm r,p}$ & $\beta_{\rm r,d}$ \\
& (kHz) & (kHz) & (kHz) && (kHz/fm$^2$)  & (kHz/fm$^2$) \\
\hline 
$0,0\! \to\! 0,1$ &   1\,314\,925\,752.929 
& $ -1.0601\times 10^{6}$ & $ -3.2126\times 10^{1}$ & 
 $ 3.9969\times 10^{-4}$ & $ -9.0991\times 10^{-1}$ & 
 $ -9.0991\times 10^{-1}$ \\
$0,0\! \to\! 1,1$ &  58\,605\,052\,163.88 & 
$ -2.3201\times 10^{7}$ & $ -7.0874\times 10^{2}$ &
 $ 1.7814\times 10^{-2}$ & $ -2.4253\times 10^{1}$ 
 & $ -2.4220\times 10^{1}$ \\
$0,3 \!\to \!9,3$ &  415\,264\,925\,502.7 
& $ -1.2580\times 10^{8}$ & $ -3.9569\times 10^{3}$ & 
 $ 1.2623\times 10^{-1}$ & $ -1.5940\times 10^{2}$ 
 & $ -1.5850\times 10^{2}$ \\
\hline\hline
\end{tabular}
\end{table*}

Useful intuition regarding the size and behavior of  some of the
coefficients $\beta_i(vL\to v'L')$ can be obtained from an analysis of
the HD$^+$ rovibrational energies within the Born-Oppenheimer
approximation for the X$^2\Sigma^+$ electronic ground state and the
harmonic approximation of its potential $V_X(R)$ around the equilibrium
separation $R_{\rm e}$.  That is, $V_X(R)\approx   V_0 +
\frac{1}{2}\kappa (R-R_{\rm e})^2$ with dissociation energy $V_0$ and
spring constant $\kappa$.  Moreover, 
\begin{equation}
    \kappa= d_{\kappa} \frac{hcR_\infty}{a_0^2}
\end{equation}
and 
\begin{equation}
  R_{\rm e}= d_{\rm e} a_0,
\end{equation}
where $d_{\kappa}$ and $d_{\rm e}$ are  dimensionless constants of order
1, and
\begin{equation}
hcR_\infty=\frac{\hbar^2}{2m_{\rm e} a_0^2} \, ,
\end{equation}
where $a_0$ is the Bohr radius.  Approximate rovibrational
energies are then 
\begin{eqnarray}
   E^{\rm approx}(v,L) &=& V_0+\hbar \omega_{\rm e} (v+1/2) \\
    && \quad  + \frac{\hbar^2}{2\mu_{\rm pd} R_{\rm e}^2}  L(L+1)+\cdots
    \nonumber 
\end{eqnarray}
with harmonic frequency $\omega_{\rm e}=\sqrt{\kappa/\mu_{\rm pd}}$ and  also
\begin{eqnarray}
   E^{\rm approx}(v,L) &=& 
   V_0+hcR_{\infty}\sqrt{\frac{m_{\rm e} }{\mu_{\rm pd}}} 
   \sqrt{2d_\kappa}  (v+1/2)\nonumber  \\
   && \ \    + hcR_{\infty}\frac{m_{\rm e} }{\mu_{\rm pd}} 
   \frac{L(L+1)}{d_{\rm e}^2}+\cdots \,.
    \label{eq:Rinftyme}
\end{eqnarray}
Note that the energy differences between vibrational states $v$ is much
larger than those between rotational states within the same $v$ as
$m_{\rm e}\ll\mu_{\rm pd}$.

A corollary of Eq.~(\ref{eq:Rinftyme}) is that the partial derivative of
approximate transition frequencies with respect to $\lambda_{\rm pd}$ is
\begin{equation}
    \frac{\partial f^{\rm approx}}{\partial \lambda_{\rm pd}}= 
    - \eta \frac{1}{\lambda_{\rm pd}} f^{\rm approx} 
    \label{eq:lpd}
\end{equation}
at fixed $cR_\infty$. Here, $\eta=1$ for rotational transitions within
the same $v$ and 1/2 for vibrational transitions.  A numerical
evaluation of the right-hand side of  Eq.~(\ref{eq:lpd}) using the
reference values $\lambda_{\rm pd, ref}$ and $f_{\rm ref}$  agrees with
$\beta_{\lambda,\rm pd}$ in Table \ref{tab:betacoeff} within 1\,\% for
the $0,0\to 0,1$ transition and within 5\,\% for the $0,0\to 1,1$
transition. For the overtone $0,3\to 9,3$ transition the agreement is
worse as anharmonic corrections to $V_X(R)$ become important.  The
partial derivative of approximate transition frequencies with respect to
$\lambda_{\rm d}$ is zero at fixed $\lambda_{\rm pd}$ and $cR_\infty$.
This explains the small value for $\beta_{\lambda,\rm d}$ relative to
that of $\beta_{\lambda,\rm pd}$.  Finally, we have
\begin{equation}
   \frac{\partial f^{\rm approx}}{\partial c R_\infty} =
      \frac{1}{c R_\infty}f^{\rm approx}\,.
\end{equation}
The right-hand side is an exact representation of $\beta_\infty$ for all
transitions.

\begin{table}[h]
\def\vb{\vbox to 10 pt {}}
\caption[Ionization energies for selected atoms]{Ionization energies for $^1$H, $^3$H, $^3$He, $^4$He, $^{12}$C, and $^{28}$Si. 
The full description of unit m$^{-1}$ is cycles or periods per meter.  
Covariances among the data in this table have not been included in the
adjustment. See text for explanation.
}
\label{tab:be}
\begin{tabular}{cD{.}{.}{3.15}@{~~}cD{.}{.}{3.15}}
\hline
\hline
  &  
\multicolumn{1}{c}{$E_{\rm I}/hc~(10^7 \ {\rm m}^{-1})$} 
&  & 
\multicolumn{1}{c}{$E_{\rm I}/hc~(10^7 \ {\rm m}^{-1})$\vb} 
        \\
\hline
$^1$H     &  1.096\,787\ldots \vbox to 10 pt {}  \\
$^3$H  &  1.097\,185\ldots    \\

$^3$He$^+$       &  4.388\,891\,939(2) \\
$^4$He           &  1.983\,106\,6637(20)  &
$^4$He$^+$        &  4.389\,088\,788\,400(80) \\
$^{12}$C     &  0.908\,203\,480(90) &
$^{12}$C$^+$     &  1.966\,6331(10)  \\
$^{12}$C$^{2+}$     &  3.862\,410(20) &
$^{12}$C$^{3+}$     &  5.201\,753(15)  \\
$^{12}$C$^{4+}$     &  31.624\,2330(20)   &
$^{12}$C$^{5+}$     &  39.520\,617\,464(18)  \\
$^{28}$Si  &  0.657\,4776(25)   &
$^{28}$Si$^+$  &  1.318\,3814(30)  \\
$^{28}$Si$^{2+}$  &  2.701\,3930(70) &
$^{28}$Si$^{3+}$  &  3.640\,9310(60)  \\
$^{28}$Si$^{4+}$  &  13.450\,70(25)   &
$^{28}$Si$^{5+}$  &  16.556\,90(40)  \\
$^{28}$Si$^{6+}$  &  19.8873(40) &
$^{28}$Si$^{7+}$  &  24.4864(42)\\
$^{28}$Si$^{8+}$  &  28.3330(50)  &
$^{28}$Si$^{9+}$  &  32.3735(34)   \\
$^{28}$Si$^{10+}$  &  38.4140(15)  &
$^{28}$Si$^{11+}$  &  42.216\,30(60) \\
$^{28}$Si$^{12+}$  &  196.610\,389(16) &
$^{28}$Si$^{13+}$  &  215.606\,3427(14) \\
\hline
\hline
\end{tabular}
\end{table}
\section{Atomic hydrogen and deuterium transition energies} 
\label{sec:ryd}

Comparison of theory and experiment for electronic transition energies
in atomic hydrogen and deuterium is currently the most precise way to
determine the Rydberg constant, or equivalently the Hartree energy, and
to a lesser extent the charge radii of the proton and deuteron.  Here,
we summarize the theory of and the experimental input data
on H and D energy levels in Secs.~\ref{ssec:hdel} and \ref{sec:tfhd},
respectively.  

The charge radii of the proton and deuteron are also constrained by data
and theory on muonic hydrogen and muonic deuterium.  These data are
discussed in Sec.~\ref{sec:mulsrprd}.

The electronic eigenstates of H and D are labeled by $n\ell_j$, where
$n=1$, 2, $\dots$ is the principal quantum number,
$\ell=0,1,\dots,{n-1}$ is the quantum number for the nonrelativistic
electron orbital angular momentum $\bf L$, and $j=\ell\pm1/2$ is the
quantum number of the total electronic angular momentum $\bf J$.

Theoretical values for the energy levels of H and D are determined by
the Dirac eigenstate energies, QED effects such as self-energy and
vacuum-polarization corrections, as well as nuclear size and recoil
effects.  The energies satisfy
\begin{eqnarray} 
        E &=& -\frac{E_{\rm h}}{2n^2}
        \left(1 + {\cal F}\right)
        = -\frac{R_\infty hc}{n^2} \left(1 + {\cal F}\right)  \,,
\end{eqnarray}
where $E_{\rm h}=\alpha^2 m_{\rm e} c^2=2R_{\infty}hc$ is the Hartree
energy, $R_\infty$ is the Rydberg constant, $\alpha$ is the
fine-structure constant, and $m_{\rm e}$ is the electron mass.  The
dimensionless function ${\cal F}$ is small compared to one and is
determined by QED, recoil corrections, etc.  Consequently, the measured
H and D transition energies determine $E_{\rm h}$ and $R_\infty$ as $h$
and $c$ are exact in the SI.  The transition energy between states $i$
and $i^\prime$ with energies $E_i$ and $E_{i^\prime}$ is given by
\begin{eqnarray}
        \Delta {\cal E}_{ii^\prime}&=&E_{i^\prime} - E_i \,.
\end{eqnarray}
Alternatively, we write $ \Delta {\cal E}_{ii^\prime}= \Delta {\cal
E}(i\shyphen i^\prime)$.

\subsection{Theory of hydrogen and deuterium energy levels}
\slabel{ssec:hdel}

This section describes the theory of hydrogen and deuterium energy
levels.  References to literature cited in earlier CODATA reviews are
generally omitted; they may be found in \citet{1990020};
\citet{2001057}; \citet{2005204}; \citet{2007353}; \citet{2015084};
\citet{2019003}; and in earlier CODATA reports listed in
Sec.~\ref{sec:intro}.  References to new developments are given where
appropriate.  

Theoretical contributions from hyperfine structure due to nuclear
moments are not included here.  The theory of nuclear moments is limited
by the incomplete understanding of nuclear structure effects.
Hyperfine structure corrections are discussed, for example, by
\citet{1967004}; \citet{2005204}; \citet{2006127}; \citet{2010178}; and
\citet{2016038}.

Various contributions to the energies are discussed in the following
nine subsections.  Each contribution has ``correlated'' and/or
``uncorrelated'' uncertainties due to limitations in the calculations.
An important correlated uncertainty is where a contribution to the
energy has the form $C/n^3$ with a coefficient $C$ that is the same for
states with the same $\ell$ and $j$. The uncertainty in $C$ leads to
correlations among energies of states with the same $\ell$ and $j$.
Such uncertainties are denoted as uncertainty type $u_0$ in the text.
Uncorrelated uncertainties, i.e.,~those that depend on $n$, are denoted
as type $u_n$.  Other correlations are those between corrections for the
same state in different isotopes, where the difference in the correction
is only due to the difference in the masses of the isotopes.
Calculations of the uncertainties of the energy levels and the
corresponding correlation coefficients are further described in
Sec.~\ref{par:teu}.

\subsubsection{Dirac eigenvalue and mass corrections}
\slabel{par:dev}

The largest contribution to the electron energy, including its rest
mass, is the Dirac eigenvalue for an electron bound to an infinitely
heavy point nucleus, which is given by 
\begin{eqnarray} 
E_{\rm D} = f(n,\kappa)\, m_{\rm e}c^2 \ , 
\elabel{eq:diracen}
\end{eqnarray} 
with
\begin{eqnarray} 
f(n,\kappa) &=& \left[ 1+\frac{(Z\alpha)^2}{(n-\delta)^2} \right]
^{-\frac{1}{2}}
\end{eqnarray}
where $n$ is the principal quantum number, $\kappa =
(-1)^{j-\ell+1/2}(j+\frac{1}{2})$ is the Dirac angular-momentum-parity
quantum number, $j = |\kappa|-1/2$, $\ell = |\kappa +1/2|-1/2$, and
$\ell_j={\rm S}_{1/2}$, P$_{1/2}$, P$_{3/2}$, D$_{3/2}$, and D$_{5/2}$
states correspond to $\kappa = -1,~1,-2$, $2$, $-3$, respectively, and
$\delta = |\kappa|-\sqrt{\kappa^2-(Z\alpha)^2}$.  States with the same
$n$ and $j$ have degenerate eigenvalues, and we retain the atomic number
$Z$ in the equations in order to help indicate the nature of the
contributions.

For a nucleus with a finite mass $m_{N}$, we have
\begin{eqnarray} 
E_M({\rm H}) &=& Mc^2 +[f(n,\kappa)-1]m_{\rm r}c^2 
-[f(n,\kappa)-1]^2\frac{m_{\rm
r}^2c^2}{2M} \nonumber\\ 
&&+\, \frac{1-\delta_{\ell 0}}{\kappa(2\ell+1)} 
\frac{(Z\alpha)^4m_{\rm r}^3c^2}{2 n^3 m_{N}^2} +\cdots
\elabel{eq:relred}
\end{eqnarray} 
for hydrogen and 
\begin{eqnarray} 
E_M({\rm D}) &=& Mc^2 +[f(n,\kappa)-1]m_{\rm r}c^2 
-[f(n,\kappa)-1]^2\frac{m_{\rm
r}^2c^2}{2M} \nonumber\\ 
&&+\, \frac{1}{\kappa(2\ell+1)} \frac{(Z\alpha)^4m_{\rm r}^3c^2}
{2 n^3 m_{N}^2} +\cdots
\elabel{eq:relredd}
\end{eqnarray}
for deuterium, where $\delta_{\ell \ell'}$ is the Kronecker delta, $M =
m_{\rm e} + m_{N}$, and $m_{\rm r} = m_{\rm e}m_{N}/(m_{\rm e}+m_{ N})$
is the reduced mass.  In these equations the energy of $n$S$_{1/2}$
states differs from that of $n$P$_{1/2}$  states.

Equations~(\ref{eq:relred}) and (\ref{eq:relredd}) follow a slightly
different classification of terms than that used by \citet{2015084} and
\citet{2019003}.  The difference between the sum of either
Eqs.~(\ref{eq:relred}) and (\ref{eq:relredd}) and the relativistic
recoil corrections given in the following section and the corresponding
sum of terms given by \citet{2015084} and \citet{2019003} is negligible
at the current level of accuracy.

\subsubsection{Relativistic recoil}

The leading relativistic-recoil correction, to lowest order in $Z\alpha$
and all orders in $m_{\rm e}/m_{N}$, is \cite{1977002,1990020}
\begin{eqnarray} 
E_{\rm S} &=& \frac{m_{\rm r}^3}{m_{\rm e}^2 m_{N}}
\frac{(Z\alpha)^5}{\rmpi n^3} \,
m_{\rm e} c^2 \label{eq:salp}\\ 
&&\times \bigg\{\frac{1}{3}\delta_{\ell 0}\ln (Z\alpha)^{-2} -\frac{8}{3}\ln 
k_0(n,\ell) -\frac{1}{9}\delta_{\ell 0}-\frac{7}{3}a_n \nonumber\\ 
&&- \, \frac{2}{m_{N}^2-m_{\rm e}^2}\,\delta_{\ell 0} \left[m_{N}^2\ln 
\Big(\frac{m_{\rm e}}{m_{\rm r}}\Big) 
- m_{\rm e}^2\ln\Big(\frac{m_{N}}{m_{\rm r}} 
\Big)\right]\bigg\}\, ,  \nonumber
\end{eqnarray}
where
\begin{eqnarray}
a_n &=& \left(-2\ln{\frac{2}{n}} -2 + \frac{1}{n}
- 2 \sum_{i=1}^n\,\frac{1}{i}
\right)\delta_{\ell0}
\nonumber\\[0 pt]&&
+\frac{1-\delta_{\ell0}}{\ell(\ell+1)(2\ell+1)} \, .
\end{eqnarray}
Values  for the Bethe logarithms $\ln k_0(n,\ell)$ are given in
Table~\ref{tab:bethe}.  Equation~(\ref{eq:salp}) has been derived only
for a spin 1/2 nucleus.  We assume the uncertainty in using it for
deuterium is negligible.

\begin{table}
\def\vb{\vbox to 8pt{}}
\caption[Values of Bethe logarithms]{Relevant values of the Bethe
logarithms $\ln k_0(n,\ell)$.  Missing entries are for states for
which no experimental measurement is included.}
\tlabel{tab:bethe} 
\begin{tabular}{rccc} 
\hline
\hline
$n$ & S\vb & P & D \\
\hline
1 & $ 2.984\,128\,556$ & \vb & \\
2 & $ 2.811\,769\,893$ & $ -0.030\,016\,709$ & \\
3 & $ 2.767\,663\,612$ &                  &                  \\
4 & $ 2.749\,811\,840$ & $ -0.041\,954\,895$ & $ -0.006\,740\,939$ \\
6 & $ 2.735\,664\,207$ &                  & $ -0.008\,147\,204$ \\
8 & $ 2.730\,267\,261$ &                  & $ -0.008\,785\,043$ \\
12&                  &                  & $ -0.009\,342\,954$ \\
\hline
\hline
\end{tabular} 
\end{table} 

\begin{table}
\def \vb{\vbox to 8pt {}}
\caption[Values of the function $G_{\rm REC}({\alpha})$]{Values of the
function $\rmpi \times G_{\rm REC}({x=\alpha})$ from
\citet{2015098,2016019}. Numbers in parentheses are the
one-standard-deviation uncertainty in the last digit of the value.  [The
definitions of $G_{\rm REC}(x)$ in this adjustment and that of
\citet{2015098,2016019} differ by a factor $\rmpi$.]  Missing entries
are states for which data are not available from these references.}
\label{tab:GREL}
\begin{tabular}{r@{~~~}c@{~~~}c@{~~~}c}
\hline
\hline
\vb
$n$ & S & P$_{1/2}$ & P$_{3/2}$ \\
\hline
1 &    9.720(3) &&\\
2 &   14.899(3) & 1.5097(2) & $-$2.1333(2) \\
3 &   15.242(3) &&\\
4 &   15.115(3) &&\\
5 &   14.941(3) &&\\
\hline
\hline
\end{tabular}
\end{table}

Contributions to first order in the mass ratio but of higher order in
$Z\alpha$ are \cite{1995100, 1996109}
\begin{eqnarray}
E_{\rm R} &=&
 \frac{(Z\alpha)^6}{n^3}
\frac{m_{\rm e}}{m_{N}}\,
 m_{\rm e}c^2
 \bigg\{
\left(4\ln{2}-\frac{7}{2}\right)\delta_{\ell0}
\nonumber\\[8 pt]&&
+\left[3-\frac{\ell(\ell+1)}{n^2}\right]
\frac{2(1-\delta_{\ell0})}{(2\ell-1)(2\ell+1)(2\ell+3)}
\nonumber\\[8 pt]&&+Z\alpha \,
G_{\rm REC}(Z\alpha)
\bigg\}
\end{eqnarray}
Only the leading term $G_{\rm REC}(Z\alpha) =
-11/(60\rmpi)$ $\delta_{\ell0} \ln^2{(Z\alpha)^{-2}} + \dots$ is known
analytically.  We use the numerically computed $G_{\rm REC}(Z\alpha)$ of
\citet{2015098,2016019} for $n{\rm S}$ states with $n=1,\dots, 5$ and
for the $2{\rm P}_{1/2}$ and $2{\rm P}_{3/2}$ states.  For $Z=1$, these
values and uncertainties are reproduced in Table~\ref{tab:GREL}.  For
$n{\rm S}$ states with $n=6$, 8, we extrapolate $G_{\rm REC}(\alpha)$
using $g_0+g_1/n$, where coefficients $g_0$ and $g_1$ are found from
fitting to the $n=4$ and 5 values of $G_{\rm REC}(\alpha)$. The values
are 14.8(1) and 14.7(2) for $n = 6$ and 8, with uncertainties based on
comparison to values obtained by fitting $g_0 + g_1/n + g_2/n^2$ to the
$n = 3,4$, and 5 values. For the other states with $\ell>0$, we use
$G_{\rm REC}(\alpha)=0$ and an uncertainty in the relativistic-recoil
correction of $0.01 E_{\rm R}^{(1)}$.

The covariances for $E_{\rm S}+E_{\rm R}$ between pairs of states with
the same $\ell$ and $j$ follow the dominant $1/n^3$ scaling of the
uncertainty, i.e., are type $u_0$.

\subsubsection{Self energy}
\slabel{par:selfen}

The one-photon self energy of an electron bound to a stationary point
nucleus is
\begin{eqnarray}
        E_{\rm SE}^{(2)} = \frac{\alpha}{\rmpi}{(Z\alpha)^4\over n^3} 
	\left(\frac{m_{\rm r}}{m_{\rm e}}\right)^3 \!
        F(Z\alpha) \, m_{\rm e} c^2  ,
        \label{eq:selfen}
\end{eqnarray}
where the function $F(x)$ is
\begin{eqnarray}
        F(Z\alpha) &=&
	A_{41}{\cal L}+A_{40}+\left(Z\alpha\right)A_{50}
        \nonumber\\&&
	+\left(Z\alpha\right)^2\!\big[A_{62}{\cal L}^2
        +A_{61}{\cal L}
        +G_{\rm SE}(Z\alpha)\big] \, , \quad
        \label{eq:sepow}
\end{eqnarray}
with ${\cal L} = \ln{[(m_{\rm e}/m_{\rm r})(Z\alpha)^{-2}]}$ and
\begin{eqnarray*}
&&A_{41} = \frac{4}{3} \, \delta_{\ell0},
\\[5 pt]
&&A_{40} = -\frac{4}{3}\,\ln k_0(n,\ell) + \frac{10}{9}\,
  \delta_{\ell0} - \frac{m_{\rm e}/m_{\rm r}}{2\kappa(2\ell+1)}
  \left(1-\delta_{\ell0}\right) , \qquad
\\[5 pt]
&&A_{50} = \left(\frac{139}{32} 
  - 2 \ln 2 \right) \rmpi \, \delta_{\ell0},
\\[5 pt]
&&A_{62} = - \delta_{\ell0} ,
\\[5 pt]
&&A_{61} = \left[4\left(1 + \frac{1}{2} + \cdots + \frac{1}{n}\right)
  +{28\over3}\ln{2}-4\ln{n}\right.
\\[5 pt]
&&\quad\left.-{601\over180} - {77\over 45n^2}\right]\delta_{\ell0}
  +\frac{n^2-1}{n^2}
  \left({2\over15}+{1\over3}\,
  \delta_{j{1\over2}}\right)\delta_{\ell1}
\\[5 pt]
&&\quad+{\left[96n^2-32\ell(\ell+1)\right] \left(1-\delta_{\ell0}\right)
  \over 3 n^2(2\ell-1)(2\ell)(2\ell+1)(2\ell+2)(2\ell+3)}\, .
\end{eqnarray*}
Values for $G_{\rm SE}(\alpha)$ in Eq.~(\ref{eq:sepow}) are listed in
Table~\ref{tab:gse}.  The uncertainty of the
self-energy contribution  is due to the uncertainty of
$G_{\rm SE}(\alpha)$ listed in the table and is taken to be type $u_n$.
See \citet{2012158} for details.

\begin{table*}
\def\sb{\hbox to 6mm{}}
\def\vb{\vbox to 8pt{}}
\caption[Values of the function $G_{\rm SE}(\alpha)$]{Values of the function $G_{\rm SE}(\alpha)$.}
\tlabel{tab:gse}
\begin{tabular}{r @{\sb} l @{\sb} l @{\sb} l @{\sb} l @{\sb} l}
\hline
\hline
\vb
$n$& \ \ \ \ S$_{1/2}$& \ \ \ \ P$_{1/2}$& \ \ \ \ P$_{3/2}$ & \ \ \ \ D$_{3/2}$& \ \ \ \ D$_{5/2}$\\
\hline
\vb
 $ 1  $&$  -30.290\,240(20) $&&&&\\ 
 $ 2  $&$  -31.185\,150(90) $&$  -0.973\,50(20) $&$  -0.486\,50(20) $&&\\
 $ 3  $&$  -31.047\,70(90) $&                &                 &                &                  \\
 $ 4  $&$  -30.9120(40) $&$  -1.1640(20) $&$  -0.6090(20) $&                &$  0.031\,63(22) $ \\
 $ 6  $&$  -30.711(47) $&               &                 &                &$  0.034\,17(26) $ \\
 $ 8  $&$  -30.606(47) $&                &                 &$  0.007\,940(90) $&$  0.034\,84(22) $ \\
 $ 12 $&                 &                &                 &$  0.009\,130(90) $&$  0.035\,12(22) $ \\
\hline
\hline
\end{tabular}
\end{table*}

\subsubsection{Vacuum polarization}
\slabel{par:vacpol} 

The stationary point nucleus second-order vacuum-polarization level
shift is
\begin{eqnarray} 
        E_{\rm VP}^{(2)} = {\alpha\over\rmpi}{(Z\alpha)^4\over n^3} 
	\left(\frac{m_{\rm r}}{m_{\rm e}}\right)^3 \!
	H(Z\alpha) \, 
        m_{\rm e} c^2  , 
        \label{eq:vacpol} 
\end{eqnarray}
where
\begin{eqnarray}
H(Z\alpha) &=& H^{(1)}(Z\alpha) + H^{({\rm R})}(Z\alpha)
\end{eqnarray}
and
\begin{eqnarray}
  H^{(1)}(Z\alpha) &=& V_{40}+Z\alpha \, V_{50} 
\nonumber\\[0 pt]&&
  +(Z\alpha)^2\big[V_{61}{\cal L}
  +  G_{\rm VP}^{(1)}(Z\alpha)\big] \,,
\\[5 pt]
H^{({\rm R})}(Z\alpha) &=& 
(Z\alpha)^2 G_{\rm VP}^{({\rm R})}(Z\alpha)
\end{eqnarray}
Here,
\begin{eqnarray*}
V_{40} &=& -\frac{4}{15} \, \delta_{\ell0}\,,
\\[0 pt]
V_{50} &=& \frac{5\rmpi}{48} \, \delta_{\ell0}\,,
\\[0 pt]
V_{61} &=& -\frac{2}{15} \, \delta_{\ell0} \,.
\end{eqnarray*}
Values of $G_{\rm VP}^{(1)}(\alpha)$ are given in Table~\ref{tab:gvp1}
and
\begin{eqnarray} 
G_{\rm VP}^{\rm (R)}(Z\alpha) &=& \frac{19}{45} 
- \frac{\rmpi^2}{27}
+ \left(\frac{1}{16} 
- \frac{31\rmpi^2}{2880}\right)\rmpi\, Z\alpha
+\cdots \qquad
\label{eq:hovp}
\end{eqnarray}
for $\ell=0$.  Higher-order and higher-$\ell$ terms are negligible.

\begin{table*}
\def\sb{\hbox to 5mm{}}
\def\vb{\vbox to 8pt{}}
\caption[Values of the function $G_{\rm VP}^{(1)}(\alpha)$]{Values of the function $G_{\rm VP}^{(1)}(\alpha)$.}
\tlabel{tab:gvp1}
\begin{tabular}{r @{\sb} l @{\sb} l @{\sb} l @{\sb} l @{\sb} l}
\hline
\hline
\vb
$n$& \ \ \ \ S$_{1/2}$& \ \ \ \ P$_{1/2}$& \ \ \ \ P$_{3/2}$ & \ \ \ \ D$_{3/2}$& \ \ \ \ D$_{5/2}$\\
\hline
\vb
 $ 1  $&$  -0.618\,724 $&&&&\\ 
 $ 2  $&$  -0.808\,872 $&$  -0.064\,006 $&$  -0.014\,132 $&&\\
 $ 3  $&$  -0.814\,530 $&                &                 &                &                  \\
 $ 4  $&$  -0.806\,579 $&$  -0.080\,007 $&$  -0.017\,666 $&                &$  -0.000\,000 $ \\
 $ 6  $&$  -0.791\,450 $&                &                 &                &$  -0.000\,000 $ \\
 $ 8  $&$  -0.781\,197 $&                &                 &$  -0.000\,000 $&$  -0.000\,000 $ \\
 $ 12 $&                 &                &                 &$  -0.000\,000 $&$  -0.000\,000 $ \\
\hline
\hline
\end{tabular}
\end{table*}

Vacuum polarization from ${\rmmu}^+{\rmmu}^-$ pairs is
\begin{eqnarray} 
 E_{{\rmssmu}{\rm VP}}^{(2)} = {\alpha\over\rmpi}{(Z\alpha)^4\over n^3} 
 \left[-\frac{4}{15}\,\delta_{\ell0}\right] 
 \left({m_{\rm e}\over m_{\rmssmu}}\right)^2 \!
 \left({m_{\rm r}\over m_{\rm e}}\right)^3\!
 m_{\rm e} c^2 \,,\qquad
 \label{eq:vacpolmu} 
\end{eqnarray}
while the hadronic vacuum polarization is given by
\begin{eqnarray}
 E_{\rm had \, VP}^{(2)} =  0.671(15) E_{{\rmssmu}{\rm VP}}^{(2)} \ .
\end{eqnarray}
Uncertainties are of type $u_0$.  The muonic and hadronic
vacuum-polarization contributions are negligible for higher-$\ell$
states.

\subsubsection{Two-photon corrections}
\slabel{par:tpc} 

The two-photon correction is 
\begin{eqnarray} 
E^{(4)} &=& \left({\alpha\over\rmpi}\right)^2 {(Z\alpha)^4\over n^3} 
  \left(\frac{m_{\rm r}}{m_{\rm e}}\right)^3  \!
  F^{(4)}(Z\alpha)\, m_{\rm e}c^2 \,,
\label{eq:total4}
\end{eqnarray} 
where
\begin{eqnarray} 
&&F^{(4)}(Z\alpha) =
B_{40} + Z\alpha\,B_{50}
+ (Z\alpha)^2\big[B_{63} {\cal L}^3
\nonumber\\ && \qquad
+ B_{62} {\cal L}^2
+ B_{61} {\cal L}
+ G^{(4)}(Z\alpha)\big] \, ,
\label{eq:total4ps}
\end{eqnarray} 
with
\begin{eqnarray*} 
&&B_{40} = \left[\frac{3\rmpi^2}{2}\ln 2
-\frac{10\rmpi^2}{27}-\frac{2179}{648}-\frac{9}{4}\zeta(3) 
\right] \delta_{\ell0} \\&& \quad
+ \left[\frac{\rmpi^2\ln 2}{2}-\frac{\rmpi^2}{12}
-\frac{197}{144}-\frac{3\zeta(3)}{4} 
\right] \frac{m_{\rm e}}{m_{\rm r}}\,
\frac{1-\delta_{\ell0}}{\kappa(2\ell+1)}  ,
\\[3 pt]
&&B_{50} = -21.554\,47(13)\,\delta_{\ell0} \,, \\
&&B_{63} = -\frac{8}{27}\,\delta_{\ell0} \,,\\
&&B_{62} = \frac{16}{9} \left[{71\over60}-\ln{2} 
+ \psi(n) +\gamma - \ln n - \frac{1}{n} + \frac{1}{4n^2}
\right] \delta_{\ell0} \\
&& \quad +\frac{4}{27} \,\frac{n^2 - 1 }{ n^2} \, \delta_{\ell1} \,,
\\[5 pt]
&&B_{61} = \bigg\{\frac{413\,581}{64\,800} +
{4N(n{\rm S})\over3} + {2027\rmpi^2\over864} - \frac{616\,\ln{2}}{135}
   \\&&\quad
- \frac{2\rmpi^2\ln{2}}{3} 
+ \frac{40\ln^2{2}}{9}
+ \zeta(3)
+\left(\frac{304}{135}-\frac{32\,\ln{2}}{9}\right)
\\&&\quad\times
\left[\frac{3}{4} + \gamma
  + \psi(n) - \ln n - \frac{1}{n} + \frac{1}{4n^2}
\right]
\\&&\quad
-\frac{43}{36} + \frac{709\rmpi^2}{3456}
\bigg\} \delta_{\ell0}
+ \bigg[\frac{4}{3}\,N(n{\rm P})
\\&&\quad
+ \frac{n^2-1}{n^2}
\bigg(\frac{31}{405}
+ \frac{1}{3}\,\delta_{j\frac{1}{2}}
-\frac{8}{27}\,\ln{2}\bigg)
\bigg]\delta_{\ell1}\,,
\nonumber
\end{eqnarray*}
where the relevant values and uncertainties for the function $N(n\ell)$
are given in Table~\ref{tab:b61n}.  The term $B_{61}$ includes an
updated value for the light-by-light contribution by \citet{2019091}.

\begin{table}
	\def\vb{\vbox to 8pt {}}
        \caption[Values of $N(n\ell)$]{Values of $N(n\ell)$ used in the 2022 adjustment and taken
        from \citet{2003118,2005212}.}
\tlabel{tab:b61n} 
\begin{tabular}{D{.}{.}{3.3}D{.}{.}{3.12}D{.}{.}{3.12}} 
\hline
\hline
\vb
 n & \multicolumn{1}{c}{$N(n{\rm S})$} & \multicolumn{1}{c}{$N(n{\rm P})$} \\
\hline
1 &  17.855\,672\,03(1) & \vb \\
2 &  12.032\,141\,58(1) &  0.003\,300\,635(1)\\
3 &  10.449\,809(1) & \\
4 &  9.722\,413(1) &  -0.000\,394\,332(1)\\
6 &  9.031\,832(1) & \\
8 &  8.697\,639(1) & \\
\hline
\hline
\end{tabular} 
\end{table} 

Before describing the next term, {\it i.e.} $B_{60}$, 
it is useful to observe that \citet{2018104} have derived that 
\begin{eqnarray*}
        B_{72}&= &\left(-\frac{427}{144}+\frac{4\ln 2}{3}
            \right)\rmpi\, \delta_{\ell 0} \,.
\end{eqnarray*}
In addition, they find the difference 
\begin{eqnarray*}
         B_{71}(n{\rm S}) &-& B_{71}(1{\rm S}) = 
\rmpi\left(\frac{427}{36} - \frac{16}{3}\,\ln{2}\right) 
\nonumber\\
&\times&\left[\frac{3}{4}-\frac{1}{n}+\frac{1}{4n^2}
+\psi(n)+\gamma-\ln{n}\right] 
\label{eq:dB71}
\end{eqnarray*}
for S states, and
\begin{eqnarray*}
      B_{71}(n{\rm P}) &=&
                       \rmpi
        \left( \frac{427}{432}-\frac{4\ln 2}{9}
                       \right)
               \left (1-\frac{1}{n^2}\right)
\end{eqnarray*}
for P states, and $B_{71}(n\ell)=0$ for states with $\ell>1$.

\begin{table*}
	\def\vb{\vbox to 8 pt {}}
\def\spa{\hbox to 18 pt{}}
\caption[Values of $B_{60}$ and $B_{71}(n{\rm S}_{1/2})$]{Values of $B_{60}$ and $B_{71}(n{\rm S}_{1/2})$
used in the 2018 adjustment.  The uncertainties of $B_{60}$ are
explained in the text.}
\tlabel{tab:b60} 
\begin{tabular}{c@{\spa}c@{\spa}c@{\spa}c@{\spa}c@{\spa}c@{\spa}c@{\spa}c}
\hline
\hline
  $n$  & $B_{60}(n$S$_{1/2})$  &  $B_{60}(n$P$_{1/2})$ &
  $B_{60}(n$P$_{3/2})$ & $B_{60}(n$D$_{3/2})$ &
  $B_{60}(n$D$_{5/2})$ & $ B_{71}(n$S$_{1/2})$\vb\\
\hline
1&$ -78.7( 0.3)( 9.3)$&&&&& $ -116(12)$\\
2&$ -63.6( 0.3)( 9.3)$&$ -1.8(3)$&$ -1.8(3)$&&&$ -100(12)$\\
3&$ -60.5( 0.6)( 9.3)$&&&&&$ -94(12)$\\
4&$ -58.9( 0.8)( 9.3)$&$ -2.5(3)$&$ -2.5(3)$&&$ 0.178(2)$&$ -91(12)$\\
6&$ -56.9( 0.8)( 9.3)$&&&&$ 0.207(4)$&$ -88(12)$\\
8&$ -55.9( 2.0)( 9.3)$&&&$ 0.245(5)$&$ 0.221(5)$&$ -86(12)$\\
12&&&&$ 0.259(7)$&$ 0.235(7)$\\
\hline
\hline
\end{tabular} 
\end{table*} 

We determine
the coefficients $B_{60}(1{\rm S})$ and $B_{71}(1{\rm S})$ 
by combining the analytical expression for $B_{72}$ 
and the values and uncertainties for the remainder
\begin{eqnarray}
 G^{(4)}(Z\alpha) &=&
 B_{60} 
 + Z\alpha\big[B_{72}\ln^2{(Z\alpha)^{-2}} 
\nonumber\\[5 pt]&&
 + B_{71}\ln{(Z\alpha)^{-2}}
+\dots \big]
\label{eq:GQED2}
\end{eqnarray}
for the 1S state extrapolated to $x\le 2\alpha$ by \citet{2019003}
from numerical calculations of $G_{\rm QED2}(x)$ as a function of $x$
for ${x=Z\alpha}$ with $Z\ge 15$ given by \citet{2008148,2009263,2018046}.
Specifically,  the remainder has three contributions.  The largest by far
has been evaluated at ${x=0}$ and $\alpha$. The remaining two are available for
${x=\alpha}$ and $2\alpha$. Fits to each of the three contributions give
corresponding contributions to $B_{60}(1{\rm S})$ and $B_{71}(1{\rm
S})$.  We assign a type-$u_0$ state-independent standard uncertainty of 9.3 for
$B_{60}(1{\rm S})$ and a 10\,\% type-$u_0$ uncertainty to
$B_{71}(1{\rm S})$.  The difference $B_{60}(n{\rm S})-B_{60}(1{\rm S})$,
given by \citet{2005212}, is then used to obtain $B_{60}(n{\rm S})$
for $n>1$ and adds an additional small state-dependent uncertainty.
Similarly, the expression for $B_{71}(n{\rm S}) - B_{71}(1{\rm S})$
in Eq.~(\ref{eq:dB71}) is used to determine $B_{71}(n{\rm S})$.

Values for $B_{60}$ for $n$P and $n$D states with $n=1,\dots,6$ are
those published by \citet{2005212} and \citet{2006316}, but using
in place of the results in Eqs.~(A3) and (A6) of the latter paper
the corrected results given in Eqs.~(24) and (25) by \citet{2019003}.
For $n>6$, we use $B_{60}=g_0+g_1/n$ with $g_0$ and $g_1$ determined
from the values and uncertainties of $B_{60}$ at $n=5$ and 6.

Relevant values and uncertainties for $B_{60}(n\ell)$ and $B_{71}(1{\rm
S})$ are listed in Table~\ref{tab:b60}.  For the $B_{60}$ of S states,
the first number in parentheses is the state-dependent uncertainty of
type $u_n$, while the second number in parentheses is the
state-independent uncertainty of type $u_0$.  Note that the
extrapolation procedure for $n$S states is by no means unique. In fact,
\citet{2019003} used a different approach that leads to consistent and
equally accurate values for $B_{60}(n{\rm S})$.  See also
\citet{2019093,2022039}.  For $B_{71}(1{\rm S})$ and $B_{60}(n\ell)$
with $\ell>0$, the uncertainties are of type $u_0$.

\subsubsection{Three-photon corrections}
\slabel{par:thpc}

The three-photon contribution in powers of $Z\alpha$ is 
\begin{eqnarray}
E^{(6)} &=& \left({\alpha\over\rmpi}\right)^3 {(Z\alpha)^4\over n^3}
  \left(\frac{m_{\rm r}}{m_{\rm e}}\right)^3  \!
  F^{(6)}(Z\alpha)\, m_{\rm e}c^2 \,,
 \elabel{eq:total6}
\end{eqnarray} 
where
\begin{eqnarray}
	F^{(6)}(Z\alpha) &=& 
        C_{40} + Z\alpha\,C_{50} + (Z\alpha)^2\big[C_{63} {\cal L}^3
        \nonumber\\&&
        + C_{62} {\cal L}^2 
	+ C_{61} {\cal L} + C_{60} \big]
+ \cdots \, . \qquad
\elabel{eq:total6ps}
\end{eqnarray} 
The leading term $C_{40}$ is 
\begin{eqnarray*}
   C_{40} &=& \bigg[
   -{{568\,{a_4}}\over{9}}+{{85\,\zeta(5)}\over{24}}
   -{{121\,\rmpi^{2}\,\zeta(3)}\over{72}}
   -{{84\,071\,\zeta(3)}\over{2304}}
   \\&&
   -{{71\,\ln ^{4}2}\over{27}}
   -{{239\,\rmpi^{2}\,\ln^{2}2}\over{135}}
   +{{4787\,\rmpi^{2}\,\ln 2}\over{108}}
   \\&&
   +{{1591\,\rmpi^{4}}\over{3240}}
   -{{252\,251\,\rmpi^{2}}\over{9720}}
   +{679\,441\over93\,312}
   \bigg] \delta_{\ell0} 
   \\&&
   + \bigg[
   -{{100\,{a_4}}\over{3}}+{{215\,\zeta(5)}\over{24}}
   -{{83\,\rmpi^{2}\,\zeta(3)}\over{72}}
   -{{139\,\zeta(3)}\over{18}} 
   \\&&
   -{{25\,\ln ^{4}2}\over{18}}
   +{{25\,\rmpi^{2}\,\ln ^{2}2}\over{18}}
   +{{298\,\rmpi^{2}\,\ln 2}\over{9}}
   \\&&
   +{{239\,\rmpi^{4}}\over{2160}}
   -{{17\,101\,\rmpi^{2}}\over{810}}-{28\,259\over5184}
   \bigg] \frac{m_{\rm e}}{m_{\rm r}}\,
   \frac{1 - \delta_{\ell0}}{\kappa(2\ell+1)} \ ,
   \label{eq:c40}
\end{eqnarray*}
where $a_4 = \sum_{n=1}^\infty 1/(2^n\,n^4) = 0.517\,479\,061\dots$.
An estimate for the complete value has been
given by \citet{2019094,2019079} who obtain
\begin{eqnarray}
C_{50} &=& -3.3(10.5)\,\delta_{\ell0} \, ,
\nonumber
\end{eqnarray}
which reduces the uncertainty of this term by a factor of three compared
to the value used in CODATA 2018.  The uncertainty is taken to be type
$u_0$.

\citet{2018104} derived that
\[
        C_{63}=0
\]
and
\[
      C_{62} = -\frac{2}{3} \left(
              -\frac{2179}{648}-\frac{10\rmpi^2}{27}+\frac{3}{2}\rmpi^2\ln 2
                              -\frac{9}{4}\zeta(3)
        \right)
        \delta_{\ell 0} \,.
\]
They also gave an expression for the difference
$C_{61}(n{\rm S})-C_{61}(1{\rm S})$ as well as
\begin{eqnarray*}
        C_{61}(n{\rm P})&=&\frac{2}{9}\frac{n^2-1}{n^2}
        \left(
              -\frac{2179}{648}-\frac{10\rmpi^2}{27}+\frac{3}{2}\rmpi^2\ln 2
              \right.
              \\
              &&\quad\quad \quad\quad\quad\left.
                              -\frac{9}{4}\zeta(3)
                      \right) \,,
\end{eqnarray*}
and $C_{61}(n\ell)=0$ for $\ell>1$.  We do not use the expression
for the difference.  Instead, we assume that $C_{61}(n{\rm S}) =
0$ with an uncertainty of 10 of type $u_n$.  Finally, we set
$C_{60}=0$ with uncertainty 1 of type $u_n$ for P and higher-$\ell$
states. For S states we also use $C_{60}=0$, but do not need to
specify an uncertainty as the uncertainty of their three-photon
correction is determined by the uncertainties of $C_{50}$ and
$C_{61}$.

The contribution from four photons is negligible at the level of
uncertainty of current interest, as shown by \citet{2020002}.

\subsubsection{Finite nuclear size and polarizability} \label{par:nucsize}

Finite-nuclear-size and nuclear-polarizability corrections are ordered
by powers of $\alpha$, following \citet{2019003}, rather than
by finite size and polarizability. Thus,
we write for the total correction
\begin{equation}
        E_{\rm nucl}= \sum_{i=4}^\infty E^{(i)}_{\rm nucl} \,,
\end{equation}
where index $i$ indicates the order in $\alpha$.
The first and lowest-order contribution is 
\begin{equation}
        E^{(4)}_{\rm nucl}= \frac{2}{3} m_{\rm e}c^2 \frac{(Z\alpha)^4}{n^3} 
        \left(\frac{m_{\rm r}}{m_{\rm e}}\right)^3 
        \left(\frac{r_{N}}{\lbar_{\rm C}}\right)^2 
        \delta_{\ell0}
\end{equation}
and is solely due to the finite rms charge radius $r_N$ of nucleus $N$.
Here, $\lbar_{\rm C}=\hbar/m_{\rm e}c$ is the reduced Compton wavelength
of the electron.

The $\alpha^5$ correction has both nuclear-size and polarizability
contributions and has been computed by \citet{2019065}. For hydrogen, the
correction is parametrized as
\begin{equation}
        E^{(5)}_{\rm nucl}({\rm H})= -\frac{1}{3} m_{\rm e}c^2 
        \frac{(Z\alpha)^5}{n^3} 
         \left(\frac{m_{\rm r}}{m_{\rm e}}\right)^3 
	 \left(\frac{r_{{\rm pF}}}{\lbar_{\rm C}}\right)^3 \delta_{\ell0}
         \label{eq:Zemach}
\end{equation}
with effective Friar radius for the proton 
\begin{equation}
	r_{{\rm pF}}=1.947(75)\ {\rm fm} \,.
\end{equation}
The functional form of Eq.~(\ref{eq:Zemach}) is inspired by the results
of \citet{1979028} and his definition of the third Zemach moment.

For deuterium, the $\alpha^5$ correction is parametrized as \cite{2019003}
\begin{eqnarray}
        \lefteqn{ E^{(5)}_{\rm nucl}({\rm D})= -\frac{1}{3} m_{\rm e}c^2 \frac{(Z\alpha)^5}{n^3} 
        \left(\frac{m_{\rm r}}{m_{\rm e}}\right)^3 }
        \\
        &&    \times \left[ Z 
                 \left(\frac{r_{{\rm pF}}}{\lbar_{\rm C}}\right)^3 
                  +(A-Z)
                 \left(\frac{r_{{\rm nF}}}{\lbar_{\rm C}}\right)^3 
         \right]
         \delta_{\ell0}
         +E^{(5)}_{\rm pol}({\rm D})
         \nonumber
\end{eqnarray}
with atomic number $A$, effective Friar radius for the neutron
\begin{equation}
        r_{{\rm nF}}=1.43(16)\ {\rm fm} \,,
\end{equation}
and two-photon polarizability
\begin{equation}
        E^{(5)}_{\rm pol}({\rm D})/h = -21.78(22) \frac{\delta_{\ell0}}{n^3}\ {\rm kHz}
        \,.
\end{equation}

In principle, the effective Friar radius for the proton might be
different in hydrogen and deuterium. Similarly, the Friar radius
of the neutron extracted from electron-neutron scattering can be
different from that in a deuteron. We assume that such changes in
the Friar radii are smaller than the quoted uncertainties.

The $\alpha^6$ correction has finite-nuclear-size, nuclear-polarizability,
and radiative finite-nuclear-size contributions and can thus be written as $E^{(6)}_{\rm
nucl}=E^{(6)}_{\rm fns}+E^{(6)}_{\rm pol}+E^{(6)}_{\rm rad}$.
The finite-nuclear-size and nuclear-polarizability contributions are
given by \citet{2018056}. The  finite-nuclear-size contribution is
\begin{eqnarray}
        E^{(6)}_{\rm fns}&=&
        m_{\rm e}c^2
        \frac{(Z\alpha)^6}{n^3} 
         \left(\frac{m_{\rm r}}{m_{\rm e}}\right)^3 
        \left(\frac{r_{N}}{\lbar_{\rm C}}\right)^2 
          \left\{
                  -\frac{2}{3}
                  \left[
                      \frac{9}{4 n^2}-3-\frac{1}{n}
                     \right.\right. \nonumber\\
               && \quad  \left. 
                      +\,2\gamma 
                     -\ln(n/2) +\psi(n)
                      +\ln\left(\frac{m_{\rm r}}{m_{\rm e}}
                      \frac{r_{N2}}{\lbar_{\rm C}} Z\alpha\right)
              \right] \delta_{\ell 0} 
              \nonumber \\
              &&\quad\quad \quad \left.
              +\, \frac{1}{6}\left( 1-\frac{1}{n^2}\right)
              \delta_{\kappa 1}
          \right\} \,,
\end{eqnarray}
and the polarization contribution for hydrogen is
\begin{eqnarray}
        E^{(6)}_{\rm pol}({\rm H})/h=0.393 \frac{\delta_{\ell0}}{n^3}
        \ {\rm kHz}
\end{eqnarray}
with a 100\,\% uncertainty
and for deuterium
\begin{eqnarray}
        E^{(6)}_{\rm pol}({\rm D})/h=-0.541 \frac{\delta_{\ell0}}{n^3}
        \ {\rm kHz}
\end{eqnarray}
with a 75\,\% uncertainty.  The model-dependent effective radius
$r_{N2}$ describes high-energy contributions and is given by
\begin{eqnarray}
        r_{N2}=1.068\,497 \, r_N \,.
\end{eqnarray}

The radiative finite-nuclear-size contribution
of order $\alpha^6$ is
\numcite{2001057}
\begin{equation}
        E^{(6)}_{\rm rad}=
        \frac{2}{3} m_{\rm e}c^2 
        \frac{\alpha(Z\alpha)^5}{n^3} 
         \left(\frac{m_{\rm r}}{m_{\rm e}}\right)^3 
        \left(\frac{r_{N}}{\lbar_{\rm C}}\right)^2 
         (4\ln2 -5)\delta_{\ell0} \,.
\end{equation}
Next-order radiative finite-nuclear-size corrections of order $\alpha^7$
also have logarithmic dependencies on $Z\alpha$; see \citet{2011021}.
In fact, for $n$S states we have
\begin{eqnarray}
        E^{(7)}_{\rm nucl} & = &
        \frac{2}{3} m_{\rm e}c^2 
        \frac{\alpha(Z\alpha)^6}{\rmpi n^3} 
         \left(\frac{m_{\rm r}}{m_{\rm e}}\right)^3 
        \left(\frac{r_{N}}{\lbar_{\rm C}}\right)^2 \\
         && \quad \times \left[
         -\frac{2}{3} \ln^2(Z\alpha)^{-2}
   + \ln^2\left(\frac{m_{\rm r}}{m_{\rm e}} \frac{r_N}{\lbar_{\rm C}}  \right)\right] \,. \nonumber
\end{eqnarray}
We assume a zero value with uncertainty $1$  for the uncomputed
coefficient of  $\ln (Z\alpha)^{-2}$ inside the square brackets, {\it
i.e.,} that the coefficient is equal to 0(1).  For $n$P$_j$ states we
have
\begin{eqnarray}
        \lefteqn{
        E^{(7)}_{\rm nucl}  = 
        \frac{1}{6} m_{\rm e}c^2 
        \frac{\alpha(Z\alpha)^6}{\rmpi n^3} 
         \left(\frac{m_{\rm r}}{m_{\rm e}}\right)^3 
        \left(\frac{r_{N}}{\lbar_{\rm C}}\right)^2
\left(1-\frac{1}{n^2} \right) } \\
         && \times \left[
         \frac{8}{9} \ln(Z\alpha)^{-2}
         - \frac{8}{9}\ln 2 + \frac{11}{27} + \delta_{\kappa1}
	+ \frac{4n^2}{n^2-1} N(n{\rm P}) \right]
 \nonumber
\end{eqnarray}
with a zero value for the uncomputed coefficient of
$Z\alpha$ inside the square brackets with an
uncertainty of 1.
[This equation fixes a typographical error in Eq.~(64) of \citet{2019003}.
See also Eq.~(31) of \citet{2003118}.]
We assume a zero value for states with $\ell>1$.

Uncertainties in this subsection are of type $u_0$.

\subsubsection{Radiative-recoil corrections} \slabel{par:rrc}

Corrections for radiative-recoil effects are 
\begin{eqnarray} 
        E_{\rm RR} &=& 
        {m_{\rm r}^3\over m_{\rm e}^2m_{N}} {\alpha(Z\alpha)^5\over
\rmpi^2 \, n^3} m_{\rm e}c^2 \delta_{\ell0} \nonumber\\
&&\times\bigg[
 6\,\zeta(3) -2\,\rmpi^2\ln{2} + {35\,\rmpi^2\over 36} - {448\over27}
\nonumber\\
&& \qquad + {2\over3}\rmpi(Z\alpha)\,\ln^2(Z\alpha)^{-2} +
\cdots \bigg] \ . \qquad \elabel{eq:radrec} 
\end{eqnarray} 
We assume a zero value for the uncomputed coefficient of
$(Z\alpha)\ln(Z\alpha)^{-2}$ inside the square brackets with an
uncertainty of 10 of type $u_0$ and 1 for type $u_n$.  Corrections for
higher-$\ell$ states vanish at the order of $\alpha(Z\alpha)^5$.

\subsubsection{Nucleus self energy} \slabel{par:nse}

The nucleus self-energy correction  is 
\begin{eqnarray} 
 E_{\rm SEN} &=& {4Z^2\alpha(Z\alpha)^4 \over 3 \rmpi n^3} {m_{\rm r}^3\over
m_{N}^2}c^2 \nonumber\\ 
&& \times\left[ \ln{\left({m_{N}\over
m_{\rm r}(Z\alpha)^2}\right)}\delta_{\ell0} -\ln k_0(n,\ell) \right]
, \qquad \elabel{eq:nucse} 
\end{eqnarray} 
with an uncertainty of $0.5$ for S states in the constant 
($\alpha$-independent) term in
square brackets.  This uncertainty is of type $u_0$ and  given by
Eq.~(\ref{eq:nucse}) with the factor in the square brackets replaced
by 0.5. For higher $\ell$ states, the correction is negligibly small
compared to current experimental uncertainties.

\subsection{Total theoretical energies and uncertainties}
\slabel{par:teu}

The theoretical energy of centroid $E_n(L)$ of a relativistic level
${L=n\ell_j}$ is the sum of the contributions given in
Secs.~\ref{par:dev}$-$\ref{par:nse}, with atom $X = {\rm H}$ or D.
Uncertainties in the adjusted constants that enter the theoretical
expressions are found by the least-squares adjustment.  The most
important adjusted constants are $R_{\infty}=\alpha^2 m_{\rm e}c^2/2hc$,
$\alpha$, $r_{\rm p}$, and $r_{\rm d}$.

The uncertainty in the theoretical energy is taken into account by
introducing additive corrections to the energies.  Specifically, we
write
\[
        E_X(L) \to E_X(L) +  \delta_{\rm th}(X,L) 
\]
for relativistic levels $L=n\ell_j$ in atom $X$.  The energy
$\delta_{\rm th}(X,L)$ is treated as an adjusted constant and we include
$\delta_X(L)$ as an input datum with zero value and an uncertainty that
is the square root of the sum of the squares of the uncertainties of the
individual contributions. That is,
\begin{eqnarray} 
        u^2[\delta_X(L)] = \sum_i \left[u_{0i}^2(X,L) + u_{ni}^2(X,L)\right] \,,
\end{eqnarray} 
where energies $u_{0i}(X,L)$ and $u_{ni}(X,L)$ are type $u_0$ and $u_n$
uncertainties of contribution $i$.  The observational equation is
$\delta_X(L)\doteq\delta_{\rm th}(X,L)$.

Covariances among the corrections $\delta_X(L)$ are accounted for in the
adjustment. We assume that nonzero covariances for a given
atom $X$ only occur between states with the same $\ell$ and $j$.
We then have
\begin{eqnarray}
u\left[\delta_X(n_1\ell_j), \delta_X(n_2\ell_j)\right] = \sum_i
u_{0i}(X,n_2\ell_j)u_{0i}(X,n_1\ell_j) \,,  \nonumber
\end{eqnarray}
when $n_1\neq n_2$ and only uncertainties of type $u_0$ are present.
Covariances between the corrections $\delta$ for hydrogen and deuterium 
in the same electronic state $L$ are 
\begin{eqnarray} 
        \lefteqn{
        u\left[\delta_{\rm H}(L),\delta_{\rm D}(L)\right] }
        \nonumber\\
        & =& \sum_{i = \{i_{\rm c}\}} \left[u_{0i}({\rm H},L)u_{0i}({\rm
 D},L)
+ u_{ni}({\rm H},L)u_{ni}({\rm D},L)\right] \nonumber
\end{eqnarray} 
and for $n_1\neq n_2$ 
\begin{eqnarray}
        u\left[\delta_{\rm
H}(n_1\ell_j),\delta_{\rm D}(n_2\ell_j)\right] = \sum_{i =
\{i_{\rm c}\}} u_{0i}({\rm H},n_1\ell_j)u_{0i}({\rm D},n_2\ell_j)  ,
\nonumber
\end{eqnarray}
where the summation over $i$ is only over the uncertainties common to
hydrogen and deuterium.  This excludes, for example, contributions that
depend on the nuclear-charge radii.

Values and standard uncertainties of $\delta_X(L)$ are given in
Table~\ref{tab:deltaRyd} and the covariances greater than $0.0001$ of
the corrections $\delta$ are given as correlation coefficients in
Table~\ref{tab:cchydrogen}.

\subsection{Experimentally determined transition energies in hydrogen
and deuterium}
\slabel{sec:tfhd}

\begin{table*}
\def\vsp{\vbox to 8pt{}}

\caption[Input data for H and D transition frequencies]{ Summary of
measured transition frequencies $\Delta {\cal E}_X(i\shyphen
i^\prime)/h$ between states $i$ and $i^\prime$ for electronic hydrogen
($X={\rm H}$) and electronic deuterium ($X={\rm D}$) considered as input
data for the determination of the Rydberg constant $R_\infty$.  The
label in the first column is used in Table \ref{tab:cchydrogen} to list
correlation coefficients among these data and in Table
\ref{tab:pobseqsa} for observational equations.  Columns two and three
give the reference and an abbreviation of the name of the laboratory in
which the experiment has been performed.  See Sec.~\ref{sec:nom} for an
extensive list of abbreviations.}

\label{tab:rydfreq}

\begin{tabular}{c|l@{\ }l@{\ }l@{\ }l@{\ }l}
\hline
\hline
& \multicolumn{1}{c}{Reference}  & Lab.
  & \multicolumn{1}{c}{Energy interval(s)} & \quad Reported value
 & Rel. stand. \\ 
 & & &  & 
\multicolumn{1}{c}{$\Delta {\cal E}/h$ (kHz)} & uncert. $u_{\rm r}$ \\
\hline

A1 & \vsp\citet{1995159}  & MPQ & $\Delta{\cal E}_{\rm H}({\rm
2S_{1/2}}\shyphen{\rm 4S_{1/2}}) - \frac{1}{4}\Delta{\cal E}_{\rm
H}({\rm 1S_{1/2}}\shyphen{\rm 2S_{1/2}})$ & $ 4\,797\,338(10)$ &
$ 2.1\times 10^{-6}$ \\

A2 & &    & $\Delta{\cal E}_{\rm H}({\rm 2S_{1/2}}\shyphen{\rm
4D_{5/2}}) - {1\over4}\Delta{\cal E}_{\rm H}({\rm 1S_{1/2}}\shyphen{\rm
2S_{1/2}})$ & $ 6\,490\,144(24)$ & $ 3.7\times 10^{-6}$ \\

A3 & && $\Delta{\cal E}_{\rm D}({\rm 2S_{1/2}}\shyphen{\rm 4S_{1/2}}) -
{1\over4}\Delta{\cal E}_{\rm D}({\rm 1S_{1/2}}\shyphen{\rm 2S_{1/2}})$ &
$ 4\,801\,693(20)$ & $ 4.2\times 10^{-6}$ \\

A4 & &    & $\Delta{\cal E}_{\rm D}({\rm 2S_{1/2}}\shyphen{\rm
4D_{5/2}}) - {1\over4}\Delta{\cal E}_{\rm D}({\rm 1S_{1/2}}\shyphen{\rm
2S_{1/2}})$ & $ 6\,494\,841(41)$ & $ 6.3\times 10^{-6}$ \\

A5 & \vsp\citet{2010083} & MPQ & $\Delta{\cal E}_{\rm D}({\rm 1S_{1/2}}
\shyphen{\rm 2S_{1/2}}) - \Delta{\cal E}_{\rm H}({\rm 1S_{1/2}} \shyphen
{\rm 2S_{1/2}})$ & $ 670\,994\,334.606(15)$ & $ 2.2\times 10^{-11}$ \\

A6 & \citet{2011221} & MPQ & $\Delta{\cal E}_{\rm H}({\rm
1S_{1/2}}\shyphen{\rm 2S_{1/2}})$ & $ 2\,466\,061\,413\,187.035(10)$ & $ 4.2\times 10^{-15}$
\\ 

A7 & \citet{2013045} & MPQ & $\Delta{\cal E}_{\rm H}({\rm
1S_{1/2}}\shyphen{\rm 2S_{1/2}})$ & $ 2\,466\,061\,413\,187.018(11)$ & $ 4.4\times 10^{-15}$
\\ 

A8 & \vsp\citet{2017071} & MPQ & $\Delta{\cal E}_{\rm H}({\rm
2S_{1/2}}\shyphen{\rm 4P})$ & $ 616\,520\,931\,626.8(2.3)$ & $ 3.7\times 10^{-12}$
\\ 

A9 & \vsp\citet{2020077} & MPQ & $\Delta{\cal E}_{\rm H}({\rm
1S_{1/2}}\shyphen{\rm 3S_{1/2}})$ & $ 2\,922\,743\,278\,665.79(72)$ &
$ 2.5\times 10^{-13}$ \\ 

A10 & \vsp\citet{1997001} & LKB/ & $\Delta{\cal E}_{\rm H}({\rm
2S_{1/2}}\shyphen{\rm 8S_{1/2}})$ & $ 770\,649\,350\,012.0(8.6)$ & $ 1.1\times 10^{-11}$ \\ 

A11 & & ~SYRTE & $\Delta{\cal E}_{\rm H}({\rm 2S_{1/2}}\shyphen{\rm
8D_{3/2}})$ & $ 770\,649\,504\,450.0(8.3)$ & $ 1.1\times 10^{-11}$ \\ 

A12 & & & $\Delta{\cal E}_{\rm H}({\rm 2S_{1/2}}\shyphen{\rm 8D_{5/2}})$
& $ 770\,649\,561\,584.2(6.4)$ & $ 8.3\times 10^{-12}$ \\ 

A13 & & & $\Delta{\cal E}_{\rm D}({\rm 2S_{1/2}}\shyphen{\rm 8S_{1/2}})$
& $ 770\,859\,041\,245.7(6.9)$ & $ 8.9\times 10^{-12}$ \\ 

A14 & & & $\Delta{\cal E}_{\rm D}({\rm 2S_{1/2}}\shyphen{\rm 8D_{3/2}})$
& $ 770\,859\,195\,701.8(6.3)$ & $ 8.2\times 10^{-12}$ \\ 

A15 & & & $\Delta{\cal E}_{\rm D}({\rm 2S_{1/2}}\shyphen{\rm 8D_{5/2}})$
& $ 770\,859\,252\,849.5(5.9)$ & $ 7.7\times 10^{-12}$ \\ 

A16 & \vsp\citet{1999072e} & LKB/ & $\Delta{\cal E}_{\rm H}({\rm
2S_{1/2}}\shyphen{\rm 12D_{3/2}})$ & $ 799\,191\,710\,472.7(9.4)$ &
$ 1.2\times 10^{-11}$ \\ 

A17 & &  ~SYRTE& $\Delta{\cal E}_{\rm H}({\rm 2S_{1/2}}\shyphen{\rm
12D_{5/2}})$ & $ 799\,191\,727\,403.7(7.0)$ & $ 8.7\times 10^{-12}$ \\ 

A18 & & & $\Delta{\cal E}_{\rm D}({\rm 2S_{1/2}}\shyphen{\rm
12D_{3/2}})$ & $ 799\,409\,168\,038.0(8.6)$ & $ 1.1\times 10^{-11}$ \\ 

A19 & & & $\Delta{\cal E}_{\rm D}({\rm 2S_{1/2}}\shyphen{\rm
12D_{5/2}})$ & $ 799\,409\,184\,966.8(6.8)$ & $ 8.5\times 10^{-12}$ \\ 

A20 & \vsp \citet{1996001} & LKB & $\Delta{\cal E}_{\rm H}({\rm
2S_{1/2}}\shyphen{\rm 6S_{1/2}}) - {1\over4}\Delta{\cal E}_{\rm H}({\rm
1S_{1/2}}\shyphen{\rm 3S_{1/2}})$ & $ 4\,197\,604(21)$ & $ 4.9\times 10^{-6}$
\\ 

A21 & && $\Delta{\cal E}_{\rm H}({\rm 2S_{1/2}}\shyphen{\rm 6D_{5/2}}) -
{1\over4}\Delta{\cal E}_{\rm H}({\rm 1S_{1/2}}\shyphen{\rm 3S_{1/2}})$ &
$ 4\,699\,099(10)$ & $ 2.2\times 10^{-6}$ \\ 

A22 & \vsp\citet{2018042} & LKB & $\Delta{\cal E}_{\rm H}({\rm
1S_{1/2}}\shyphen{\rm 3S_{1/2}})$ & $ 2\,922\,743\,278\,671.5(2.6)$ &
$ 8.9\times 10^{-13}$ \\ 

\color{blue}
A23 & \vsp\citet{2022002} & CSU & $\Delta{\cal E}_{\rm H}({\rm
2S_{1/2}}\shyphen{\rm 8D_{5/2}})$ & $ 770\,649\,561\,570.9(2.0)$ &
$ 2.6\times 10^{-12}$ \\ 
\color{black}

A24 & \vsp\citet{1995138} & Yale & $\Delta{\cal E}_{\rm H}({\rm
2S_{1/2}}\shyphen{\rm 4P_{1/2}}) - {1\over4}\Delta{\cal E}_{\rm H}({\rm
1S_{1/2}}\shyphen{\rm 2S_{1/2}})$ & $ 4\,664\,269(15)$ &
$ 3.2\times 10^{-6}$ \\ 

A25 & && $\Delta{\cal E}_{\rm H}({\rm 2S_{1/2}}\shyphen{\rm 4P_{3/2}}) -
{1\over4}\Delta{\cal E}_{\rm H}({\rm 1S_{1/2}}\shyphen{\rm 2S_{1/2}})$ &
$ 6\,035\,373(10)$ & $ 1.7\times 10^{-6}$ \\ 

A26 & \vsp\citet{1979001} & Sussex & $\Delta{\cal E}_{\rm H}({\rm
2P_{1/2}}\shyphen{\rm 2S_{1/2}})$ & $ 1\,057\,862(20)$ & $ 1.9\times 10^{-5}$
\\

A27 & \vsp\citet{1981005} & Harvard & $\Delta{\cal E}_{\rm H}({\rm
2P_{1/2}}\shyphen{\rm 2S_{1/2}})$ & $ 1\,057\,845.0(9.0)$ & $ 8.5\times 10^{-6}$
\\ 

A28 & \vsp\citet{1994090} & Harvard & $\Delta{\cal E}_{\rm H}({\rm
2S_{1/2}}\shyphen{\rm 2P_{3/2}})$ & $ 9\,911\,200(12)$ & $ 1.2\times 10^{-6}$
\\ 

A29 & \vsp\citet{2019095} & York & $\Delta{\cal E}_{\rm H}\big({\rm
2P}_{1/2}\shyphen{\rm 2S}_{1/2}\big)$ &  $ 1\,057\,829.8(3.2)$ &
$ 3.0\times 10^{-6}$ \\

\hline
\hline
\end{tabular}
\end{table*}

Table~\ref{tab:rydfreq} gives the measured transition frequencies in
hydrogen and deuterium used as input data in the 2022 adjustment.  All
but two data are the same as in the 2018 report. The new results in
hydrogen are reviewed in the next two subsections.  The new frequencies
for the 1S$-$3S and 2S$-$8D$_{5/2}$ transitions were measured at the
Max-Planck-Institut f\"ur Quantenoptik (MPQ), Garching, Germany and at
Colorado State University, Fort Collins, Colorado, USA, respectively.
Observational equations for the data are given in
Table~\ref{tab:pobseqsa}.

\begin{table}
\caption[Input data for additive corrections of H and D energy levels]{Summary of input data for the additive energy
corrections to account for missing contributions to the theoretical
description of the electronic hydrogen (H) and deuterium (D) energy
levels.  These correspond to 25 additive corrections
$\delta_{\rm H,D}(n\ell_j)$ for the centroids of levels $n\ell_j$.
The label in the first column is used in Table \ref{tab:cchydrogen} to list correlation
coefficients among these data and in Table \ref{tab:pobseqsa} for observational equations.
Relative uncertainties are with respect
to the binding energy. 
        \label{tab:deltaRyd} }
        \begin{tabular}{l|@{~~}ll@{~~}l}
        \hline\hline
           &  Input datum &  \multicolumn{1}{c}{Value} & Rel. stand. \\
           &              &  \multicolumn{1}{c}{(kHz)}     & uncert. $u_{\rm r}$\\
        \hline
 B1   & $\delta_{\rm H}({\rm 1S_{1/2}})/h$ & $ 0.0(1.6)$  & $ 4.9\times 10^{-13}$ \\
 B2   & $\delta_{\rm H}({\rm 2S_{1/2}})/h$ & $ 0.00(20)$  & $ 2.4\times 10^{-13}$ \\
 B3   & $\delta_{\rm H}({\rm 3S_{1/2}})/h$ & $ 0.000(59)$  & $ 1.6\times 10^{-13}$ \\
 B4   & $\delta_{\rm H}({\rm 4S_{1/2}})/h$ & $ 0.000(25)$  & $ 1.2\times 10^{-13}$ \\
 B5   & $\delta_{\rm H}({\rm 6S_{1/2}})/h$ & $ 0.000(12)$  & $ 1.3\times 10^{-13}$ \\
 B6   & $\delta_{\rm H}({\rm 8S_{1/2}})/h$ & $ 0.0000(51)$  & $ 9.9\times 10^{-14}$ \\
 B7   & $\delta_{\rm H}({\rm 2P_{1/2}})/h$ & $ 0.0000(39)$  & $ 4.8\times 10^{-15}$ \\
 B8   & $\delta_{\rm H}({\rm 4P_{1/2}})/h$ & $ 0.0000(16)$  & $ 7.6\times 10^{-15}$ \\
 B9   & $\delta_{\rm H}({\rm 2P_{3/2}})/h$ & $ 0.0000(39)$  & $ 4.8\times 10^{-15}$ \\
 B10  & $\delta_{\rm H}({\rm 4P_{3/2}})/h$ & $ 0.0000(16)$  & $ 7.6\times 10^{-15}$ \\
 B11  & $\delta_{\rm H}({\rm 8D_{3/2}})/h$ & $ 0.000\,000(13)$  & $ 2.6\times 10^{-16}$ \\
 B12  & $\delta_{\rm H}({\rm 12D_{3/2}})/h$ & $ 0.000\,0000(40)$  & $ 1.8\times 10^{-16}$ \\
 B13  & $\delta_{\rm H}({\rm 4D_{5/2}})/h$ & $ 0.000\,00(17)$  & $ 8.2\times 10^{-16}$ \\
 B14  & $\delta_{\rm H}({\rm 6D_{5/2}})/h$ & $ 0.000\,000(58)$  & $ 6.3\times 10^{-16}$ \\
 B15  & $\delta_{\rm H}({\rm 8D_{5/2}})/h$ & $ 0.000\,000(22)$  & $ 4.2\times 10^{-16}$ \\
 B16  & $\delta_{\rm H}({\rm 12D_{5/2}})/h$ & $ 0.000\,0000(64)$  & $ 2.8\times 10^{-16}$ \\
 B17   & $\delta_{\rm D}({\rm 1S_{1/2}})/h$ & $ 0.0(1.5)$  & $ 4.5\times 10^{-13}$ \\
 B18  & $\delta_{\rm D}({\rm 2S_{1/2}})/h$ & $ 0.00(18)$  & $ 2.2\times 10^{-13}$ \\
 B19  & $\delta_{\rm D}({\rm 4S_{1/2}})/h$ & $ 0.000(23)$  & $ 1.1\times 10^{-13}$ \\
 B20  & $\delta_{\rm D}({\rm 8S_{1/2}})/h$ & $ 0.0000(49)$  & $ 9.6\times 10^{-14}$ \\
 B21  & $\delta_{\rm D}({\rm 8D_{3/2}})/h$ & $ 0.000\,0000(95)$  & $ 1.8\times 10^{-16}$ \\
 B22  & $\delta_{\rm D}({\rm 12D_{3/2}})/h$ & $ 0.000\,0000(28)$  & $ 1.2\times 10^{-16}$ \\
 B23  & $\delta_{\rm D}({\rm 4D_{5/2}})/h$ & $ 0.000\,00(15)$  & $ 7.5\times 10^{-16}$ \\
 B24  & $\delta_{\rm D}({\rm 8D_{5/2}})/h$ & $ 0.000\,000(19)$  & $ 3.8\times 10^{-16}$ \\
 B25  & $\delta_{\rm D}({\rm 12D_{5/2}})/h$ & $ 0.000\,0000(58)$  & $ 2.5\times 10^{-16}$ \\
        \hline\hline
\end{tabular}
\end{table}
\begin{table*}
\def\vsp{\vbox to 10pt{}}
        \caption[Correlation coefficients for data in  Tables~\ref{tab:rydfreq} and \ref{tab:deltaRyd}]{Correlation coefficients $r(x_i,x_j)>0.0001$ among the input data for the hydrogen and deuterium energy levels given in 
        Tables~\ref{tab:rydfreq} and \ref{tab:deltaRyd}. Coefficients $r$ are strictly zero between input data A$n$ and B$m$ for positive integers $n$ and $m$.}
        \label{tab:cchydrogen}
        \begin{tabular}{l@{~~~~~}l@{~~~~~}l@{~~~~~}l@{~~~~~}l}
                \hline\hline
\vsp $r$(A1,A2)$=$ 0.1049 & $r$(A1,A3)$=$ 0.2095 & $r$(A1,A4)$=$ 0.0404 & $r$(A2,A3)$=$ 0.0271 & $r$(A2,A4)$=$ 0.0467 \\
$r$(A3,A4)$=$ 0.0110 & $r$(A6,A7)$=$ 0.7069 & $r$(A10,A11)$=$ 0.3478 & $r$(A10,A12)$=$ 0.4532 & $r$(A10,A13)$=$ 0.1225 \\
$r$(A10,A14)$=$ 0.1335 & $r$(A10,A15)$=$ 0.1419 & $r$(A10,A16)$=$ 0.0899 & $r$(A10,A17)$=$ 0.1206 & $r$(A10,A18)$=$ 0.0980 \\
$r$(A10,A19)$=$ 0.1235 & $r$(A10,A20)$=$ 0.0225 & $r$(A10,A21)$=$ 0.0448 & $r$(A11,A12)$=$ 0.4696 & $r$(A11,A13)$=$ 0.1273 \\
$r$(A11,A14)$=$ 0.1387 & $r$(A11,A15)$=$ 0.1475 & $r$(A11,A16)$=$ 0.0934 & $r$(A11,A17)$=$ 0.1253 & $r$(A11,A18)$=$ 0.1019 \\
$r$(A11,A19)$=$ 0.1284 & $r$(A11,A20)$=$ 0.0234 & $r$(A11,A21)$=$ 0.0466 & $r$(A12,A13)$=$ 0.1648 & $r$(A12,A14)$=$ 0.1795 \\
$r$(A12,A15)$=$ 0.1908 & $r$(A12,A16)$=$ 0.1209 & $r$(A12,A17)$=$ 0.1622 & $r$(A12,A18)$=$ 0.1319 & $r$(A12,A19)$=$ 0.1662 \\
$r$(A12,A20)$=$ 0.0303 & $r$(A12,A21)$=$ 0.0602 & $r$(A13,A14)$=$ 0.5699 & $r$(A13,A15)$=$ 0.6117 & $r$(A13,A16)$=$ 0.1127 \\
$r$(A13,A17)$=$ 0.1512 & $r$(A13,A18)$=$ 0.1229 & $r$(A13,A19)$=$ 0.1548 & $r$(A13,A20)$=$ 0.0282 & $r$(A13,A21)$=$ 0.0561 \\
$r$(A14,A15)$=$ 0.6667 & $r$(A14,A16)$=$ 0.1228 & $r$(A14,A17)$=$ 0.1647 & $r$(A14,A18)$=$ 0.1339 & $r$(A14,A19)$=$ 0.1687 \\
$r$(A14,A20)$=$ 0.0307 & $r$(A14,A21)$=$ 0.0612 & $r$(A15,A16)$=$ 0.1305 & $r$(A15,A17)$=$ 0.1750 & $r$(A15,A18)$=$ 0.1423 \\
$r$(A15,A19)$=$ 0.1793 & $r$(A15,A20)$=$ 0.0327 & $r$(A15,A21)$=$ 0.0650 & $r$(A16,A17)$=$ 0.4750 & $r$(A16,A18)$=$ 0.0901 \\
$r$(A16,A19)$=$ 0.1136 & $r$(A16,A20)$=$ 0.0207 & $r$(A16,A21)$=$ 0.0412 & $r$(A17,A18)$=$ 0.1209 & $r$(A17,A19)$=$ 0.1524 \\
$r$(A17,A20)$=$ 0.0278 & $r$(A17,A21)$=$ 0.0553 & $r$(A18,A19)$=$ 0.5224 & $r$(A18,A20)$=$ 0.0226 & $r$(A18,A21)$=$ 0.0449 \\
$r$(A19,A20)$=$ 0.0284 & $r$(A19,A21)$=$ 0.0566 & $r$(A20,A21)$=$ 0.1412 & $r$(A24,A25)$=$ 0.0834 & \\

$r$(B1,B2)$=$ 0.9946 & $r$(B1,B3)$=$ 0.9937 & $r$(B1,B4)$=$ 0.9877 & $r$(B1,B5)$=$ 0.6140 & $r$(B1,B6)$=$ 0.6124 \\
$r$(B1,B17)$=$ 0.9700 & $r$(B1,B18)$=$ 0.9653 & $r$(B1,B19)$=$ 0.9575 & $r$(B1,B20)$=$ 0.5644 & $r$(B2,B3)$=$ 0.9937 \\
$r$(B2,B4)$=$ 0.9877 & $r$(B2,B5)$=$ 0.6140 & $r$(B2,B6)$=$ 0.6124 & $r$(B2,B17)$=$ 0.9653 & $r$(B2,B18)$=$ 0.9700 \\
$r$(B2,B19)$=$ 0.9575 & $r$(B2,B20)$=$ 0.5644 & $r$(B3,B4)$=$ 0.9869 & $r$(B3,B5)$=$ 0.6135 & $r$(B3,B6)$=$ 0.6119 \\
$r$(B3,B17)$=$ 0.9645 & $r$(B3,B18)$=$ 0.9645 & $r$(B3,B19)$=$ 0.9567 & $r$(B3,B20)$=$ 0.5640 & $r$(B4,B5)$=$ 0.6097 \\
$r$(B4,B6)$=$ 0.6082 & $r$(B4,B17)$=$ 0.9586 & $r$(B4,B18)$=$ 0.9586 & $r$(B4,B19)$=$ 0.9704 & $r$(B4,B20)$=$ 0.5605 \\
$r$(B5,B6)$=$ 0.3781 & $r$(B5,B17)$=$ 0.5959 & $r$(B5,B18)$=$ 0.5959 & $r$(B5,B19)$=$ 0.5911 & $r$(B5,B20)$=$ 0.3484 \\
$r$(B6,B17)$=$ 0.5944 & $r$(B6,B18)$=$ 0.5944 & $r$(B6,B19)$=$ 0.5896 & $r$(B6,B20)$=$ 0.9884 & $r$(B11,B12)$=$ 0.6741 \\
$r$(B11,B21)$=$ 0.9428 & $r$(B11,B22)$=$ 0.4803 & $r$(B12,B21)$=$ 0.4782 & $r$(B12,B22)$=$ 0.9428 & $r$(B13,B14)$=$ 0.2061 \\
$r$(B13,B15)$=$ 0.2391 & $r$(B13,B16)$=$ 0.2421 & $r$(B13,B23)$=$ 0.9738 & $r$(B13,B24)$=$ 0.1331 & $r$(B13,B25)$=$ 0.1352 \\
$r$(B14,B15)$=$ 0.2225 & $r$(B14,B16)$=$ 0.2253 & $r$(B14,B23)$=$ 0.1128 & $r$(B14,B24)$=$ 0.1238 & $r$(B14,B25)$=$ 0.1258 \\
$r$(B15,B16)$=$ 0.2614 & $r$(B15,B23)$=$ 0.1309 & $r$(B15,B24)$=$ 0.9698 & $r$(B15,B25)$=$ 0.1459 & $r$(B16,B23)$=$ 0.1325 \\
$r$(B16,B24)$=$ 0.1455 & $r$(B16,B25)$=$ 0.9692 & $r$(B17,B18)$=$ 0.9955 & $r$(B17,B19)$=$ 0.9875 & $r$(B17,B20)$=$ 0.5821 \\
$r$(B18,B19)$=$ 0.9874 & $r$(B18,B20)$=$ 0.5821 & $r$(B19,B20)$=$ 0.5774 & $r$(B21,B22)$=$ 0.3407 & $r$(B23,B24)$=$ 0.0729 \\
$r$(B23,B25)$=$ 0.0740 & $r$(B24,B25)$=$ 0.0812 & & & \\
        \hline\hline
\end{tabular}
\end{table*}

\subsubsection{Measurement of the hydrogen 2S$-$8D$_{5/2}$ transition}

\citet{2022002} have measured the frequency of the 2S$-$8D$_{5/2}$
transition in hydrogen with a relative uncertainty of
$2.6\times10^{-12}$.  The same transition had been measured earlier by
\citet{1997001} at LKB/SYRTE and that measurement and the recent
measurement differ by 13.3(6.7) kHz.  The more recent result has an
uncertainty that is more than three times smaller than the earlier
result.

\subsubsection{Measurement of the hydrogen two-photon 1S$-$3S transition}
 \slabel{ssec:lkbh1s3s}

The hydrogen 1S$-$3S transition energy was measured by \citet{2016034}
at the MPQ and \citet{2018042} at the LKB, as discussed in the CODATA
2018 publication.  The earlier LKB measurement by \citet{2010184},
listed among the 2018 data, is not included in Table~\ref{tab:rydfreq}.
More recently, \citet{2020077} at the MPQ measured this transition with
an uncertainty more than 20 times smaller than the previous MPQ
measurement.  Their result differs by a combined standard deviation of
2.1 from the LKB result.  These data are items A9 and A22 in
Table~\ref{tab:rydfreq}.  The difference between these results is not
currently understood.  \citet{2023004} give a more recent discussion of
the experiment where they note a newly discovered systematic effect due
to a stray accumulation of atoms in the vacuum chamber.  However, they
feel that this effect is too small to explain the difference between the
LKB and MPQ results.

The researchers at the LKB used two-photon spectroscopy.  In this
technique, the first-order Doppler shift is eliminated by having
room-temperature atoms simultaneously absorb photons from
counter-propagating laser beams.  The measured transition energy has a
five times smaller uncertainty than two older measurements of the same
transition energy.  \citet{2017088} and \citet{2019005} give more
information about the LKB measurement.  A history of Doppler-free
spectroscopy is given by \citet{2019106}.

The development of a continuous-wave laser source at 205\,nm for the
two-photon excitation by \citet{2015083} contributed significantly to
the fivefold uncertainty reduction by improving the signal-to-noise
ratio compared to previous LKB experiments with a chopped laser source.
The frequency of the 205\,nm laser was determined with the help of a
transfer laser, several Fabry-Perot cavities, and a femtosecond frequency
comb whose repetition rate was referenced to a Cs-fountain 
frequency standard.

The laser frequency was scanned to excite the $1{\rm S}_{1/2}({f =
1})-3{\rm S}_{1/2}({f = 1})$ transition and the resonance was detected
from the 656\,nm radiation emitted by the atoms when they decay from the
3S to the 2P level. The well-known 1S and 3S hyperfine splittings 
were used to obtain the final transition energy between the 
hyperfine centroids with $u(\Delta {\cal E}/h)=2.6$\,kHz
and $u_{\rm r} = 8.9 \times10^{-13}$.

The distribution of velocities of the atoms in the room-temperature
hydrogen beam led to a second-order Doppler shift of roughly
$-140$\,kHz, or 500 parts in $10^{13}$, and was the largest systematic
effect in the experiment. To account for this shift, the velocity
distribution of the hydrogen atoms was mapped out by applying a small
magnetic flux density ${\bm B}$ perpendicular to the hydrogen beam. In
addition to Zeeman shifts, the flux density leads to Stark shifts of 3S
hyperfine states by mixing with the nearby 3P$_{1/2}$ level via the
motional electric field in the rest frame of the atoms.  Both this
motional Stark shift and the second-order Doppler shift have a
quadratic dependence on velocity.  Then the LKB researchers fit
resonance spectra obtained at different $\bm B$ to a line-shape model
averaged over a modified Maxwellian velocity distribution of an effusive
beam.  The fit gives the temperature of the hydrogen beam, distortion
parameters from a Maxwellian distribution, and a line position with the
second-order Doppler shift removed.

Finally, the observed line position was corrected for light shifts
due to the finite 205\,nm laser intensity and pressure shifts due to
elastic collisions with background hydrogen molecules.  Light shifts
increase the apparent transition energy by up to $h\times 10$\, kHz
depending on the laser intensity in the data runs, while 
pressure shifts decrease this energy by slightly less than $h\times
1$\,kHz/($10^{-5}$\,hPa). Pressures up to $20\times 10^{-5}$\, hPa
were used in the experiments.  Quantum interference effects, mainly from
the 3D state, are small for the $1{\rm S}-3{\rm S}$ transition and led
to a correction of $h\times 0.6(2)$\, kHz.

\section{Muonic atoms and muonic ions}
\label{sec:mulsrprd}

\subsection{Theory and experiment}
\label{ssec:mute}

The muonic atoms ${\rmmu}$H and ${\rmmu}$D and muonic atomic ions
$\rmmu^3$He$^+$ and $\rmmu^4$He$^+$ are ``simple'' systems consisting of
a negatively charged muon bound to a positively charged nucleus.
Because the mass of a muon is just over 200 times greater than that of
the electron, the muonic Bohr radius is 200 times smaller than the
electronic Bohr radius, making the muon charge density for S states at
the location of the nucleus more than a million times larger than those
for electronic hydrogenic atoms and ions.  Consequently, the muonic Lamb
shift, the energy difference $E_{\rm L}(X)=E_{2{\rm
P}_{1/2}}(X)-E_{2{\rm S}_{1/2}}(X)$ between the $nL_j=2{\rm S_{1/2}}$
and $2{\rm P_{1/2}}$ states of muonic atoms or ions, is more sensitive
to the rms charge radius $r_{X}$ of the nucleus.  Here
$X=$p, d, h, or $\rmalpha$.  In fact, measuring the Lamb shift of muonic
atoms and ions is a primary means of determining $r_X$.

For this 2022 adjustment, measurements of the Lamb shift are available
for $\rmmu$H, $\rmmu$D, and $\rmmu^4$He$^+$ from \citet{2013011},
\citet{2016037}, and \citet{2021028}, respectively.  The value for
$\rmmu^4$He$^+$ is a new input datum for this adjustment and will be
discussed in more detail below.  An experimental input datum for
$\rmmu^3$He$^+$ is not available.  We can compare the measurement
results with equally accurate theoretical estimates of the Lamb shift
for the four systems derived over the past 25 years and summarized  by
\citet{Pachucki2022}. These comparisons help determine the rms charge
radii $r_X$ of the proton, deuteron, and $\rmalpha$ particle.  For the
2022 CODATA adjustment, we follow \citet{Pachucki2022}  and summarize
the Lamb-shift calculations with 
\begin{equation}
 E^{\rm (th)}_{\rm L}(X) =  
 E_{\rm QED}(X)  + {\cal C}_X r^2_{X} + E_{\rm NS}(X)
 \label{eq:muonicHtheory}
\end{equation}
for $X=$p, d, h, and $\rmalpha$. The values and (uncorrelated)
uncertainties for  coefficients $E_{\rm QED}(X)$, ${\cal C}_X$, and
$E_{\rm NS}(X)$ are given  in Table \ref{tab:muonicatoms}.  The
theoretical coefficients for $\rmmu^3$He$^+$ are listed here for
the sake of completeness. 

The coefficient $E_{\rm QED}(X)$ contains 19 QED contributions starting
with the one-loop electron vacuum-polarization correction, contributing
about 99.5\,\% to $E_{\rm QED}(X)$ at order $O(\alpha(Z\alpha)^2
m_{\rmssmu}c^2)$, up to and including corrections of order $O(\alpha^6
m_{\rmssmu}c^2)$.  The  uncertainty  $u(E_{\rm QED}(X))$ is $\sim
10^{-6} E_{\rm QED}(X)$ and is dominated by that of the one-loop
hadronic vacuum-polarization correction.  The energy ${\cal C}_X r^2_X$
is the finite nuclear size contribution containing all contributions
that depend on nuclear structure proportional to  $r_X^2$. Three terms
contribute and  uncertainties in the calculation of ${\cal C}_X$ do not
affect the determination of the rms charge radii at the current level of
our theoretical understanding as well as measurement uncertainties.

The third term in Eq.~(\ref{eq:muonicHtheory}), $E_{\rm NS}(X)$, is the
nuclear structure contribution and includes effects from higher-order
moments in the nuclear charge and magnetic moment distribution of a
nucleus  in its nuclear ground state  as well as polarizability
contributions when the nucleus is virtually excited by the muon. Again
multiple terms contribute, the largest by far being the
two-photon-exchange contribution.  For the four muonic atoms, the
uncertainty of the two- and three-photon-exchange contributions
determine the corresponding uncertainty of the theoretical value of the
Lamb shift.

For $\rmmu$H, the two-photon-exchange contribution to $E_{\rm NS}(X)$ is
conventionally split into multiple terms.  The largest of these terms,
contributing about 70\,\% of the total value, is the Friar contribution
and is related to a cubic moment of a product of the  ground-state
proton  charge distribution and is part of the two-photon exchange
contribution. The uncertainty of $E_{\rm NS}(\rmmu{\rm p})$, however, is
dominated by the ``subtraction'' term related to the magnetic dipole
polarizability of the proton.  For $\rmmu$D, \citet{Pachucki2022}
computed the two-photon exchange contribution in three different ways,
one based on chiral effective field theory, one based on pion-less
effective field theory, and one based on nuclear theory with an
effective Hamiltonian for the interactions among nucleons in the
presence of an electromagnetic field. The three approaches are
consistent and \citet{Pachucki2022} chose, as the best value, the mean
of the three values with an uncertainty set by the approach with the
largest uncertainty.  For levels of electronic H and D, described in
Sec.~\ref{sec:ryd}, the Friar contribution is negligible compared to the
final theoretical uncertainty. Therefore, at the current state of
theory, we do not need to account for correlations between the energy
levels of H and $\rmmu$H beyond those due to $r_{\rm p}$. Similarly,
there is no correlation between energy levels of D and $\rmmu$D. For
$\rmmu^3$He$^+$ and $\rmmu^4$He$^+$, two-photon exchange contributions
are computed from nuclear theory.

The relevant observational equations for the 2022 adjustment are  
\begin{equation} 
  E^{\rm (exp)}_{\rm L}(X)\doteq 
  E^{\rm (th)}_{\rm L}(X)+\delta_{\rm th}(\rmmu{}X)
\end{equation}
and 
\begin{equation}
  \delta_{\rm L}(\rmmu X)\doteq \delta_{\rm th}(\rmmu{}X)
\end{equation}
with adjusted constants $r_X$ and  $\delta_{\rm th}(\rmmu{}X)$.  The
input data $\delta_{\rm L}(\rmmu X)=0$ with standard uncertainty
$\sqrt{ u^2(E_{\rm QED}(X))+u^2(E_{\rm NS}(X))}$ account for the
uncertainty from uncomputed terms in the theoretical expression for the
muonic Lamb shift.

We finish this section with a brief description of the experiment of
\citet{2021028} measuring the Lamb shift of $\rmmu^4$He$^+$. The
experiment follows the techniques of \citet{2013011} and
\citet{2016037}.  About 500 negatively charged muons per second with a
kinetic energy of a few keV are stopped in a room temperature $^4$He gas
at a pressure of 200 Pa.  In the last collision with a $^4$He atom, the
muon ejects an electron and gets captured by $^4$He in a highly excited
Rydberg state.  In an Auger process the remaining electron is ejected.
The resulting highly excited $\rmmu^4$He$^+$ relaxes by radiative decay
to the ground $nL_j=1{\rm S}_{1/2}$ or metastable 2S$_{1/2}$ state.  The
approximately 1\,\% $\rmmu ^4{\rm He}^+$ ions in the 2S state are then
resonantly excited to the $2{\rm P}_{1/2}$ or $2{\rm P}_{3/2}$ states by
a pulsed titanium:sapphire-based laser  with a frequency bandwidth of
0.1 GHz and an equally accurately characterized frequency. The presence
of 8.2 keV Lyman-$\alpha$ x-ray photons from the radiative decay of the
2P states indicates the successful excitation. These x-ray photons  were
counted by large-area avalanche photodiodes.  Finally, the two 2S to 2P
transition frequencies were measured with an accuracy of $\approx 15$
GHz, mostly due to statistics from the limited number of events. The
theoretical value for the $2{\rm P}_{3/2}$ to $2{\rm P}_{1/2}$
fine-structure splitting is far more accurate than the experimental
uncertainties of the 2S to 2P transition frequencies, and the two data
points were combined to lead to the value in Table
\ref{tab:muonicatoms}.

\def\vb{\vbox to 8 pt {}}
 \LTcapwidth=\textwidth
\begin{longtable*}{@{~}l@{~~}|@{~~}l@{\qquad~~}l@{\qquad~~}l@{\qquad~~}l
@{\quad~~}l}
\caption[Input data for muonic hydrogen, deuterium and $^4$He$^+$.]
{ Input data that
determine the radii of the proton, deuteron, and $\alpha$ particle.
Sec.~\ref{sec:mulsrprd}.
\label{tab:muhdata}} \\
\hline
\hline
& Input datum  &  Value &  Rel. stand.  & Lab. & Reference(s) \\
& & & unc.  $u_{\rm r}$ &  & \\
\hline

C1  & $ E^{({\rm exp})}_{\rm L}(\rmmu{\rm H}) $    & $
 202.3706(23)$ meV  & $ 1.1\times 10^{-5}$ & CREMA-13 & \citet{2013011}
 \\

C2  & $ ~\delta_{\rm L}(\rmmu{\rm H}) $    & $
 0.0000(25)$ meV  & $[ 1.2\times 10^{-5}]$ & theory &
\citet{Pachucki2022}  \\

C3  & $ E^{({\rm exp})}_{\rm L}(\rmmu{\rm D}) $    & $
 202.8785(34)$ meV  & $ 1.7\times 10^{-5}$ & CREMA-16 & \citet{2016037}
 \\

C4  & $ ~\delta_{\rm L}(\rmmu{\rm D}) $    & $
 0.0000(200)$ meV  & $[ 1.0\times 10^{-4}]$ & theory &
\citet{Pachucki2022}  \\

C5  & $ E^{({\rm exp})}_{\rm L}(\rmmu {}^4{\rm He}^+) $    & $
 1378.521(48)$ meV  & $ 3.5\times 10^{-5}$ & CREMA-21 &
\citet{2021028}  \\

C6  & $ ~\delta_{\rm L}(\rmmu {}^4{\rm He}^+) $    & $
 0.000(433)$ meV  & $[ 3.1\times 10^{-4}]$ & theory
& \citet{Pachucki2022} \\

 \hline
 \hline
\end{longtable*}

\begin{table}
\caption{Values and standard uncertainties for theoretical coefficients
$E_{\rm QED}(X)$, ${\cal C}_X$, and $E_{\rm NS}(X)$ for nuclei $X={\rm
p}$, d, and $\rmalpha$ that determine the theoretical  Lamb shift for
$\rmmu$H, $\rmmu$D, and $\rmmu^4$He$^+$, respectively.
\label{tab:muonicatoms}}
\begin{tabular}{c|@{\quad}D{.}{.}{4.7}@{\qquad}D{.}{.}{4.4}@{\qquad}D{.}{.}{2.9}}
\hline\hline 
  $X$  & \multicolumn{1}{c}{$E_{\rm QED}(X)$} 
  & \multicolumn{1}{c}{${\cal C}_X$} 
  & \multicolumn{1}{c}{$E_{\rm NS}(X)$} \\  
  & \multicolumn{1}{c}{  (meV) } 
  & \multicolumn{1}{c}{(meV\, fm$^{-2}$)}  
  &  \multicolumn{1}{c}{  (meV) }  \\
 \hline
  p  &   206.034\,4(3) &  -5.2259 &  0.0289(25) \\
  d  &  228.744\,0(3) &  -6.1074 &  1.7503(200) \\
  $\rmalpha$ &  1\,668.491(7) &  -106.209 &  9.276(433) \\
\hline \hline  
\end{tabular}
\end{table}

\subsection{Values of $\bm{r_{\rm p}}$, $\bm{r_{\rm d}}$, and
$\bm{r_{\rmssalpha}}$ from hydrogen, deuterium, and {\textbf He}$\bm{^+}$
transition energies} \label{ssec:vrads}

Finite nuclear size and polarizability contributions to the theoretical
expressions for hydrogen and deuterium energy levels are discussed in
Sec.~\ref{par:nucsize}.  A number of these contributions depend on
$r_N$, the rms charge radius of the nucleus $N$ of the atom, which for
hydrogen is denoted by $r_{\rm p}$, for deuterium by $r_{\rm d}$, and
for $^4{\rm He}$ by $r_{\rmssalpha}$.  Although the complete theoretical
expression for an energy level in hydrogen (or deuterium or the ion
$^4{\rm He}^+$) is lengthy, a simplified form can be derived that
depends directly on $R_\infty$ and contains a term which is the product
of a coefficient and $r_N^2$ [see, for example, Eq.~(1) of the paper by
\citet{2017071}].  There are other constants in the expression,
including the fine-structure constant $\alpha$ and the mass ratio
$m_{\rm e}/m_N$, but these are obtained from other experiments and in
this context are adequately known.  Thus, in principle, two measured
transition energies in the same atom and their theoretical expressions
can be combined to obtain values of the two unknowns $R_\infty$ and
$r_N$.

In the least-squares adjustment that determines the 2022 recommended
values of the constants, the theoretical and experimental muonic data in
Tables~\ref{tab:muhdata} and \ref{tab:muonicatoms} of this section are
used as input data together with the theoretical and experimental
hydrogen and deuterium transition energies data discussed in
Sec.~\ref{sec:ryd}.  As discussed in Sec.~\ref{sec:2022crv} the
uncertainties of all of these input data are multiplied by an expansion
factor of 1.7 to reduce the inconsistencies among the transition energy
data to an acceptable level.  With this in mind, we compare in
Table~\ref{tab:radcomp} the values of $R_\infty$, $r_{\rm p}$, $r_{\rm
d}$, and $r_{\rmssalpha}$ obtained in different ways and from which the
following three conclusions can be drawn.

(i) The muonic data has a significant impact on the recommended value of
$R_\infty$, as a comparison of the values in columns two and three of
Table~\ref{tab:radcomp} shows.  Including the muonic data lowers the
value of $R_\infty$ by 2.7 times the uncertainty of the value that
results when the muonic data are omitted and reduces the uncertainty of
that value of $R_\infty$ by a factor of 3.5.

(ii) The value of $r_{\rm p}$ and $r_{\rm d}$ from the H and D
transition energies alone, which are in column three of the table, differ
significantly from their corresponding muonic-data values in column four.
For both $r_{\rm p}$ and $r_{\rm d}$ the H-D alone value exceeds the
muonic value by $2.8\,\sigma$, where as usual $\sigma$ is the standard
uncertainty of the difference.

(iii)  Including the ${\rmmu}$H and ${\rmmu}$D data in the 2022
adjustment leads to a recommended value of $R_\infty$ with $u_{\rm r} =
1.1\times10^{-12}$ compared to $1.9\times10^{-12}$ for the 2018
recommended value (${\rmmu}$H and ${\rmmu}$D data are also included in
that adjustment but have been improved since then).  However, the lack
of good agreement between the H-D alone transition-energy values of
$r_{\rm p}$ and $r_{\rm d}$ and the ${\rmmu}$H and ${\rmmu}$D Lamb-shift
values is unsatisfactory and needs both experimental and theoretical
investigation.

\def\vh{\vbox to 10 pt {}}
\begin{longtable*}{@{\quad}c@{\quad}@{\quad}c@{\quad}@{\quad}c@{\quad}@{\quad}c@{\quad}}
\caption[]{Results of various least-squares
adjustments.\label{tab:radcomp}} \\
\hline
\hline
\vh Constant & Complete & No muonic atom data & No electronic atom data \\
\hline
\vh $R_\infty/{\rm m}^{-1}$ &
$10\,973\,731.568\,157(12)$ \ $[1.1\times10^{-12}]$& 
$10\,973 731.568\,276(44)$ \ $[3.9\times10^{-12}]$ & Not applicable \\
\vh $r_{\rm p}/$fm &
$0.840\,75(64)$ \ $[7.6\times10^{-4}]$ &
$0.8529(43)$ \ $[51\times10^{-4}]$ &
$0.840\,60(66)$ \ $[7.8\times10^{-4}]$ \\
\vh $r_{\rm d}/$fm &
$2.127\,78(27)$ \  $[1.3\times10^{-4}]$ &
$2.1326(17)$ \ $[8.1\times10^{-4}]$ &
$2.12643(133)$ \ $[6.2\times10^{-4}]$ \\
\vh $r_{\rmssalpha}$/fm	&
$1.6785(21)$ \ $[12\times10]^{-4}$ & 
No data available &
$1.6785(21)$ \ $[12\times10^{-4}]$ \\[3 pt]
\hline
\end{longtable*}

In Table~\ref{tab:radcomp}, the uncertainty of the radius of the
deuteron $r_{\rm d}$ appears to be anomalously small compared to the
value obtained by combining the no-$\rmmu$ data with the $\rmmu$ data.
The apparent combined relative uncertainty is $4.9\times10^{-4}$ which
may be compared to the $1.3\times10^{-4}$ for the complete least-squares
adjustment given in the table.  The seeming disparity is due to a
phenomenon of the least-squares adjustment which takes into account
relations between data that may not be apparent.  In particular, the
isotope shift in electronic atoms (item A5 in Table~\ref{tab:rydfreq})
provides a link between the deuteron and proton radii which translates
to a link between the muonic hydrogen and muonic deuterium theory.  This
link takes advantage of the fact that the muonic hydrogen theory is
nearly an order-of-magnitude more accurate than the muonic deuterium
theory and serves to provide an independent source of information about
muonic deuterium theory.  This phenomenon has been confirmed by running
the complete least-squares adjustment with the exclusion of the
electronic isotope-shift measurement.  The result is $r_{\rm d} =
2.1266(13)$ with $u_{\rm r} = 6.1\times10^{-4}$.  This result is just
slightly more accurate than the $\rmmu$-only data.  The reason for this
is that without the isotope shift data, the electronic-only value for
the deuteron radius is $r_{\rm d} = 2.1362(63)$ with $u_{\rm r} =
29\times10^{-4}$.  When combined with the $\rmmu$-only data, this gives
an uncertainty of $u_{\rm r} = 6.1\times10^{-4}$ which is consistent
with the no-isotope combined result.  Finally, one sees that the isotope
shift improves the electron-only value for the deuteron radius
significantly, because much of the information about the electron
deuteron radius comes from measurements on electron hydrogen combined
with the isotope shift measurement which links this information to the
deuteron radius.

The deuteron-proton squared-radius difference $r^{2}_{\rm d} -
r^{2}_{\rm p}$ is somewhat constrained by the $\rmmu$H and $\rmmu$D
Lamb-shift measurements, but mainly by the measurement of the isotope
shift of the 1S$-$2S transition in H and D by \citet{2010083}, item A5
in Table~\ref{tab:rydfreq}.  The 2022 CODATA value is
\begin{eqnarray}
        r_{\rm d}^2 - r_{\rm p}^2 &=& 3.820\,36(41)\ \mbox{fm}^2 \, .
        \elabel{eq:drcodata}
\end{eqnarray}

\begin{figure*} \includegraphics[angle=0,scale=0.5, trim=0 0 0
10,clip]{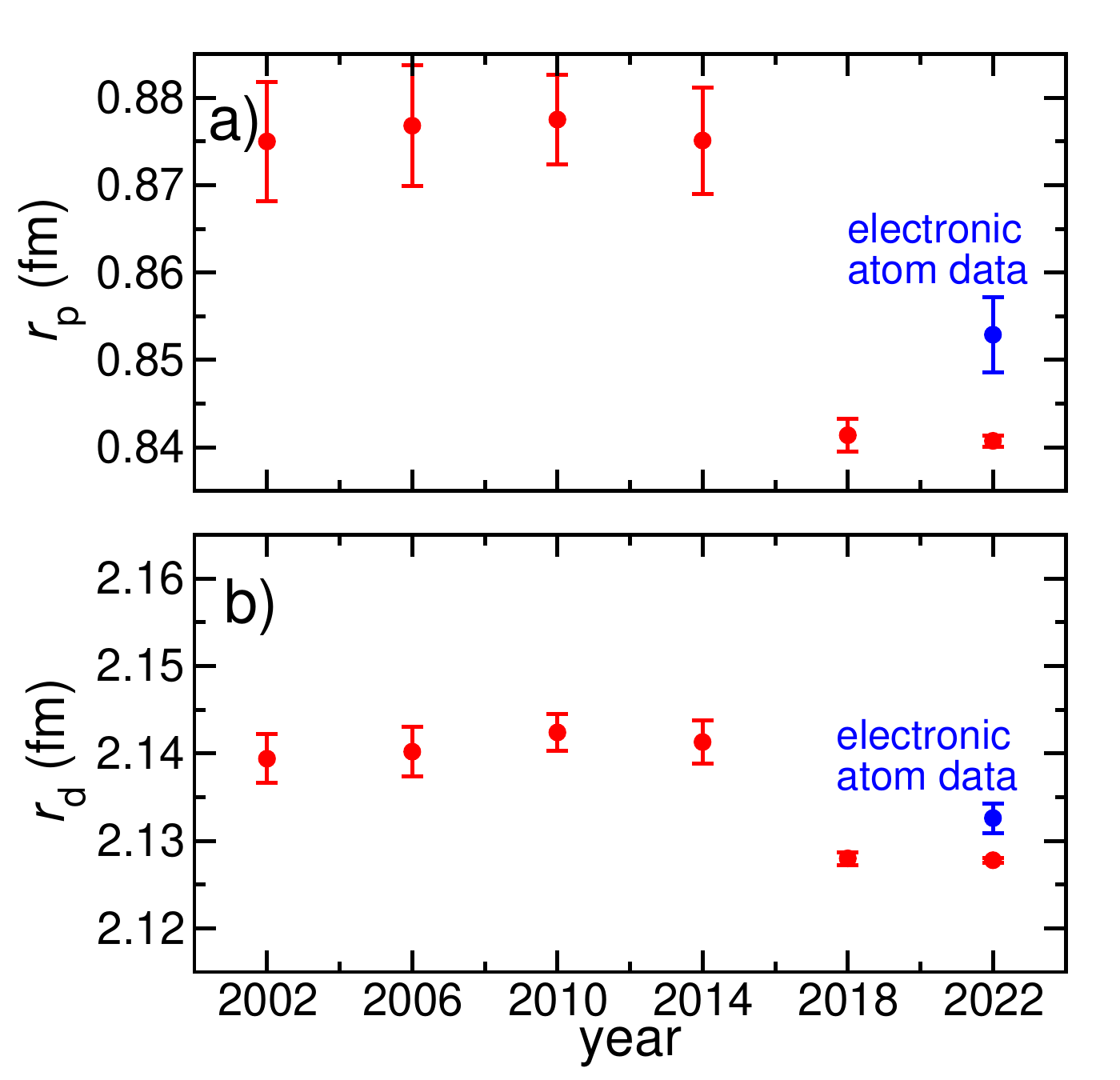} 
\caption[Comparison of the recommended values of the root-mean-square
(rms) charge radii of the proton and deuteron]{Comparison of the
recommended value of the rms charge radii of the proton $r_{\rm p}$ and
of the deuteron $r_{\rm d}$ from the 2022 and previous five CODATA
adjustments (in red).  Values from the 2022 adjustment are given in
Table~\ref{tab:radcomp}.}
\label{fig:history}
\end{figure*}

We conclude this section with Fig.~\ref{fig:history}, which shows how
the recommended values of $r_{\rm p}$ and $r_{\rm d}$ have evolved over
the past 20 years.  (The 2002 adjustment was the first that provided
recommended values for these radii.) Measurements of the Lamb shift in
the muonic atoms $\rmmu$H and $\rmmu$D as a source of information for
determining the radii are discussed starting with the 2010 CODATA
report, but only the e-p and e-d scattering values and electronic
spectroscopic data were used in the 2002, 2006, 2010, and 2014
adjustments. The recommended values have not varied greatly over this
period because the scattering and H-D spectroscopic values have not
varied much.  The 2018 adjustment was the first to include $\rmmu$H and
$\rmmu$D Lamb-shift input data, and it also included scattering values.
Nevertheless, the large shifts in the 2018 recommended values of $r_{\rm
p}$ and $r_{\rm d}$ are due to both new and more accurate H-D
spectroscopic data and improved muonic atom theory. As can be seen from
Secs.~\ref{sec:ryd} and \ref{ssec:mute} of this report, these advances
have continued, and as discussed in Sec.~\ref{ssec:ova}, the scattering
data are not included in the 2022 adjustment. The values of $r_{\rm p}$
and of $r_{\rm d}$ ``electronic atom data'' (in blue) are given in
Table~\ref{tab:radcomp}.

\begin{table*}
\caption[Observational equations for (muonic-)H, D, and $^4$He
spectroscopy]{Observational equations for input data on H, D, $^4$He
spectroscopy of muonic-H, -D, and  $^4$He Lamb shifts given in
Tables~\ref{tab:rydfreq}, \ref{tab:deltaRyd}, and \ref{tab:muhdata} as
functions of adjusted constants.  Labels in the first column correspond
to those defined in the tables with input data. Subscript $X$ is H, D,
or $^4$He.  Energy levels of hydrogenic atoms are discussed in
Sec.~\ref{ssec:hdel}.  Here, the symbol $\Gamma_X$ represents the six
adjusted constants $R_\infty,\alpha,A_{\rm r}({\rm e}),m_{\rm
e}/m_{\rmssmu}, A_{\rm r}({N})$, and $r_{N}$ such that when $X={\rm H}$
the nucleus $N={\rm p}$ is the proton, when $X={\rm D}$ the nucleus
$N={\rm d}$ is the deuteron, and when $X=^4$He the nucleus
$N={\rmalpha}$ is the $\alpha$ particle.  The Lamb shift for  muonic
atoms, $\Delta {\cal E}_{\rm LS}(\rmmu X)$, is discussed in
Sec.~\ref{sec:mulsrprd}.  \label{tab:pobseqsa}}

\begin{tabular}{l@{\qquad}rcl}
\hline
\hline
\multicolumn{1}{c}{Input data} & 
\multicolumn{3}{c}{Observational equation} \\
\hline

 A1--A4, A20, A21,
&$ \nu_{X}({n_1{\ell_1}_{j_1} \!-\! n_2{\rm \ell_2}_{j_2}}) 
-\fr{1}{4}\nu_{X}({n_3{\ell_3}_{j_3}\! -\! n_4{\ell_4}_{j_4}})$&$\doteq$&$ 
\Big\{E_{X}\big(n_2{\ell_2}_{j_2};\Gamma_X\big)
        +\delta_{X}(n_2{\ell_2}_{j_2}) $  \\[1 pt]
~~A24, A25&&&$ \quad -E_{X}\big(n_1{\ell_1}_{j_1};\Gamma_X\big)
   -\delta_{X}(n_1{\ell_1}_{j_1}) $  \\[1 pt]

&&&$ \quad\quad-\fr{1}{4} \big[E_{X}\big(n_4{\ell_4}_{j_4};\Gamma_X\big)
  +\delta_{X}(n_4{\ell_4}_{j_4}) $  \\[1 pt]
&&&$\quad \quad\quad\quad -E_{X}\big(n_3{\ell_3}_{j_3};\Gamma_X\big)
   -\delta_{X}(n_3{\ell_3}_{j_3})\big]\Big\}/h $  \\[3 pt]

 A5 
&$ \nu_{\rm D}(1{\rm S}_{1/2}\! - \!2{\rm S}_{1/2}) 
-\nu_{\rm H}(1{\rm S}_{1/2}\! -\! 2{\rm S}_{1/2})$&$\doteq$&$ 
\Big\{E_{\rm D}\big(2{\rm S}_{1/2};\Gamma_{\rm D}\big)
        +\delta_{\rm D}(2{\rm S}_{1/2}) $  \\[3 pt]
        &&&$ \quad -E_{\rm D}\big(1{\rm S}_{1/2};\Gamma_{\rm D}\big)
         -\delta_{\rm D}(1{\rm S}_{1/2}) $  \\[1 pt]
&&&$ \quad\quad - \big[E_{\rm H}\big(2{\rm S}_{1/2};\Gamma_{\rm H}\big)
     +\delta_{\rm H}(2{\rm S}_{1/2}) $  \\[0 pt]
&&&$ \quad\quad\quad -E_{\rm H}\big(1{\rm S}_{1/2};\Gamma_{\rm H}\big)
     -\delta_{\rm H}(1{\rm S}_{1/2})\big]\Big\}/h $  \\[1 pt]

 A6, A7, A9--A19, 
 & $ \nu_{X}(n_1{\ell_1}_{j_1}\! -\! n_2{\ell_2}_{j_2}) $&$\doteq$&$ 
\big[E_{X}\big(n_2{\ell_2}_{j_2};\Gamma_X
        \big) +\delta_{X}(n_2{\ell_2}_{j_2}) $  \\[1 pt]
 ~~A22, A23, A26--A29 &&&$\quad -E_{X}\big(n_1{\ell_1}_{j_1};\Gamma_X \big)-
\delta_{X}(n_1{\ell_1}_{j_1})\big]/h $  \\[1 pt]

 A8 &  $\nu_{\rm H}({\rm 2S_{1/2}}\shyphen{\rm 4P,\,centroid})$ & $\doteq$& 
     $  \Big\{ \big(E_{\rm H}\big(4{\rm P}_{1/2};\Gamma_{\rm H}\big) 
     +  \delta_{\rm H}(4{\rm P}_{1/2})\big)/3$ \\[1 pt]
       &&&$\quad 
    + 2\, \big(E_{\rm H}\big(4{\rm P}_{3/2};\Gamma_{\rm H}\big) 
    + \delta_{\rm H}(4{\rm P}_{3/2})\big)/3 $ \\[1 pt]
       &&&$\quad \quad - E_{\rm H}\big(2{\rm S}_{1/2};\Gamma_{\rm H}\big) 
       -\delta_{\rm H}(2{\rm S}_{1/2})\Big\}/h$\\[1 pt]

 B1--B25 &$ \delta_{X}(n\ell_j) $&$\doteq$&$ \delta_{X}(n\ell_j) $ \\[5 pt]

  C1, C3, C5 & $ \Delta {\cal E}_{\rm LS}(\rmmu X) $ 
  & $\doteq$&$ {\cal E}_{0X} + {\cal E}_{2X}\, r_N^2 + 
  \delta_{\rm th}(\rmmu X)$\\[1 pt]
  
 C2, C4, C6  & $\delta E_{\rm LS}(\rmmu X)$  & $\doteq$&$\delta_{\rm
th}(\rmmu X)$\\[1 pt]

\hline
\hline
\end{tabular}
\end{table*}
\section{Electron magnetic-moment anomaly}
\label{sec:elmagmom}

The interaction of the magnetic moment of a charged lepton $\ell$ in a
magnetic flux density (or magnetic field) $\bm B$ is described by the
Hamiltonian ${\cal H}=-\bm\mu_\ell \cdot \bm B$, with
\begin{eqnarray}
        \bm\mu_\ell = g_\ell \, \frac{e}{2m_\ell}\,\bm s ,
        \elabel{eq:lgdef}
\end{eqnarray}
where $\ell={\rm e}^{\pm}$, $\rmmu^{\pm}$, or $\rmtau^{\pm}$, $g_\ell$
is the $g$-factor, with the convention that it has the same sign as the
charge of the particle, $e$ is the positive elementary charge, $m_\ell$
is the lepton mass, and $\bm s$ is its spin.  Since the spin has
projection eigenvalues of $s_{z} = \pm \hbar/2$, the magnitude of a
magnetic moment is 
\begin{eqnarray}
 \mu_{\rm \ell} = \frac{g_{\ell}}{2} \,\frac{e\hbar}{2m_\ell} \,.
 \elabel{eq:gedef}
\end{eqnarray}
The lepton magnetic-moment anomaly $a_{\ell}$ is defined by the relationship
\begin{eqnarray}
 |g_\ell| &\equiv& 2(1+a_\ell) \,,
 \elabel{eq:adef}
\end{eqnarray}
based on the Dirac $g$-value of $-2$  and $+2$ for the negatively and
positively charged lepton $\ell$, respectively.

The Bohr magneton is defined as
\begin{eqnarray}
        \mu_{\rm B} = \frac{e\hbar}{2m_{\rm e}} \,, 
        \elabel{eq:bohrmag}
\end{eqnarray}
and the theoretical expression for the anomaly of the electron $a_{\rm
e}({\rm th})$  is
\begin{eqnarray}
	a_{\rm e}({\rm th}) = a_{\rm e}({\rm QED}) + a_{\rm e}({\rm weak})
	 + a_{\rm e}({\rm had}) \ ,
	 \label{eq:aeth}
 \end{eqnarray}
where terms denoted by ``QED'', ``weak'', and ``had'' account for the
purely quantum electrodynamic, predominantly electroweak, and
predominantly hadronic (that is, strong interaction) contributions,
respectively.  

The QED contribution may be written as 
\begin{eqnarray}
a_{\rm e}({\rm QED}) 
&=& \sum_{n=1}^{\infty} C_{\rm e}^{(2n)}\left({\alpha\over\pi}\right)^n\,,
\label{eq:appaeqed}
\end{eqnarray}
where the index $n$ corresponds  to contributions with $n$ virtual photons
and
\begin{equation}
C_{\rm e}^{(2n)}= A^{(2n)}_1 + A^{(2n)}_2(x_{{\rm e}\rmssmu}) 
+ A^{(2n)}_2(x_{{\rm e}\rmsstau})+\cdots 
 \label{eq:aeqed}
\end{equation}
with mass-independent coefficients $A^{(2n)}_1$ and functions
$A^{(2n)}_2(x)$ evaluated at mass ratio $x={x_{{\rm e}X}\equiv m_{\rm
e}/m_X}\ll 1$ for lepton ${X=\rmmu}$ or $\rmtau$.  For $n=1$, we have
\begin{eqnarray}
 A_1^{(2)} &=& 1/2 , \label{eq:a12}
\end{eqnarray}
and function ${A^{(2)}_2(x)=0}$, while
for $n>1$  coefficients $A^{(2n)}_1$   include
vacuum-polarization corrections with virtual electron/positron pairs. In fact,
\begin{eqnarray}
    A_1^{(4)} &=&  -0.328\,478\,965\,579\,193\ldots \,, \label{eq:a14} \\
    A_1^{(6)} &=&  1.181\,241\,456\,587\ldots\, ,  \label{eq:a16}\\
    A_1^{(8)}&=&  -1.912\,245\,764\ldots \,, \label{eq:a18}
\end{eqnarray}
The coefficient $A^{(8)}_1$ has been evaluated by \citet{2017026}.

\citet{2019055} have published an updated value for coefficient
$A_1^{(10)}=6.737(159)$, reducing their uncertainty by $\approx 25$\,\%
from that in \citet{2018025}.  In the same year, \citet{2019124}
published an independent evaluation of the diagrams contributing to
$A_1^{(10)}$ that have no virtual lepton loops and found
$A_1^{(10)}[{\rm no\mhyphen lepton\mhyphen loops}]=6.793(90)$.  The
total coefficient $A_1^{(10)}=A_1^{(10)}[\rm no\mhyphen lepton\mhyphen
loops] + A_1^{(10)}[\rm lepton\mhyphen loops]=5.863(90)$, where
$A_1^{(10)}[\rm lepton\mhyphen loops]=-0.93042(361)$ from
\citet{2018025} contains the contributions from diagrams that have
virtual lepton loops, which have a relatively small uncertainty.  All
uncertainties are statistical from numerically evaluating
high-dimensional integrals by Monte-Carlo methods.

\begin{figure}
\includegraphics[angle=-90,scale=0.31,
trim=100 0 150 10,clip]{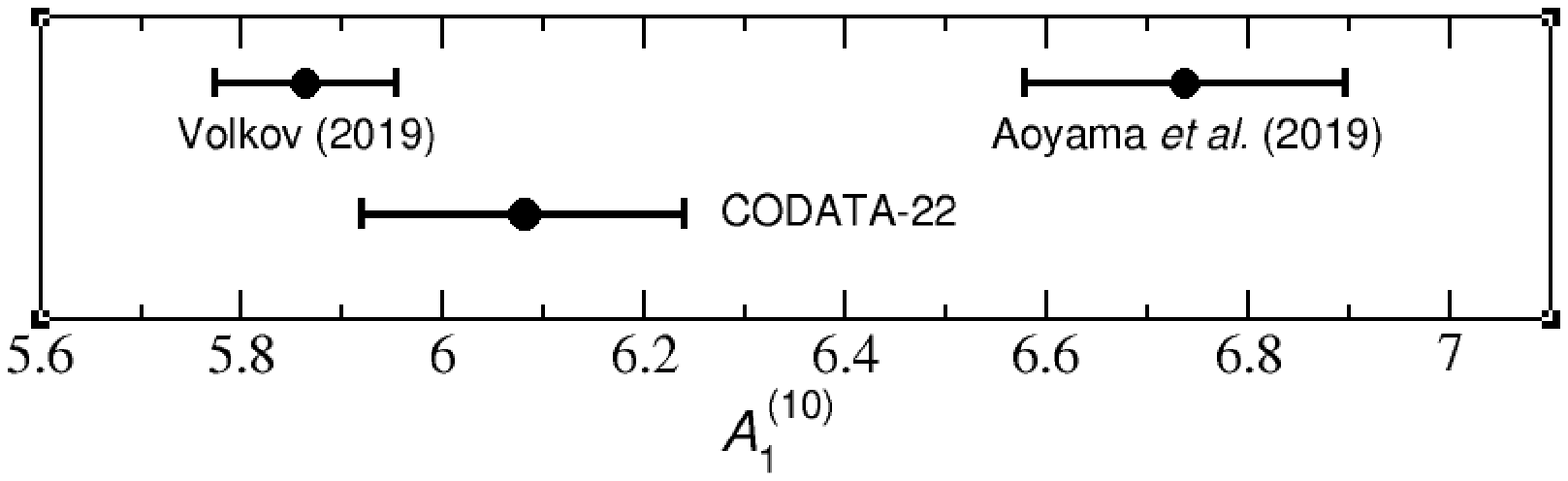}
\caption[Coefficient $A_1^{(10)}$ for the electron anomaly]{Coefficient
$A_1^{(10)}$ for the electron anomaly and its uncertainty as evaluated
by \citet{2019055} and \citet{2019124} as well as its value and
uncertainty used in the 2022 CODATA adjustment. The values of
$A_1^{(10)}$ used in the adjustment include an expansion factor of $2.1$
so that both values lie within two expanded standard deviations of the
2022 CODATA value.}
\label{fig:A10}
\end{figure}

The two values for $A_1^{(10)}$ by \citet{2019124} and
\citet{2019055} are discrepant by  $4.8\,\sigma$ as shown in
Fig.~\ref{fig:A10}, implying the need for an expansion factor for this
2022 CODATA adjustment.  A least-squares fit gives normalized residuals
$-2.4$ and $4.2$ for the values by \citet{2019124} and \citet{2019055},
respectively, and
\begin{equation} 
A^{(10)}_{1}\bigl|_{\rm CODATA\mhyphen
2022}=6.08(16)\,,
\label{eq:a110}
\end{equation}
where we have applied an expansion factor of $2.1$ to the uncertainty
used for the two input data to ensure that the values of \citet{2019124}
and \citet{2019055} lie within twice their (expanded) uncertainties of
the recommended value.

The functions $A^{(2n)}_2(x)$ for $n>1$ are vacuum-polarization
corrections due to heavier leptons.  For $x\to 0$, we have
$A^{(4)}_2(x)=x^2/45 + {\cal O}(x^4)$ and $A^{(6)}_2(x)=x^2(b_0 + b_1\ln
x ) + {\cal O}(x^4)$ with
$b_0=0.593\,274\cdots$ and $b_1=23/135$
\numcite{1993015,1993034}.  The ${\cal O}(x^4)$ contributions are known
and included in the calculations but not reproduced here.  The functions
$A^{(8)}_2(x)$ and $A^{(10)}_2(x)$ are also ${\cal O}(x^2)$ for small
$x$, but not reproduced here \cite{2014031,2015023}. Currently, terms
with $n>5$ and vacuum-polarization corrections that depend on two lepton
mass ratios can be neglected.

Table \ref{tab:anomaly} summarizes the relevant QED coefficients and
summed contributions to $C^{(2n)}_{\rm e}$ with their
one-standard-deviation uncertainties where appropriate as used in the
2022 CODATA adjustment.  Additional references to the original
literature can be found in descriptions of previous CODATA adjustments.

\begin{table*}
\def\vb{\vbox to 10pt {}}
\caption[Coefficients for QED contributions to the electron
anomaly]{Coefficients for the QED contributions to the electron anomaly.
The coefficients $A_1^{(2n)}$ and functions $A_2^{(2n)}(x)$, evaluated
at mass ratios $x_{ {\rm e}\rmssmu}=m_{\rm e}/m_{\rmssmu}$ and $x_{ {\rm
e}\rmsstau}=m_{\rm e}/m_{\rmsstau}$ for the muon and tau lepton,
respectively; summed values $C^{(2n)}_{\rm e}$, based on values for
lepton mass ratios from the 2022 CODATA adjustment, are listed as
accurately as needed for the tests described in this article.  Missing
values indicate that their contributions to the electron anomaly are
negligible.} \label{tab:anomaly}

\begin{tabular}{cl@{\ \ }l@{\ \ }l@{\ \ }l}
\hline
\hline
\vb $n$  & \multicolumn{1}{c}{$A_1^{(2n)}$} &
\multicolumn{1}{c}{$A^{(2n)}_2(x_{ {\rm e}\rmssmu})$} & 
\multicolumn{1}{c}{$A^{(2n)}_2(x_{{\rm e}\rmsstau})$} & 
\multicolumn{1}{c}{ $C_{\rm e}^{(2n)}$}  \\
\hline
1 & \multicolumn{1}{c}{$1/2$} &    \multicolumn{1}{c}{$0$}            
&  \multicolumn{1}{c}{$0$}                  & $\phantom{-} 0.5$ \\
2 & $ -0.328\,478\,965\,579\,193\ldots$ & $\phantom{-} 5.197\,386\,76(23)\times 10^{-7}$ & $\phantom{-} 1.837\,90(25)\times 10^{-9}$   
& $ -0.328\,478\,444\,00$\\
3 & $\phantom{-} 1.181\,241\,456\,587\ldots$ & $ -7.373\,941\,70(24)\times 10^{-6}$ & $ -6.582\,73(79)\times 10^{-8}$   
& $\phantom{-} 1.181\,234\,017$\\
4 & $ -1.912\,245\,764\ldots$ & $\phantom{-} 9.161\,970\,83(33)\times 10^{-4}$ & $\phantom{-} 7.428\,93(88)\times 10^{-6}$   
& $ -1.911\,322\,138\,91(88)$\\
5 & $\phantom{-} 6.080(160)$ & $ -0.003\,82(39)$ &                     
& $\phantom{-} 6.08(16)$\\
\hline
\hline
\end{tabular}
\end{table*}

The electroweak contribution is
\begin{eqnarray}
a_{\rm e}({\rm weak}) &=&  0.030\,53(23)\times 10^{-12} 
\label{eq:aeweak}
\end{eqnarray}
and is calculated as discussed in the 1998 CODATA adjustment, but with
the 2022 values of the Fermi coupling constant $G_{\rm F}/(\hbar c)^3$
and the weak mixing angle $\theta_{\rm W}$ \cite{2022048}.  

\citet{2019096} has provided updates to hadronic contributions to the
electron anomaly. See also \citet{2021024}.  Four such contributions
have been considered.  They are
\begin{eqnarray}
        a_{\rm e}({\rm had}) =&&
        a_{\rm e}^{\rm LO,VP}({\rm had})+a_{\rm e}^{\rm NLO,VP}({\rm had})
        \nonumber\\
        &&\quad
        + \,a_{\rm e}^{\rm NNLO,VP}({\rm had}) + a_{\rm e}^{\rm LL}({\rm had})
\end{eqnarray}
corresponding to leading-order (LO), next-to-leading-order (NLO), and
next-to-next-to-leading-order (NNLO) hadronic vacuum-polarization
corrections and a hadronic light-by-light (LL) scattering term,
respectively.  Contributions are determined from analyzing experimental
cross sections for electron-positron annihilation into hadrons and
tau-lepton-decay data.  The values in the 2022 adjustment are
\begin{equation}
        \begin{array}{r@{\ =\ }D{.}{.}{12}}
   a_{\rm e}^{\rm LO,VP}({\rm had})  & 1.849(11)\times 10^{-12} \,, \\
   a_{\rm e}^{\rm NLO,VP}({\rm had}) & -0.2213(12)\times 10^{-12}  \,,\\
   a_{\rm e}^{\rm NNLO,VP}({\rm had})&  0.028\,00(20)\times 10^{-12}  \,,\\
   a_{\rm e}^{\rm LL}({\rm had}) &  0.0370(50)\times 10^{-12} 
        \end{array} 
\end{equation}
leading to the total hadronic contribution 
\begin{equation}
a_{\rm e}({\rm had}) =   1.693(12)\times 10^{-12} \,.  \label{eq:aehad}
\end{equation}

A first-principle lattice quantum chromodynamics (QCD) evaluation of the
leading-order hadronic correction $a_{\rm e}^{\rm LO,VP}({\rm had})$ to
the electron anomaly was published in 2018 \cite{2018074}. The value is
\begin{equation}
  a_{\rm e}^{\rm LO,VP}({\rm had}) = 1.893(26)(56) \times 10^{-12}\,,
\end{equation}
where the first and second numbers in parentheses correspond to the
statistical and systematic uncertainty, respectively. The systematic
uncertainty is dominated by finite-volume artifacts. 
The combined uncertainty is six times larger than that
obtained by analyzing electron-positron scattering data.

The theoretical uncertainty of the electron anomaly (apart from
uncertainty in the fine-structure constant) is dominated by two
contributions: the mass-independent $n=5$ QED correction and the
hadronic contribution; its value is
\begin{eqnarray}
 u[a_{\rm e}({\rm th)}] 
=  0.000(16)\times 10^{-12} =  1.4\times 10^{-11}\, a_{\rm e} \, .
\label{eq:uncaeth}
\end{eqnarray}

In 2022, \citet{2023002} at Northwestern University measured the
electron anomaly $a_{\rm e}$ in an apparatus storing single electrons in
a homogeneous magnetic field. Their value is new for our CODATA
adjustments and has a 2.2 times smaller uncertainty than that reported
in 2008 by a Harvard research group led by the same senior researcher,
G. Gabrielse \cite{2008043}. Still the theoretical uncertainty is
significantly smaller than the $1.1\times 10^{-10}a_{\rm e}$ uncertainty
reported by \citet{2023002}. Following the recommendation by
\citet{2023002}, the 2022 measurement of $a_{\rm e}$ supersedes the
14-year-old datum.

The success of the 2022 measurements of $a_{\rm e}$ relied on a stable
magnetic field even though their frequency ratio measurement is to a
large degree independent of the actual field strength. Residual
dependencies are a consequence of the fact that the anomalous and
cyclotron frequencies of the electron are not measured simultaneously.
Relative field drifts of $2\times10^{-9}$/day, four times below that in
\citet{2008043}, allowed round-the-clock measurements and, thus,
improved statistical precision.  The researchers achieved this field
stability by supporting a 50 mK electron trap in a mixing chamber
flexibly hanging from the rest of a dilution refrigerator.

A stable magnetic field and electric fields generated by cylindrical
Penning trap electrodes with appropriately chosen relative dimensions
and potentials produce the trapping potential for the electron. The
Penning trap is also a low-loss microwave cavity that is used to inhibit
the decay rate of excited cyclotron states, here due to spontaneous
emission of synchrotron radiation, by a factor of 50 to 70. Cyclotron
excitations can then be detected before they decay. Nevertheless, the
limits on the cavity design, leading to cavity shifts in the cyclotron
frequency, and residual spontaneous emission were the two largest
uncertainties in the experiment.

For the least-squares adjustment, we use the observational equations
\begin{equation}
 a_{\rm e}({\rm exp}) \doteq a_{\rm e}({\rm th}) +\delta_{\rm th}({\rm e})
\end{equation}
and
\begin{equation}
 \delta_{\rm e}   \doteq \delta_{\rm th}({\rm e})
\end{equation}
with additive adjusted constant $\delta_{\rm th}({\rm e})$.  Input data
for $a_{\rm e}({\rm exp})$ is from \citet{2023002}, while input datum
$\delta_{\rm e}= 0$ with $u[\delta_{\rm e}] =0.018 \times
10^{-12}$ accounts for the uncertainty of the theoretical expression.
The input data is entry D$1$ in Table~\ref{tab:pdata}.
Relevant observational equations are found in Table~\ref{tab:pobseqsb}.
Atom-recoil experiments, discussed in Sec.~\ref{sec:atomrecoil}, form a
second competitive means to determine $\alpha$.

\section{Atom-recoil measurements}
\label{sec:atomrecoil}

Atom-recoil measurements with rubidium and cesium atoms from the
stimulated absorption and emission of photons are relevant for the
CODATA adjustment as they determine the electron mass, the atomic mass
constant, and the fine-structure constant
\numcite{1997036,1997175,2000035}.  This can be understood as follows. 

First and foremost, recoil measurements determine the mass $m(X)$ of a
neutral atom $X$ in kg using interferometers with atoms in
superpositions of momentum states and taking advantage of the fact that
photon energies  can be precisely measured.  Equally precise photon
momenta $p$ follow from their dispersion or energy-momentum relation
$E=pc$.  In practice, Bloch oscillations are used to transfer a large
number of photon momenta to the atoms in order to improve the
sensitivity of the measurement \numcite{2015047,2015152}.  Before the
adoption of the revised SI on 20 May 2019, these experiments only
measured the ratio $h/m(X)$, since the Planck constant $h$ was not an
exactly defined constant.

Second, atom-recoil measurements are a means to determine the atomic
mass constant, $m_{\rm u}=m(^{12}{\rm C})/12$, and the mass of the
electron, $m_{\rm e}$, in kg.  This follows, as many relative atomic masses
$A_{\rm r}(X)=m(X)/m_{\rm u}$ of atoms $X$ are well known.  For $^{87}$Rb
and $^{133}$Cs, the relative atomic masses have a relative uncertainty
smaller than $1\times 10^{-10}$ from the 2020 recommended values of the
AMDC (see Table \ref{tab:rmass20}).  The relative
atomic mass of the electron can be determined even more precisely
with spin-precession and cyclotron-frequency-ratio measurements on
hydrogenic $^{12}{\rm C}^{5+}$ and $^{28}{\rm Si}^{13+}$ as discussed
in Sec.~\ref{sec:ehcs}. We thus have 
\begin{equation}
        m_{\rm u} = m(X)/A_{\rm r}(X)
\end{equation}
and
\begin{equation}
        m_{\rm e} = \frac{A_{\rm r}({\rm e})}{A_{\rm r}(X)} m(X)
\end{equation}
from a measurement of the mass of atom $X$.

Finally, the fine-structure constant follows from the observation that
the Rydberg constant $R_{\infty} = \alpha^2m_{\rm e}c/2h$ 
has a relative standard uncertainty of $ 1.1\times 10^{-12}$
based on spectroscopy of atomic hydrogen discussed in
Sec.~\ref{sec:ryd}. The expression for $R_\infty$ can be rewritten as  
\begin{eqnarray}
 \alpha = \sqrt{\frac{2\,hcR_{\infty}}{m(X)c^2}\,
 \frac{A_{\rm r}(X)}{ A_{\rm r}(\rm e)}}
\elabel{eq:alhmx}
\end{eqnarray}
and a value of $\alpha$ with a competitive uncertainty can be
obtained from a  measurement of $m(X)$.

Two $m(X)$ or equivalently ${h/m(X)}$ measurements are input data in the
current least-squares adjustment: A mass for ${\rm^{133}Cs}$ measured at
the University of California at Berkeley, USA by \citet{2018033} and a
mass for ${\rm^{87}Rb}$ measured at the Laboratoire Kastler-Brossel
(LKB), France by \citet{2020064}.  The results are items D3 and D4 in
Table~\ref{tab:pdata} and satisfy the relevant observational equations
in Table~\ref{tab:pobseqsb}.

\begin{figure}
\includegraphics[scale=0.32,trim=0 0 50
0,clip,angle=-90,origin=c]{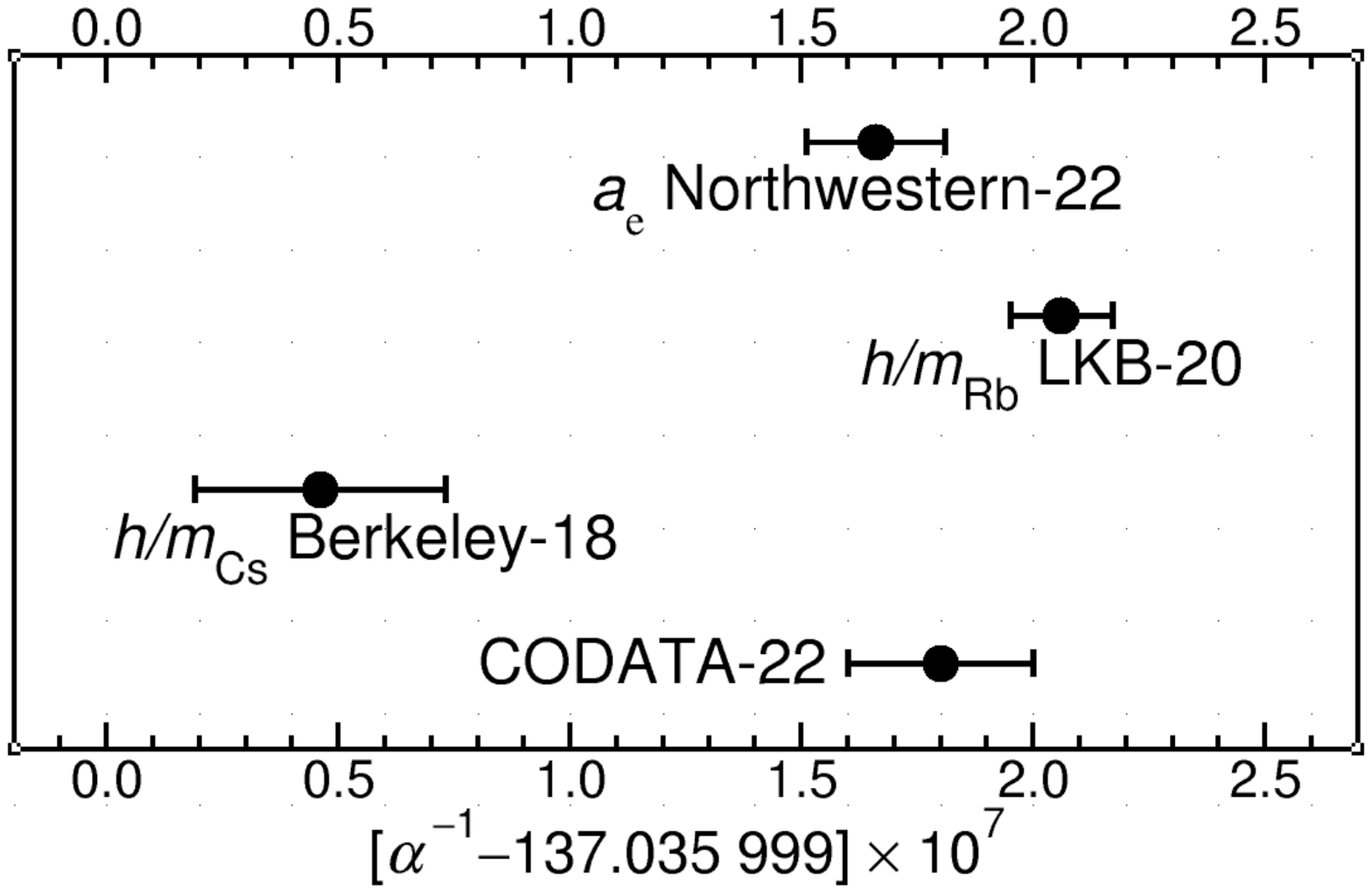} \caption[Determinations of
the value of the fine-structure constant from atom recoil
measurements]{Determinations of the value of the fine-structure constant
from atom recoil measurements and the measurement of the electron
magnetic-moment anomaly.  An expansion factor of 2.5 is applied to the
uncertainties of these data in the adjustment, so that all values lie
within two (expanded) standard deviations of the CODATA value.}
\label{fig:alpha}
\end{figure}

The values of ${\rm \alpha}$ inferred from the two atom-recoil
measurements are shown in Fig.~\ref{fig:alpha}, together with those
inferred from electron magnetic-moment anomaly $a_{\rm e}$ measurements.
Their comparison provides a useful test of the QED-based determination
of $a_{\rm e}$ and is further discussed in Sec.~\ref{ssec:cd}.

\citet{2020064} at LKB reduced by one order of magnitude several
systematic effects identified in their 2011 measurement \cite{2011014}.
Now they are able to adjust the altitude of atomic trajectories
within 100 $\rmmu$m and suppress the effect of the Earth’s rotation.
Long-term drifts in the laser beam alignment have been reduced and laser
frequencies further stabilized with a Fabry–P\'erot cavity and measured
with a frequency comb.  A lower density atomic sample was used to avoid
problems with a changing refractive index and atom–atom interactions.
Effects related to the Gouy phase and wave front curvature have been
mitigated.

In addition, \citet{2020064} identified new systematic effects
\cite{2018110,2019049}.  The most subtle one is related to the question
of how to calculate the photon momentum in a spatially distorted optical
field.  Moreover, in the expanding atomic cloud, a small phase shift due
to the variation of the intensity perceived by the atoms is present and
had to be mitigated.  Finally, a correction for a frequency ramp used to
compensate Doppler shifts induced by gravity had to be made.
\citet{2020064} think that these new systematic effects could explain
the $2.4\,\sigma$ discrepancy with the measurement in \citet{2011014}.
Unfortunately, they could not retroactively evaluate these contributions
for their 2011 measurement. 

The determinations of the fine-structure constant from both the recoil
measurements and the electron magnetic-moment anomaly are shown in
Fig.~\ref{fig:alpha}.  The recommended value includes an expansion
factor of 2.5 on the uncertinties of both the measurement input data and
the QED theory of the electron anomaly.

\section{Muon magnetic-moment anomaly}
\label{sec:mmma}

The muon magnetic-moment anomaly $a_{\rmssmu}$ and thus the muon
$g$-factor $g_{\rmssmu}=-2(1+a_{\rmssmu})$ were first measured in 2006,
and a recent measurement was made at the Fermi National Accelerator
Laboratory~\cite{2021012}.  A review of the theoretical value for
$a_{\rmssmu}$ is available from the PDG~\cite{2022048}.

\subsection{Theory of the muon anomaly}
\label{ssec:amuth}

The muon magnetic-moment anomaly theory is reviewed and summarized by A.
H\"ocker (CERN) and W.J. Marciano (BNL) on p.~796 of the PDG
report~\cite{2022048}.  These authors conclude that there is a
$4.2\,\sigma$ discrepancy between theory and experiment.  Active
searches for explanations for the discrepancy, which may indicate
deviations from the standard model of QCD, are underway and possible
such effects are discussed by H\"ocker and Marciano.

In view of the discrepancy and ongoing work seeking an explanation, the
TGFC has decided not to include theoretical input for the recommended
value of the muon magnetic-moment anomaly, so it is based on the
experimental values only.

The muon magnetic-moment anomaly can be expressed as
\begin{equation}
a_{\rmssmu}({\rm th}) = a_{\rmssmu}({\rm QED}) + a_{\rmssmu}({\rm weak})
 + a_{\rmssmu}({\rm had}) \, ,
\elabel{eq:appamth}
\end{equation}
where terms denoted by ``QED'', ``weak'', and ``had'' account for the
purely quantum electrodynamic, predominantly electroweak, and
predominantly hadronic (that is, with hadrons in intermediate states)
contributions, respectively.

Here, we update the QED contribution based on the CODATA 2022
recommended values of the fine-structure constant and the relevant mass
ratios.  It may be written as
\begin{eqnarray}
a_{\rmssmu}({\rm QED}) &=& 
 \sum_{n=1}^{\infty} C_{\rmssmu}^{(2n)}\left({\alpha\over\rmpi}\right)^n
 \,,
\label{eq:appamqed}
\end{eqnarray}
where
\begin{eqnarray}
 C_{\rmssmu}^{(2n)}&=& A^{(2n)}_1 + A^{(2n)}_2(x_{{\rmssmu}{\rm e}}) 
  + A^{(2n)}_2(x_{{\rmssmu}\rmsstau})
  \nonumber\\[5 pt]&&
  + A^{(2n)}_3(x_{{\rmssmu}\rm e}, x_{{\rmssmu}\rmsstau})
\end{eqnarray}
with mass-independent coefficients $A^{(2n)}_1$ given by
Eqs.~(\ref{eq:a12})-(\ref{eq:a110}) and functions $A^{(2n)}_2(x)$ and
$A^{(2n)}_3(x,y)$ evaluated at mass ratios $m_{\rmssmu}/m_{X}$ for
lepton ${X=\rm e}$ or $\rmtau$.  The expression for the QED contribution
has the same functional form as that for the electron anomaly described
in Sec.~\ref{sec:elmagmom}, except that the mass-dependent terms
$A^{(2n)}_2(x)$ are evaluated at different mass ratios, while
contributions due to $A^{(2n)}_3(x,y)$ are negligibly small for the
electron anomaly.  Contributions from the mass-dependent terms are
generally more important for the muon anomaly.
\begin{table*}
\def\vb{\vbox to 10 pt {}}
\caption[Coefficients for QED contributions to the muon
anomaly]{Mass-dependent functions $A^{(2n)}_2(x)$, $A^{(2n)}_3(x,y)$,
and summed $C_{\rmssmu}^{(2n)}$ coefficients for the QED contributions
to the muon anomaly based on the 2018 recommended values of lepton mass
ratios.  The functions are evaluated at mass ratios $x_{ \rmssmu{\rm
e}}\equiv m_{\rmssmu}/m_{\rm e} $ and/or $x_{\rmssmu\rmsstau}\equiv
m_{\rmssmu}/m_{\rmsstau}$.  }
\label{tab:muonanomaly}
\begin{tabular}{c@{\ \ }l@{\ \ \ }l@{\ \ \ }l@{\ \ \ }l}
\hline
\hline
\vb $n$  & \multicolumn{1}{c}{$A^{(2n)}_2(x_{ \rmssmu{\rm e}})$} & 
\multicolumn{1}{c}{$A^{(2n)}_2(x_{ \rmssmu\rmsstau})$} & 
\multicolumn{1}{c}{$A^{(2n)}_3(x_{ \rmssmu{\rm e}},x_{\rmssmu\rmsstau})$} & 
\multicolumn{1}{c}{ $C_{\rmssmu}^{(2n)}$}  \\
\hline
1 &     \multicolumn{1}{c}{$0$}   &  \multicolumn{1}{c}{$0$}  
 & \multicolumn{1}{c}{$0$}  & \phantom{75}$ 0.5$ \\
2 &  $\phantom{74} 1.094\,258\,3093(72)$ & $\phantom{-} 0.000\,078\,076(10)$  
 & \multicolumn{1}{c}{$0$} & $\phantom{75} 0.765\,857\,420(10)$\\
3 &  $\phantom{7} 22.868\,379\,98(17)$ & $\phantom{-} 0.000\,360\,599(86)$  
 & $ 0.000\,527\,738(71)$ & $\phantom{7} 24.050\,509\,77(16)$\\
4 &  $ 132.6852(60)$ & $\phantom{-} 0.042\,4928(40)$  
 & $ 0.062\,72(1)$  & $ 130.8782(60)$\\
5 &  $ 742.32(86)$ & $ -0.066(5)$  
 & $ 2.011(10)$  & $ 750.35(87)$\\
\hline
\hline
\end{tabular}
\end{table*}

The mass-dependent functions $A^{(2)}_2(x)$, $A^{(2)}_3(x)$, and
$A^{(4)}_3(x,y)$ are zero. The remaining nonzero mass-dependent
coefficients computed at the relevant mass ratios are given in
Table~\ref{tab:muonanomaly}. Their fractional contributions to the muon
anomaly are given in Table \ref{tab:fracmuonanomaly}. Only four of the
mass-dependent QED corrections contribute significantly to the
theoretical value for the muon anomaly.  Finally,  $a_{\rmssmu}({\rm
QED})$ based on the 2022 recommended value of $\alpha$ and lepton mass
ratios is
\begin{equation}
    a_{\rmssmu}({\rm QED}) =  0.001\,165\,847\,188\,14(84) \ [ 7.2\times 10^{-10}] \,.
\label{eq:amuqed}
\end{equation}

\begin{table}
\def\vb{\vbox to 10 pt {}}
\caption[Fractional values for QED contributions to the muon
anomaly]{Fractional contribution of mass-dependent functions
$A^{(2n)}_2(x)$ and $A^{(2n)}_3(x,y)$ for the QED contributions to the
muon anomaly based on the 2018 recommended values for $\alpha$ and
lepton mass ratios.  Fractional contributions are defined as
$A_{j}^{(2n)} \!\times\! (\alpha/\rmpi)^n/a_{\rmssmu}({\rm th})$ for
$j=2$ or $3$ and the relative standard uncertainty of $a_{\rmssmu}({\rm
th})$ is $3.3\times 10^{-7}$.  The functions are evaluated at mass
ratios $x_{\rmssmu{\rm e}}\equiv m_{\rmssmu}/m_{\rm e}$ and/or
$x_{\rmssmu\rmsstau}\equiv m_{\rmssmu}/m_{\rmsstau}$.  }
\label{tab:fracmuonanomaly}
\begin{tabular}{c@{\ \ \ }c@{\ \ \ }c@{\ \ \ }l}
\hline
\hline
\vb
$n$  & \multicolumn{1}{c}{$A^{(2n)}_2(x_{\rmssmu{\rm e}})$} & 
\multicolumn{1}{c}{$A^{(2n)}_2(x_{\rmssmu\rmsstau})$} & 
\multicolumn{1}{c}{$A^{(2n)}_3(x_{\rmssmu{\rm e}},x_{\rmssmu\rmsstau})$} \\
\hline
2 &  $ 5.06\times 10^{-3}$ & $\  3.61\times 10^{-7}$  
 & \vbox to 10 pt {}\ \\
3 &  $ 2.46\times 10^{-4}$ & $\  3.88\times 10^{-9}$ 
 & $ 5.67\times 10^{-9}$ \\
4 &  $ 3.31\times 10^{-6}$ & $\  1.06\times 10^{-9}$  
 & $ 1.57\times 10^{-9}$  \\
5 &  $ 4.31\times 10^{-8}$ & $ -3.80\times 10^{-12}$  
& $ 1.17\times 10^{-10}$  \\
\hline
\hline
\end{tabular}
\end{table}

For the electroweak and hadronic corrections to the muon anomaly, we
follow the recommendations of the PDG PDG~\cite{2022048}.  The primarily
electroweak contribution is
\begin{eqnarray}
a_{\rmssmu}{({\rm weak})} &=&  153.6(1.0)\times 10^{-11} 
\label{eq:amw}
\end{eqnarray}
and contains both the leading term and also some higher-order
corrections.
The PDG value of the hadronic corrections is
\begin{eqnarray}
a_{\rmssmu}({\rm had}) &=&  6937(44) \, ,
\label{eq:mhadtotal}
\end{eqnarray}
which yields a total theoretical value of
\begin{eqnarray}
a_{\rmssmu}({\rm th}) &=& 1.165\,918\,09(44) \times 10^{-3} \, .
\label{eq:amuth}
\end{eqnarray}

\subsection{2006 Measurement of the muon anomaly at Brookhaven
National Laboratory}
\label{ssec:amb}

The 2006 determination of the muon anomaly \cite{2006132} at Brookhaven
National Laboratory (BNL), USA in a 1.45 T superconducting storage ring
magnet with an approximate diameter of 7 m has been discussed in past
CODATA adjustment reports.  The relativistic velocity of the muons in
the ring is chosen such that the dependence on the static electric field
used for transverse confinement is tuned out. The spread in their
velocities leads to a negligible contribution to the uncertainty budget.
The quantity measured is the anomaly difference frequency $\omega_{\rm
a}=\omega_{\rm s}-\omega_{\rm c}$, where $\omega_{\rm s} =
|g_{\rmssmu}|(e/2m_{\rmssmu})B$ is the muon spin-flip (or precession)
frequency in  applied magnetic flux density $B$ and $\omega_{\rm
c}=(e/m_{\rmssmu})B$ is the muon cyclotron frequency.

The magnetic flux density was eliminated from these expressions by
determining its value from a measurement of the precession frequency of
the shielded proton in a spherical sample of pure H$_2$O in the same
apparatus, which depends on the temperature of the H$_2$O.  

\citet{2006132} in Table XV give as the result of their
experiment $R_{\rmssmu} = \omega_a/\omega_{\rm p} = 0.003\,707\,2063(20)
~[5.4\times 10^{-7}]$, where $\omega_{\rm p}$ is the precession frequency
of the {\it free} proton. This value, to which consistent measurements
on both positively and negatively charged muons contributed, was used as
the input datum for the BNL experiment in the 2006, 2010, 2014, and 2018
CODATA adjustments. Nevertheless, it was decided for the 2022 adjustment
to treat the BNL result in the same way other data involving the
shielded proton are treated, including the result reported in 2021 from
the similar experiment carried out at the Fermi National Accelerator
Laboratory (FNAL) in Batavia, Illinois, USA. It is discussed in the next
section.

\citet{2006132} (see section IV.A.1 of their paper) obtained the above
free proton value by converting their (unreported) result to the free
proton value based on the measurements of the shielding by
\citet{1977016} and the temperature dependence measured by
\citet{1984035}.

In previous CODATA adjustments, except for the value of $R_{\rmssmu}$,
all input data involving the precession frequency of protons in a
spherical pure H$_2$O sample obtained at a temperature other than
$25\,^\circ$C are converted to the reference temperature $25\,^\circ$C
using the expression obtained experimentally by Petley and Donaldson
given in Eq.~(\ref{eq:tconv}) in the next section.  Further, the
adjusted constants that depend on the shielded proton are taken to be
for protons in a spherical sample of pure H$_2$O at $25\,^\circ$C. For
such converted input data and adjusted constants, no temperature is
specified but a prime alone is used. For example, $\omega_{\rm
p}^\prime$ and $\mu_{\rm p}^\prime$ are the shielded proton precession
frequency and the shielded proton magnetic moment for protons in a
spherical sample of pure H$_2$O at $25\,^\circ$C, respectively.

Converting the \citet{2006132} reported value to $25\,^\circ$C yields
\begin{eqnarray}
&&R_{\rmssmu}^\prime({\rm
BNL}) =\omega_a/\omega^\prime_{\rm p}
\nonumber\\&& \qquad\qquad=
0.003\,707\,3015(20) ~[5.4\times 10^{-7}] \, .
\label{eq:id}
\end{eqnarray}
The relevant observational equation is
\begin{eqnarray}
  R^\prime_{\rmssmu} = \frac{e\hbar/(2m_{\rmssmu})}{\mu^\prime_{\rm p}}
 a_{\rmssmu} &\doteq&  
 -\frac{a_{\rmssmu}}{1+a_{\rm e}({\rm th})+\delta_{\rm th}({\rm e})} 
  \frac{m_{\rm e}}{m_{\rmssmu}}   \frac{\mu_{\rm e}}{\mu^\prime_{\rm p}}
  \, , \qquad
  \label{eq:rprime}
\end{eqnarray}
where the adjusted constants are $a_{\rmssmu}$, $m_{\rm e}/m_{\rmssmu}$,
and $\mu_{\rm e}/\mu^\prime_{\rm p}$, as well as the additive correction
$\delta_{\rm th}({\rm e})$ for the theoretical anomaly of the electron.

\subsection{2021 Measurement of the muon anomaly at the Fermi
National Accelerator Laboratory (FNAL)}
\label{ssub:fnal}

New for the 2022 CODATA adjustment is a measurement of the anomaly  of
the positively charged muon $a_{\rmssmu}$ from FNAL \cite{2021012}.  It
uses the same 1.45 T superconducting storage ring magnet but has a 2.5
times improved uniformity of the magnetic flux density. Moreover,
metrology for beam properties and the characterization of positrons, a
decay product of the muon, have been improved.  The apparatus operates
in a pulsed mode, where every 1.4 s a compact bunch of about 5000  muons
are prepared and stored in the ring.  The velocity of the muons is again
tuned such that the dependence on electric fields, used for transverse
confinement, cancels in the expression for the spin precession
frequency.  The result of the FNAL experiment run 1 as reported by
\citet{2021012} is 
\begin{eqnarray}
&&R_{\rmssmu}^\prime(34.7\,^\circ{\rm C})_{\rm
FNAL} =\omega_a/\omega^\prime_{\rm p}(34.7\,^\circ {\rm C})
\nonumber\\&& \qquad\qquad=
0.003\,707\,3003(17) ~[4.6\times 10^{-7}] \,,
\label{eq:fnalrep}
\end{eqnarray}
where the reference temperature of $34.7\,^\circ{\rm C}$ reported for
the magnetic field calibration is the same as that used by
\citet{1977016}.  Conversion of this value to the reference value of
$25\,^\circ {\rm C}$ yields
\begin{eqnarray}
&&R^\prime_{\rmssmu}({\rm FNAL}) = \omega_a/\omega^\prime_{\rm p}
\nonumber\\&& \qquad\qquad=
 0.003\,707\,2999(17) \quad [ 4.6\times 10^{-7}]
   \label{eq:rprime21} \, .
   \label{eq:fnal}
\end{eqnarray}
Its observational equation is Eq.~(\ref{eq:rprime}), the same as for
$R^\prime_{\rmssmu}({\rm BNL})$.  

The  input data and observational equation for both muon-anomaly
experiments can also be found in Table~\ref{tab:pdata} and
Table~\ref{tab:pobseqsb}, respectively.  

The quantity $R_{\rmssmu}^\prime({\rm FNAL})$, which is input datum D34
in Table \ref{tab:pdata}, is one of five input data whose determination
involved measuring the precession frequency of shielded protons in a
sample of pure H$_2$O. The other four are D33, $R_{\rmssmu}^\prime({\rm
BNL})$; D43, $\mu_{\rm e}(\mbox{H})/\mu_{\rm p}^\prime$; D44, $\mu_{\rm
h}(^3\mbox{He})/\mu_{\rm p}^\prime$ (former symbol $\mu_{\rm
h}^\prime/\mu_{\rm p}^\prime$); and D46, $\mu_{\rm n}/\mu_{\rm
p}^\prime$.  For D33, D34, and D43 the temperature reported for the
sample was $34.7\,^\circ {\rm C}$, for D44 it was $25\,^\circ$C and
hence no conversion is necessary, and for D46 the temperature was
$22\,^\circ$C. The conversion factor is obtained from the experimentally
derived expression of \citet{1984035} valid over the temperature range
$5\,^\circ {\rm C}$ to $45\,^\circ {\rm C}$:
\begin{eqnarray}
\frac{\mu_{\rm p}^\prime(t)}{\mu_{\rm p}^\prime} &=& 1 -
10.36(30)\times10^{-9}\,
\frac{(t-
25\,^\circ\mbox{C})}{^\circ\mbox{C}} \, ,
\label{eq:tconv}
\end{eqnarray}
with $t = 34.7\,^\circ {\rm C}$ or $t = 22\,^\circ {\rm C}$.
Thus
\begin{eqnarray}
\frac{\mu_{\rm p}^\prime(34.7\,^\circ {\rm C})}{\mu_{\rm p}^\prime}
&=& 0.999\,999\,8995(29) \,[2.9 \times 10^{-9}] \, ,
\nonumber\\
\frac{\mu_{\rm p}^\prime(22\,^\circ {\rm C})}{\mu_{\rm p}^\prime}
&=& 1.000\,000\,031\,08(90) \,[9.0 \times 10^{-10}] \, .
\qquad
\end{eqnarray}

The relative uncertainties of D33, D34, D43, and D46 before conversion
are $5.4 \times 10^{-7}$, $4.6 \times 10^{-7}$, $1.0 \times 10^{-8}$,
and $2.4 \times 10^{-7}$, respectively.  For D33 and D34, the $2.9
\times 10^{-9}$ uncertainty of its conversion factor is more than two
orders of magnitude smaller than their relative uncertainties and is
entirely negligible. This is also the case for the uncertainty of D46
and the $9.0 \times 10^{-10}$ uncertainty of its conversion factor.
However, the $2.9 \times 10^{-9}$ uncertainty of the conversion factor
for D43 is not negligible compared to its $1.0 \times 10^{-8}$
uncertainty and increases its relative uncertainty from $1.0 \times
10^{-8}$ to $1.1 \times 10^{-8}$.

We conclude this section by noting that on 11 August 2023, some seven
months after the 31 December 2022 closing date for new data for the 2022
adjustment, the FNAL Muon $g-2$ Collaboration posted the preprint
arXiv:2308.06230v1 on the archive server at https://arxiv.org.  It not
only reports new values of $R_{\rmssmu}^\prime(34.7\,^\circ$C) obtained
from runs 2 and 3, but in reference 23 updates the run-1 value in Eq.
(\ref{eq:fnalrep}) by the fractional amount $+0.28 \times 10^{-7}$.
However, this correction is small compared with the $4.6 \times 10^{-7}$
relative uncertainty of the run-1 value and only increases the last
digit of that value from 3 to 4. The preprint, subsequently published
\cite{2023025}, also gives as the combined result from all three runs as
$0.003\,707\,300\,82(75) \  [2.0 \times 10^{-7}]$. The value in
Eq.~(\ref{eq:fnalrep}), which is the basis of the input datum in
Eq.~(\ref{eq:fnal}), is in excellent agreement with this result but its
uncertainty is 2.3 times larger. Nevertheless, it means that the
numerical value of the 2022 recommended experimental value of
$a_{\rmssmu}$ should not be significantly different from that which
would result if the new three-run result had been available by the
closing date.

\subsection{Comparison of experiment and theory}
\label{sec:amuet}

The combined experimental value for $a_{\rmssmu}$(exp) is
\begin{eqnarray}
a_{\rmssmu}{\rm (exp)} &=& 0.001\,165\,920\,62(41)\times10^{-3} \, ,
\end{eqnarray}
which is the recommended value.  Comparison to Eq.~(\ref{eq:amuth})
yields
\begin{eqnarray}
a_{\rmssmu}{\rm (exp)} - a_{\rmssmu}{\rm (th)} 
= 253(60)\times10^{-11} \, ,
\end{eqnarray}
corresponding to the $4.2\,\sigma$ disagreement mentioned at the beginning
of Sec.~\ref{ssec:amuth} above.

\section{Atomic $g$-factors in hydrogenic $^{\mathbf 12}$C and $^{\mathbf 28}$Si ions} 
\label{sec:ehcs}

The most accurate value for the relative atomic mass of the
electron is obtained from measurements of the ratio of spin-precession
and cyclotron frequencies in hydrogenic carbon and silicon and
theoretical expressions for the $g$-factors of their bound electron.
See, for example, the recent analysis by \citet{2017032}.
These measurements also play an important role in determining the
fine-structure constant using atom-recoil experiments discussed in
Sec.~\ref{sec:atomrecoil}.

For a hydrogenic ion $X$ in its electronic ground state 1S$_{1/2}$ and
with a spinless nucleus, the Hamiltonian in an applied magnetic flux
density $\bm B$ is 
\begin{eqnarray}
 {\cal H} &=& - g_{\rm e}(X) \, \frac{e}{2m_{\rm e}}\,{\bm J} \cdot {\bm B}\, ,
\end{eqnarray}
where ${\bm J}$ is the electron angular momentum and $g_{\rm e}(X)$ is
the bound-state $g$-factor for the electron.  The electron angular
momentum projection is $J_z = \pm \hbar/2$ along the direction of $\bm
B$, so the energy splitting between the two levels is
\begin{eqnarray}
  \Delta E = |g_{\rm e}(X)|  \, \frac{e\hbar}{2m_{\rm e}}\, B ,
\elabel{eq:deemf}
\end{eqnarray}
and the spin-flip precession frequency is
\begin{eqnarray}
\omega_{\rm s} = \frac{\Delta E}{\hbar} =  |g(X)|  \, \frac{eB}{2m_{\rm e}} .
\end{eqnarray}
The ion's cyclotron frequency is
\begin{eqnarray}
\omega_{\rm c} &=& \frac{q_X B}{m_X}\,,
\end{eqnarray}
where $q_X = (Z-1)e$, $Z$, and $m_X$ are its net charge, atomic number,
and mass, respectively.  The frequency ratio $\omega_{\rm s}/\omega_{\rm
c}$ is then independent of $\bm B$ and satisfies 
\begin{equation}
\frac{\omega_{\rm s}}{\omega_{\rm c}} = 
\frac{|g_{\rm e}(X)|}{2(Z-1)}\,\frac{m_X}{m_{\rm e}}
=\frac{|g_{\rm e}(X)|}{2(Z-1)}\,\frac{A_{\rm r}(X)}{A_{\rm r}({\rm e})} \, ,
\label{eq:ratbsgf}
\end{equation}
where $A_{\rm r}(X)$ is the relative atomic mass of the ion.

We summarize the theoretical computations of the $g$-factor in
Sec.~\ref{ssec:thbegf} and describe the experimental input data and
observational equations in Secs.~\ref{ssec:bsgfexps} and
\ref{ssec:bsgfobs}.

\subsection{Theory of the bound-electron $g$-factor}
\label{ssec:thbegf}

The bound-electron $g$-factor is given by
\begin{eqnarray}
  g_{\rm e}(X) = g_{\rm D} + \Delta g_{\rm rad} + \Delta g_{\rm rec} 
+ \Delta g_{\rm ns} + \cdots \ ,
\elabel{eq:gsumdef}
\end{eqnarray}
where the individual terms on the right-hand side are the Dirac value,
radiative corrections, recoil corrections, and nuclear-size corrections,
and the dots represent possible additional corrections not already
included.

The Dirac value is \cite{1928001}
\begin{eqnarray}
g_{\rm D} &=& - \frac{2}{3}\left[1+2\sqrt{1-(Z\alpha)^2}\right]
        \label{eq:diracg}\\
&=& - 2\left[1-\frac{1}{3}(Z\alpha)^2 -\frac{1}{12}(Z\alpha)^4 
-\frac{1}{24}(Z\alpha)^6
 + \cdots\right] \,,
\nonumber
\end{eqnarray}
where the only uncertainty is due to that in $\alpha$.

The radiative correction is given by the series
\begin{equation}
\Delta g_{\rm rad} = \sum_{n=1}^\infty \Delta g^{(2n)} \,,
\end{equation}
where
\begin{eqnarray}
        \Delta g^{(2n)} &=& 
-2\, C_{\rm e}^{(2n)}(Z\alpha)\left({\alpha\over\rmpi}\right)^n
\, .
\end{eqnarray}
The first or one-photon coefficient in the series has self-energy (SE)
and vacuum-polarization (VP) contributions, i.e., $C_{\rm
e}^{(2)}(Z\alpha)=C_{\rm e,SE}^{(2)}(Z\alpha)+C_{\rm
e,VP}^{(2)}(Z\alpha)$.  The self-energy coefficient is \cite{1970020,
1970019, 1971019, 2004078e, 2005079}
\begin{eqnarray}
 C_{\rm e,SE}^{(2)}(Z\alpha) 
&=& \frac{1}{2}\,\bigg\{1 + \frac{(Z\alpha)^2}{6}
+(Z\alpha)^4\left[
        \frac{32}{9}\,\ln(Z\alpha)^{-2}
\right.
\nonumber\\ 
&&\left.   + \frac{247}{216}
-\frac{8}{9}\,\ln{k_0}
- \frac{8}{3}\, \ln{k_3}
\right] 
\nonumber\\
&&+ (Z\alpha)^5\,R_{\rm SE}(Z\alpha) \bigg\} \ ,
\label{eq:pachetal}
\end{eqnarray}
where
\begin{eqnarray}
\ln{k_0} &=&  2.984\,128\,556 \, ,
\label{eq:lnk0}
\\
\ln{k_3} &=&  3.272\,806\,545 \, ,
\label{eq:lnk3}
\\
R_{\rm SE}(6\alpha) &=&  22.1660(10) \, ,
\label{eq:rsec}
\\
R_{\rm SE}(14\alpha) &=&  21.000\,5(1) \, .
\label{eq:rsesi}
\end{eqnarray}
Values for the remainder function $R_{\rm SE}(Z\alpha)$ for carbon and
silicon have been taken from \citet{2017033}.  \citet{2017079} have
shown that
\begin{equation}
 R_{\rm SE}(0)=\pi\left\{\frac{89}{16}+\frac{8}{3} \ln 2\right\} \,.
 \label{eq:rse0}
\end{equation}
Finally, we have
\begin{equation}
 \begin{array}{rcl}
C_{\rm e,SE}^{(2)}(6\alpha) &=&  0.500\,183\,607\,131(80) \, , \\
C_{\rm e,SE}^{(2)}(14\alpha) &=&  0.501\,312\,638\,14(56) \, .
\end{array}
\label{eq:yerokgcsi}
\end{equation}

The lowest-order vacuum-polarization coefficient $C_{\rm e,
VP}^{(2)}(Z\alpha)$ has a wave-function and a potential contribution,
each of which can be separated into a lowest-order Uehling-potential
contribution and a higher-order Wichmann-Kroll contribution.  The
wave-function correction is \cite{2000153, 2000111, 2000037, 2001061,
2001218}
\begin{equation}
        \begin{array}{rcl}
C_{\rm e,VPwf}^{(2)}(6\alpha) &=&  -0.000\,001\,840\,3431(43) \, , \\
C_{\rm e,VPwf}^{(2)}(14\alpha) &=&  -0.000\,051\,091\,98(22) \, .
\end{array}
\label{eq:gvpwfcsi}
\end{equation}
For the potential correction, the Uehling contribution vanishes
\cite{2000111}, and for the Wichmann-Kroll part we take the value of
\citet{2005090}, which has a negligible uncertainty from omitted binding
corrections for the present level of uncertainty. This leads to
\begin{equation}
 \begin{array}{rcl}
C_{\rm e,VPp}^{(2)}(6\alpha) &=&  0.000\,000\,008\,201(11)  \,, \\
C_{\rm e,VPp}^{(2)}(14\alpha) &=&  0.000\,000\,5467(11)  \,,
\end{array}
\label{eq:gvppcsi} 
\end{equation}
and for the total lowest-order vacuum-polarization coefficient 
\begin{equation}
        \begin{array}{rcl}
C_{\rm e,VP}^{(2)}(6\alpha) &=& -0.000\,001\,832\,142(12) \, , \\
C_{\rm e,VP}^{(2)}(14\alpha) &=& -0.000\,050\,5452(11) \, .  
\end{array}
\label{eq:cvp2csi}
\end{equation}
Moreover, we have
\begin{eqnarray}
C_{\rm e}^{(2)}(6\alpha) &=&
C_{\rm e,SE}^{(2)}(6\alpha)+ C_{\rm e,VP}^{(2)}(6\alpha) \nonumber\\
&=&  0.500\,181\,774\,989(81)\, ,
\nonumber\\
C_{\rm e}^{(2)}(14\alpha) &=&
C_{\rm e,SE}^{(2)}(14\alpha)+ C_{\rm e,VP}^{(2)}(14\alpha) \nonumber\\
&=&  0.501\,262\,0929(12) \, .
\elabel{eq:c2csi}
\end{eqnarray}
 
The two-photon $n=2$ correction factor for the ground S state is
\cite{2005079,2006041}
\begin{eqnarray}
C_{\rm e}^{(4)}(Z\alpha) &=&
\left(1 + {(Z\alpha)^2\over 6} \right) C_{\rm e}^{(4)}
+(Z\alpha)^4
\nonumber\\
&&\times\bigg[ \frac{14}{9}\,\ln(Z\alpha)^{-2} 
+ \frac{991\,343}{155\,520} - \frac{2}{9}\,\ln{k_0}
\nonumber\\
&&- \frac{4}{3}\,\ln{k_3}
+ \frac{679\,\rmpi^2}{12\,960}
- \frac{1441\,\rmpi^2}{720}\,\ln{2} 
\nonumber\\&&+
\frac{1441}{480}\,\zeta({3})
+  \frac{16-19\rmpi^2}{216}
\bigg] + (Z\alpha)^5 
\nonumber\\ &&\times
 \bigg[ {\frac{14\rmpi}{135} \,\ln(Z\alpha)^{-2}}
 + \frac{1}{2}\, R^{(4)}(Z\alpha)\bigg] \, ,
\label{eq:c4g}
\end{eqnarray}
where $C_{\rm e}^{(4)} =  -0.328\,478\,444\,00\ldots$.  The last term in square
brackets for the contribution of order $(Z\alpha)^4$ is a light-by-light
scattering contribution \numcite{2016093}.   The first term in square
brackets for the contribution of order $(Z\alpha)^5$, absent in the
previous adjustment, is also a light-by-light scattering contribution
\numcite{2020062}.

The term $(Z\alpha)^5R^{(4)}(Z\alpha)$ in Eq.~(\ref{eq:c4g}) is the
contribution of order $(Z\alpha)^5$ and higher from diagrams with zero,
one, or two vacuum-polarization loops.  \citet{2013101} have performed
nonperturbative calculations for many of the vacuum-polarization
contributions to this function, denoted here by $R^{(4)}_{\rm
VP}(Z\alpha)$, with the results
\begin{equation}
R^{(4)}_{\rm VP}(6\alpha) = 14.28(39)\,,
\quad R^{(4)}_{\rm VP}(14\alpha) = 12.72(4)
\end{equation}
for the two ions.  These vacuum-polarization values are the sum of three
contributions.  The first, denoted with subscript SVPE, is from
self-energy vertex diagrams with a free-electron vacuum-polarization
loop included in the photon line and magnetic interactions on the
bound-electron line.  This calculation involves severe numerical
cancellations when lower-order terms are subtracted for small $Z$. The
results
\begin{equation}
R^{(4)}_{\rm SVPE}(6\alpha) = 0.00(15)\,,
\quad
R^{(4)}_{\rm SVPE}(14\alpha) = -0.152(43)
\end{equation}
were extrapolated from results for $Z\ge20$.  The second  contribution,
denoted with subscript SEVP, is from screening-like diagrams with
separate self-energy and vacuum-polarization loops.  The
vacuum-polarization loop includes the higher-order Wichmann-Kroll terms
and magnetic interactions are only included in the bound-electron line.
This set gives
\begin{equation}
R^{(4)}_{\rm SEVP}(6\alpha) = 7.97(36) \,,
\quad
R^{(4)}_{\rm SEVP}(14\alpha) = 7.62(1) \,.
\end{equation}
The third contribution, denoted with subscript VPVP, comes from
twice-iterated vacuum-polarization diagrams and from the
K\"all\'en-Sabry corrections with free-electron vacuum-polarization
loops, all with magnetic interactions on the bound-electron line.  This
set gives
\begin{equation}
R^{(4)}_{\rm VPVP}(6\alpha) = 6.31 \,,
\quad
R^{(4)}_{\rm VPVP}(14\alpha) = 5.25 \,.
\end{equation}
The results for this latter contribution are consistent with a
perturbative result at $Z\alpha=0$ given by \numcite{2009089} 
\begin{eqnarray}
R^{(4)}_{\rm VPVP}(0) &=& \left(\frac{1\,420\,807}{238\,140} +
\frac{832}{189}\,\ln{2} - \frac{400}{189}\,\rmpi\right)\rmpi
 \nonumber\\
 &=&  7.4415\dots \,.
\end{eqnarray}

\citet{2018016} performed perturbative calculations at $Z\alpha=0$ for a
complementary set of diagrams contributing to $R^{(4)}(Z\alpha)$. These
calculations include self-energy diagrams without vacuum-polarization
loops, with the combined result
\begin{equation}
\Delta R^{(4)}(0) = 4.7304(9)\,.
\end{equation}
This value has three contributions.  One is from self-energy diagrams
without vacuum-polarization loops given by
\begin{equation}
R^{(4)}_{\rm SE}(0) = 0.587\,35(9)\,\rmpi^2\,.
\end{equation}
The second set has light-by-light diagrams with nuclear interactions in
a vacuum-polarization loop inserted into the photon line in a
self-energy diagram, which gives
\begin{equation}
R^{(4)}_{\rm LBL}(0) = -0.172\,452\,6(1)\,\rmpi^2 \,.
\end{equation}
The remaining contribution with external magnetic-field coupling to a
virtual-electron loop is given by
\begin{eqnarray}
R^{(4)}_{\rm ML}(0) &=& \left(-\frac{101\,698\,907}{3\,402\,000} +
\frac{92\,368}{2\,025}\,\ln{2} - \frac{7\,843}{16\,200}\,\rmpi\right)\rmpi
 \nonumber \\
& = &0.064\,387\dots\rmpi^2 \,.
\end{eqnarray}

The results by \citet{2013101} and \citet{2018016} can be combined to
give
\begin{eqnarray}
R^{(4)}(Z\alpha) &=& R^{(4)}_{\rm VP}(Z\alpha) + \Delta R^{(4)}(0) \,,
\end{eqnarray}
which has uncertainty computed in quadrature from that of $R^{(4)}_{\rm
VP}(Z\alpha)$ and, following \citet{2018016},
\begin{equation}
u\left[\Delta R^{(4)}(0)\right] = \left|  Z\alpha\, \ln^3{(Z\alpha)^2} \right|
\end{equation}
taken to be on the order of the contribution of the next-order term.
For $Z=6$ and $14$, this uncertainty is approximately twice $\Delta
R^{(4)}(0)$.  Finally, we have for the two-photon coefficients 
\begin{equation}
        \begin{array}{rcl}
C_{\rm e}^{(4)}(6\alpha) &=&  -0.328\,579\,27(86) 
		     \\
C_{\rm e}^{(4)}(14\alpha)&=&  -0.329\,171(54) .
	\end{array} 
\label{eq:c4csi}
\end{equation}

For $n>2$ contributions $\Delta g^{(2n)}$ to the radiative correction,
it is sufficient to use the observations of \citet{1997162} and
\citet{2000193}, who showed that 
\begin{eqnarray}
C_{\rm e}^{(2n)}(Z\alpha) = 
\left(1 + {(Z\alpha)^2\over 6} + \cdots \right) 
	C_{\rm e}^{(2n)}
\elabel{eq:egbinding}
\end{eqnarray}
for all $n$. The values for constants $C_{\rm e}^{(2n)}$ for $n=1$
through 5 are given in Table~\ref{tab:anomaly}. This dependence for
$n=1$ and 2 can be recognized in Eqs.~(\ref{eq:pachetal}) and
(\ref{eq:c4csi}), respectively.  For $n=3$ we use
\begin{equation}
 \begin{array}{rcl}
C_{\rm e}^{(6)}(Z\alpha) &=& 
 1.181\,611\ldots \quad {\rm for}~Z = 6,
\\ &=&
 1.183\,289\ldots \quad {\rm for}~Z = 14 \, ,
\end{array}
\elabel{eq:c6csi}
\end{equation}
while for $n=4$ we have
\begin{equation}
       \begin{array}{rcl}
C_{\rm e}^{(8)}(Z\alpha) &=& 
-1.911\,933\dots  \ {\rm for}~Z = 6,
\\ &=&
-1.914\,647\dots
  \ {\rm for}~Z = 14 \, ,
\end{array}
\elabel{eq:c8csi}
\end{equation}
and, finally, for $n=5$
\begin{equation}
        \begin{array}{rcl}
C_{\rm e}^{(10)}(Z\alpha) &=& 
 6.08(16) \quad {\rm for}~Z = 6
\\ &=&
 6.09(16) \quad {\rm for}~Z = 14 \, .
\end{array}
\elabel{eq:c10csi}
\end{equation}

Recoil of the nucleus gives a correction $\Delta g_{\rm rec}$
proportional to the electron-nucleus mass ratio and can be written as
$\Delta g_{\rm rec} = \Delta g_{\rm rec}^{(0)} + \Delta g_{\rm
rec}^{(2)} + \dots$\,, where the two terms are zero and first order in
$\alpha/\rmpi$, respectively.  The first term is \cite{2002003,1997162}
\begin{eqnarray}
\Delta g_{\rm rec}^{(0)}  &=&
\bigg\{- (Z\alpha)^2 + {(Z\alpha)^4\over
3[1+\sqrt{1-(Z\alpha)^2}]^2}
\label{eq:grr0}\\
&& -(Z\alpha)^5\,P(Z\alpha)\bigg\}{m_{\rm e}\over m_{N}}
+(1+Z)(Z\alpha)^2\left({m_{\rm e}\over m_{N}}\right)^{\!2} \,,
\nonumber
\end{eqnarray}
where $m_{N}$ is the mass of the nucleus.  Mass ratios, based on the
current adjustment values of the constants, are ${m_{\rm e}/ m(^{12}{\rm
C}^{6+})} =  0.000\,045\,727\,5\ldots$ and ${m_{\rm e}/ m(^{28}{\rm Si}^{14+})} =
 0.000\,019\,613\,6\ldots$.  For carbon $P(6\alpha) = 10.493\,96(1)$,
and for silicon $P(14\alpha) = 7.162\,26(1)$~\cite{2020006}.

For $\Delta g_{\rm rec}^{(2)}$ we have
\begin{eqnarray}
\Delta g_{\rm rec}^{(2)}  &=&
{\alpha\over\rmpi}{(Z\alpha)^2\over3}
{m_{\rm e}\over m_{N}} +\cdots \,.
\label{eq:grr2}
\end{eqnarray}
The uncertainty in $\Delta g_{\rm rec}^{(2)}$ is negligible compared to
that of $\Delta g_{\rm rad}^{(2)}$.

\citet{2002058} have calculated the nuclear-size correction $\Delta
g_{\rm ns, LO}$ within lowest-order perturbation theory based on a
homogeneous-sphere nuclear-charge distribution and Dirac wave functions
for the electron bound to a point charge.  To good approximation, the
correction is \numcite{2000037}
\begin{equation}
   - {8\over3}(Z\alpha)^4 
      \left({R_{N}\over \lbar_{\rm C}}\right)^2 ,
\elabel{eq:gnsg}
\end{equation}
where $R_{N}$ is the rms nuclear charge radius and $\lbar_{\rm C}$ is
the reduced Compton wavelength of the electron.  In the CODATA
adjustment, we scale the values of \citet{2002058} with the squares of
updated values for the nuclear radii $R_{N} =  2.4702(22)$~fm and $R_{N}
=  3.1224(24)$~fm from the compilation of \citet{2013161} for $^{12}$C
and $^{28}$Si, respectively.

Recently, higher-order contributions of the nuclear-size correction have
been computed by \citet{2018020}.  They are
\begin{equation}
   \Delta g_{\rm ns, NLO} = -\left(\frac{2}{3} Z\alpha  {R_{N}\over \lbar_{\rm C}} C_{\rm ZF}+ \frac{\alpha}{4\rmpi} \right) \Delta g_{\rm ns, LO} \,,
   \label{eq:gnsNLO}
\end{equation} 
where $C_{\rm ZF}=3.3$ is the ratio of the Zemach or Friar moment
\numcite{1997168} to $R_{N}^3$ for a homogeneous-sphere nuclear-charge
distribution. We assume that $\Delta g_{\rm ns, NLO}$ has a 10\,\%
uncertainty.

The sum of the scaled nuclear-size correction of \citet{2002058} and
Eq.~(\ref{eq:gnsNLO}) yields 
\begin{equation}
\begin{array}{rcl}
\Delta g_{\rm ns} &=&  -0.000\,000\,000\,407(1) \quad {\rm for}~^{12}{\rm C}^{5+} \, ,
\\
\Delta g_{\rm ns} &=&  -0.000\,000\,020\,48(3) \quad\,\ {\rm for}~^{28}{\rm Si}^{13+} 
\end{array}
\elabel{eq:gns}
\end{equation}
for the total nuclear-size correction.

\begin{table}[t]
\def\va{\vbox to 12 pt {}}
\def\vb{\vbox to 8 pt {}}
\def\m{\phantom{-}}
\caption[Contributions to and value of the $g$-factor of hydrogenic carbon 12]{Theoretical contributions and total value for the $g$-factor of
	hydrogenic $^{12}$C$^{5+}$ based on the 2022 recommended values of the
constants.} \tlabel{tab:gfactthc}
\begin{center}
\begin{tabular}{c@{\quad}l@{\quad}c}
\hline
\hline
\vb
	Contribution   &  \multicolumn{1}{c}{Value} & Source \\
\hline
\vb
Dirac $g_{\rm D}$                 & $   -1.998\,721\,354\,392\,8(4)  $ & Eq.~(\ref{eq:diracg}) \\
 $\Delta g^{(2)}_{\rm SE}       $ & $   -0.002\,323\,672\,436\,6(5)$ & Eq.~(\ref{eq:yerokgcsi}) \\
 $\Delta g^{(2)}_{\rm VP}       $ & $\m 0.000\,000\,008\,511$ & Eq.~(\ref{eq:cvp2csi})  \\
 $\Delta g^{(4)}                $ & $\m 0.000\,003\,545\,6925(93)  $ & Eq.~(\ref{eq:c4csi})  \\
 $\Delta g^{(6)}                $ & $   -0.000\,000\,029\,618  $ & Eq.~(\ref{eq:c6csi})  \\
 $\Delta g^{(8)}                $ & $\m 0.000\,000\,000\,111  $ & Eq.~(\ref{eq:c8csi})  \\
 $\Delta g^{(10)}               $ & $   -0.000\,000\,000\,001  $ & Eq.~(\ref{eq:c10csi})  \\
 $\Delta g_{\rm rec}            $ & $ -0.000\,000\,087\,629 $ & Eqs.~(\ref{eq:grr0}) and (\ref{eq:grr2}) \\
 $\Delta g_{\rm ns}             $ & $   -0.000\,000\,000\,407(1)  $ & Eq.~(\ref{eq:gns})  \\
 $g(^{12}{\rm C}^{5+})          $ & $ -2.001\,041\,590\,1691(94) \va $ & Eq.~(\ref{eq:gcsi}) \\
\hline
\hline
\end{tabular}
\end{center}
\end{table}
\begin{table}[t]
\def\va{\vbox to 12 pt {}}
\def\vb{\vbox to 8 pt {}}
\def\m{\phantom{-}}
\caption[Contribution to and value of the $g$-factor of
hydrogenic silicon 28]{Theoretical contributions and total value for the $g$-factor of
	hydrogenic $^{28}$Si$^{13+}$ based on the 2022 recommended values of the
constants.} \tlabel{tab:gfactthsi}
\begin{center}
\begin{tabular}{c@{\qquad}l@{\qquad}c}
\hline
\hline
\vb
	Contribution   &  \multicolumn{1}{c}{Value} & Source \\
\hline
\vb
Dirac $g_{\rm D}$                 & $   -1.993\,023\,571\,561(2)  $ & Eq.~(\ref{eq:diracg}) \\
 $\Delta g^{(2)}_{\rm SE}       $ & $   -0.002\,328\,917\,507(3)$ & Eq.~(\ref{eq:yerokgcsi}) \\
 $\Delta g^{(2)}_{\rm VP}       $ & $\m 0.000\,000\,234\,81(1)$ & Eq.~(\ref{eq:cvp2csi})  \\
 $\Delta g^{(4)}                $ & $\m 0.000\,003\,552\,08(58)  $ & Eq.~(\ref{eq:c4csi})  \\
 $\Delta g^{(6)}                $ & $   -0.000\,000\,029\,66  $ & Eq.~(\ref{eq:c6csi})  \\
 $\Delta g^{(8)}                $ & $\m 0.000\,000\,000\,11  $ & Eq.~(\ref{eq:c8csi})  \\
 $\Delta g^{(10)}               $ & $   -0.000\,000\,000\,00  $ & Eq.~(\ref{eq:c10csi})  \\
 $\Delta g_{\rm rec}            $ & $ -0.000\,000\,205\,88 $ & Eqs.~(\ref{eq:grr0})-(\ref{eq:grr2}) \\
 $\Delta g_{\rm ns}             $ & $   -0.000\,000\,020\,48(3)  $ & Eq.~(\ref{eq:gns})  \\
 $g(^{28}{\rm Si}^{13+})        $ & $ -1.995\,348\,958\,08(58) \va $ & Eq.~(\ref{eq:gcsi}) \\
\hline
\hline
\end{tabular}
\end{center}
\end{table}

Tables~\ref{tab:gfactthc} and \ref{tab:gfactthsi} list the contributions
discussed above to $g_{\rm e}(X)$ for $X=^{12}$C$^{5+}$ and
$^{28}$Si$^{13+}$, respectively. The final values are
\begin{equation}
  \begin{array}{rcl}
    g_{\rm e}(^{12}{\rm C}^{5+}) &=&  -2.001\,041\,590\,1691(94)\, , \\
    g_{\rm e}(^{28}{\rm Si}^{13+}) &=&  -1.995\,348\,958\,08(58) 
  \end{array}
\label{eq:gcsi}
\end{equation}
with uncertainties that are dominated by that of the two-photon
radiative correction $\Delta g^{(4)}$.  This uncertainty is dominated by
terms proportional to $(Z\alpha)^6$ multiplying various powers of
$\ln[(Z\alpha)^{-2}]$.  We shall assume that the uncertainties for this
contribution have a correlation coefficient of
\begin{eqnarray}
r &=&  0.80 
\label{eq:gfthcov}
\end{eqnarray}
for the two hydrogenic ions.  The derived value for the electron mass
depends only weakly on this assumption; the value for the mass changes
by only 2 in the last digit and the uncertainty varies by 1 in its last
digit.

\subsection{Measurements of precession and cyclotron frequencies of
$^{12}$C$^{5+}$ and $^{28}$Si$^{13+}$}
\label{ssec:bsgfexps}

The experimentally determined quantities  are ratios of the electron
spin-precession (or spin-flip) frequency in hydrogenic carbon and
silicon ions to the cyclotron frequency of the ions, both in the same
magnetic flux density.  The input data used in the 2022 adjustment for
hydrogenic carbon and silicon are
\begin{equation}
\frac{\omega_{\rm s}\left(^{12}{\rm C}^{5+}\right)}{
 \omega_{\rm c}\left(^{12}{\rm C}^{5+}\right)} =
 4376.210\,500\,87(12) 
\quad [ 2.8\times 10^{-11}]  
\elabel{eq:rfsfcc14}
\end{equation}
and
\begin{equation}
\frac{\omega_{\rm s}\left(^{28}{\rm Si}^{13+}\right)}{
 \omega_{\rm c}\left(^{28}{\rm Si}^{13+}\right)} =
 3912.866\,064\,84(19)
\quad [ 4.8\times 10^{-11}] 
\elabel{eq:rfsfcsi13}
\end{equation}
with correlation coefficient
\begin{eqnarray}
r\left[\frac{\omega_{\rm s}\left(^{12}{\rm C}^{5+}\right)}{
 \omega_{\rm c}\left(^{12}{\rm C}^{5+}\right)}\,,\,
 \frac{\omega_{\rm s}\left(^{28}{\rm Si}^{13+}\right)}{
  \omega_{\rm c}\left(^{28}{\rm Si}^{13+}\right)}\right]
  &=&  0.347,
\end{eqnarray}
both obtained at the Max-Planck-Institut f\"ur Kernphysik, Heidelberg,
Germany (MPIK) using a multi-zone cylindrical Penning trap operating at
$B = 3.8$~T and in thermal contact with a liquid helium bath
\numcite{2013046,pc15ss,2014006,2015052}.  The development of this trap
and associated measurement techniques has occurred over a number of
years, leading to the current relative uncertainties below 5 parts in
$10^{11}$.  A detailed discussion of the uncertainty budget and
covariance  and additional references can be found in the 2014 CODATA
adjustment. We identify the results in Eqs.~(\ref{eq:rfsfcc14}) and
(\ref{eq:rfsfcsi13}) by MPIK-15.

\subsection{Observational equations for 
$^{12}$C$^{5+}$ and $^{28}$Si$^{13+}$ experiments}
\label{ssec:bsgfobs}

The observational equations that apply to the frequency-ratio
experiments on hydrogenic carbon and silicon and theoretical
computations of their $g$-factors follow from Eq.~(\ref{eq:ratbsgf})
when it is expressed in terms of the adjusted constants. That is,
\begin{eqnarray}
\frac{\omega_{\rm s}\left(^{12}{\rm C}^{5+}\right)}{
 \omega_{\rm c}\left(^{12}{\rm C}^{5+}\right)} &\doteq&
 -\frac{g_{\rm e}\left(^{12}{\rm C}^{5+}\right)+\delta_{\rm th}({\rm
 C})}{ 10 A_{\rm r}({\rm e})}
\nonumber\\&&\hbox to -2 cm {} \times
\left[12-5A_{\rm r}({\rm e}) 
+ \frac{\alpha^2 A_{\rm r}({\rm e})}{2R_\infty} 
 \frac{\Delta E_{\rm B}\left(^{12}{\rm
C^{5+}}\right) }{ hc}\right] 
\label{eq:rfsfccoe}
\end{eqnarray}
for $^{12}$C$^{5+}$ using $A_{\rm r}(^{12}{\rm C}) \equiv 12$,
Eq.~(\ref{eq:araxn}), and Eq.~(\ref{eq:ebxn}). Similarly,
\begin{equation}
{\omega_{\rm s}\left(^{28}{\rm Si}^{13+}\right)\over
 \omega_{\rm c}\left(^{28}{\rm Si}^{13+}\right)} \doteq
 -\frac{g_{\rm e}\left(^{28}{\rm Si}^{13+}\right)
 +\delta_{\rm th}({\rm
 Si})}{ 26 A_{\rm r}({\rm e})} \,
A_{\rm r}\!\left(^{28}{\rm Si}^{13+}\right)
\label{eq:rfsfcooe}
\end{equation}
for $^{28}$Si$^{13+}$. In these two equations,  $\alpha$, $R_\infty$,
the relative atomic masses $A_{\rm r}({\rm e})$ and $A_{\rm r}(^{28}{\rm
Si}^{13+}$), binding energy $\Delta E_{\rm B}\left(^{12}{\rm
C^{5+}}\right)$, and additive corrections $\delta_{\rm th}({\rm C})$ and
$\delta_{\rm th}({\rm Si})$ to the theoretical $g$-factors of $^{12}{\rm
C^{5+}}$ and $^{28}{\rm Si}^{13+}$ are adjusted constants.  Of course,
the observational equation
\begin{eqnarray}
A_{\rm r}\left(^{28}{\rm Si}\right) 
&\doteq&
A_{\rm r}\left(^{28}{\rm Si}^{13+}\right) 
+13A_{\rm r}({\rm e}) \nonumber\\
&& \quad
- \frac{\alpha^2 A_{\rm r}({\rm e})}{2R_\infty} \frac{
\Delta E_{\rm B}\left(^{28}{\rm Si^{13+}}\right)
}{hc} 
\label{eq:aroobseq}
\end{eqnarray}
relates the relative atomic mass of the silicon ion to that of the input
datum of the neutral atom and $\Delta E_{\rm B}\left(^{28}{\rm
Si^{13+}}\right)$ is an adjusted constant.

The theoretical expressions for $g$-factors $g_{\rm e}\left(^{12}{\rm
C}^{5+}\right)$ and $g_{\rm e}\left(^{28}{\rm Si}^{13+}\right)$ are
functions of adjusted constant $\alpha$.  The observational equations
for the additive corrections $\delta_{\rm th}({\rm C})$ and $\delta_{\rm
th}({\rm Si})$ for these $g$-factors are
\[
 \delta_X \doteq \delta_{\rm th}(X)
\]
for $X={\rm C}$ and Si with input data
\begin{equation}
 \begin{array}{rcl}
\delta_{\rm C} &=&  0.00(94)\times 10^{-11} \,,
\\
\delta_{\rm Si} &=&  0.00(58)\times 10^{-9} \,,
\end{array}
\label{eq:cgcsith}
\end{equation}
and $u(\delta_{\rm C},\delta_{\rm Si})=  0.4\times 10^{-20}$ from
Eqs.~(\ref{eq:gcsi}) and (\ref{eq:gfthcov}).

The input data are summarized as entries D7 through D13 in Table
\ref{tab:pdata} and observational equations can be found in Table
\ref{tab:pobseqsb}.

\section{Electron-to-muon mass ratio and muon-to-proton magnetic-moment ratio}
\label{sec:muhfs}

Muonium (Mu) is an atom consisting of a (positively charged) antimuon
and a (negatively charged) electron.  Measurements of two muonium
ground-state hyperfine transition energies in a strong magnetic
flux density combined with theoretical expressions for these energies
provide information on the electron-to-muon mass ratio, $m_{\rm
e}/m_{\rmssmu}$, as well as the antimuon-to-proton magnetic-moment
ratio, $\mu_{\rmssmu^+}/\mu_{\rm p}$.  Here, the proton magnetic
moment only appears because the applied magnetic field or flux
density is found by ``replacing'' the muonium with a proton in the
experimental apparatus and measuring the transition frequency
$\omega_{\rm p}$ of its precessing spin.  (More precisely, replacing
muonium  with a liquid-water sample, measuring the proton spin-precession
frequency in water, and accounting for a shielding correction.)

In the remainder of this section, we summarize the theoretical
determination of the zero-flux-density muonium hyperfine splitting
(HFS) and the experimental measurements at field fluxes between one
to two tesla.  Results of relevant calculations and measurements
are given along with references to new work; references to the
original literature used in earlier CODATA adjustments are not
repeated.  We finish with an analysis of the data.

\subsection{Theory of the muonium ground-state hyperfine splitting}
\label{ssec:muhfs}

The theoretical expression for the muonium hyperfine energy splitting absent
a magnetic field may be factored into
\begin{eqnarray}
\Delta E_{\rm Mu}({\rm th}) 
&=&\Delta E_{\rm F}\,\cdot \, {\cal F}
\label{eq:hfsth}
\end{eqnarray}
with the Fermi energy formula
\begin{eqnarray} 
\Delta E_{\rm F} 
&=& {16\over 3} hc R_\infty Z^3\alpha^2 {m_{\rm e}\over m_{\rmssmu}} 
\left( 1+{m_{\rm e}\over m_{\rmssmu}} \right)^{-3}  \,,
\label{eq:dfermi}\\
\nonumber 
\end{eqnarray} 
which contains the main dependence on fundamental constants,
and a function ${\cal F}=1+\alpha/\rmpi+\cdots$
that only depends weakly on them. 
(Recall $E_{\rm h}=2R_\infty hc=\alpha^2m_{\rm e}c^2$.)
The charge of the antimuon is specified by $Ze$ rather than $e$
in order to identify the source of terms contributing to $\Delta E_{\rm
Mu}({\rm th})$.

The Fermi formula in Eq.~(\ref{eq:dfermi}) is
expressed in terms of our adjusted constants $R_\infty$, $\alpha$,
and $m_{\rm e}/m_{\rmssmu}$.  The relative uncertainties of $R_\infty$
and $\alpha$ are much smaller than those for the measured $\Delta E_{\rm
Mu}$. Hence, a measurement of $\Delta E_{\rm Mu}$ determines the
electron-to-muon mass ratio.

The general expression for the hyperfine splitting and thus also $\cal F$ is
\begin{eqnarray} 
\Delta E_{\rm Mu} (\hbox{th}) 
&=& \Delta E_{\rm D} + \Delta E_{\rm rad} + \Delta E_{\rm rec} 
\elabel{eq:dmuth}
\\
&&\quad +\, \Delta E_{\rm r\hbox{-}r} 
+ \Delta E_{\rm weak} + \Delta E_{\rm had} \, , 
\nonumber
\end{eqnarray} 
where subscripts  D, rad, rec, r-r, weak, and had denote the
Dirac, radiative, recoil, radiative-recoil, electroweak, and hadronic
contributions to the hyperfine splitting, respectively. 

The Dirac equation yields
\begin{eqnarray}
\Delta E_{\rm D} &=& 
\Delta E_{\rm F}(1 + a_{\rmssmu})
\left[1+\frac{3}{2}(Z\alpha)^2 + \frac{17}{8}(Z\alpha)^4
+ \cdots \ \right] ,
\nonumber\\
\elabel{eq:dbreit}
\end{eqnarray}
where $a_{\rmssmu}$ is the muon magnetic-moment anomaly.
Radiative corrections are
\begin{eqnarray}
\Delta E_{\rm rad} &=& 
\Delta E_{\rm F}(1 + a_{\rmssmu})
       \sum_{n=1}^\infty D^{(2n)}(Z\alpha) \left({\alpha\over\rmpi}\right)^n
        \,,
\elabel{eq:drad}
\end{eqnarray}
where functions $D^{(2n)}(Z\alpha)$ are contributions from $n$
virtual photons.  The leading term is
\begin{eqnarray}
&&  D^{(2)}(Z\alpha) =
  A_{1}^{(2)} + \left(\ln 2 - \frac{5}{2}\right)\rmpi \,Z\alpha
   + \bigg[-\frac{2}{3}\ln^2(Z\alpha)^{-2} 
\nonumber\\
&&\qquad+ \left(\frac{281}{360}-\frac{8}{3}\ln 2\right)\ln(Z\alpha)^{-2}
 + 16.9037\ldots\bigg] (Z\alpha)^2
\nonumber\\
&&\qquad+ \bigg[\left(\frac{5}{2}\ln 2 - \frac{547}{96}\right)
\ln(Z\alpha)^{-2}\bigg]\rmpi \, (Z\alpha)^3 
\nonumber\\
&&\qquad + G(Z\alpha)\, (Z\alpha)^3 \,,
\elabel{eq:mud2}
\end{eqnarray}
where $A_{1}^{(2)} = 1/2$, as in Eq.~(\ref{eq:a12}).  The function $G(Z\alpha)$
accounts for all higher-order contributions in powers of $Z\alpha$; it can
be divided into self-energy (SE) and vacuum-polarization (VP) contributions,
$G(Z\alpha) = G_{\rm SE}(Z\alpha) + G_{\rm VP}(Z\alpha)$.  \citet{2008088,2010017} and
\citet{1999173,2000031} have calculated the one-loop self-energy and
vacuum-polarization contributions for the muonium HFS with $Z=1$.
Their results are
\begin{eqnarray}
G_{\rm SE}(\alpha) = - 13.8308(43) 
\end{eqnarray}
and
\begin{eqnarray}
G_{\rm VP}(\alpha) = 7.227(9)\, ,
\end{eqnarray}
where the latter uncertainty is meant to account for neglected
higher-order Uehling-potential terms; it corresponds to energy
uncertainties less than $h\times 0.1$ Hz, and is thus negligible.

For $D^{(4)}(Z\alpha)$, we have
\begin{eqnarray}
&& D^{(4)}(Z\alpha) =
A_{1}^{(4)} + 0.770\,99(2) \rmpi \,Z\alpha
+ \Big[-\frac{1}{3}\ln^2(Z\alpha)^{-2}
\nonumber\\ && 
\quad  -0.6390 \ldots\ln (Z\alpha)^{-2}
+10(2.5) \Big]\,(Z\alpha)^2
+ \cdots \,,
\elabel{eq:mud4}
\end{eqnarray}
where $A_1^{(4)}$ is given in Eq.~(\ref{eq:a14}).
The next term is
\begin{eqnarray}
D^{(6)}\!(Z\alpha) = A_{1}^{(6)} + \cdots \ ,
\elabel{eq:mud6}
\end{eqnarray}
where the leading contribution $A_{1}^{(6)}$ is given in
Eq.~(\ref{eq:a16}), but only partial results of relative order $Z\alpha$
have been calculated \cite{2007138}.  Higher-order functions
$D^{(2n)}\!(Z\alpha)$ with $n>3$ are expected to be negligible.

The recoil contribution is
\begin{eqnarray} 
\elabel{eq:hfsrec} 
\Delta E_{\rm rec} &=& 
\Delta E_{\rm F}\,\frac{m_{\rm e}}{m_{\rmssmu}}
\Bigg(-{3\over1-\left(m_{\rm e}/ m_{\rmssmu}\right)^2}
\ln \Big({m_{\rmssmu}\over m_{\rm e}}\Big){Z\alpha\over\rmpi} \nonumber\\ 
&& +\, {1\over\left(1+m_{\rm e}/m_{\rmssmu}\right)^2} 
\bigg\{\ln {(Z\alpha)^{-2}}-8\ln 
2 + {65\over 18}
\nonumber\\&&
+\bigg[{9\over 2\rmpi^2}
\ln^2\left({m_{\rmssmu}\over m_{\rm e}}\right)
+\left({27\over 2\rmpi^2}-1\right)
\ln\left({m_{\rmssmu}\over m_{\rm e}}\right)
\nonumber\\&&
+{93\over4\rmpi^2} + {33\zeta(3)\over\rmpi^2}
-{13\over12}-12\ln2 \bigg] 
{m_{\rm e}\over m_{\rmssmu}}
\bigg\}(Z\alpha)^2 \nonumber\\ 
&& +\, \bigg\{-{3\over 2} \ln{\Big({m_ 
{\rmssmu}\over m_{\rm e}}\Big)} \ln(Z\alpha)^{-2} 
-{1\over6}\ln^2 {(Z\alpha)^{-2}} 
\nonumber\\ && 
+\left({101\over18} - 10 \ln 2\right)\ln(Z\alpha)^{-2}
\nonumber\\&&\qquad\qquad
+40(10)\bigg\}
{(Z\alpha)^3\over\rmpi}
\Bigg)
 + \cdots \,.
\elabel{eq:drec}
\end{eqnarray} 

The leading-order ${\cal O}(\Delta E_{\rm F}\alpha^2)$ radiative-recoil
contribution is
\begin{eqnarray} 
\elabel{eq:hfsradrec} 
\Delta E_{\rm r\hbox{-}r} &=& \Delta E_{\rm F} 
\left({\alpha\over \rmpi}\right)^2 {m_{\rm e}\over m_{\rmssmu}} 
\bigg\{\bigg[ -2\ln^2\Big({m_{\rmssmu}\over m_{\rm e}}\Big) +{13\over
12}\ln{\Big({m_{\rmssmu}\over m_{\rm e}}\Big)} \nonumber\\ 
&& +\, {21\over2}\zeta(3)+{\rmpi^2\over6}+{35\over9}\bigg] 
+\bigg[\, {4\over3} \ln^2\alpha^{-2}
\nonumber\\
&&+\left({16\over3} \ln 2 - {341\over180}\right) \ln\alpha^{-2}
-40(10) \bigg]\rmpi\alpha \nonumber\\ 
&&+\bigg[-{4\over3}\ln^3 \Big({m_{\rmssmu}\over 
m_{\rm e}}\Big) 
+{4\over3}\ln^2 \Big({m_{\rmssmu}\over m_{\rm e}}\Big)
\bigg] {\alpha\over \rmpi}\bigg\} 
\nonumber\\ 
&& - \Delta E_{\rm F}\alpha^2\!\left(m_{\rm e}\over m_{\rmssmu}\right)^{\!2}
\left(6\ln{2} + {13\over6}\right) + \cdots \,,
\elabel{eq:dradrec}
\end{eqnarray} 
where, for simplicity, the explicit dependence on $Z$ is not shown.
Single-logarithmic and nonlogarithmic three-loop radiative-recoil
corrections of ${\cal O}(\Delta E_{\rm F}\alpha^3)$ are \numcite{2014159} 
\begin{eqnarray}
            \Delta E_{\rm F} \left(\frac{\alpha}{\rmpi}\right)^3
\frac{m_{\rm e}}{m_{\rmssmu}}  
 \bigg\{
        \left[-6\rmpi^2\ln{2}
        +\frac{\pi^2}{3}+\frac{27}{8} \right] \ln{\frac{m_{\rmssmu}}{m_{\rm e}}}
	\quad
\elabel{eq:essum}
        \\
+\, 68.507(2) \bigg\} = - h\times 30.99\ {\rm Hz}\,.
\nonumber
\end{eqnarray}
Uncalculated remaining terms of the same order as those included
in Eq.~(\ref{eq:essum}) have been estimated by \citet{2014159} to be about
$h\times10$~Hz to $h\times15$~Hz.  Additional radiative-recoil
corrections have been calculated, but are negligibly small, less
than $h\times0.5$ Hz.

The electroweak contribution due to the exchange
of a Z$^0$ boson is \cite{1996059}
\begin{eqnarray} 
\Delta E_{\rm weak}/h &=& -65 \ {\rm Hz} \ ,
\elabel{eq:dweak}
\end{eqnarray} 
while for the hadronic vacuum-polarization contribution
we have 
\begin{eqnarray}
\Delta E_{\rm had}/h &=& 
         237.7(1.5) \ {\rm Hz}\,.
\elabel{eq:dhad}
\end{eqnarray} 
This hadronic contribution combines the result of \citet{2013139}
with a newly computed $h\times 4.97(19)$~Hz contribution from \citet{2018131}.
A negligible contribution ($\approx h\times 0.0065$ Hz) from the hadronic
light-by-light correction has been given by \citet{2008206}.

The uncertainty of  $\Delta E_{\rm Mu}(\rm th)$ in
Eq.~(\ref{eq:dmuth}) is determined, from the largest to smallest
component, by those in $\Delta E_{\rm rec}$, $\Delta  E_{\rm
r\hbox{-}r}$, $\Delta E_{\rm rad}$, and $\Delta E_{\rm had}$. The
$h\times1.5$~Hz uncertainty in the latter is only of marginal interest.

For $\Delta E_{\rm rec}$, the total uncertainty is $h\times64$~Hz and has
three components. They are $h\times53$~Hz from twice the uncertainty 10
of the number 40 in Eq.~(\ref{eq:drec}) as discussed in the 2002 CODATA
adjustment, $h\times34$~Hz due to a possible recoil correction of order
$\Delta E_{\rm F}({m_{\rm e}/ m_{\rmssmu}}) \times(Z\alpha)^3\,\ln({m_{\rm
e}/ m_{\rmssmu}})$, and, finally, $h\times6$~Hz to reflect a possible
recoil term  of order $\Delta E_{\rm F}({m_{\rm e}/ m_{\rmssmu}})
\times(Z\alpha)^4\,\ln^2(Z\alpha)^{-2}$.

The  uncertainty in $\Delta  E_{\rm r\hbox{-}r}$ is $h\times55$~Hz,
with $h\times53$~Hz due to twice the uncertainty 10 of the number $-40$
in Eq.~(\ref{eq:dradrec}) as above, and $h\times15$~Hz as discussed in
connection with Eq.~(\ref{eq:essum}).  The uncertainty in $\Delta E_{\rm
rad}$ is $h\times5$~Hz and consists of two components: $h\times4$~Hz from
an uncertainty of $1$ in $G_{\rm VP}(\alpha)$ due to the uncalculated
contribution of order $\alpha (Z\alpha)^3$, and $h\times3$~Hz from the
uncertainty 2.5 of the number 10 in the function $D^{(4)}(Z\alpha)$.

The final uncertainty in $\Delta  E_{\rm Mu} {\rm(th)}$ is then
\begin{eqnarray}
u[\Delta  E_{\rm Mu} {\rm(th)}]/h =  85 \ {\rm Hz}. 
\elabel{eq:runcmuhfs}
\end{eqnarray}
For the least-squares adjustment, we use the 
observational equations
\begin{eqnarray}
        \Delta E_{\rm Mu} &\doteq& \Delta  E_{\rm Mu} ({\rm th})
        + \delta_{\rm th}({\rm Mu})
\label{eq:dmuc}
\end{eqnarray}
and
\begin{equation}
        \delta_{\rm Mu} \doteq \delta_{\rm th}({\rm Mu}) \,,
\end{equation}
where $\delta_{\rm th}({\rm Mu})$ accounts for the uncertainty of
the theoretical expression and is taken to be an adjusted constant.
Based on Eq.~(\ref{eq:runcmuhfs}), its corresponding input datum
in the 2018 adjustment is $\delta_{\rm Mu} = 0(85)$\,Hz. The input
data $\Delta E_{\rm Mu}$ are discussed later.  The theoretical
hyperfine splitting $\Delta E_{\rm Mu} ({\rm th})$ is mainly a
function of the adjusted constants $R_\infty$, $\alpha$, and $m_{\rm
e}/m_{\rmssmu}$. Finally, the covariance between $\Delta E_{\rm
Mu}$ and $\delta_{\rm Mu}$ is zero.

\subsection{Measurements of muonium transition energies}
\label{sssec:mufreqs}

The two most precise determinations of muonium hyperfine transition
energies were carried out by researchers at the Meson Physics Facility
at Los Alamos (LAMPF), New Mexico, USA and published in 1982 and 1999,
respectively.  These transition energies are compared to differences
between eigenvalues of the Breit-Rabi Hamiltonian
\numcite{1931002,1938002} modified for muonium using a magnetic flux
density determined from the free-proton nuclear magnetic resonance (NMR)
frequency measured in the apparatus.  The experiments were reviewed in
the 1998 CODATA adjustment.

Data reported in 1982 by \citet{1982003,pc81mar} are
\begin{eqnarray}
        \Delta E_{\rm Mu}/h &=&  4\,463\,302.88(16) \ {\rm kHz} \quad [ 3.6\times 10^{-8}]
\elabel{eq:nupL82}
\end{eqnarray}
for the hyperfine splitting and
\begin{eqnarray}
E(\omega_{\rm p})/h &=&  627\,994.77(14) \ {\rm kHz} \quad [ 2.2\times 10^{-7}]
\elabel{eq:numL82}
\end{eqnarray}
for the difference of two transition energies
with correlation coefficient
\begin{eqnarray}
        r[\Delta E_{\rm Mu},E(\omega_{\rm p})] &=&  0.227  \,.
\elabel{eq:rnupm82}
\end{eqnarray}
In fact, $\Delta E_{\rm Mu}$ and $E(\omega_{\rm p})$ are the sum
and difference of two measured transition energies, $\hbar \omega_{\rm
p}=2\mu_{\rm p}B$ is the free-proton NMR transition energy,
and only $E(\omega_{\rm p})$ depends on $\omega_{\rm p}$.  In this
experiment, $\hbar\omega_{\rm p}=h\times 57.972\,993$ MHz at its 1.3616 T
magnetic flux density.

The data reported in 1999 by \numcite{1999002} are
\begin{eqnarray}
        \Delta E_{\rm Mu}/h &=&  4\,463\,302\,765(53) \ {\rm Hz} \quad [ 1.2\times 10^{-8}]
\, ,
\elabel{eq:nupL99}\\
E(\omega_{\rm p})/h &=&  668\,223\,166(57) \ {\rm Hz} \quad [ 8.6\times 10^{-8}]
\elabel{eq:numL99}
\end{eqnarray}
with correlation coefficient
\begin{eqnarray}
r[\Delta E_{\rm Mu},E(\omega_{\rm p})] =  0.195
\elabel{eq:rnupm99}
\end{eqnarray}
and $\hbar\omega_{\rm p}=h\times 72.320\,000$ MHz for the proton
transition energy in a flux density of approximately $1.7$ T.

The observational equations are Eq.~(\ref{eq:dmuc}) and
\begin{eqnarray}
        \lefteqn{E(\omega_{\rm p}) \doteq 
-(W_{\rm e^-}+W_{\rmssmu^+})} 
        \label{eq:freqp}\\
&& +\sqrt{\bigl(\Delta E_{\rm Mu}({\rm th})+\delta_{\rm th}({\rm Mu})\bigr)^2+\left(W_{\rm e^-}-W_{\rmssmu^+}\right)^2}
        \,,
\nonumber 
\end{eqnarray}
where $W_\ell =-[\mu_{\ell}({\rm Mu})/\mu_{\rm p}] \hbar \omega_{\rm p}$.
Explicitly expressing $W_{\rm e^-}$ and $W_{\rmssmu^+}$  in terms of adjusted constants then yields
\begin{equation}
    W_{\rm e^-}= -  \frac{g_{\rm e}({\rm Mu})}{g_{\rm e}} \frac{\mu_{\rm e^-}}{\mu_{\rm p}} \hbar\omega_{\rm p}
    \label{eq:welectron}
\end{equation}
and
\begin{equation}
    W_{\rmssmu^+}
    =   \frac{g_{\rmssmu}({\rm Mu})}{g_{\rmssmu}} \frac{1+a_{\rmssmu}}{1+a_{\rm e}({\rm th})+\delta_{\rm th}({\rm e})}
    \frac{m_{\rm e}}{m_{\rmssmu}} \frac{\mu_{\rm e^-}}{\mu_{\rm p}} \hbar\omega_{\rm p} \,.
    \label{eq:wmuon}
\end{equation}
Here, we have used the fact that $\mu_\ell({\rm Mu})=g_\ell({\rm Mu})e\hbar/4m_\ell$
for the magnitude of the magnetic moment of lepton $\ell$ in muonium (see
Secs.~\ref{sec:elmagmom} and \ref{ssec:thbfrats}), $|g_\ell|=2(1+a_\ell)$,
and crucially $g_{\rmssmu^+}=-g_{\rmssmu^-}$.  The  $g$-factor ratios $g_{{\rm
e}}({\rm Mu})/ g_{{\rm e}}$ and $g_{{\rmssmu}}({\rm Mu})/ g_{{\rmssmu}}$
are given in Table \ref{tab:gfactrat}.

The adjusted constants in Eq.~(\ref{eq:dmuc}) and
Eqs.~(\ref{eq:freqp})-(\ref{eq:wmuon})
are the magnetic-moment anomaly $a_{\rmssmu}$, mass ratio $m_{\rm
e}/m_{\rmssmu}$, magnetic-moment ratio $\mu_{\rm e^-}/\mu_{\rm p}$, and
additive constants $\delta_{\rm th}({\rm Mu})$ and $\delta_{\rm th}({\rm
e})$. The latter two constants account for uncomputed theoretical
contributions to $\Delta E_{\rm Mu}({\rm th})$ and $a_{\rm e}({\rm
th})$, respectively.  Finally, $\Delta E_{\rm Mu}({\rm th})$ is mainly
a function of adjusted constants $m_{\rm e}/m_{\rmssmu}$, $R_\infty$,
and $\alpha$;  $a_{\rm e}({\rm th})$ is mainly a function of $R_\infty$
and $\alpha$.  The accurately measured or computed $\hbar\omega_{\rm
p}$ and ratios $g_\ell({\rm Mu})/g_\ell$ are treated as exact in our
least-squares adjustment.

In Eq.~(\ref{eq:freqp}) the energy $W_{\rm e^-}>0$, and at the flux
densities used in the experiments $|W_{\rm e^-}| \approx \Delta E_{\rm
Mu}({\rm th})$ and $|W_{\rmssmu^+}|\ll |W_{\rm e^-}|$. Consequently, the
right-hand side of Eq.~(\ref{eq:freqp}) only has  a weak dependence on
$\Delta E_{\rm Mu}({\rm th})$ and the corresponding input datum does not
significantly constrain $\Delta E_{\rm Mu}({\rm th})$ and thus $m_{\rm
e}/m_{\rmssmu}$ in the adjustment.

For ease of reference, the experimental and theoretical input data
for muonium hyperfine splittings are summarized in Table~\ref{tab:pdata}
and given labels D35 through D39. Observational equations are summarized
in Table~\ref{tab:pobseqsb}. 

\subsection{Analysis of the muonium hyperfine splitting and mass ratio $m_{\rmssmu}/m_{\rm e}$ }
\label{ssec:muhfsvalues}

The 2022 recommended value for the muonium hyperfine splitting is
\begin{eqnarray}
        \lefteqn{    \Delta E_{\rm Mu}({\rm th}) + \delta_{\rm th}({\rm Mu}) =} \nonumber\\
        &&h \times 4\,463\,302\,776(51) \ {\rm Hz}
\quad [ 1.1\times 10^{-8}] \,,
\elabel{eq:dnmup}
\end{eqnarray}
which is consistent both in value and uncertainty with the most
accurately measured value of Eq.~(\ref{eq:nupL99}). More importantly,
the prediction $\delta_{\rm th}({\rm Mu})/h=-4(83)$ Hz for the
additive constant falls well inside the 85 Hz theoretical uncertainty.
As $\delta_{\rm th}({\rm Mu})$ is a measure of uncomputed terms in
the theory, the value implies that the theory is sufficiently
accurate given the current constraints.  \citet{2019068} gave an
alternative prediction for the uncertainty of the recommended muonium
hyperfine splitting.

The 2022 recommended value for the muon-to-electron mass ratio is
\begin{equation}
        m_{\rmssmu}/m_{\rm e} =  206.768\,2827(46)
\end{equation}
and has a relative standard uncertainty of $ 2.2\times 10^{-8}$ that
is nearly twice that of the 1999 measurement of $\Delta E_{\rm Mu}$
in Eq.~(\ref{eq:nupL99}).  This increase simply reflects
the fact that the square of the relative standard uncertainty for
$m_{\rmssmu}/m_{\rm e}$ to good approximation satisfies
\begin{equation}
        u^2_{\rm r}(m_{\rmssmu}/m_{\rm e}) = u^2_{\rm r}\left(\Delta E_{\rm Mu}({\rm th})\right)
               + u^2_{\rm r}(\Delta E_{\rm Mu}) \,,
\end{equation}
which follows from error propagation with Eqs.~(\ref{eq:hfsth}) and
(\ref{eq:dmuc}). The relative standard uncertainties in the theory for
and measurement of the hyperfine splitting are almost the same.

New data on the muonic hyperfine splitting by the MuSEUM collaboration
at the J-PARC Muon Science Facility are expected in the near future
\numcite{2019025}.

\LTcapwidth=\columnwidth    %

\begin{table}
\caption[List of Adjusted constants]{Fifty-five of the eighty
adjusted constants in the 2022 CODATA least-squares adjustment.  The
other 25 adjusted constants are given in Table \ref{tab:deltaRyd}.
} \label{tab:adjother}
\begin{tabular}{ll}
\hline
\hline
\multicolumn{1}{c}{Adjusted constant} & Symbol \\
\hline
fine-structure constant & $\alpha$ \\
Rydberg constant & $R_\infty$ \\
proton rms charge radius & $r_{\rm p}$ \\
deuteron rms charge radius & $r_{\rm d}$ \\
alpha particle rms charge radius & $r_{\rmssalpha}$ \\
Newtonian constant of gravitation & $ G $ \\
electron relative atomic mass & $A_{\rm r}({\rm e})$ \\
proton relative atomic mass & $A_{\rm r}({\rm p})$ \\
neutron relative atomic mass & $A_{\rm r}({\rm n})$ \\
deuteron relative atomic mass & $A_{\rm r}({\rm d})$ \\
triton relative atomic mass & $A_{\rm r}({\rm t})$ \\
helion relative atomic mass & $A_{\rm r}({\rm h})$ \\
alpha particle relative atomic mass & $A_{\rm r}(\rmalpha)$ \\
$^{28}$Si$^{13+}$ relative atomic mass & $A_{\rm r}(^{28}{\rm Si}^{13+})$ \\
$^{87}$Rb relative atomic mass & $A_{\rm r}(^{87}{\rm Rb})$ \\
$^{133}$Cs relative atomic mass & $A_{\rm r}(^{133}{\rm Cs})$ \\
H$_2^+$ electron ionization energy & $\Delta E_{\rm I}({\rm H}_2^+)$ \\
HD$^+$ electron ionization energy & $\Delta E_{\rm I}({\rm HD}^+)$ \\
$^3$He$^+$ electron ionization energy & $\Delta E_{\rm I}(^3{\rm He}^+)$ \\
$^{12}$C$^{4+}$ electron removal energy & $\Delta E_{\rm B}(^{12}{\rm C}^{4+})$ \\
$^{12}$C$^{5+}$ electron removal energy & $\Delta E_{\rm B}(^{12}{\rm C}^{5+})$ \\
$^{12}$C$^{6+}$ electron removal energy & $\Delta E_{\rm B}(^{12}{\rm C}^{6+})$ \\
$^{28}$Si$^{13+}$ electron removal energy & $\Delta E_{\rm B}(^{28}{\rm Si}^{13+})$ \\
additive correction to $f_{\rm SA}^{\rm th}(0,\!0\!\rightarrow0,\!1)$ 
& $\delta_{\rm HD^+}^{\rm th}(0,\!0\!\rightarrow0,\!1)$ \\
additive correction to $f_{\rm SA}^{\rm th}(0,\!0\!\rightarrow1,\!1)$ 
& $\delta_{\rm HD^+}^{\rm th}(0,\!0\!\rightarrow1,\!1)$ \\
additive correction to $f_{\rm SA}^{\rm th}(0,\!3\!\rightarrow9,\!3)$ 
& $\delta_{\rm HD^+}^{\rm th}(0,\!3\!\rightarrow9,\!3)$ \\
additive correction to $a_{\rm e}({\rm th})$ & $\delta_{\rm th}({\rm e})$ \\
additive correction to $g_{\rm C}(\rm th)$ & $\delta_{\rm th}({\rm C})$ \\
additive correction to $g_{\rm Si}(\rm th)$ & $\delta_{\rm th}({\rm Si})$ \\
additive correction to $\Delta\nu_{\rm Mu}({\rm th})$ & $\delta_{\rm th}({\rm Mu})$ \\
additive correction to $\rmmu$-H Lamb shift & $\delta_{\rm th}(\rmmu{\rm H})$ \\
additive correction to $\rmmu$-D Lamb shift & $\delta_{\rm th}(\rmmu{\rm D})$ \\
additive correction to $\rmmu$-$^4$He$^+$ Lamb shift & $\delta_{\rm
th}(\rmmu^4\mbox{He}^+)$ \\
electron-muon mass ratio & $m_{\rm e}/m_{\rmssmu}$ \\
muon magnetic moment & $a_{\rmssmu}$ \\
deuteron-electron magnetic-moment ratio & $\mu_{\rm d}/\mu_{\rm e^-}$ \\
electron-proton magnetic-moment ratio & $\mu_{\rm e^-}/\mu_{\rm p}$ \\
electron to shielded proton & $\mu_{\rm e^-}/\mu^\prime_{\rm p}$ \\
\b                 magnetic-moment ratio  & \\
bound helion to shielded proton & $\mu_{\rm h}(^3{\rm He})/\mu^\prime_{\rm p}$ \\
\b                 magnetic-moment ratio  & \\
neutron to shielded proton & $\mu_{\rm n}/\mu_{\rm p}^\prime$ \\
\b                 magnetic-moment ratio  & \\
triton to proton magnetic-moment ratio & $\mu_{\rm t}/\mu_{\rm p}$ \\
shielding difference of d and p in HD & $\sigma_{\rm dp}$ \\
shielding difference of t and p in HT & $\sigma_{\rm tp}$ \\
$d_{220}$ of an ideal natural Si crystal & $d_{220}$ \\
$d_{220}$ of Si crystal ILL & $d_{220}({\rm {\scriptstyle ILL}})$ \\
$d_{220}$ of Si crystal MO$^*$ & $d_{220}({\rm {\scriptstyle MO^*}})$ \\
$d_{220}$ of Si crystal N & $d_{220}({\rm {\scriptstyle N}})$ \\
$d_{220}$ of Si crystal NR3 & $d_{220}({\rm {\scriptstyle NR3}})$ \\
$d_{220}$ of Si crystal NR4 & $d_{220}({\rm {\scriptstyle NR4}})$ \\
$d_{220}$ of Si crystal WASO 04 & $d_{220}({\rm {\scriptstyle W04}})$ \\
$d_{220}$ of Si crystal WASO 17 & $d_{220}({\rm {\scriptstyle W17}})$ \\
$d_{220}$ of Si crystal WASO 4.2a & $d_{220}({\rm {\scriptstyle W4.2a}})$ \\
Copper K$\alpha_1$ x unit & ${\rm xu}({\rm CuK}\alpha_{\rm 1})$ \\
\AA{}ngstrom star & \AA$^{\ast}$ \\
Molybdenum K$\alpha_1$ x unit & ${\rm xu}({\rm MoK}\alpha_{\rm 1})$ \\
\hline
\hline
\end{tabular}
\end{table}

\section{Magnetic-moment ratios of light atoms and molecules}
\label{sec:mmrmemr}

The CODATA Task Group recommends values for the free-particle magnetic
moments of leptons, the neutron, and light nuclei.  The most precise
means to determine the free magnetic moments of the electron, muon, and
proton are discussed in Secs.~\ref{sec:elmagmom}, \ref{sec:mmma}, and
\ref{sec:ppm}, respectively.  In this section, we describe the
determination of the neutron, deuteron, triton, and helion magnetic
moments.  The magnetic moment of the $^4$He nucleus or $\rmalpha$
particle is zero.

The determination of the ratio of the neutron magnetic moment to the
moment of the proton in water is discussed in Sec.~III.C.8 of
\citet{2000035}.

Nuclear magnetic moments may be determined from hydrogen and deuterium
maser experiments and NMR experiments on atoms and molecules.  Both
types of experiments measure ratios of magnetic moments to remove the
need to know the strength of the applied magnetic field.  We rely on NMR
measurements for ratios of nuclear magnetic moments in the HD and HT
molecules as well as the ratio of the magnetic moment of the neutron and
the helion in $^3$He with respect to that of the proton in H$_2$O.  For
these molecules, the electronic ground state is an electron spin
singlet.

The magnetic moment of a nucleus or electron in an atom or molecule,
however, differs from that of a free nucleus or electron and theoretical
binding corrections are used to relate bound moments to free moments.
In the remainder of this section, we give the relevant theoretical
binding corrections to magnetic-moment ratios and describe experimental
input data.  We also describe the binding corrections for
magnetic-moment ratios of an antimuon and electron bound in muonium
(Mu).  These are relevant in the determination of the electron-to-muon
mass ratio in Sec.~\ref{sec:muhfs}.

\subsection{Definitions of bound-state and free $\bm g$-factors}
\label{ssec:Zeeman}

The Hamiltonian for a magnetic moment $\bm \mu$ in a magnetic flux
density $\bm B$ is ${\cal H}=-\bm\mu\cdot\bm B$. For a charged lepton
$\ell$, the magnetic moment is
\begin{eqnarray}
\bm \mu_{\ell}=g_{\ell} \,\frac{e}{2m_{\ell}}\, \bm s \, ,
\end{eqnarray}
where $g_{\ell}$, $m_{\ell}$, and $\bm s$ are its $g$-factor, mass, and
spin, respectively.  The magnitude of the moment is
\begin{equation}
        \mu_\ell =  g_\ell \,\frac{e}{2m_{\ell}}\,\frac{\hbar}{2}  \,.
\end{equation}

By convention, the magnetic moment of a neutron or nucleus with spin
$\bm I$ is denoted by
\begin{equation}
\bm\mu = g \, \frac{e}{2m_{\rm p}} \, \bm I \,,
\label{eq:ngdef}
\end{equation}
where $g$ is the $g$-factor of the neutron or nucleus. The charge and
mass of the proton $m_{\rm p}$ appear in the definition, regardless
whether or not the particle in question is a proton.  The magnitude of
the moment is
\begin{equation}
    \mu = g \,\mu_{\rm N} \,i \,, \label{eq:ngdeff}
\end{equation}
where $\mu_{\rm N} = e\hbar/2m_{\rm p}$ is the nuclear magneton and
integer or half-integer $i$ gives the maximum positive spin projection of
$\bm I$ as $i\,\hbar$.

When electrons bind with nuclei to form ground-state atoms or molecules,
the effective $g$-factors change. For atomic H and D in their electronic
ground state, the non-relativistic Hamiltonian is
\begin{equation}
{\cal H} =
\frac{\Delta\omega_{X}}{\hbar}\,\bm s \cdot \bm I
- g_{\rm e}(X)\,\frac{e}{2m_{\rm e}}\, \bm s \cdot \bm B
- g_{N}(X)\,\frac{e}{2m_{\rm p}}\, \bm I\cdot \bm B \,,
\end{equation}
where $(X, N)$=(H,p) or (D,d) and the coefficients $g_{\rm e}(X)$ and
$g_{N}(X)$ are bound-state $g$-factors.  For muonium, an atom where an
electron is bound to an antimuon, the corresponding Hamiltonian is
\begin{eqnarray}
{\cal H}_{\rm Mu} &=&
\frac{\Delta\omega_{\rm Mu}}{\hbar}\,\bm s_{\rm e} \cdot \bm s_{\rmssmu}
- g_{\rm e}({\rm Mu})\,\frac{e}{2m_{\rm e}}\, \bm s_{\rm e} \cdot \bm B
	\\
	&&
- g_{\rmssmu}({\rm Mu})\,\frac{e}{2m_{\rmssmu}}\, \bm s_{\rmssmu} \cdot \bm B
\,.
\label{eq:ZeeMu}
\end{eqnarray}

\subsection{Theoretical ratios of $\bm g$-factors in H, D, $^3$He, and muonium}
\label{ssec:thbfrats}

Theoretical binding corrections to $g$-factors in the relevant atoms and
muonium have already been discussed  in previous CODATA reports.
Relevant references can be found there as well.  Here, we only give the
final results.  For atomic hydrogen, we have
\begin{eqnarray}
\frac{g_{\rm e}({\rm H})}{g_{\rm e}}
&=& 1 -\frac{1}{3}(Z\alpha)^2 - \frac{1}{12}(Z\alpha)^4
+ \frac{1}{4}(Z\alpha)^2 \frac{\alpha}{\rmpi} 
\nonumber \\
&& +\, \frac{1}{2}(Z\alpha)^2\frac{m_{\rm e}}{m_{\rm p}} 
 + \frac{1}{2}\left(A_1^{(4)}-\frac{1}{4}\right)(Z\alpha)^2
\left(\frac{\alpha}{\rmpi}\right)^2 
\nonumber \\
 && -\,\frac{5}{12}(Z\alpha)^2
\frac{\alpha}{\rmpi} 
\frac{m_{\rm e}}{m_{\rm p}} + \cdots 
\elabel{eq:ehgrat} 
\end{eqnarray} 
and 
\begin{eqnarray} \frac{g_{\rm
p}({\rm H})}{g_{\rm p}}  &=& 1 - \frac{1}{3}\alpha(Z\alpha) -
\frac{97}{108}\alpha(Z\alpha)^3 
\nonumber\\[5 pt]&&
+\frac{1}{6} \alpha(Z\alpha) \,
\frac{m_{\rm e} }{m_{\rm p}} \, \frac{3+4a_{\rm p}}{1+a_{\rm p}} +
\cdots , \elabel{eq:phgrat} 
\end{eqnarray} 
where $A_1^{(4)}$ is given in
Eq.~(\ref{eq:a14}) and the proton magnetic-moment anomaly is $a_{\rm
p} = \mu_{\rm p}/\left(e\hbar/2m_{\rm p}\right)-1 \approx 1.793$.
For deuterium, we have
\begin{eqnarray}
\frac{g_{\rm e}({\rm D})}{g_{\rm e}}
&=& 1 -\frac{1}{3}(Z\alpha)^2 - \frac{1}{12}(Z\alpha)^4 +
\frac{1}{4}(Z\alpha)^2 \frac{\alpha}{\rmpi} 
\nonumber\\[5 pt]&&+
\frac{1}{2}(Z\alpha)^2\frac{m_{\rm e}}{m_{\rm d}}
 + \frac{1}{2}\left(A_1^{(4)}-\frac{1}{4}\right)(Z\alpha)^2
\left(\frac{\alpha}{\rmpi}\right)^2 
\nonumber\\[5 pt]&& -\frac{5}{12}(Z\alpha)^2
\frac{\alpha}{\rmpi} 
\frac{m_{\rm e}}{m_{\rm d}} + \cdots 
\elabel{eq:edgrat} 
\end{eqnarray}
and
\begin{eqnarray}
  \frac{g_{\rm d}({\rm D})}{g_{\rm d}} 
  &=& 1 - \frac{1}{3}\,\alpha(Z\alpha)
- \frac{97}{108}\,\alpha(Z\alpha)^3 
\nonumber\\[5 pt]&&
+\frac{1}{6}\, \alpha(Z\alpha) \,
\frac{m_{\rm e}}{m_{\rm d}} \,\frac{3+4a_{\rm d}}{1+a_{\rm d}} + \cdots
, \elabel{eq:ddgrat} 
\end{eqnarray} 
where the deuteron magnetic-moment
anomaly is $a_{\rm d} = \mu_{\rm d}/ \left(e\hbar/ m_{\rm d}\right) -
1 \approx -0.143$.  

For helium-3, there are new results for the screening correction
$\sigma(X)$, defined by
\begin{eqnarray}
\mu_{\rm h}(X) = \mu_{\rm h}\left[1 - \sigma(X)\right]
\end{eqnarray}
from \citet{2021044}, both for the
neutral atom and the singly charged ion.  For the atom, we have 
\begin{eqnarray} 
\frac{\mu_{\rm h}(^3{\rm He})}{\mu_{\rm h}} &=& 1 - 
 59.967\,029(23)\times 10^{-6} 
\label{eq:3helium}
\end{eqnarray}
and for the singly charged ion
\begin{eqnarray} 
\frac{\mu_{\rm h}(^3{\rm He}^+)}{\mu_{\rm h}} &=& 1 -
 35.507\,434(9)\times 10^{-6} \, .
\label{eq:3heliump}
\end{eqnarray}
There is also an independent value by \citet{2022029} given as
\begin{eqnarray} 
\frac{\mu_{\rm h}(^3{\rm He}^+)}{\mu_{\rm h}} &=& 1 -
 35.507\,38(3)\times 10^{-6} \, .
\label{eq:3heliumpa}
\end{eqnarray}
These screening factors may be used to deduce the bare nuclear moment
from measurements made on the atoms.

Finally, for muonium we have
\begin{eqnarray}
&&\frac{g_{\rm e}({\rm Mu})}{g_{\rm e}}
= 1 -\frac{1}{3}(Z\alpha)^2 - \frac{1}{12}(Z\alpha)^4 
+ \frac{1}{4}(Z\alpha)^2
\frac{\alpha}{\rmpi}
\nonumber \\ &&\quad
+ \frac{1}{2}(Z\alpha)^2\frac{m_{\rm e}}{m_{\rmssmu}}
+ \frac{1}{2}\left(A_1^{(4)}-\frac{1}{4}\right)(Z\alpha)^2
\left(\frac{\alpha}{\rmpi}\right)^2 
\elabel{eq:emugrat}
 \\ &&\quad 
-\frac{5}{12}(Z\alpha)^2
	\frac{\alpha}{\rmpi}
\frac{m_{\rm e}}{m_{\rmssmu}}
- \frac{1}{2}(1+Z)(Z\alpha)^2
\left(\frac{m_{\rm e}}{m_{\rmssmu}}\right)^{\!2}
+ \cdots \qquad\nonumber
\end{eqnarray} 
and
\begin{eqnarray}
&&\frac{g_{{\rmssmu}}({\rm Mu})}{g_{{\rmssmu}}} =
1 - \frac{1}{3}\alpha(Z\alpha) - \frac{97}{108}\,\alpha(Z\alpha)^3
+ \frac{1}{2} \alpha(Z\alpha) \, \frac{m_{\rm e}}{m_{\rmssmu}} 
\nonumber \\ &&
+ \frac{1}{12} \, \alpha(Z\alpha) 
        \frac{\alpha}{\rmpi}
\,\frac{m_{\rm e}}{m_{\rmssmu}}
- \frac{1}{2}(1+Z)\alpha(Z\alpha) 
\left(\frac{m_{\rm e}}{m_{\rmssmu}}\right)^2
 + \cdots \,.  \qquad
\elabel{eq:mumugrat}
\end{eqnarray}

Numerical values for the corrections in Eqs.~(\ref{eq:ehgrat}) to
(\ref{eq:mumugrat}) based on 2022 recommended values for $\alpha$, mass
ratios, etc. are listed in Table \ref{tab:gfactrat}; uncertainties are
negligible.  See \citet{2009018} for a negligible additional term.

\begin{table}
\caption[Values for various bound-particle to free-particle $g$-factor
ratios]{Theoretical values for various bound-particle to free-particle
$g$-factor ratios based on the 2022
recommended values of the constants.}
\tlabel{tab:gfactrat}
\begin{center}
\begin{tabular}{l@{\qquad}l}
\hline
\hline
\quad Ratio   & \qquad Value \\
\hline
 $g_{\rm e}({\rm H})/g_{\rm e}$      & $  1  -17.7054\times 10^{-6} $ \T \\
 $g_{\rm p}({\rm H})/g_{\rm p}$          & $  1  -17.7354\times 10^{-6} $ \\
 $g_{\rm e}({\rm D})/g_{\rm e}$      & $  1  -17.7126\times 10^{-6} $ \\
 $g_{\rm d}({\rm D})/g_{\rm d}$          & $  1  -17.7461\times 10^{-6} $ \\
 $g_{\rm e}({\rm Mu})/g_{\rm e}$     & $ 1  -17.5926\times 10^{-6} $ \\
 $g_{\rmssmu}({\rm Mu})/g_{\rmssmu}$ & $ 1  -17.6254\times 10^{-6} $ \B \\
\hline
\hline
\end{tabular}
\end{center}
\end{table}

\subsection{Theoretical ratios of nuclear $\bm g$-factors in HD and HT}
\label{ssec:thbfratsmol}

Bound-state corrections to the magnitudes of nuclear magnetic moments in
the diatomic molecules HD and HT are expressed as
\begin{equation}
\mu_N(X) = \left[1-\sigma_N(X)\right] \mu_N\,, 
\end{equation}
for nucleus $N$ in molecule $X$. Here, $\mu_N$ is the magnitude of the
magnetic moment of the free nucleus and $\sigma_N(X)$ is the nuclear
magnetic shielding correction.   In fact, $|\sigma_N(X)|\ll 1$.

NMR experiments for these molecules measure the ratio
\begin{equation}
        \frac{\mu_N(X)}{\mu_{N'}(X)}= \left[ 1 + \sigma_{N'N}
        + {\cal O}(\sigma^2) \right] \frac{\mu_N}{\mu_{N'}}
\end{equation}
for nuclei $N$ and $N'$ in molecule $X={\rm HD}$ or HT and $\sigma_{
N'N} = \sigma_{N'}(X) - \sigma_{N}(X)$ is the shielding difference of
molecule $X$.  In the adjustment, corrections of  ${\cal O}(\sigma^2)$,
quadratic in $\sigma_N(X)$, are much smaller than the uncertainties in
the experiments and are omitted.

The theoretical values for shielding differences in HD and HT are
$\sigma_{\rm dp} =  19.877(1)\times 10^{-9}$ and $\sigma_{\rm tp} =  23.945(2)\times 10^{-9}$,
respectively, as reported by \citet{2022013}. The values are more
accurate than those used in the 2018 CODATA adjustment and are also
listed as items D51 and D52 in Table~\ref{tab:pdata}.  The two shielding
differences are taken as adjusted constants with observational equations
$\sigma_{\rm dp}\doteq \sigma_{\rm dp}$ and $\sigma_{\rm tp}\doteq
\sigma_{\rm tp}$, respectively.

\subsection{Ratio measurements in atoms and molecules}
\label{ssec:exps}

Ten atomic and molecular magnetic-moment ratios obtained with H and D
masers and NMR experiments are used as input data in the 2022
adjustment, and determine the magnetic moments of the neutron, deuteron,
triton, and helion.  For ease of reference, these experimental frequency
ratios are summarized in Table~\ref{tab:pdata} and given labels D41
through D50.  There are no correlation coefficients among these data
greater than 0.0001.  Observational equations are summarized in
Table~\ref{tab:pobseqsb}.

We note that the primed magnetic moment $\mu_{\rm p}^\prime$ appearing
in three input data in Table~\ref{tab:pdata} indicates that the proton
is bound in a H$_2$O molecule in a spherical sample of liquid water at
25~$^\circ{\rm C}$ surrounded by vacuum.  The shielding factor for the
proton in water is not known theoretically and, thus, these measurements
cannot be used to determine the free-proton magnetic moment. The
relationships among these three input data, however, help determine
other magnetic moments as well as the shielding factor of the proton in
water.

The adjusted constants for the determination of the relevant magnetic
moments are $\mu_{\rm d}/\mu_{\rm e}$, $\mu_{\rm e}/\mu_{\rm p}$,
$\mu_{\rm e}/\mu^\prime_{\rm p}$, $\mu_{\rm h}(^3{\rm
He})/\mu^\prime_{\rm p}$, $\mu_{\rm n}/\mu^\prime_{\rm p}$, $\mu_{\rm
t}/\mu_{\rm p}$, $\sigma_{\rm dp}$, and $\sigma_{\rm tp}$.

The ratio $\mu_{\rm p}({\rm HD}) /\mu_{\rm d}({\rm HD})$ obtained by
\citet{2012162}, item D49 in Table~\ref{tab:pdata}, is a relatively old
result that was not included in the 2014 adjustment, but is included in
the current adjustment.  We rely on three determinations of ${\mu}_{\rm
p}({\rm HD})/{\mu}_{\rm d}({\rm HD})$ in the 2022 CODATA adjustment.
The values are from \citet{2012223}, researchers at the University of
Warsaw, Poland; and from \citet{2003262} and \citet{2012162},
researchers in Saint Petersburg, Russia, who have a long history of NMR
measurements in atoms and molecules.  (The remaining experimental input
data have been reviewed in previous CODATA reports and are not discussed
further.)

\citet{2012162} describe a complex set of experiments to determine the
free-helion to free-proton magnetic-moment ratio.  We had previously
overlooked their frequency ratio measurements on HD, which satisfy 
\begin{equation}
        \frac{ {\omega}_{\rm p}({\rm HD})}{{\omega}_{\rm d}({\rm HD})}= 2\,
                \frac{{\mu}_{\rm
                p}({\rm HD})}{{\mu}_{\rm d}({\rm HD})} \,,
\end{equation}
where the factor two appears because the spins of the proton and
deuteron are 1/2 and 1, respectively.  The statistical relative
uncertainty of the frequency ratio is given as 7.7 parts in $10^{10}$.
The lineshape fits by \citet{2012162}, however, visibly disagree with
the experimental data and, thus, systematic effects are present. We
account for these effects by increasing the uncertainty by a factor of
4.0 consistent with determining the NMR frequency of d in HD to
approximately one-tenth of the full-width-half-maximum of the Lorentzian
line.

\section{Magnetic moments}

\subsection{Proton magnetic moment in nuclear magnetons}
\label{sec:ppm}

The 2017 measurement of the proton magnetic moment in nuclear magnetons,
${\mu}_{\rm p}/{\mu}_{\rm N}$ was obtained using a single proton in a
double Penning trap at the University of Mainz, Germany
\numcite{2017113}.  The ratio was determined by measuring its spin-flip
transition frequency $\omega_{\rm s} = 2\mu_{\rm p}B/{\hbar}$ and its
cyclotron frequency $\omega_{\rm c} = eB/m_{\rm p}$ in a magnetic flux
density $\bm B$.  As $\bm B$ is the same in both measurements, 
\begin{equation}
 \frac{ \omega_{\rm s}}{\omega_{\rm c}} = 
 \frac{{\mu}_{\rm p}}{{\mu}_{\rm N}}
\end{equation}
independent of $\bm B$ and where $\mu_{\rm N}=e\hbar/2m_{\rm p}$
is the nuclear magneton.

The Mainz value 
\begin{equation}
\frac{ \omega_{\rm s}}{\omega_{\rm c}} =   2.792\,847\,344\,62(82)
\quad [ 2.9\times 10^{-10}]
\end{equation}
is consistent with but supersedes the 2014 result by the same research
group \numcite{2014060}.  Improvements in the apparatus led to a
relative uncertainty that is more than an order of magnitude smaller
than in 2014.  The linewidth of the resonant Lorentzian signal was
narrowed by reducing magnetic-field inhomogeneity, and an improved
detector for the cyclotron frequency doubled the data acquisition rate.
The relative uncertainty of the new result comprises 2.7 and 1.2 parts
in $10^{10}$ from statistical and systematic effects, respectively.  The
two largest components contributing to the systematic uncertainty are
due to limits on line-shape fitting and on the characterization of a
relativistic shift and have been added linearly to account for
correlations.  The total correction from systematic effects is $-1.3$
parts in $10^{10}$.

The observational equation for $\omega_{\rm s}/\omega_{\rm c}$ is
\begin{eqnarray}
\frac{ \omega_{\rm s}}{\omega_{\rm c}} &\doteq& -
\left[1+a_{\rm e}({\rm th})+\delta_{\rm th}({\rm e})\right]
\frac{A_{\rm r}({\rm p})}{A_{\rm r}({\rm e})}\,\frac{\mu_{\rm p}}{\mu_{\rm
e}} 
\end{eqnarray}
using the definition of $\mu_{\rm e}$ in Eq.~(\ref{eq:gedef}).  The
quantities $\delta_{\rm th}({\rm e})$, $A_{\rm r}({\rm e})$, $A_{\rm
r}({\rm p})$, and $\mu_{\rm e}/\mu_{\rm p}$ are adjusted constants.  The
theoretical expression for the electron anomaly $a_{\rm e}({\rm th})$ is
mainly a function of adjusted constant $\alpha$.

The input datum has identifier UMZ-17 and is item D40 in Table
\ref{tab:pdata}. Its observational equation can be found in Table
\ref{tab:pobseqsb}.

\subsection{Direct measurement of the $^3$He$^+$ magnetic moment}
\label{ssec:dmhe3p}

\citet{2022029} have measured the magnetic moment of the $^3$He$^+$ ion
in a Penning trap.  The combined hyperfine and Zeeman effect leads to a
splitting of the 1S electronic ground state into four magnetic
sublevels, as described by the Breit–Rabi formula up to first-order
perturbation theory in the magnetic field strength $B$.  At the level of
experimental precision, second-order corrections in $B$ have to be taken
into account.  These include the quadratic Zeeman shift, which is
identical for all four levels involved and has no influence on the
transition frequencies.  There is a shielding correction which means
that the measurement determines the shielded moment, and the relation to
the unshielded moment can be calculated by theory.  The magnetic field
strength is determined by measurement of the free cyclotron frequency of
the ion.  This eliminates the need for an absolute field calibration.
The result is given for the bound $g$-factor as
\begin{eqnarray}
g^\prime_I(^3{\rm He}^+) = -4.255\,099\,6069(30)_{\rm stat}(17)_{\rm sys} \, .
\end{eqnarray}
Because the field calibration is based on the ion cyclotron frequency,
the quoted value for the $g$-factor is proportional to the proton-helion
mass ratio times the measured frequency ratio.  However, the uncertainty
of the mass ratio is approximately 20 times smaller than the quoted
uncertainty, and has changed by less since the 2018 adjustment, hence no
correction is needed at this time.  For the 2022 adjustment, we use the
ratio
\begin{eqnarray}
\frac{\mu_{\rm h}(^3{\rm He}^+)}{\mu_{\rm N}} = \frac{g^\prime_I(^3{\rm
He}^+)}{2}
\end{eqnarray}
as the input datum, with the observational equation
\begin{eqnarray}
\frac{\mu_{\rm h}(^3{\rm He}^+)}{\mu_{\rm N}} &\doteq&
\frac{1-\sigma_{\rm h}(^3{\rm He^+})}{1-\sigma_{\rm h}(^3{\rm He})}\,
\frac{\mu_{\rm h}(^3{\rm He})}{\mu_{\rm p}^\prime}\,
\frac{\mu_{\rm p}^\prime}{\mu_{\rm N}}
\nonumber\\[10 pt] &=& -
\frac{1-\sigma_{\rm h}(^3{\rm He^+})}{1-\sigma_{\rm h}(^3{\rm He})}\,
\frac{\mu_{\rm h}(^3{\rm He})}{\mu_{\rm p}^\prime}
\nonumber\\[5 pt] &&\times
\left(1+a_{\rm e}(\rm{th}) + \delta_{\rm th}({\rm e})\right)
\,\frac{A_{\rm r}({\rm p})}{A_{\rm r}(\rm e)}\,
\frac{\mu_{\rm p}^\prime}{\mu_{\rm e}} \, ,
\qquad
\end{eqnarray}
where the screening corrections $\sigma_{\rm h}(^3{\rm He^+})$ and
$\sigma_{\rm h}(^3{\rm He})$ are given in
Eqs.~(\ref{eq:3helium})-(\ref{eq:3heliumpa}).  In the 2022 adjustment,
we use the weighted average of the latter two values, given by
\begin{eqnarray}
\sigma_{^3{\rm He}^+} =  35.507\,430(9)\times 10^{-6} \, .
\label{eq:wav}
\end{eqnarray}

The magnetic moment of the helion itself follows from the shielding
correction according to
\begin{eqnarray}
\frac{\mu_{\rm h}}{\mu_{\rm N}} =
\frac{g^\prime_I(^3{\rm He}^+)}{2}\,\frac{1}{1-\sigma_{^3{\rm He}^+}} \, .
\end{eqnarray}

\section{Electroweak quantities}
\label{sec:xeq}

There are a few cases in the 2022 adjustment, as in previous
adjustments, where an inexact constant is used in the analysis of input
data but not treated as an adjusted quantity, because the adjustment has
a negligible effect on its value.  Three such constants, used in the
calculation of the theoretical expression for the electron and muon
magnetic-moment anomaly $a_{\rm e}$ and $a_{\rmssmu}$, are the mass of
the tau lepton $m_{\rmsstau}$, the Fermi coupling constant $G_{\rm F}$,
and sine squared of the weak mixing angle $\sin^{2}{\theta}_{\rm W}$.
These are electroweak quantities with values obtained from the most
recent report of the PDG \numcite{2022048}:
\begin {eqnarray}
&&m_{\rmsstau}c^{2} =  1776.86(12) \ {\rm MeV}
\quad [ 6.8\times 10^{-5}]\, ,
\elabel{eq:mtaumev}
\\ 
&&\frac{G_{\rm F}}{(\hbar c)^{3}} =  1.166\,3787(6)\times 10^{-5} \ {\rm GeV}^{-2}
\quad [  5.1\times 10^{-7}]\, , \qquad
\nonumber\\
\elabel{eq:gf} 
\\ 
&&\sin^2{\theta}_{\rm W} =  0.223\,05(23)
\quad [ 1.0\times 10^{-3}] \, .
\elabel{eq:sin2thw}
\end{eqnarray}

We use $\sin^{2}{\theta}_{\rm W} = 1 - (m_{\rm W}/m_{\rm Z})^{2}$, where
$m_{\rm W}$ and $m_{\rm Z}$ are the masses of the ${\rm W}^{\pm}$ and
${\rm Z}^{0}$ bosons, respectively.  The PDG value $m_{\rm W}/m_{\rm Z}
=  0.881\,45(13)$ leads to the given value of $\sin^{2}{\theta}_{\rm W}$.

\LTcapwidth=\columnwidth    %

\def\vb{\vbox to 8 pt {}}
\newcounter{mycounter}
\setcounter{mycounter}{1}
\newcommand{\mynext}{D\themycounter \stepcounter{mycounter}}
 \LTcapwidth=\textwidth
\begin{longtable*}{l@{~}|l@{~~}l@{~~}l@{~~}l@{~}l@{~~}l}
\caption[Input data for masses and magnetic moments of the electron,
muon, and light nuclei.]{ Input data for the 2022 CODATA adjustment to
determine the fine-structure constant, the relative atomic masses of the
electron, muon, and nuclei with $Z\le2$, and magnetic-moment ratios among
these nuclei as well as those of leptons.  Relative standard
uncertainties in square brackets are relative to the value of the
theoretical quantity to which the additive correction corresponds.  The
label in the first column is used to specify correlation coefficients
among these data and in Table~\ref{tab:pobseqsb} for observational
equations.  Columns five and six give the reference, an abbreviation of
the name of the laboratory in which the experiment has been performed,
and the year of publication.  An extensive list of abbreviations can be
found at the end of this report. Correlations among these data are given
in Table \ref{tab:ccpdata}.}
\label{tab:pdata}\\
\hline
\hline
\vb
& Input datum  &  Value
&  Rel. stand.
& Lab. & Reference(s)& Sec. \\
& & & unc.
$u_{\rm r}$ &  & &\\
\hline
\endfirsthead

\caption{continued} \\
\hline
\vb
& Input datum  &  Value
&  Rel. stand.
& Lab. & Reference(s)& Sec. \\
& & & unc.
$u_{\rm r}$ &  & &\\
\hline
\endhead
\hline
\endfoot
\endlastfoot
& \multicolumn{6}{c}{Input data relevant for the fine-structure constant
and the electron mass}\\

\mynext & $a_{\rm e}({\rm exp})$ & $  1.159\,652\,180\,59(13)\times 10^{-3}$ & $ 1.1\times 10^{-10}$
& NW-23 & \citet{2023002} & \ref{sec:elmagmom} \\

\mynext & $\delta_{\rm e}$ & $  0.000(16)\times 10^{-12}$ & [$ 1.4\times 10^{-11}$] &
theory & &\ref{sec:elmagmom} \\

\mynext  & $h/m({\rm ^{87}Rb})$ & $  4.591\,359\,258\,90(65)\times 10^{-9}$
& $ 1.4\times 10^{-10}$ & LKB-20 & \citet{2020064} & \ref{sec:atomrecoil}
\\

\mynext  & $h/m({\rm ^{133}Cs})$ & $  3.002\,369\,4721(12)\times 10^{-9}$
& $ 4.0\times 10^{-10}$ & UCB-18 & \citet{2018033} & \ref{sec:atomrecoil}
\\
&                      &   \multicolumn{1}{r}{
m$^2$\,s$^{-2}\,{\rm Hz}^{-1}$}  &                     &        &
&    \\

\mynext  & $A_{\rm r}(^{87}{\rm Rb})$   &    86.909\,180\,5291(65)  &
$ 7.5\times 10^{-11}$  & AMDC-20 & \citet{2021032,2021033} &
\ref{sec:ram}\\

\mynext  & $A_{\rm r}(^{133}{\rm Cs})$    &    132.905\,451\,9585(86) &
$ 6.5\times 10^{-11}$ & AMDC-20 & \citet{2021032,2021033}&
\ref{sec:ram}\\

\mynext   & $\omega_{\rm s}/\omega_{\rm c}$ for $^{12}{\rm C}^{5+}$  & $
 4376.210\,500\,87(12)$ & $ 2.8\times 10^{-11}$ & MPIK-15 & \citet{2015052} &
\ref{ssec:bsgfexps} \\

\mynext  & $\Delta E_{\rm B}(^{12}{\rm C}^{5+})/hc $     & $
 43.563\,233(25)\times10^7$ m$^{-1}$  & $ 5.8\times 10^{-7}$ &
ASD-22 &  & \ref{sec:ram} \\

\mynext   & $\delta_{\rm C}$ & $  0.00(94)\times 10^{-11}$   & [$ 0.5\times 10^{-11}$] &
theory & &\ref{ssec:bsgfobs} \\

\mynext   & $\omega_{\rm s}/\omega_{\rm c}$ for $^{28}{\rm Si}^{13+}$  &
$  3912.866\,064\,84(19)$ & $ 4.8\times 10^{-11}$ & MPIK-15 &
\citet{2013046,pc15ss} & \ref{ssec:bsgfexps}  \\

\mynext  & $A_{\rm r}(^{28}{\rm Si})$       &    27.976\,926\,534\,42(55)  &
$ 2.0\times 10^{-11}$ & AMDC-20 & \citet{2021032,2021033} & \ref{sec:ram}
\\

\mynext   & $\Delta E_{\rm B}(^{28}{\rm Si}^{13+})/hc $     & $
 420.6467(85)\times10^7$ m$^{-1}$  & $ 2.0\times 10^{-5}$ &
ASD-22 &  & \ref{sec:ram} \\

\mynext  &   $\delta_{\rm Si}$ & $  0.00(58)\times 10^{-9}$ & [$ 2.9\times 10^{-10}$]
& theory & & \ref{ssec:bsgfobs}\\
\hline 

& \multicolumn{6}{c}{Input data relevant for masses of light nuclei}\\

\mynext &  $\eta_{\rm d}$  &  $ 2.904\,302\,45(49)\times 10^{-3}$ m &
$ 1.7\times 10^{-7}$  &  NIST-98  & \citet{1999052}  & \ref{sec:ram}\\

\mynext &  $\omega_{\rm c}(^{12}{\rm C}^{6+})/\omega_{\rm c}(\rm p)$ &
$ 0.503\,776\,367\,670(17)$ & $ 3.3\times 10^{-11}$ & MPIK-19 & \citet{2019074}
& \ref{sec:ram}\\

\mynext &  $\omega_{\rm c}(^{12}{\rm
C}^{6+})/\omega_{\rm c}(\rm d)$ & $ 1.007\,052\,737\,9117(85)$ &
$ 8.4\times 10^{-12}$ & MPIK-20 & \citet{2020060}  & \ref{sec:ram}\\

\mynext &  $\omega_{\rm c}({\rm H}_2^{+})/\omega_{\rm c}(\rm d)$ &
$ 0.999\,231\,660\,0030(43)$ & $ 4.3\times 10^{-12}$ & FSU-21 & \citet{2021042}
& \ref{sec:ram}\\

\mynext &  $\omega_{\rm c}(^{12}{\rm C}^{4+})/\omega_{\rm c}({\rm
HD}^+)$ & $ 1.007\,310\,263\,905(20)$ & $ 2.0\times 10^{-11}$ & MPIK-20 &
\citet{2020060}  & \ref{sec:ram}\\

\mynext & $\omega_{\rm c}(\rm HD^+)/\omega_{\rm c}(^{3}{\rm He}^{+})$  &
$  0.998\,048\,085\,122(23)$ & $ 2.3\times 10^{-11}$ & FSU-17 &
\citet{2017099} & \ref{sec:ram} \\ 

\mynext & $\omega_{\rm c}(\rm t)/\omega_{\rm c}(^{3}{\rm He}^{+})$  & $
 0.999\,993\,384\,997(24)$ & $ 2.4\times 10^{-11}$ & FSU-15 & \citet{2015002}
& \ref{sec:ram} \\

\mynext & $\omega_{\rm c}(^{4}{\rm He}^{2+})/\omega_{\rm c}(^{12}{\rm
C}^{6+})$ &  0.999\,349\,502\,360(16) & $ 1.6\times 10^{-11}$  & UWash-06
& \citet{2006036} & \ref{sec:ram}  \\

\mynext  & $ E_{\rm I}(^{3}{\rm He}^{+})/hc $     & $
 4.388\,891\,939(2)\times 10^7$ m$^{-1}$  & $ 4.6\times 10^{-10}$ &
ASD-22 &  & \ref{sec:ram}  \\

\mynext  & $\Delta E_{\rm B}(^{12}{\rm C}^{4+})/hc $     & $
 11.939\,000(25)\times 10^7$ m$^{-1}$  & $ 2.1\times 10^{-6}$ &
ASD-22 & &  \ref{sec:ram} \\

\mynext  & $\Delta E_{\rm B}(^{12}{\rm C}^{6+})/hc $     & $
 83.083\,850(25)\times 10^7$ m$^{-1}$  & $ 3.0\times 10^{-7}$ &
ASD-22 & &  \ref{sec:ram} \\

\mynext  & $ E_{\rm I}({\rm H}_2^{+})/hc $     & $  1.310\,581\,219\,937(6) \times
10^7$ m$^{-1}$  & $ 4.6\times 10^{-12}$ & theory & \citet{2017094} &
\ref{sec:ram}  \\

\mynext  & $ E_{\rm I}({\rm HD}^{+})/hc $     &
$  1.312\,246\,841\,650(6) \times 10^7$ m$^{-1}$  & $ 4.6\times 10^{-12}$ & theory &
\citet{2017094} & \ref{sec:ram}  \\

\mynext  & $f_{\rm SA}^{\rm exp}(0,\!0\! \to\! 0,\!1)$ &  1\,314\,925\,752.978(48)
kHz   & $ 3.7\times 10^{-11}$ & HHU-20& \citet{2020029}; &
\ref{sec:HDplus}\\ 
& &   &  & & \citet{2023021}\\

\mynext & $f_{\rm SA}^{\rm exp}(0,\!0\! \to\! 1,\!1)$ &  58\,605\,052\,164.14(56)
kHz  & $ 9.6\times 10^{-12}$ & HHU-21& \citet{2021049}; & \ref{sec:HDplus} \\ 
& &   &  & & \citet{2023021}\\

\mynext & $f_{\rm SA}^{\rm exp}(0,\!3\! \to\! 9,\!3)$ &  415\,264\,925\,501.3(1.6)
kHz & $ 3.9\times 10^{-12}$ & VUA-20 & \citet{2020061}; & \ref{sec:HDplus}\\ 
& & & & & \citet{2023021}\\

\mynext & $ \delta^{\rm th}_{\rm HD^+}(0,\!0\! \to\! 0,\!1)$ &
 0.000(0.019)  kHz  && & \citet{2023021} & \ref{sec:HDplus}\\

\mynext & $ \delta^{\rm th}_{\rm HD^+}(0,\!0\! \to\! 1,\!1)$ &
 0.00(0.49)  kHz  && &\citet{2023021} & \ref{sec:HDplus}\\

\mynext & $ \delta^{\rm th}_{\rm HD^+}(0,\!3\! \to\! 9,\!3)$ &
 0.0(3.2)  kHz  && &\citet{2023021} & \ref{sec:HDplus}\\ \hline 

   \hline
 & \multicolumn{6}{c}{Input data relevant for the muon anomaly}\\

\mynext   & $R^\prime_{\rmssmu}$ & $ 0.003\,707\,3015(20)$ &
$ 5.4\times 10^{-7}$ & BNL-06 & \citet{2006132} &
\ref{ssec:amb}  \\

\mynext   & $R^\prime_{\rmssmu}$ & $ 0.003\,707\,2999(17)$ & $ 4.6\times 10^{-7}$ & FNAL-21 & \citet{2021012} & \ref{ssub:fnal}  \\

                \hline
 &    \multicolumn{6}{c}{Input data relevant for the muon mass and muon magnetic moment}\\
\mynext   & $E(58~{\rm MHz})/h$ & $ 627\,994.77(14)$ kHz & $ 2.2\times 10^{-7}$ & LAMPF-82 & \citet{pc81mar,1982003} & \ref{sssec:mufreqs}  \\
\mynext   & $E(72~{\rm MHz})/h$ & $ 668\,223.166(57)$ kHz & $ 8.6\times 10^{-8}$ & LAMPF-99 & \citet{1999002} &\ref{sssec:mufreqs}  \\
\mynext & $\Delta E_{\rm Mu}/h$ & $ 4\,463\,302.88(16)$ kHz & $ 3.6\times 10^{-8}$ & LAMPF-82 &  \citet{pc81mar,1982003} & \ref{sssec:mufreqs}  \\
\mynext & $\Delta E_{\rm Mu}/h$ & $ 4\,463\,302.765(53)$ kHz & $ 1.2\times 10^{-8}$ & LAMPF-99 & \citet{1999002} &\ref{sssec:mufreqs}  \\
\mynext   & $\delta_{\rm Mu}/h$ & $  0(85)$ Hz & [$ 1.9\times 10^{-8}$] & theory & &\ref{ssec:muhfs} \\
    \hline
 &    \multicolumn{6}{c}{Input data relevant for the magnetic moments of light nuclei}\\
\mynext   & $\mu_{\rm p}/\mu_{\rm N}$ & $  2.792\,847\,344\,62(82)$ & $ 2.9\times 10^{-10}$ & UMZ-17 &  \citet{2017113} &\ref{sec:ppm}  \\
\mynext   & $\mu_{\rm e}({\rm H})/\mu_{\rm p}({\rm H})$ & $ -658.210\,7058(66)$ & $ 1.0\times 10^{-8}$ & MIT-72 & Sec.~III.C.3 of \citet{2000035} & \ref{ssec:exps} \\
\mynext   & $\mu_{\rm d}({\rm D})/\mu_{\rm e}({\rm D})$ & $ -4.664\,345\,392(50)\times 10^{-4}$ & $ 1.1\times 10^{-8}$ & MIT-84 & Sec.~III.C.4 of \citet{2000035}  & \ref{ssec:exps} \\
\mynext   & $\mu_{\rm e}({\rm H})/\mu_{\rm p}^\prime$ & $ -658.215\,9430(72)$ & $ 1.1\times 10^{-8}$ & MIT-77 & Sec.~III.C.6 of \citet{2000035}  &\ref{ssec:exps} \\
\mynext   & $\mu_{\rm h}(^3{\rm He})/\mu_{\rm p}^\prime$ & $ -0.761\,786\,1313(33)$ & $ 4.3\times 10^{-9}$ & NPL-93 & \citet{1993113} & \ref{ssec:exps}  \\
\mynext   & $\mu_{\rm h}(^3{\rm He^+})/\mu_{\rm N}$ & $ -2.127\,549\,8035(17)$ & $ 8.1\times 10^{-10}$ & MPIK-22 & \citet{2022029} & \ref{ssec:dmhe3p}  \\
\mynext   & $\mu_{\rm n}/\mu_{\rm p}^\prime$ & $ -0.684\,996\,94(16)$ & $ 2.4\times 10^{-7}$ & ILL-79 & Sec.~III.C.8 of \citet{2000035} & \ref{ssec:exps} \\
\mynext   & $\mu_{\rm p}({\rm HD})/\mu_{\rm d}({\rm HD})$ & $  3.257\,199\,531(29)$ & $ 8.9\times 10^{-9}$ & StPtrsb-03 & \citet{2003262} &\ref{ssec:exps} \\
\mynext   & $\mu_{\rm p}({\rm HD})/\mu_{\rm d}({\rm HD})$ & $  3.257\,199\,514(21)$ & $ 6.6\times 10^{-9}$ & WarsU-12 & \citet{2012223} &\ref{ssec:exps}   \\
\mynext   & $\mu_{\rm p}({\rm HD})/\mu_{\rm d}({\rm HD})$ & $  3.257\,199\,516(10)$ & $ 3.1\times 10^{-9}$ & StPtrsb-12 & \citet{2012162} &\ref{ssec:exps}  \\
\mynext   & $\mu_{\rm t}({\rm HT})/\mu_{\rm p}({\rm HT})$ & $  1.066\,639\,8933(21)$ & $ 2.0\times 10^{-9}$ & StPtrsb-11 & \citet{2011216} & \ref{ssec:exps} \\
\mynext   & $\sigma_{\rm dp}$ & $  19.877(1)\times 10^{-9}$ & &  &\citet{2022013} & \ref{ssec:thbfratsmol} \\
\mynext   & $\sigma_{\rm tp}$ & $  23.945(2)\times 10^{-9}$ & & &\citet{2022013} & \ref{ssec:thbfratsmol} \\
 \hline
 \hline
\end{longtable*}

\begin{table}[h]
        \caption[Correlation coefficients for Table~\ref{tab:pdata}]{Correlation coefficients $r(x_i,x_j)>0.0001$ among the input data 
                        in Table~\ref{tab:pdata}. }
        \label{tab:ccpdata}
        \begin{tabular}{l@{~~~}l}
                \hline\hline
$r$(D5,D6)$=$ 0.1032 & $r$(D5,D11)$=$ 0.0678 \\
$r$(D6,D11)$=$ 0.0630 & $r$(D7,D10)$=$ 0.3473 \\
$r$(D8,D23)$=$ 0.9968 & $r$(D8,D24)$=$ 1.0000 \\
$r$(D9,D13)$=$ 0.7957 & $r$(D23,D24)$=$ 0.9968 \\
$r$(D27,D28)$=$ 0.0004 & $r$(D27,D29)$=$ 0.0064 \\
$r$(D28,D29)$=$ 0.0074 & $r$(D30,D31)$=$ 0.9957 \\
$r$(D30,D32)$=$ 0.9573 & $r$(D31,D32)$=$ 0.9800 \\
$r$(D35,D37)$=$ 0.2267 & $r$(D36,D38)$=$ 0.1946 \\
        \hline\hline
\end{tabular}
\end{table}

\section{Lattice spacings of silicon crystals}
\label{sec:msc}

\begin{table*}
\def\vb{\vbox to 8 pt {}}
        \caption[Input data for Si lattice constant and X-ray units]{Input data for the determination of the
2022 recommended values of the lattice spacings of an ideal natural
Si crystal and x-ray units.  The label in the first column is used
in Table \ref{tab:ccXray} to list correlation coefficients among
the data and in Table~\ref{tab:pobseqsb} for observational equations.
The uncertainties are not those as originally published, but corrected
according the considerations in Sec.~III.I of \citet{2000035}.  For
additional information about the uncertainties of data published
after the closing date of the 1998 CODATA adjustment, see also the
corresponding text in this and other CODATA publications.  Columns
four and five give the reference and an abbreviation of the name
of the laboratory in which the experiment has been performed, and
year of publication.  See Sec.~\ref{sec:nom} for an extensive list
of abbreviations.
}
        \label{tab:Xray}
        \begin{tabular}{l@{~}|l@{\quad}l@{\quad}l@{\qquad}l@{~}l}
                \hline
                \hline
                \vb
                & Input datum  &  \multicolumn{1}{c}{Value}
                & \multicolumn{1}{c}{Relat. stand.}
                & Laboratory  &Reference(s)\\
                & & & \multicolumn{1}{c}{uncert. $u_{\rm r}$}  && \\
                \hline
\vb
                E1  & $1-d_{220}({\rm {\scriptstyle{W17}}})/d_{220}({\rm {\scriptstyle ILL}})$  & $  -8(22)\times 10^{-9}$ & & NIST-99 & \citet{2000068}  \\
                E2  & $1-d_{220}({\rm {\scriptstyle{MO^*}}})/d_{220}({\rm {\scriptstyle ILL}})$  & $  86(27)\times 10^{-9}$ & & NIST-99 & \citet{2000068} \\
                E3  & $1-d_{220}({\rm {\scriptstyle{NR3}}})/d_{220}({\rm {\scriptstyle ILL}})$  & $  33(22)\times 10^{-9}$ & & NIST-99 & \citet{2000068}  \\
                E4  & $1-d_{220}({\rm {\scriptstyle{N}}})/d_{220}({\rm {\scriptstyle W17}})$  & $   7(22)\times 10^{-9}$ & & NIST-97 & \citet{1997078}\\
                E5  & $d_{220}({\rm {\scriptstyle{W4.2a}}})/d_{220}({\rm {\scriptstyle W04}})-1$  & $  -1(21)\times 10^{-9}$ & & PTB-98 & \citet{1998127}  \\
                E6 & $d_{220}({\rm {\scriptstyle{W17}}})/d_{220}({\rm {\scriptstyle W04}})-1$  & $  22(22)\times 10^{-9}$ & & PTB-98 & \citet{1998127}  \\
                E7 & $d_{220}({\rm {\scriptstyle{W17}}})/d_{220}({\rm {\scriptstyle W04}})-1$  & $  11(21)\times 10^{-9}$ & & NIST-06 & \citet{2005010} \\
                E8  & $d_{220}({\rm {\scriptstyle{MO^*}}})/d_{220}({\rm {\scriptstyle W04}})-1$  & $  -103(28)\times 10^{-9}$ & & PTB-98 & \citet{1998127}  \\
                E9 & $d_{220}({\rm {\scriptstyle{NR3}}})/d_{220}({\rm {\scriptstyle W04}})-1$  & $  -23(21)\times 10^{-9}$ & & PTB-98 & \citet{1998127}  \\
                E10 & $d_{220}({\rm {\scriptstyle{NR3}}})/d_{220}({\rm {\scriptstyle W04}})-1$  & $  -11(21)\times 10^{-9}$ & & NIST-06 & \citet{2005010} \\
                E11  & $d_{220}/d_{220}({\rm {\scriptstyle W04}})-1$  & $  10(11)\times 10^{-9}$ & & PTB-03 & \citet{2003137} \\
                E12  & $d_{220}({\rm {\scriptstyle{NR4}}})/d_{220}({\rm {\scriptstyle W04}})-1$  & $   25(21)\times 10^{-9}$ & & NIST-06 & \citet{2005010} \\
                E13 &  $d_{220}({\rm {\scriptstyle{ILL}}})/d_{220}({\rm {\scriptstyle W04}})-1$  & $   -20(22)\times 10^{-9}$ &  & NIST-17 & \citet{2017085}  \\
                E14 & $d_{220}({\rm {\scriptstyle{MO^*}}})$ & $  192\,015.5508(42)$ fm & $ 2.2\times 10^{-8}$ & INRIM-08 & \citet{2008194}\\
                E15 & $d_{220}({\rm {\scriptstyle{W04}}})$ & $  192\,015.5702(29)$ fm & $ 1.5\times 10^{-8}$ & INRIM-09 & \citet{2009113} \\
                E16 & $d_{220}({\rm {\scriptstyle{W4.2a}}})$ & $  192\,015.5691(29)$ fm & $ 1.5\times 10^{-8}$ & INRIM-09 & \citet{2009062}\\
                E17 & $d_{220}({\rm {\scriptstyle{W4.2a}}})$ & $  192\,015.563(12)$ fm & $ 6.2\times 10^{-8}$ & PTB-81 & \citet{1981017}; \\
                E18 & $\lambda({\rm Cu\,K\alpha_1})/d_{220}({\rm {\scriptstyle{W4.2a}}})$ & $  0.802\,327\,11(24)$ & $ 3.0\times 10^{-7}$ & FSUJ/PTB-91 & \citet{1990055}; \\
                    &                                                                    &                        &                          &             & \quad\citet{1991096} \\
                E19 & $\lambda({\rm Cu\,K\alpha_1})/d_{220}({\rm {\scriptstyle{N}}})$ & $  0.802\,328\,04(77)$ & $ 9.6\times 10^{-7}$ & NIST-73 &\citet{1973019}  \\
                E20  & $\lambda({\rm W\,K\alpha_1})/d_{220}({\rm {\scriptstyle{N}}})$ & $  0.108\,852\,175(98)$ & $ 9.0\times 10^{-7}$ & NIST-79 & \citet{1979013}  \\
                E21  & $\lambda({\rm Mo\,K\alpha_1})/d_{220}({\rm {\scriptstyle{N}}})$ & $  0.369\,406\,04(19)$ & $ 5.3\times 10^{-7}$ & NIST-73 &\citet{1973019}  \\
                \hline
                \hline
        \end{tabular}

\end{table*}

\begin{table*}

\caption[Correlation coefficients for data in
Table~\ref{tab:Xray}]{Correlation coefficients $r(x_i,x_j)>0.0001$ among
the input data for the lattice spacing of an ideal natural Si crystal
and x-ray units given in Table~\ref{tab:Xray}. }

        \label{tab:ccXray}
        \begin{tabular}{l@{~~~~~}l@{~~~~~}l@{~~~~~}l@{~~~~~}l}
                \hline\hline
                $r($E1,E2$)= 0.4214$   &  $r($E1,E3$)= 0.5158$   &  $r($E1,E4$)= -0.2877$   &  $r($E1,E7$)= -0.3674$   &  $r($E1,E10$)= 0.0648$\\
                $r($E1,E12$)= 0.0648$   &  $r($E2,E3$)= 0.4213$   &  $r($E2,E4$)= 0.0960$   &  $r($E2,E7$)= 0.0530$   &  $r($E2,E10$)= 0.0530$\\
                $r($E2,E12$)= 0.0530$   &  $r($E3,E4$)= 0.1175$   &  $r($E3,E7$)= 0.0648$   &  $r($E3,E10$)= -0.3674$   &  $r($E3,E12$)= 0.0648$\\
                $r($E4,E7$)= 0.5037$   &  $r($E4,E10$)= 0.0657$   &  $r($E4,E12$)= 0.0657$   &  $r($E5,E6$)= 0.4685$   &  $r($E5,E8$)= 0.3718$\\
                $r($E5,E9$)= 0.5017$   &  $r($E6,E8$)= 0.3472$   &  $r($E6,E9$)= 0.4685$   &  $r($E7,E10$)= 0.5093$   &  $r($E7,E12$)= 0.5093$\\
                $r($E8,E9$)= 0.3718$   &  $r($E10,E12$)= 0.5093$   &  $r($E14,E15$)= 0.0230$   &  $r($E14,E16$)= 0.0230$   &  $r($E15,E16$)= 0.0269$\\
        \hline\hline
\end{tabular}
\end{table*}

In this section, we summarize efforts to determine the lattice spacing
of an ideal (or nearly perfect) natural-silicon single crystal.  We also
give values for several historical x-ray units in terms of the SI unit
meter.  Three stable isotopes of silicon exist in nature. They are
$^{28}{\rm Si}$, $^{29}{\rm Si}$, and $^{30}{\rm Si}$ with
amount-of-substance fractions $x(^A{\rm Si})$ of approximately 0.92,
0.05, and 0.03, respectively.  Highly enriched silicon single crystals
have $x(^{28}{\rm Si})\approx 0.999\,96$.

The quantities of interest are the \{220\} crystal lattice spacing
$d_{220}(X)$ in meters of a number of different crystals $X$ using a
combined x-ray and optical interferometer (XROI) as well as the
fractional differences
\begin{equation}
          \frac{d_{220}(X)- d_{220}(Y)}{d_{220}(Y)}
\end{equation}
for single crystals $X$ and $Y$, determined using a lattice
comparator based on x-ray double-crystal nondispersive diffractometry.

Data on eight natural Si crystals, in the literature denoted by
WASO\,4.2a, WASO\,04, WASO\,17, NRLM3, NRLM4, MO*, ILL, and N, are
relevant for the 2022 CODATA adjustment.  Their lattice spacings
$d_{220}(X)$ are adjusted constants in the least-squares adjustment.
The simplified notations W4.2a, W04, W17, NR3, and NR4 are used in
quantity symbols and tables for the first five crystals.  The lattice
spacing for the ideal natural-silicon single crystal $d_{220}$ is an
adjusted constant.

Lattice-spacing data included in this adjustment are items E1-E17
in Table \ref{tab:Xray} and quoted at a temperature of 22.5\,$^\circ$C and
in vacuum.  All data were already included in the 2018 adjustment.

The copper K${\rmalpha}_1$ x~unit with symbol xu(CuK${\rmalpha}_1$),
the molybdenum K${\rmalpha}_1$ x~unit with symbol xu(MoK${\rmalpha}_1$),
and the {\aa}ngstr\"om star with symbol ${\rm \AA}^*$ are historic
x-ray units that are still of current interest. They are defined
by assigning an exact, conventional value to the wavelength of the
CuK${\rmalpha}_1$, MoK${\rmalpha}_1$, and WK${\rmalpha}_1$ x-ray
lines.  These assigned wavelengths for $\lambda$(CuK${\rmalpha}_1$),
$\lambda$(MoK${\rmalpha}_1$), and $\lambda$(WK${\rmalpha}_1$) are
1537.400~xu(CuK${\rmalpha}_1$), 707.400~xu(MoK${\rmalpha}_1$),
and 0.209\,010\,0\,${\rm \AA}^*$, respectively. The four relevant
experimental input data are the measured ratios of CuK${\rmalpha}_1$,
MoK${\rmalpha}_1$, and  WK${\rmalpha}_1$  wavelengths to the \{220\}
lattice spacings of crystals WASO\,4.2a and N and are items E18-E21
in Table~\ref{tab:Xray}.  In the least-squares calculations, the units
xu(CuK${\rmalpha}_1$), xu(MoK${\rmalpha}_1$), and ${\rm \AA}^*$ are
adjusted constants.

The correlation coefficients among the data on lattice spacings and
x-ray units are given in Table~\ref{tab:ccXray}. Discussions of these
correlations can be found in previous adjustments.  Observational
equations may be found in Table~\ref{tab:pobseqsc}.

\LTcapwidth=\textwidth      %
\begin{table*}
\caption[Observational equations excluding those for H and D
spectroscopy]{Observational equations for input data in
Table~\ref{tab:Xray} as functions of the adjusted
constants. Sec.~\ref{sec:msc}.}
\label{tab:pobseqsc}
\begin{tabular}{l@{\quad}rcl@{\qquad\qquad\qquad}l@{\quad}rcl}
\hline
\hline
Input data & \multicolumn{3}{l}{\quad Observational equation}  &
Input data & \multicolumn{3}{c}{Observational equation}  \\
\hline
E1-E4&$ 1-\tfrac{d_{220}({{\scriptstyle Y}}) }{ d_{220}({{\scriptstyle
X}})}  $&$\doteq$& $1-\tfrac{d_{220}({{\scriptstyle Y}}) }{
d_{220}({{\scriptstyle X}})} $ \vbox to 15 pt {} &

E18,E19&$\tfrac{\lambda({\rm Cu\,K\rmalpha_1})}{d_{220}({{\scriptstyle
X}})} $&$\doteq$&$ \tfrac{\rm 1\,537.400
~xu(Cu\,K\rmalpha_1)}{d_{220}({{\scriptstyle X}})}$  \\[8 pt]

E5-E13&$ \tfrac{d_{220}({{\scriptstyle X}}) }{ d_{220}({{\scriptstyle
Y}})} - 1 $&$\doteq$& $\tfrac{d_{220}({{\scriptstyle X}}) }{
d_{220}({{\scriptstyle Y}})}-1 $  &

E20&$\tfrac{\lambda({\rm W\,K\rmalpha_1})}{d_{220}({\rm {\scriptstyle
N}})} $&$\doteq$&$ \tfrac{\rm 0.209\,010\,0 ~\AA^*}{d_{220}({\rm
{\scriptstyle N}})} $  \\[8 pt]

E14-E17&$ d_{220}({{\scriptstyle X}}) $&$\doteq$&$
d_{220}({{\scriptstyle X}}) $  &

E21&$\tfrac{\lambda({\rm Mo\,K\rmalpha_1})}{d_{220}({\rm {\scriptstyle
N}})} $&$\doteq$&$ \tfrac{\rm 707.831 ~xu(Mo\,K\rmalpha_1)}{d_{220}({\rm
{\scriptstyle N}})} $ \\[10 pt] \hline \hline

\end{tabular}
\end{table*}

\section{Newtonian constant of gravitation}
\label{sec:ncg}

While various efforts continue towards the determination of the
Newtonian constant of gravitation, $G$, there is no new relevant input
datum for the 2022 adjustment. Of note is a novel dynamic method using
the gravitational coupling between resonating beams by \citet{2022022};
however, their combined relative standard uncertainty of
$1.7\times10^{-2}$ is not competitive.  As with the 2018 adjustment,
Table \ref{tab:bg} summarizes the 16 measured values of $G$ as input
data for the 2022 adjustment.  See section \ref{sec:2022crv} for the
treatment of the input data.

\begin{table*}[t]
\def\hsp{\hbox to 14 pt{}}
\def\vb{\vbox to 10 pt{}}
\caption[Input data for the Newtonian constant of gravitation]{
Input data for the Newtonian constant of gravitation $G$ relevant to the 
2022 adjustment. 
The first two columns give the reference and an abbreviation
of the name of the laboratory in which the experiment has been performed,
and year of publication.
The data are uncorrelated except for three cases with correlation coefficients
$r({\rm NIST}\mhyphen 82,{\rm LANL}\mhyphen 97)=\protect 0.351$,
$r({\rm HUST}\mhyphen 05,{\rm HUST}\mhyphen 09)=\protect 0.134$, and
$r({\rm HUST\mhyphen 09},{\rm HUST}_{\rm T}\mhyphen 18)=\protect 0.068$.
}
\label{tab:bg}
\begin{tabular}{l@{\hsp}l@{\quad}l@{\hsp}l@{\qquad}l}
\hline
\hline
\vb
Source & Identification%
	& Method & \multicolumn{1}{l}{$G\,\, (10^{-11}\,{\rm kg}^{-1}$ }       & Rel. stand. \\
       &                     &        & \multicolumn{1}{c}{$~\times\, {\rm m}^3\,{\rm s}^{-2})$ }  & uncert. $u_{\rm r}$ \\
\hline
\vb
  \citet{1982013}  & NIST-82  & Fiber torsion balance,    & $ 6.672\,48(43)$     & $ 6.4\times 10^{-5}$  \\
                         &           & \quad dynamic mode              &                 &                     \\

  \citet{1996199}  & TR\&D-96  & Fiber torsion balance,    & $ 6.672\,9(5)$     & $ 7.5\times 10^{-5}$  \\
                         &           & \quad dynamic mode              &                 &                     \\

  \citet{1997025}      & LANL-97   & Fiber torsion balance,    & $ 6.673\,98(70)$     & $ 1.0\times 10^{-4}$  \\
                             &           & \quad dynamic mode              &                 &                     \\

  \citet{2000088}; & UWash-00    & Fiber torsion balance,            & $ 6.674\,255(92)$   & $ 1.4\times 10^{-5}$  \\
\quad \citet{pc02gm}                    &           & \quad dynamic compensation      &                 &   \\[10 pt]

  \citet{2001089}      & BIPM-01   & Strip torsion balance,    & $ 6.675\,59(27)$     & $ 4.0\times 10^{-5}$  \\
                             &    & \quad compensation mode,         &        &                     \\
                             &    & \quad static deflection          &        &                     \\

  \citet{02kleinevoss,pc02kmph} &  UWup-02   & Suspended body,           & $ 6.674\,22(98)$    & $ 1.5\times 10^{-4}$  \\
                                      &           & \quad displacement              &                 &                     \\

  \citet{2003219}      & MSL-03    & Strip torsion balance,    &  $ 6.673\,87(27)$     &  $ 4.0\times 10^{-5}$ \\
                             &           & \quad compensation mode         &                 &                     \\

  \citet{2005292} & HUST-05  & Fiber torsion balance, & $ 6.672\,22(87)$& $ 1.3\times 10^{-4}$  \\
                        &           & \quad dynamic mode              &                 &                     \\

  \citet{2006238}      & UZur-06   & Stationary body,          & $ 6.674\,25(12)$    & $ 1.9\times 10^{-5}$  \\
                             &           & \quad weight change             &                 &                     \\

  \citet{2009099,2010122}  & HUST-09   & Fiber torsion balance, & $ 6.673\,49(18)$    & $ 2.7\times 10^{-5}$  \\
    &           & \quad dynamic mode             &                 &                     \\

  \citet{2013067e,2014120}   & BIPM-14   & Strip torsion balance,    & $ 6.675\,54(16)$    & $ 2.4\times 10^{-5}$  \\
                            &           & \quad compensation mode,            &                 &                     \\
                            &           &  \quad static deflection             &                 &                     \\

  \citet{2014074,2014122}   & LENS-14   & Double atom interferometer, & $ 6.671\,91(99)$    & $ 1.5\times 10^{-4}$  \\
                           &           & \quad  gravity gradiometer           &                 &                     \\

  \citet{2014121}   & UCI-14   & Cryogenic torsion balance, & $ 6.674\,35(13)$    & $ 1.9\times 10^{-5}$  \\
                             &           & \quad dynamic mode             &                 &                     \\

\citet{2018109}   & HUST$_{\rm T}$-18  & Fiber torsion balance, & $ 6.674\,184(78)$    & $ 1.2\times 10^{-5}$  \\
                        &           & \quad dynamic mode             &                 &                     \\

\citet{2018109}   & HUST$_{\rm A}$-18   & Fiber torsion balance, & $ 6.674\,484(77)$    & $ 1.2\times 10^{-5}$  \\
                        &           & \quad dynamic compensation       &                 &                     \\

  \citet{2010144e}   & JILA-18   & Suspended body, & $ 6.672\,60(25)$    & $ 3.7\times 10^{-5}$  \\
      &           &\quad  displacement             &                 &                     \\
\hline
\hline
\end{tabular}
\end{table*}

\begin{center}
\begin{figure*}[t]
 \includegraphics[width=0.85\textwidth]{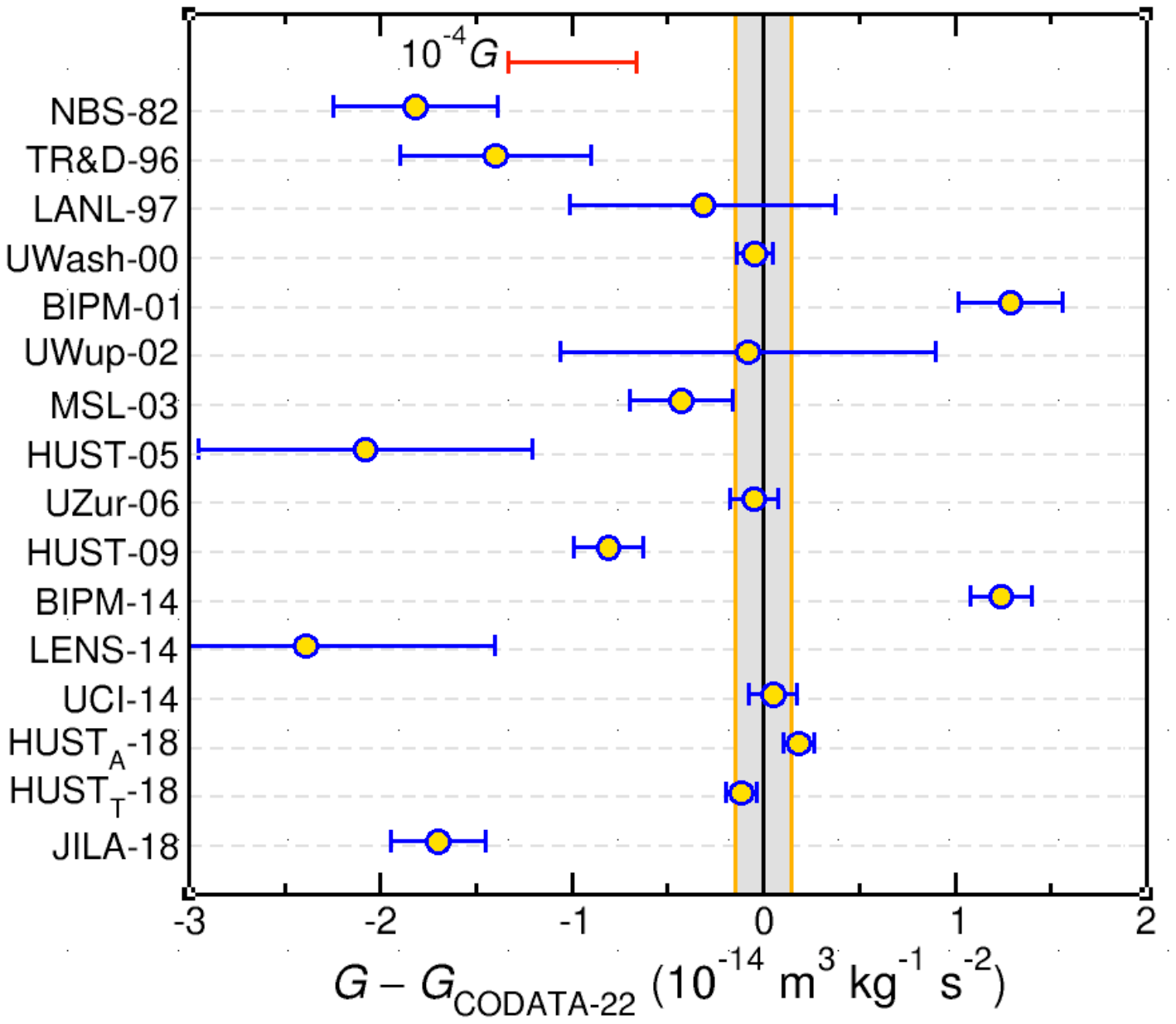}
 \caption[Input data for the Newtonian constant of gravitation]
 {The 16 input data determining
   the Newtonian constant of gravitation $G$ ordered by publication
   year.  The 2022 recommended value for $G$ has been subtracted.
   Error bars correspond to one-standard-deviation uncertainties
   as reported in Table~\ref{tab:bg}.  
The uncertainties after applying the 3.9 multiplicative expansion
   factor to determine the 2022 recommended value are not shown.
   Labels on the left side of the figure denote the laboratories
   and the last two digits of the year in which the data were
   reported.  See Table \ref{tab:bg} for details.  The gray band
   corresponds to the one-standard-deviation uncertainty of the
   recommended value.
    }
\label{fig:BigG} \end{figure*}
\end{center}

\clearpage

\LTcapwidth=\textwidth      %
\begin{table*}
\caption[Observational equations excluding those for H and D
spectroscopy]{Observational equations for input data in
Tables~\ref{tab:pdata} and \ref{tab:Xray} as functions of the adjusted
constants.}
\label{tab:pobseqsb}
\begin{tabular}{l@{\quad}rcl@{\quad}l}
\hline
\hline
\multicolumn{1}{c}{Input data} & &
\multicolumn{2}{l}{Observational equation}& Sec. \quad\\
D1 &$ a_{\rm e}({\rm exp}) $&$\doteq$&$ a_{\rm e}({\rm th}) 
+ \delta_{\rm th}({\rm e})$ & \ref{sec:elmagmom} \\

D2 &$ \delta_{\rm e}$&$\doteq$&$ \delta_{\rm th}({\rm e})$ 
& \ref{sec:elmagmom}\\

D3, D4&$ \tfrac{h}{m(X)} $&$\doteq$& $ 
\tfrac{ A_{\rm r}({\rm e}) } {  A_{\rm r}(X)}  $ 
$ \tfrac{ c\alpha^2 } { 2 R_\infty } $ &\ref{sec:atomrecoil} \\

D5, D6 &$ A_{\rm r}(X) $&$\doteq$&$ A_{\rm r}(X)$ & \ref{sec:ram} \\

D7&$ \tfrac{\omega_{\rm s}(^{12}{\rm C}^{5+})}{\omega_{\rm c}(^{12}{\rm
C}^{5+})} $&$\doteq$&$ -\tfrac{g_{\rm e}(^{12}{\rm C}^{5+})+\delta_{\rm
th}({\rm C})}{10 A_{\rm r}({\rm e})} \left[12-5A_{\rm r}({\rm e}) +
\Delta E_{\rm B}(^{12}{\rm C}^{5+})\alpha^2A_{\rm r}({\rm e})/2R_\infty
hc\right] $ & \ref{ssec:bsgfobs} \\

D8, D12, D23, D24 &$ \Delta E_{\rm B}(X^{n+}) $&$\doteq$&$ \Delta E_{\rm
B}(X^{n+}) $& \ref{sec:ram} \\

D9&$ \delta_{\rm C}$&$\doteq$&$ \delta_{\rm th}({\rm C})$ &
\ref{ssec:bsgfobs} \\

D10&$ \tfrac{\omega_{\rm s}(^{28}{\rm Si}^{13+})}{\omega_{\rm
c}(^{28}{\rm Si}^{13+})} $&$\doteq$&$ -\tfrac{g_{\rm e}(^{28}{\rm
Si}^{13+})+\delta_{\rm th}({\rm Si})}{26 A_{\rm r}({\rm e})} A_{\rm
r}(^{28}{\rm Si}^{13+}) $ & \ref{ssec:bsgfobs} \\

D11 &$ A_{\rm r}(^{28}{\rm Si}) $&$\doteq$&$ A_{\rm r}(^{28}{\rm
Si}^{13+}) + 13 A_{\rm r}({\rm e}) - \Delta E_{\rm B}(^{28}{\rm
Si^{13+}})\,\alpha^2 A_{\rm r}({\rm e})/2R_\infty h c$  & \ref{sec:ram}
\\

D13&$ \delta_{\rm Si}$&$\doteq$&$ \delta_{\rm th}({\rm Si})$ &
\ref{ssec:bsgfobs} \\

D14 &   $\eta_{\rm d}$  &  $ \doteq $  & $
\tfrac{\alpha^2}{R_\infty}\,\tfrac{1}{d_{220}({\rm {\scriptstyle ILL}})}
\,\tfrac{A_{\rm r}({\rm e}) \, (A_{\rm r}({\rm n})+A_{\rm r}({\rm p}))}
{(A_{\rm r}({\rm n})+A_{\rm r}({\rm p}))^2-A_{\rm r}({\rm d})^2}$ &
\ref{sec:ram}\\

D15 &  $\tfrac{\omega_{\rm c}(^{12}{\rm C}^{6+})}{\omega_{\rm c}({\rm
p})}$&$\doteq$& $\tfrac{6 A_{\rm r}({\rm p})}{12-6 A_{\rm r}({\rm e})+
\Delta E_{\rm B}(^{12}{\rm C}^{6+})\,\alpha^2 A_{\rm r}({\rm
e})/2R_\infty h c}$ & \ref{sec:ram}\\

D16 &   $ \tfrac{\omega_{\rm c}(^{12}{\rm C}^{6+})}{\omega_{\rm
c}({\rm d})} $ & $\doteq$ & $\tfrac{6 A_{\rm r}({\rm d})}{ 12-6 A_{\rm
r}({\rm e})+ \Delta E_{\rm B}(^{12}{\rm C}^{6+})\,\alpha^2 A_{\rm
r}({\rm e})/2R_\infty h c}$ & \ref{sec:ram} \\ 

D17 & $\tfrac{\omega_{\rm c}({\rm H}^{+}_2)}{\omega_{\rm c}({\rm
d})}$&$\doteq$& $ \tfrac{ A_{\rm r}({\rm d})}{2 A_{\rm r}({\rm p})
+A_{\rm r}({\rm e}) -  E_{\rm I}({\rm H}_2^+) \,\alpha^2 A_{\rm r}({\rm
e})/2R_\infty h c}$ &  \ref{sec:ram} \\ %

D18 &   $ \tfrac{\omega_{\rm c}(^{12}{\rm C}^{4+})}{\omega_{\rm
c}({\rm HD}^+)} $ & $\doteq$ & $\tfrac{4 [A_{\rm r}({\rm p})+A_{\rm
r}({\rm d})+A_{\rm r}({\rm e})- \Delta E_{\rm I}({\rm HD^+})\,\alpha^2
A_{\rm r}({\rm e})/2R_\infty h c]  }{12-4 A_{\rm r}({\rm e})+ \Delta
E_{\rm B}(^{12}{\rm C}^{4+})\,\alpha^2 A_{\rm r}({\rm e})/2R_\infty h c}
$ & \ref{sec:ram}\\ 

D19&$\tfrac{\omega_{\rm c}({\rm HD^+})}{\omega_{\rm c}(^3{\rm
He}^+)}$&$\doteq$&$ \tfrac{A_{\rm r}({\rm h})+A_{\rm r}({\rm e})- E_{\rm
I}(^3{\rm He^+})\,\alpha^2 A_{\rm r}({\rm e})/2R_\infty h c}{A_{\rm
r}({\rm p})+A_{\rm r}({\rm d})+A_{\rm r}({\rm e})-  E_{\rm I}({\rm
HD^+})\,\alpha^2 A_{\rm r}({\rm e})/2R_\infty h c} $ & \ref{sec:ram}
\\

D20 & $\tfrac{\omega_{\rm c}({\rm t})}{\omega_{\rm c}(^{3}{\rm
He}^{+})}$&$\doteq$& $\tfrac{A_{\rm r}({\rm h})+A_{\rm r}({\rm e}) -
E_{\rm I}(^3{\rm He^+}) \,\alpha^2 A_{\rm r}({\rm e})/2R_\infty h c} {
A_{\rm r}({\rm t})}$ & \ref{sec:ram} \\ %

D21 & $\tfrac{ \omega_{\rm c}(^4{\rm He}^{2+})}{ \omega_{\rm
c}(^{12}{\rm C}^{6+})}$ & $\doteq$ & $ \tfrac{12-6 A_{\rm r}({\rm e})+
\Delta E_{\rm B}(^{12}{\rm C}^{6+})\,\alpha^2 A_{\rm r}({\rm
e})/2R_\infty h c}{3 A_{\rm r}(\rmalpha)}$ & \ref{sec:ram} \\

D22, D25, D26 &$ E_{\rm I}(X^+) $&$\doteq$&$ E_{\rm I}(X^+) $&
\ref{sec:ram} \\

D27-D29 & $f^{\rm exp}_{\rm SA}(vL\to v'L')$ &$\doteq$& $f^{\rm
th}_{\rm SA}(vL\to v'L')+\delta_{\rm HD^+}^{\rm th}(vL\to v'L')$  &
\ref{sec:HDplus} \\[2 pt]

D30-D32 & $\delta_{\rm HD^+}(vL\to v'L')$ &$\doteq$& 
$\delta_{\rm HD^+}^{\rm th}(vL\to v'L')$  &
\ref{sec:HDplus} \\

D33, D34  & $ R^\prime_{\rmssmu} $ & $\doteq$&
$-\tfrac{a_{\rmssmu}}{1+a_{\rm e}({\rm th})+\delta_{\rm th}({\rm e})}
\tfrac{m_{\rm e}}{m_{\rmssmu}} \tfrac{\mu_{\rm e}}{\mu^\prime_{\rm p}} $
& \ref{ssub:fnal}\\

D35, D36 &$ E(\omega_{\rm p}) $&$\doteq$&$ E\!\left(\omega_{\rm
p};R_\infty,\alpha,\tfrac{m_{\rm e}}{m_{\rmssmu}},
a_{\rmssmu},\tfrac{\mu_{\rm e}}{\mu_{\rm p}},\delta_{\rm th}({\rm e}),
\delta_{\rm th}({\rm Mu})\right) $ &\ref{sssec:mufreqs} \\

D37, D38 &$ \Delta E_{\rm Mu} $&$\doteq$&$ \Delta E_{\rm
Mu}\!\!\left({\rm th};R_\infty,\alpha,\tfrac{m_{\rm e}}{m_{\rmssmu}},
a_{\rmssmu}\right) + \delta_{\rm th}({\rm Mu})$&
\ref{ssec:muhfs} \\

D39 &$ \delta_{\rm Mu}$&$\doteq$&$ \delta_{\rm th}({\rm Mu})$ &
\ref{ssec:muhfs}\\

D40 &$ \tfrac{\mu_{\rm p}}{\mu_{\rm N}} $&$\doteq$&$ -\left(1+a_{\rm
e}({\rm th}) + \delta_{\rm th}({\rm e})\right)\,\tfrac{A_{\rm r}({\rm
p})}{A_{\rm r}({\rm e})}\,\tfrac{\mu_{\rm p}}{\mu_{\rm e}}$ &
\ref{sec:ppm} \\

D41 &$ \tfrac{\mu_{\rm e}({\rm H})}{\mu_{\rm p}({\rm H})} $&$\doteq$&$
\tfrac{g_{\rm e}({\rm H})}{g_{\rm e}} \left(\tfrac{g_{\rm p}({\rm
H})}{g_{\rm p}}\right)^{-1} \tfrac{\mu_{\rm e}}{\mu_{\rm p}} $ &
\ref{ssec:exps} \\

D42 &$ \tfrac{\mu_{\rm d}({\rm D})}{\mu_{\rm e}({\rm D})} $&$\doteq$&$
\tfrac{g_{\rm d}({\rm D})}{{g_{\rm d}}} \left(\tfrac{g_{\rm e}({\rm
D})}{g_{\rm e}}\right)^{-1} \tfrac{\mu_{\rm d}}{\mu_{\rm e}} $ &
\ref{ssec:exps} \\

D43 &$ \tfrac{\mu_{\rm e}({\rm H})}{\mu_{\rm p}^\prime} $&$\doteq$&$
\tfrac{g_{\rm e}({\rm H})}{g_{\rm e}} \tfrac{\mu_{\rm e}}{\mu_{\rm
p}^\prime} $ & \ref{ssec:exps} \\

D44 &$ \tfrac{\mu_{\rm h}(^3{\rm He})}{\mu_{\rm p}^\prime} $&$\doteq$&$
\tfrac{\mu_{\rm h}(^3{\rm He})}{\mu_{\rm p}^\prime} $ & \ref{ssec:exps} 
\\

D45 &$ \tfrac{\mu_{\rm h}(^3{\rm He^+})}{\mu_{\rm N}} $&$\doteq$&$
-\tfrac{1-\sigma_{\rm h}(^3{\rm He^+})}{1-\sigma_{\rm h}(^3{\rm He})}\,
\tfrac{\mu_{\rm h}(^3{\rm He})}{\mu_{\rm p}^\prime}
\left(1+a_{\rm e}(\rm{th}) + \delta_{\rm th}({\rm e})\right)
\,\tfrac{A_{\rm r}({\rm p})}{A_{\rm r}(\rm e)}\,
\tfrac{\mu_{\rm p}^\prime}{\mu_{\rm e}}
$ & \ref{ssec:dmhe3p}\\

D46 &$ \tfrac{\mu_{\rm n}}{\mu_{\rm p}^\prime} $&$\doteq$&$
\tfrac{\mu_{\rm n}}{\mu_{\rm p}^\prime} $& \ref{ssec:exps} \\

D47-D49 &$ \tfrac{\mu_{\rm p}({\rm HD})}{\mu_{\rm d}({\rm HD})}
$&$\doteq$&$ \left[1 + \sigma_{\rm dp} \right] \tfrac{\mu_{\rm
p}}{\mu_{\rm e}} \tfrac{\mu_{\rm e}}{\mu_{\rm d}} $ & \ref{ssec:exps}
\\

D50 &$ \tfrac{\mu_{\rm t}({\rm HT})}{\mu_{\rm p}({\rm HT})} $&$\doteq$&$
\tfrac{1}{1 + \sigma_{\rm tp}}  \tfrac{\mu_{\rm t}}{\mu_{\rm p}} $ &
\ref{ssec:exps} \\

D51, D52 &$ \sigma_{NN'}$&$\doteq$&$ \sigma_{NN'}$ &
\ref{ssec:thbfratsmol} \\ %

\end{tabular}
\end{table*}

\section{The 2022 CODATA recommended values}
\label{sec:2022crv}

\subsection{Calculational details}
\label{ssec:cd}

The focus of this section is the treatment of the data discussed in the
previous sections to obtain the 2022 CODATA recommended values.  In this
regard, we recall that when the same expansion factor is applied to the
a priori assigned uncertainties of both members of a correlated pair of
input data to reduce data inconsistencies to an acceptable level, its
square is also applied to their covariance so that their correlation
coefficient is unchanged.  For the same reason, when an expansion factor
is only applied to the uncertainty of one member of a correlated pair,
the expansion factor is applied to their covariance.

We begin with the 16 measurements of the Newtonian constant of
gravitation $G$ in Table~\ref{tab:bg} and the three correlation
coefficients in the table caption.  As indicated in Sec.~\ref{sec:ncg},
there have been no changes in or additions to these data since the 31
December 2018 closing date of the 2018 CODATA adjustment.  Because $G$
is independent of all other constants, it can be determined in a
separate least-squares adjustment, which is simply a calculation of
their weighted mean.  Since the data are unchanged, the same 3.9
expansion factor is used in this adjustment and the 2022 recommended
value of $G$ is the same as the 2018 value.

The factor 3.9 reduces all normalized residuals to 2 or less, a
long-established requirement of CODATA adjustments.  For this
adjustment, $\chi^2=12.9$, $p(12.9|15)=0.61$, and $R_{\rm B}=0.93$.
(Here $p$ is the probability of $\chi^2=12.9$ for $\nu = N-M=16-1=15$
degrees of freedom occurring by chance and $R_{\rm B}=\sqrt{\chi^2/\nu}$
is the Birge ratio.)  The two values of $G$ that had the largest
residuals prior to expansion were BIPM-14 and JILA-18 in
Table~\ref{tab:bg}, which were $7.75$ and $-6.80$, respectively.  In
some past adjustments, input data with a self-sensitivity coefficient
$S_{\rm c}$ less than $0.01$ were eliminated because they contributed
less than 1\,\% to the determination of their own adjusted value.
Although six of the sixteen values of $G$ have such small values of
$S_{\rm c}$, they are retained because of the significant disagreements
among the data and the desirability of having the recommended value
reflect all of the data.

We next consider the silicon lattice spacing data in
Table~\ref{tab:Xray} with correlation coefficients in
Table~\ref{tab:ccXray}  and observational equations in
Table~\ref{tab:pobseqsc}.  Because no new data of this type have become
available in the past four years, like the $G$ data, they are the same as
used in the 2018 adjustment.  Also like the $G$ data, the silicon
lattice spacing data were treated separately in the 2018 adjustment
since they too were independent of all other input data.  However,
because there were no inconsistencies among them, no expansion factors
were required.

The situation for silicon lattice spacings is somewhat different in the
2022 adjustment.  As discussed in Sec.~\ref{sec:nmnc}, rather than using
the 2020 AME value for $A_{\rm r}$(n), the TGFC decided to determine
$A_{\rm r}$(n) from the original experimental data using
Eq.~(\ref{eq:etad}) as the observational equation with the
experimentally determined quantity $\eta_{\rm d}$ as the input datum.
That equation not only contains the adjusted constant $A_{\rm r}$(n) but
also the silicon lattice spacing adjusted constant $d_{220}$(ILL), which
couples $\eta_{\rm d}$ to the lattice spacing data in
Tables~\ref{tab:Xray} and \ref{tab:ccXray}.  Thus, in the 2022
adjustment, these data are not treated separately but are included as
part of the calculation that uses all the input data except the $G$
data.  However, because the adjusted constant $A_{\rm r}$(n) is not used
in any other observational equation, the input datum $\eta_{\rm d}$ does
not contribute to the determination of $d_{220}$(ILL) and hence the 2022
recommended values of the six silicon lattice spacing and x-ray related
data given in Table~\ref{tab:xrayvalues} are identical to those in the
corresponding Table XXXIII of the 2018 CODATA report~\cite{2021035}.

The following is a concise summary of  the data that determine the 2022
CODATA recommended values other than $G$.  For input data with
correlation coefficients, their covariances are included in all
calculations.
\begin{itemize}

\item Table~\ref{tab:rydfreq}, A1 to A29, H and D experimentally
determined transition energies in kHz.

\item Table~\ref{tab:deltaRyd}, B1 to B25, additive energy corrections
in kHz to the theoretical expressions for these transitions.  The
correlation coefficients for the data in these two tables are given in
Table~\ref{tab:cchydrogen}.

\item Table~\ref{tab:muhdata}, C1 to C6, muonic H, D, and $^4$He$^+$
experimentally determined Lamb-shift transition energies in meV and
additive energy corrections in meV to the theoretical expressions for
these transitions. (The data in Table~\ref{tab:muonicatoms} are used in
the theoretical expressions.)

\item Table~\ref{tab:pdata}, D1 to D52, a wide variety of input data
ranging from the experimental value of the electron magnetic-moment
anomaly $a_{\rm e}$(exp) to the theoretically calculated
magnetic-shielding difference of the triton t and proton p in the HT
molecule $\sigma_{\rm tp}$.  The correlation coefficients among these 52
data are given in Table~\ref{tab:ccpdata}.

\item Table~\ref{tab:Xray}, E1 to E21, experimentally determined lattice
spacings, lattice spacing differences, and ratios of reference x-ray
wave lengths to lattice spacings of a number of silicon crystals.  The
correlation coefficients among these 21 data are given in
Table~\ref{tab:ccXray}.

\item Values of quantities the uncertainties of which are so small in
the context in which they are used that they can be assumed to be exact.
These include the reference constants in Table~\ref{tab:HDptheory-input}
and coefficients in Table~\ref{tab:betacoeff} for the theoretical
expressions for HD$^+$ transition frequencies; Bethe logarithms in
Table~\ref{tab:bethe}; theoretical values for various bound-particle to
free-particle $g$-factor ratios in Table~\ref{tab:gfactrat}; and the
magnetic shielding corrections $\sigma_{\rm h}(^3{\rm He})$ and
$\sigma_{\rm h}(^3{\rm He}^+)$ in
Eqs.~(\ref{eq:3helium})-(\ref{eq:3heliumpa}) and (\ref{eq:wav}).

\end{itemize}

The 79 adjusted constants used in the observational equations for the
133 input data may be found in Tables~\ref{tab:deltaRyd} and
\ref{tab:adjother} and the observational equations in
Tables~\ref{tab:pobseqsc}, \ref{tab:pobseqsa}, and \ref{tab:pobseqsb}.
The degrees of freedom for adjustments with this dataset is $\nu =
N-M=133-79=54$.  For the initial least-squares adjustment without
expansion factors, $\chi^2=109.6$, $p(109.6|54)=0.001\,\%$, and $R_{\rm
B}=1.42$.  This large value of $\chi^2$ is mainly due to the following
eight input data, each with a normalized residual greater than 2: In
Table~\ref{tab:rydfreq}, data A12 through A15 and A22 and A23, which are
measured energy-level transitions in H and D; and in
Table~\ref{tab:pdata}, data D3 and D4, which are measurements of
$h/m(^{87}{\rm Rb})$ and $h/m(^{133}{\rm Cs})$.  The residuals of these
eight data are 3.1, 2.5, 2.5, 3.1, 2.7, 3.4, $-$2.3, and 4.7,
respectively.

The expansion factors of the uncertainties of the data to reduce these
residuals to 2 or less are: (i) 1.7 for the 29 measured transition
energies in Table~\ref{tab:rydfreq} (A1 through A29) and the 25 additive
corrections to the theoretical expressions for the transitions in
Table~\ref{tab:deltaRyd} (B1 through B25); (ii) 1.7 for the six muonic
H, D, and $^4$He$^+$ Lamb-shift data in Table~\ref{tab:muhdata} (C1
through C6); and (iii) 2.5 for the six data related to the determination
of $\alpha$ at the beginning of Table~\ref{tab:pdata} (D1 through D6).
The expansion factor 1.7 is applied to the uncertainties of all the data
in Tables~\ref{tab:rydfreq} and \ref{tab:deltaRyd} to maintain their
original relative weights and correlations and is also applied to the
six muonic atom Lamb-shift data for the same reason.  Doing so ensures
that all of the data that determine the Rydberg constant $R_\infty$ and
the proton, deuteron, and alpha particle radii $r_{\rm p}$, $r_{\rm d}$,
and $r_{\rmssalpha}$ contribute in the same proportion that they would
have if no expansion factor was needed.

As expected, the final least-squares adjustment that determines the 2022
CODATA recommended values of the constants using the input data modified
in this way yields the quite satisfactory results $\chi^2=44.2$,
$p(44.2|54)=0.83$, and $R_{\rm B}=0.90$.  As was the case for the 16
values of $G$ discussed above, a number of the 133 input data have
self-sensitivity coefficients $S_{\rm c}$ less than 0.01.  Nevertheless,
they are retained for the same reason, namely because of the significant
disagreements among some of the data and the desirability of having the
recommended values reflect all of the relevant data.

\subsection{Tables of recommended values}
\label{ssec:tov}

The six tables \ref{tab:abbr} through \ref{tab:enconv2}, which have
the same form as those in previous CODATA reports, give the 2022
recommended values of the basic constants and conversion factors of
physics and chemistry and related quantities.

They range from Table~\ref{tab:abbr}, which gives an abbreviated list of
constants, to Tables~\ref{tab:enconv1} and \ref{tab:enconv2} which give
energy equivalents.  Although some of the 79 adjusted constants are
recommended values, most are obtained from combinations of them taking
into account their covariances.  However, a few are based on the
Particle Data Group values of the Fermi coupling constant $G_{\rm F}$,
the weak mixing angle $\sin^2{\theta_{\rm w}}$, and the mass of the tau
lepton \cite{2022048}.  The value of the free helion magnetic moment
$\mu_{\rm h}$ is obtained from the adjusted constant $\mu_{\rm h}(^3{\rm
He})/\mu^\prime_{\rm p}$ with the aid of the theoretically calculated
value of the magnetic shielding constant $\mu_{\rm h}(^3{\rm
He})/\mu_{\rm h}$ in Eq.~(\ref{eq:3helium}) based on the expression
$\mu_{\rm h}(^3{\rm He})= \mu_{\rm h}\left[1-\sigma_{\rm h}(^3{\rm
He})\right]$.

All of the values given in these tables are available on the website of
the Fundamental Constants Data Center of the NIST Physical Measurement
Laboratory at physics.nist.gov/constants.  In fact, like its
predecessors, this electronic version of the 2022 CODATA recommended
values of the constants enables users to obtain the correlation
coefficient of any two constants listed in the tables.  It also allows
users to automatically convert the value of an energy-related quantity
expressed in one unit to the corresponding value expressed in another
unit (in essence, an automated version of Tables~\ref{tab:enconv1} and
\ref{tab:enconv2}).

\begin{table*}[h]

\caption[Abrreviated list of fundamental constants]{An abbreviated list
of the CODATA recommended values of the fundamental constants of physics
and chemistry based on the 2022 adjustment.} \label{tab:abbr}

\renewcommand{\thefootnote}{\fnsymbol{footnote}}
\begin{center}
\begin{tabular}{lllll}
\hline\hline
&&&& \multicolumn{1}{c}{Relative std.}    \\
\multicolumn{1}{c}{Quantity}& \multicolumn{1}{c}{Symbol} 
& \multicolumn{1}{c}{Value}  &  \multicolumn{1}{c}{Unit}   
& \multicolumn{1}{c}{uncert. $u_{\rm r}$ } \\
\hline
speed of light in vacuum & $ c $ &  299\,792\,458 & m~s$^{-1}$ & exact \vbox to 12 pt {} \\
Newtonian constant of gravitation~~ & $ G $ & $ 6.674\,30(15)\times 10^{-11}$ & m$^{3}$~kg$^{-1}$~s$^{-2}$ & $ 2.2\times 10^{-5}$ \\
Planck constant\footnotemark[1]   & $ h $ & $ 6.626\,070\,15\times 10^{-34}$ & J~Hz$^{-1}$ & exact \\
                & $\hbar$ & $ 1.054\,571\,817\ldots\times 10^{-34}$ & J~s & exact \\
elementary charge & $ e $ & $ 1.602\,176\,634\times 10^{-19}$ & C & exact \\
vacuum magnetic permeability {$4\rmpi \alpha \hbar/e^2c$} & $\mu_0$ & $ 1.256\,637\,061\,27(20)\times 10^{-6}$ & N~A$^{-2}$ & $ 1.6\times 10^{-10}$\\
vacuum electric permittivity 1/$\mu_0c^{2}$ & $\epsilon_0$ & $ 8.854\,187\,8188(14)\times 10^{-12}$ & F~m$^{-1}$ & $ 1.6\times 10^{-10}$ \\
Josephson constant\, 2$e/h$ & $K_{\rm J}$ & $ 483\,597.848\,4\ldots\times 10^{9}$ & Hz~V$^{-1}$ & exact \\
von Klitzing constant  {$\mu_0 c/2\alpha=2\rmpi\hbar/e^2$} & $R_{\rm K}$ & $ 25\,812.807\,45\ldots$ & ${\rm \Omega}$ & exact \\
magnetic flux quantum {$2\rmpi\hbar/(2e)$} & ${\it \Phi}_0$ & $ 2.067\,833\,848\ldots\times 10^{-15}$ & Wb & exact \\
conductance quantum {$2e^2/2\rmpi \hbar$} & $G_0$ & $ 7.748\,091\,729\ldots\times 10^{-5}$ & S & exact \\
electron mass & $ m_{\rm e}$ & $ 9.109\,383\,7139(28)\times 10^{-31}$ & kg & $ 3.1\times 10^{-10}$ \\
proton mass & $ m_{\rm p}$ & $ 1.672\,621\,925\,95(52)\times 10^{-27}$ & kg & $ 3.1\times 10^{-10}$ \\
proton-electron mass ratio & $m_{\rm p}$/$m_{\rm e}$ & $ 1836.152\,673\,426(32)$ & & $ 1.7\times 10^{-11}$ \\
fine-structure constant $e^2\!/4\rmpi\epsilon_0 \hbar c$ & $\alpha$ & $ 7.297\,352\,5643(11)\times 10^{-3}$ & & $ 1.6\times 10^{-10}$ \\
\b inverse fine-structure constant & $\alpha^{-1}$ & $ 137.035\,999\,177(21)$ & & $ 1.6\times 10^{-10}$ \\
Rydberg frequency $\alpha^2m_{\rm e}c^2/2h$ &$cR_\infty$ &$ 3.289\,841\,960\,2500(36)\times 10^{15}$ & Hz & $ 1.1\times 10^{-12}$ \\
Boltzmann constant  & $k$ & $ 1.380\,649\times 10^{-23}$ & J~K$^{-1}$ & exact \\
Avogadro constant & $N_{\rm A}$ & $ 6.022\,140\,76\times 10^{23}$ & mol$^{-1}$ & exact \\
molar gas constant $N_{\rm A} k$ & $ R $ & $ 8.314\,462\,618\ldots$ & J~mol$^{-1}$~K$^{-1}$ & exact \\
Faraday constant $N_{\rm A}e$ & $ F $ & $ 96\,485.332\,12\ldots$ & C~mol$^{-1}$ & exact \\
Stefan-Boltzmann constant & & & & \\
\, ($\rmpi^2$/60)$k^4\!/\hbar^3c^2$ & $\sigma$ & $ 5.670\,374\,419\ldots\times 10^{-8}$ & W~m$^{-2}$~K$^{-4}$ & exact \\
\multicolumn {5} {c} { \vbox to 12 pt {} Non-SI units accepted for use with the SI} \\ electron volt ($e$/{\rm C}) {\rm J} & eV & $ 1.602\,176\,634\times 10^{-19}$ & J & exact \\
(unified) atomic mass unit ${1\over12}m(^{12}$C)~~ & u & $ 1.660\,539\,068\,92(52)\times 10^{-27}$ & kg & $ 3.1\times 10^{-10}$ \phantom{\Big|} \\ \\
\hline
\hline
\end{tabular}
\end{center}   %
\end{table*}

\clearpage

\renewcommand{\thefootnote}{\arabic{footnote}}
\def\vbb{\vbox to 12 pt {}}
\begin{longtable*}{lllll}
\caption[CODATA recommended values of the fundamental constants]{The
CODATA recommended values of the fundamental constants of physics and
chemistry based on the 2022 adjustment.\label{tab:constants}} \\
\hline
\hline
& & & & Relative std. \\
\s{35}Quantity & \s{-15} Symbol & \s{15} Numerical value & \s{5}Unit & uncert. $u_{\rm r}$ \\
\hline
\endfirsthead
\caption{{\it (Continued).}} \\
\hline
& & & & Relative std. \\
\s{35}Quantity & \s{-15} Symbol & \s{15} Numerical value & \s{5}Unit & uncert. $u_{\rm r}$ \\
\hline
\endhead
\hline
\endfoot
\endlastfoot
\multicolumn {5} {c} { \vbb
UNIVERSAL} \\
speed of light in vacuum & $ c $ & $ 299\,792\,458$ & m~s$^{-1}$ & exact \\
vacuum magnetic permeability {$4\rmpi\alpha \hbar/e^2c$} & $\mu_0$ & $ 1.256\,637\,061\,27(20)\times 10^{-6}$ & N~A$^{-2}$ & $ 1.6\times 10^{-10}$\\
\b\b  $\mu_0/(4\rmpi\times 10^{-7})$ && $ 0.999\,999\,999\,87(16)$ &  N~A$^{-2}$ & $ 1.6\times 10^{-10}$ \\
vacuum electric permittivity $1/\mu_0c^{2}$ & $\epsilon_0$ & $ 8.854\,187\,8188(14)\times 10^{-12}$ & F~m$^{-1}$ & $ 1.6\times 10^{-10}$ \\
characteristic impedance of vacuum $\mu_0c$ & $Z_0$ & $ 376.730\,313\,412(59)$ & ${\rm \Omega}$ & $ 1.6\times 10^{-10}$ \\
Newtonian constant of gravitation & $ G $ & $ 6.674\,30(15)\times 10^{-11}$ & m$^3$~kg$^{-1}$~s$^{-2}$ & $ 2.2\times 10^{-5}$ \\
& $G/\hbar c $ & $ 6.708\,83(15)\times 10^{-39}$ & $ ({\rm GeV}/c^2)^{-2}$ & $ 2.2\times 10^{-5}$ \\
Planck constant & $ h $ & $ 6.626\,070\,15\times 10^{-34}$ & J~Hz$^{-1}$ & exact \\
\b\b & & $ 4.135\,667\,696\ldots\times 10^{-15}$ & eV~Hz$^{-1}$ & exact \\
\b  & $\hbar$ & $ 1.054\,571\,817\ldots\times 10^{-34}$ & J~s & exact \\
\b\b & & $ 6.582\,119\,569\ldots\times 10^{-16}$ & eV~s & exact \\
\b & $\hbar c$ & $ 197.326\,980\,4\ldots$ & MeV~fm & exact \\
Planck mass~$(\hbar c/G)^{1/2}$ & $m_{\rm P}$ & $ 2.176\,434(24)\times 10^{-8}$ & kg & $ 1.1\times 10^{-5}$ \\
\b energy equivalent & $m_{\rm P}c^2$ & $ 1.220\,890(14)\times 10^{19}$ & GeV & $ 1.1\times 10^{-5}$ \\
Planck temperature~$(\hbar c^5/G)^{1/2}/k$ & $T_{\rm P}$ & $ 1.416\,784(16)\times 10^{32}$ & K & $ 1.1\times 10^{-5}$ \\
Planck length~$\hbar/m_{\rm P}c=(\hbar G/c^3)^{1/2}$ & $l_{\rm P}$ & $ 1.616\,255(18)\times 10^{-35}$ & m & $ 1.1\times 10^{-5}$ \\
Planck time $l_{\rm P}/c=(\hbar G/c^5)^{1/2}$ & $t_{\rm P}$ & $ 5.391\,247(60)\times 10^{-44}$ & s & $ 1.1\times 10^{-5}$ \\ \\
\multicolumn {5} {c} { \vbb
ELECTROMAGNETIC} \\
elementary charge & $e$ & $ 1.602\,176\,634\times 10^{-19}$ & C & exact \\
& {$e/\hbar$} & $ 1.519\,267\,447\ldots\times 10^{15}$ & A J$^{-1}$ & exact \\
magnetic flux quantum {$2\rmpi \hbar/(2e)$} & ${\it \Phi}_0$ & $ 2.067\,833\,848\ldots\times 10^{-15}$ & Wb & exact \\
conductance quantum {$2e^2\!/2\rmpi \hbar$} & $G_0$ & $ 7.748\,091\,729\ldots\times 10^{-5}$ & S & exact \\
\b inverse of conductance quantum & $G_0^{-1}$ & $ 12\,906.403\,72\ldots$ & ${\rm \Omega}$ & exact \\
Josephson constant\, 2$e/h$ & $K_{\rm J}$ & $ 483\,597.848\,4\ldots\times 10^{9}$ & Hz~V$^{-1}$ & exact \\
von Klitzing constant  {$\mu_0c/2\alpha=2\rmpi\hbar/e^2$} & $R_{\rm K}$ & $ 25\,812.807\,45\ldots$ & ${\rm \Omega}$ & exact \\
Bohr magneton $e\hbar/2m_{\rm e}$ & $\mu_{\rm B}$ & $ 9.274\,010\,0657(29)\times 10^{-24}$ & J~T$^{-1}$ & $ 3.1\times 10^{-10}$ \\
\b & & $ 5.788\,381\,7982(18)\times 10^{-5}$ & eV~T$^{-1}$ & $ 3.1\times 10^{-10}$ \\
\b & $\mu_{\rm B}/h$ & $ 1.399\,624\,491\,71(44)\times 10^{10}$ & Hz~T$^{-1}$ & $ 3.1\times 10^{-10}$ \\
\b & $\mu_{\rm B}/h c$ & $ 46.686\,447\,719(15)$ &
[m$^{-1}~$T$^{-1}$]$^{\rm a}$ & $ 3.1\times 10^{-10}$ \\
\b & $\mu_{\rm B}/k$ & $ 0.671\,713\,814\,72(21)$ & K~T$^{-1}$ & $ 3.1\times 10^{-10}$ \\
nuclear magneton $e\hbar/2m_{\rm p}$ & $\mu_{\rm N}$ & $ 5.050\,783\,7393(16)\times 10^{-27}$ & J~T$^{-1}$ & $ 3.1\times 10^{-10}$ \\
\b & & $ 3.152\,451\,254\,17(98)\times 10^{-8}$ & eV~T$^{-1}$ & $ 3.1\times 10^{-10}$ \\
\b & $\mu_{\rm N}/h$ & $ 7.622\,593\,2188(24)$ & MHz~T$^{-1}$ & $ 3.1\times 10^{-10}$ \\
\b & $\mu_{\rm N}/hc$ & $ 2.542\,623\,410\,09(79)\times 10^{-2}$ & [m$^{-1}~$T$^{-1}$]$^{\rm a}$& $ 3.1\times 10^{-10}$ \\
\b & $\mu_{\rm N}/k$ & $ 3.658\,267\,7706(11)\times 10^{-4}$ & K~T$^{-1}$ & $ 3.1\times 10^{-10}$ \\ \\
\multicolumn {5} {c} { \vbb
ATOMIC AND NUCLEAR} \\
\multicolumn {5} {c} {General} \\
fine-structure constant $e^2\!/4\rmpi\epsilon_0\hbar c$ & $\alpha$ & $ 7.297\,352\,5643(11)\times 10^{-3}$ & & $ 1.6\times 10^{-10}$ \\
\b inverse fine-structure constant & $\alpha^{-1}$ & $ 137.035\,999\,177(21)$ & & $ 1.6\times 10^{-10}$ \\
Rydberg frequency $\alpha^2m_{\rm e}c^2/2h=E_{\rm h}/2h$ &$cR_\infty$ &$ 3.289\,841\,960\,2500(36)\times 10^{15}$ & Hz & $ 1.1\times 10^{-12}$ \\
\b energy equivalent &  $hc\,R_\infty$ & $ 2.179\,872\,361\,1030(24)\times 10^{-18}$ & J & $ 1.1\times 10^{-12}$ \\
\b & & $ 13.605\,693\,122\,990(15)$ & eV & $ 1.1\times 10^{-12}$ \\
  Rydberg constant  &  $R_\infty$ &  $ 10\,973\,731.568\,157(12)$ & [m$^{-1}$]$^{\rm a}$ & $ 1.1\times 10^{-12}$ \\
Bohr radius ${\hbar/\alpha m_{\rm e}c}=4\rmpi\epsilon_0\hbar^2\!/m_{\rm e}e^2$ & $a_{\rm 0}$ & $ 5.291\,772\,105\,44(82)\times 10^{-11}$ & m & $ 1.6\times 10^{-10}$ \\
Hartree energy {$ \alpha^2m_{\rm e}c^2 =e^2\!/4\rmpi\epsilon_{\rm 0}a_{\rm 0}=2 hcR_\infty$} & $E_{\rm h}$ & $ 4.359\,744\,722\,2060(48)\times 10^{-18}$ & J & $ 1.1\times 10^{-12}$ \\
\b & & $ 27.211\,386\,245\,981(30)$ & eV & $ 1.1\times 10^{-12}$ \\
quantum of circulation & $\rmpi\hbar/m_{\rm e}$ & $ 3.636\,947\,5467(11)\times 10^{-4}$ & m$^2~$s$^{-1}$ & $ 3.1\times 10^{-10}$ \\
& $2\rmpi\hbar/m_{\rm e}$ & $ 7.273\,895\,0934(23)\times 10^{-4}$ & m$^2~$s$^{-1}$ & $ 3.1\times 10^{-10}$ \\ \\
\multicolumn {5} {c} { \vbb
Electroweak} \\
Fermi coupling constant$^{\rm b}$  & $G_{\rm F}/(\hbar c)^3$ & $ 1.166\,3787(6)\times 10^{-5}$ & GeV$^{-2}$ & $ 5.1\times 10^{-7}$ \vbox to 10 pt {}\\
weak mixing angle$^{\rm c}$  $\theta_{\rm W} $ (on-shell scheme) & & & & \\
\, $\sin^2\theta_{\rm W} = s^2_{\rm W} \equiv 1-(m_{\rm W}/m_{\rm Z})^2$ & $\sin^2\theta_{\rm W}$ & $ 0.223\,05(23)$ & & $ 1.0\times 10^{-3}$ \\ \\
\multicolumn {5} {c} {\vbb
Electron, e$^-$} \\
electron mass & $m_{\rm e}$ & $ 9.109\,383\,7139(28)\times 10^{-31}$ & kg & $ 3.1\times 10^{-10}$ \\
\b\b & & $ 5.485\,799\,090\,441(97)\times 10^{-4}$ & u & $ 1.8\times 10^{-11}$ \\
\b energy equivalent & $m_{\rm e}c^2$ & $ 8.187\,105\,7880(26)\times 10^{-14}$ & J & $ 3.1\times 10^{-10}$ \\
\b\b & & $ 0.510\,998\,950\,69(16)$ & MeV & $ 3.1\times 10^{-10}$ \\
electron-muon mass ratio & $m_{\rm e}/m_{\rmssmu}$ & $ 4.836\,331\,70(11)\times 10^{-3}$ & & $ 2.2\times 10^{-8}$ \\
electron-tau mass ratio & $m_{\rm e}/m_{\rmsstau}$ & $ 2.875\,85(19)\times 10^{-4}$ & & $ 6.8\times 10^{-5}$ \\
electron-proton mass ratio & $m_{\rm e}/m_{\rm p}$ & $ 5.446\,170\,214\,889(94)\times 10^{-4}$ & & $ 1.7\times 10^{-11}$ \\
electron-neutron mass ratio & $m_{\rm e}/m_{\rm n}$ & $ 5.438\,673\,4416(22)\times 10^{-4}$ & & $ 4.0\times 10^{-10}$ \\
electron-deuteron mass ratio & $m_{\rm e}/m_{\rm d}$ & $ 2.724\,437\,107\,629(47)\times 10^{-4}$ & & $ 1.7\times 10^{-11}$ \\
electron-triton mass ratio & $m_{\rm e}/m_{\rm t}$ & $ 1.819\,200\,062\,327(68)\times 10^{-4}$ & & $ 3.8\times 10^{-11}$ \\
electron-helion mass ratio & $m_{\rm e}/m_{\rm h}$ & $ 1.819\,543\,074\,649(53)\times 10^{-4}$ & & $ 2.9\times 10^{-11}$ \\
electron to alpha particle mass ratio & $m_{\rm e}/m_{\rmssalpha}$ & $ 1.370\,933\,554\,733(32)\times 10^{-4}$ & & $ 2.4\times 10^{-11}$ \\
electron charge to mass quotient & $-e/m_{\rm e}$ & $ -1.758\,820\,008\,38(55)\times 10^{11}$ & C~kg$^{-1}$ & $ 3.1\times 10^{-10}$ \\
electron molar mass $N_{\rm A}m_{\rm e}$& $M({\rm e}),M_{\rm e}$ & $ 5.485\,799\,0962(17)\times 10^{-7}$ & kg mol$^{-1}$ & $ 3.1\times 10^{-10}$ \\
 reduced Compton wavelength $\hbar/m_{\rm e}c=\alpha a_{0}$ & $\lbar_{\rm C}$ & $ 3.861\,592\,6744(12)\times 10^{-13}$ & m & $ 3.1\times 10^{-10}$ \\
\b  Compton wavelength  &  $\lambda_{\rm C}$  & $ 2.426\,310\,235\,38(76)\times 10^{-12}$ & [m]$^{\rm a}$ & $ 3.1\times 10^{-10}$ \\
classical electron radius $\alpha^2a_{\rm 0}$ & $r_{\rm e}$ & $ 2.817\,940\,3205(13)\times 10^{-15}$ & m & $ 4.7\times 10^{-10}$ \\
Thomson cross section (8$\rmpi/3)r^2_{\rm e}$ & $\sigma_{\rm e}$ & $ 6.652\,458\,7051(62)\times 10^{-29}$ & m$^2$ & $ 9.3\times 10^{-10}$ \\
electron magnetic moment & $\mu_{\rm e}$ & $ -9.284\,764\,6917(29)\times 10^{-24}$ & J~T$^{-1}$ & $ 3.1\times 10^{-10}$ \\
\b to Bohr magneton ratio & $\mu_{\rm e}/\mu_{\rm B}$ & $ -1.001\,159\,652\,180\,46(18)$ & & $ 1.8\times 10^{-13}$ \\
\b to nuclear magneton ratio & $\mu_{\rm e}/\mu_{\rm N}$ & $ -1838.281\,971\,877(32)$ & & $ 1.7\times 10^{-11}$ \\
electron magnetic-moment & & & & \\
\, anomaly $|\mu_{\rm e}|/\mu_{\rm B}-1$ & $a_{\rm e}$ & $ 1.159\,652\,180\,46(18)\times 10^{-3}$ & & $ 1.6\times 10^{-10}$ \\
electron $g$-factor $-2(1+a_{\rm e})$ & $g_{\rm e}$ & $ -2.002\,319\,304\,360\,92(36)$ & & $ 1.8\times 10^{-13}$ \\
electron-muon magnetic-moment ratio & $\mu_{\rm e}/\mu_{\rmssmu}$ & $ 206.766\,9881(46)$ & & $ 2.2\times 10^{-8}$ \\
electron-proton magnetic-moment ratio & $\mu_{\rm e}/\mu_{\rm p}$ & $ -658.210\,687\,89(19)$ & & $ 3.0\times 10^{-10}$ \\
electron to shielded proton magnetic-& & & & \\
\, moment ratio (H$_2$O, sphere, 25 $^\circ$C) & $\mu_{\rm e}/\mu^\prime_{\rm p}$ & $ -658.227\,5856(27)$ & & $ 4.1\times 10^{-9}$ \\
electron-neutron magnetic-moment ratio & $\mu_{\rm e}/\mu_{\rm n}$ & $ 960.920\,48(23)$ & & $ 2.4\times 10^{-7}$ \\
electron-deuteron magnetic-moment ratio & $\mu_{\rm e}/\mu_{\rm d}$ & $ -2143.923\,4921(56)$ & & $ 2.6\times 10^{-9}$ \\
electron to bound helion magnetic-& & & & \\
\, moment ratio  & $\mu_{\rm e}/\mu_{\rm h}(^3{\rm He})$ & $ 864.058\,239\,86(70)$ & & $ 8.1\times 10^{-10}$ \\
electron gyromagnetic ratio $2|\mu_{\rm e}|/\hbar$ & $\gamma_{\rm e}$ & $ 1.760\,859\,627\,84(55)\times 10^{11}$ & s$^{-1}~$T$^{-1}$ & $ 3.1\times 10^{-10}$ \\
                     & & $ 28\,024.951\,3861(87)$ & MHz~T$^{-1}$ & $ 3.1\times 10^{-10}$ \\ \\
\multicolumn {5} {c} { \vbb
Muon, ${\rmmu}^-$} \\
muon mass & $m_{\rmssmu}$ & $ 1.883\,531\,627(42)\times 10^{-28}$ & kg & $ 2.2\times 10^{-8}$ \\
& & $ 0.113\,428\,9257(25)$ & u & $ 2.2\times 10^{-8}$ \\
\b energy equivalent & $m_{\rmssmu}c^2$ & $ 1.692\,833\,804(38)\times 10^{-11}$ & J & $ 2.2\times 10^{-8}$ \\
& & $ 105.658\,3755(23)$ & MeV & $ 2.2\times 10^{-8}$ \\
muon-electron mass ratio & $m_{\rmssmu}/m_{\rm e}$ & $ 206.768\,2827(46)$ & & $ 2.2\times 10^{-8}$ \\
muon-tau mass ratio & $m_{\rmssmu}/m_{\rmsstau}$ & $ 5.946\,35(40)\times 10^{-2}$ & & $ 6.8\times 10^{-5}$ \\
muon-proton mass ratio & $m_{\rmssmu}/m_{\rm p}$ & $ 0.112\,609\,5262(25)$ & & $ 2.2\times 10^{-8}$ \\
muon-neutron mass ratio & $m_{\rmssmu}/m_{\rm n}$ & $ 0.112\,454\,5168(25)$ & & $ 2.2\times 10^{-8}$ \\
muon molar mass $N_{\rm A}m_{\rmssmu}$& $M({\rmmu}),M_{\rmssmu}$ & $ 1.134\,289\,258(25)\times 10^{-4}$ & kg mol$^{-1}$ & $ 2.2\times 10^{-8}$ \\
reduced muon Compton wavelength $\hbar/m_{\rmssmu}c$ & $\lbar_{{\rm C},{\rmssmu}}$ & $ 1.867\,594\,306(42)\times 10^{-15}$ & m & $ 2.2\times 10^{-8}$ \\
\b  muon Compton wavelength & $\lambda_{{\rm C},{\rmssmu}}$   &
$ 1.173\,444\,110(26)\times 10^{-14}$ & [m]$^{\rm a}$ & $ 2.2\times 10^{-8}$ \\
muon magnetic moment & $\mu_{\rmssmu}$ & $ -4.490\,448\,30(10)\times 10^{-26}$ & J~T$^{-1}$ & $ 2.2\times 10^{-8}$ \\
\b to Bohr magneton ratio & $\mu_{\rmssmu}/\mu_{\rm B}$ & $ -4.841\,970\,48(11)\times 10^{-3}$ & & $ 2.2\times 10^{-8}$ \\
\b to nuclear magneton ratio & $\mu_{\rmssmu}/\mu_{\rm N}$ & $ -8.890\,597\,04(20)$ & & $ 2.2\times 10^{-8}$ \\
muon magnetic-moment anomaly & & & & \\
\, $|\mu_{\rmssmu}|/(e\hbar/2m_{\rmssmu})-1$ & $a_{\rmssmu}$ & $ 1.165\,920\,62(41)\times 10^{-3}$ & & $ 3.5\times 10^{-7}$ \\
muon $g$-factor $-2(1+a_{\rmssmu}$) & $g_{\rmssmu}$ & $ -2.002\,331\,841\,23(82)$ & & $ 4.1\times 10^{-10}$ \\
muon-proton magnetic-moment ratio & $\mu_{\rmssmu}/\mu_{\rm p}$ & $ -3.183\,345\,146(71)$ & & $ 2.2\times 10^{-8}$ \\ \\
\multicolumn {5} {c} { \vbb
Tau, ${\rmtau}^-$} \\
tau mass$^{\rm d}$  & $m_{\rmsstau}$ & $ 3.167\,54(21)\times 10^{-27}$ & kg & $ 6.8\times 10^{-5}$ \\
& & $ 1.907\,54(13)$ & u & $ 6.8\times 10^{-5}$ \\
\b energy equivalent & $m_{\rmsstau}c^2$ & $ 2.846\,84(19)\times 10^{-10}$ & J & $ 6.8\times 10^{-5}$ \\
& & $ 1776.86(12)$ & MeV & $ 6.8\times 10^{-5}$ \\
tau-electron mass ratio & $m_{\rmsstau}/m_{\rm e}$ & $ 3477.23(23)$ & & $ 6.8\times 10^{-5}$ \\
tau-muon mass ratio & $m_{\rmsstau}/m_{\rmssmu}$ & $ 16.8170(11)$ & & $ 6.8\times 10^{-5}$ \\
tau-proton mass ratio & $m_{\rmsstau}/m_{\rm p}$ & $ 1.893\,76(13)$ & & $ 6.8\times 10^{-5}$ \\
tau-neutron mass ratio & $m_{\rmsstau}/m_{\rm n}$ & $ 1.891\,15(13)$ & & $ 6.8\times 10^{-5}$ \\
tau molar mass $N_{\rm A}m_{\rmsstau}$& $M({\rmtau}),M_{\rmsstau}$ & $ 1.907\,54(13)\times 10^{-3}$ & kg mol$^{-1}$ & $ 6.8\times 10^{-5}$ \\
reduced tau Compton wavelength $\hbar/m_{\rmsstau}c$ & $\lbar_{{\rm C},{\rmsstau}}$ & $ 1.110\,538(75)\times 10^{-16}$ & m & $ 6.8\times 10^{-5}$ \\
\b   tau Compton wavelength  &  $\lambda_{{\rm C},{\rmsstau}}$  & $ 6.977\,71(47)\times 10^{-16}$ & [m]$^{\rm a}$ & $ 6.8\times 10^{-5}$ \\ \\
\multicolumn {5} {c} {\vbb Proton, p} \\
proton mass & $m_{\rm p}$ & $ 1.672\,621\,925\,95(52)\times 10^{-27}$ & kg & $ 3.1\times 10^{-10}$ \\
& & $ 1.007\,276\,466\,5789(83)$ & u & $ 8.3\times 10^{-12}$ \\
\b energy equivalent & $m_{\rm p}c^2$ & $ 1.503\,277\,618\,02(47)\times 10^{-10}$ & J & $ 3.1\times 10^{-10}$ \\
& & $ 938.272\,089\,43(29)$ & MeV & $ 3.1\times 10^{-10}$ \\
proton-electron mass ratio & $m_{\rm p}/m_{\rm e}$ & $ 1836.152\,673\,426(32)$ & & $ 1.7\times 10^{-11}$ \\
proton-muon mass ratio & $m_{\rm p}/m_{\rmssmu}$ & $ 8.880\,243\,38(20)$ & & $ 2.2\times 10^{-8}$ \\
proton-tau mass ratio & $m_{\rm p}/m_{\rmsstau}$ & $ 0.528\,051(36)$ & & $ 6.8\times 10^{-5}$ \\
proton-neutron mass ratio & $m_{\rm p}/m_{\rm n}$ & $ 0.998\,623\,477\,97(40)$ & & $ 4.0\times 10^{-10}$ \\
proton charge to mass quotient & $e/m_{\rm p}$ & $ 9.578\,833\,1430(30)\times 10^{7}$ & C kg$^{-1}$ & $ 3.1\times 10^{-10}$ \\
proton molar mass $N_{\rm A}m_{\rm p}$& $M$(p), $M_{\rm p}$ & $ 1.007\,276\,467\,64(31)\times 10^{-3}$ & kg mol$^{-1}$ & $ 3.1\times 10^{-10}$ \\
reduced proton Compton wavelength  $\hbar/m_{\rm p}c$ & $\lbar_{\rm C,p}$ & $ 2.103\,089\,100\,51(66)\times 10^{-16}$ & m & $ 3.1\times 10^{-10}$ \\
\b  proton Compton wavelength & $\lambda_{\rm C,p}$  & $ 1.321\,409\,853\,60(41)\times 10^{-15}$ & [m]$^{\rm a}$ & $ 3.1\times 10^{-10}$ \\
proton rms charge radius & $r_{\rm p}$ & $ 8.4075(64)\times 10^{-16}$ & m & $ 7.6\times 10^{-4}$ \\
proton magnetic moment & $\mu_{\rm p}$ & $ 1.410\,606\,795\,45(60)\times 10^{-26}$ & J~T$^{-1}$ & $ 4.3\times 10^{-10}$ \\
\b to Bohr magneton ratio & $\mu_{\rm p}/\mu_{\rm B}$ & $ 1.521\,032\,202\,30(45)\times 10^{-3}$ & & $ 3.0\times 10^{-10}$ \\
\b to nuclear magneton ratio & $\mu_{\rm p}/\mu_{\rm N}$ & $ 2.792\,847\,344\,63(82)$ & & $ 2.9\times 10^{-10}$ \\
proton $g$-factor $2\mu_{\rm p}/\mu_{\rm N}$ & $g_{\rm p}$ & $ 5.585\,694\,6893(16)$ & & $ 2.9\times 10^{-10}$ \\
proton-neutron magnetic-moment ratio & $\mu_{\rm p}/\mu_{\rm n}$ & $ -1.459\,898\,02(34)$ & & $ 2.4\times 10^{-7}$ \\
shielded proton magnetic moment & & & & \\
\, (H$_{2}$O, sphere, 25 $^\circ$C) & $\mu^\prime_{\rm p}$ & $ 1.410\,570\,5830(58)\times 10^{-26}$ & J~T$^{-1}$ & $ 4.1\times 10^{-9}$ \\
\b to Bohr magneton ratio & $\mu^\prime_{\rm p}/\mu_{\rm B}$ & $ 1.520\,993\,1551(62)\times 10^{-3}$ & & $ 4.1\times 10^{-9}$ \\
\b to nuclear magneton ratio & $\mu^\prime_{\rm p}/\mu_{\rm N}$ & $ 2.792\,775\,648(11)$ & & $ 4.1\times 10^{-9}$ \\
proton magnetic shielding correction& & & & \\
\, $1-\mu^\prime_{\rm p}/\mu_{\rm p}$ \ (H$_{2}$O, sphere, 25 $^\circ$C) & $\sigma^\prime_{\rm p}$ & $ 2.567\,15(41)\times 10^{-5}$ & &  $ 1.6\times 10^{-4}$ \\
proton gyromagnetic ratio $2\mu_{\rm p}/\hbar$ & $\gamma_{\rm p}$ & $ 2.675\,221\,8708(11)\times 10^{8}$ & s$^{-1}~$T$^{-1}$ & $ 4.3\times 10^{-10}$ \\
       &  & $ 42.577\,478\,461(18)$ & MHz~T$^{-1}$ & $ 4.3\times 10^{-10}$ \\
shielded proton gyromagnetic ratio& & & & \\
\b$2\mu^\prime_{\rm p}/\hbar$ \ (H$_{2}$O, sphere, 25 $^\circ$C)& $\gamma^\prime_{\rm p}$ & $ 2.675\,153\,194(11)\times 10^{8}$ & s$^{-1}~$T$^{-1}$ & $ 4.1\times 10^{-9}$ \\
 &  & $ 42.576\,385\,43(17)$ & MHz~T$^{-1}$ & $ 4.1\times 10^{-9}$ \\ \\
\multicolumn {5} {c} {\vbb Neutron, n} \\
neutron mass & $m_{\rm n}$ & $ 1.674\,927\,500\,56(85)\times 10^{-27}$ & kg & $ 5.1\times 10^{-10}$ \\
& & $ 1.008\,664\,916\,06(40)$ & u & $ 4.0\times 10^{-10}$ \\
\b energy equivalent & $m_{\rm n}c^2$ & $ 1.505\,349\,765\,14(76)\times 10^{-10}$ & J & $ 5.1\times 10^{-10}$ \\
& & $ 939.565\,421\,94(48)$ & MeV & $ 5.1\times 10^{-10}$ \\
neutron-electron mass ratio & $m_{\rm n}/m_{\rm e}$ & $ 1838.683\,662\,00(74)$ & & $ 4.0\times 10^{-10}$ \\
neutron-muon mass ratio & $m_{\rm n}/m_{\rmssmu}$ & $ 8.892\,484\,08(20)$ & & $ 2.2\times 10^{-8}$ \\
neutron-tau mass ratio & $m_{\rm n}/m_{\rmsstau}$ & $ 0.528\,779(36)$ & & $ 6.8\times 10^{-5}$ \\
neutron-proton mass ratio & $m_{\rm n}/m_{\rm p}$ & $ 1.001\,378\,419\,46(40)$ & & $ 4.0\times 10^{-10}$ \\
neutron-proton mass difference & $m_{\rm n}-m_{\rm p}$ & $ 2.305\,574\,61(67)\times 10^{-30}$ & kg & $ 2.9\times 10^{-7}$  \\
& & $ 1.388\,449\,48(40)\times 10^{-3}$ & u & $ 2.9\times 10^{-7}$ \\
\b energy equivalent & ($m_{\rm n}-m_{\rm p})c^2$~~ & $ 2.072\,147\,12(60)\times 10^{-13}$ & J & $ 2.9\times 10^{-7}$ \\
& & $ 1.293\,332\,51(38)$ & MeV & $ 2.9\times 10^{-7}$ \\
neutron molar mass $N_{\rm A}m_{\rm n}$ & $M({\rm n}),M_{\rm n}$ & $ 1.008\,664\,917\,12(51)\times 10^{-3}$ & kg mol$^{-1}$ & $ 5.1\times 10^{-10}$ \\
reduced neutron Compton wavelength $\hbar/m_{\rm n}c$ & $\lbar_{\rm C,n}$ & $ 2.100\,194\,1520(11)\times 10^{-16}$ & m & $ 5.1\times 10^{-10}$ \\
\b  neutron Compton wavelength & $\lambda_{\rm C,n}$  & $ 1.319\,590\,903\,82(67)\times 10^{-15}$ & [m]$^{\rm a}$ & $ 5.1\times 10^{-10}$ \\
neutron magnetic moment & $\mu_{\rm n}$ & $ -9.662\,3653(23)\times 10^{-27}$ & J~T$^{-1}$ & $ 2.4\times 10^{-7}$ \\
\b to Bohr magneton ratio & $\mu_{\rm n}/\mu_{\rm B}$ & $ -1.041\,875\,65(25)\times 10^{-3}$ & & $ 2.4\times 10^{-7}$ \\
\b to nuclear magneton ratio & $\mu_{\rm n}/\mu_{\rm N}$ & $ -1.913\,042\,76(45)$ & & $ 2.4\times 10^{-7}$ \\
neutron $g$-factor $2\mu_{\rm n}/\mu_{\rm N}$ & $g_{\rm n}$ & $ -3.826\,085\,52(90)$ & & $ 2.4\times 10^{-7}$ \\
neutron-electron magnetic-moment ratio & $\mu_{\rm n}/\mu_{\rm e}$ & $ 1.040\,668\,84(24)\times 10^{-3}$ & & $ 2.4\times 10^{-7}$ \\
neutron-proton magnetic-moment ratio & $\mu_{\rm n}/\mu_{\rm p}$ & $ -0.684\,979\,35(16)$ & & $ 2.4\times 10^{-7}$ \\
neutron to shielded proton magnetic-& & & & \\
\, moment ratio \ (H$_2$O, sphere, 25 $^\circ$C)& $\mu_{\rm n}/\mu_{\rm p}^\prime$ & $ -0.684\,996\,94(16)$ & & $ 2.4\times 10^{-7}$ \\
neutron gyromagnetic ratio $2|\mu_{\rm n}|/\hbar$ & $\gamma_{\rm n}$ & $ 1.832\,471\,74(43)\times 10^{8}$ & s$^{-1}~$T$^{-1}$ & $ 2.4\times 10^{-7}$ \\
    & & $ 29.164\,6935(69)$ & MHz~T$^{-1}$ & $ 2.4\times 10^{-7}$ \\ \\
\multicolumn {5} {c} {\vbb Deuteron, d} \\
deuteron mass & $m_{\rm d}$ & $ 3.343\,583\,7768(10)\times 10^{-27}$ & kg & $ 3.1\times 10^{-10}$ \\
& & $ 2.013\,553\,212\,544(15)$ & u & $ 7.4\times 10^{-12}$ \\
\b energy equivalent & $m_{\rm d}c^2$ & $ 3.005\,063\,234\,91(94)\times 10^{-10}$ & J & $ 3.1\times 10^{-10}$ \\
& & $ 1875.612\,945\,00(58)$ & MeV & $ 3.1\times 10^{-10}$ \\
deuteron-electron mass ratio & $m_{\rm d}/m_{\rm e}$ & $ 3670.482\,967\,655(63)$ & & $ 1.7\times 10^{-11}$ \\
deuteron-proton mass ratio & $m_{\rm d}/m_{\rm p}$ & $ 1.999\,007\,501\,2699(84)$ & & $ 4.2\times 10^{-12}$ \\
deuteron molar mass $N_{\rm A}m_{\rm d}$& $M({\rm d}),M_{\rm d}$ & $ 2.013\,553\,214\,66(63)\times 10^{-3}$ & kg mol$^{-1}$ & $ 3.1\times 10^{-10}$ \\
deuteron rms charge radius & $r_{\rm d}$ & $ 2.127\,78(27)\times 10^{-15}$ & m & $ 1.3\times 10^{-4}$ \\
deuteron magnetic moment & $\mu_{\rm d}$ & $ 4.330\,735\,087(11)\times 10^{-27}$ & J~T$^{-1}$ & $ 2.6\times 10^{-9}$ \\
\b to Bohr magneton ratio & $\mu_{\rm d}/\mu_{\rm B}$ & $ 4.669\,754\,568(12)\times 10^{-4}$ & & $ 2.6\times 10^{-9}$ \\
\b to nuclear magneton ratio & $\mu_{\rm d}/\mu_{\rm N}$ & $ 0.857\,438\,2335(22)$ & & $ 2.6\times 10^{-9}$ \\
deuteron $g$-factor $\mu_{\rm d}/\mu_{\rm N}$ & $g_{\rm d}$ & $ 0.857\,438\,2335(22)$ & & $ 2.6\times 10^{-9}$ \\
deuteron-electron magnetic-moment ratio & $\mu_{\rm d}/\mu_{\rm e}$ & $ -4.664\,345\,550(12)\times 10^{-4}$ & & $ 2.6\times 10^{-9}$ \\
deuteron-proton magnetic-moment ratio & $\mu_{\rm d}/\mu_{\rm p}$ & $ 0.307\,012\,209\,30(79)$ & & $ 2.6\times 10^{-9}$ \\
deuteron-neutron magnetic-moment ratio & $\mu_{\rm d}/\mu_{\rm n}$ & $ -0.448\,206\,52(11)$ & & $ 2.4\times 10^{-7}$ \\ \\
\multicolumn {5} {c} {\vbb Triton, t} \\
triton mass & $m_{\rm t}$ & $ 5.007\,356\,7512(16)\times 10^{-27}$ & kg & $ 3.1\times 10^{-10}$ \\
& & $ 3.015\,500\,715\,97(10)$ & u & $ 3.4\times 10^{-11}$ \\
\b energy equivalent & $m_{\rm t}c^2$ & $ 4.500\,387\,8119(14)\times 10^{-10}$ & J & $ 3.1\times 10^{-10}$ \\
& & $ 2808.921\,136\,68(88)$ & MeV & $ 3.1\times 10^{-10}$ \\
triton-electron mass ratio & $m_{\rm t}/m_{\rm e}$ & $ 5496.921\,535\,51(21)$ & & $ 3.8\times 10^{-11}$ \\
triton-proton mass ratio & $m_{\rm t}/m_{\rm p}$ & $ 2.993\,717\,034\,03(10)$ & & $ 3.4\times 10^{-11}$ \\
triton molar mass $N_{\rm A}m_{\rm t}$& $M({\rm t}),M_{\rm t}$ & $ 3.015\,500\,719\,13(94)\times 10^{-3}$ & kg mol$^{-1}$ & $ 3.1\times 10^{-10}$ \\
triton magnetic moment & $\mu_{\rm t}$ & $ 1.504\,609\,5178(30)\times 10^{-26}$ & J~T$^{-1}$ & $ 2.0\times 10^{-9}$ \\
\b to Bohr magneton ratio & $\mu_{\rm t}/\mu_{\rm B}$ & $ 1.622\,393\,6648(32)\times 10^{-3}$ & & $ 2.0\times 10^{-9}$ \\
\b to nuclear magneton ratio & $\mu_{\rm t}/\mu_{\rm N}$ & $ 2.978\,962\,4650(59)$ & & $ 2.0\times 10^{-9}$ \\
triton $g$-factor $2\mu_{\rm t}/\mu_{\rm N}$ & $g_{\rm t}$ & $ 5.957\,924\,930(12)$ & & $ 2.0\times 10^{-9}$ \\ \\
\multicolumn {5} {c} {\vbb Helion, h} \\
helion mass & $m_{\rm h}$ & $ 5.006\,412\,7862(16)\times 10^{-27}$ & kg & $ 3.1\times 10^{-10}$ \\
& & $ 3.014\,932\,246\,932(74)$ & u & $ 2.5\times 10^{-11}$ \\
\b energy equivalent & $m_{\rm h}c^2$ & $ 4.499\,539\,4185(14)\times 10^{-10}$ & J & $ 3.1\times 10^{-10}$ \\
& & $ 2808.391\,611\,12(88)$ & MeV & $ 3.1\times 10^{-10}$ \\
helion-electron mass ratio & $m_{\rm h}/m_{\rm e}$ & $ 5495.885\,279\,84(16)$ & & $ 2.9\times 10^{-11}$ \\
helion-proton mass ratio & $m_{\rm h}/m_{\rm p}$ & $ 2.993\,152\,671\,552(70)$ & & $ 2.4\times 10^{-11}$ \\
helion molar mass $N_{\rm A}m_{\rm h}$& $M({\rm h}),M_{\rm h}$ & $ 3.014\,932\,250\,10(94)\times 10^{-3}$ & kg mol$^{-1}$ & $ 3.1\times 10^{-10}$ \\
helion magnetic moment & $\mu_{\rm h}$ & $ -1.074\,617\,551\,98(93)\times 10^{-26}$ & J~T$^{-1}$ & $ 8.7\times 10^{-10}$ \\
\b to Bohr magneton ratio & $\mu_{\rm h}/\mu_{\rm B}$ & $ -1.158\,740\,980\,83(94)\times 10^{-3}$ & & $ 8.1\times 10^{-10}$ \\
\b to nuclear magneton ratio & $\mu_{\rm h}/\mu_{\rm N}$ & $ -2.127\,625\,3498(17)$ & & $ 8.1\times 10^{-10}$ \\
helion $g$-factor $2\mu_{\rm h}/\mu_{\rm N}$ & $g_{\rm h}$ & $ -4.255\,250\,6995(34)$ & & $ 8.1\times 10^{-10}$ \\
bound helion magnetic moment & $\mu_{\rm h}(^3{\rm He})$ & $ -1.074\,553\,110\,35(93)\times 10^{-26}$ & J~T$^{-1}$ & $ 8.7\times 10^{-10}$ \\
\b to Bohr magneton ratio & $\mu_{\rm h}(^3{\rm He})/\mu_{\rm B}$ & $ -1.158\,671\,494\,57(94)\times 10^{-3}$ & & $ 8.1\times 10^{-10}$ \\
\b to nuclear magneton ratio & $\mu_{\rm h}(^3{\rm He})/\mu_{\rm N}$ & $ -2.127\,497\,7624(17)$ & & $ 8.1\times 10^{-10}$ \\
bound helion to proton magnetic-& & & & \\
\b moment ratio & $\mu_{\rm h}(^3{\rm He})/\mu_{\rm p}$ & $ -0.761\,766\,577\,21(66)$ & & $ 8.6\times 10^{-10}$ \\
bound helion to shielded proton magnetic-& & & & \\
\b moment ratio \ (H$_2$O, sphere, 25 $^\circ$C) & $\mu_{\rm h}(^3{\rm He})/\mu^\prime_{\rm p}$ & $ -0.761\,786\,1334(31)$ & & $ 4.0\times 10^{-9}$ \\
bound helion gyromagnetic ratio $2|\mu_{\rm h}(^3{\rm He})|/\hbar$ & $\gamma_{\rm h}(^3{\rm He})$ & $ 2.037\,894\,6078(18)\times 10^{8}$ & s$^{-1}$ T$^{-1}$ & $ 8.7\times 10^{-10}$ \\ 
   &  & $ 32.434\,100\,033(28)$ & MHz~T$^{-1}$ & $ 8.7\times 10^{-10}$ \\ \\
\multicolumn {5} {c} {\vbb Alpha particle, ${\rmalpha}$} \\
alpha particle mass & $m_{\rmssalpha}$ & $ 6.644\,657\,3450(21)\times 10^{-27}$ & kg & $ 3.1\times 10^{-10}$ \\
& & $ 4.001\,506\,179\,129(62)$ & u & $ 1.6\times 10^{-11}$ \\
\b energy equivalent & $m_{\rmssalpha}c^2$ & $ 5.971\,920\,1997(19)\times 10^{-10}$ & J & $ 3.1\times 10^{-10}$ \\
& & $ 3727.379\,4118(12)$ & MeV & $ 3.1\times 10^{-10}$ \\
alpha particle to electron mass ratio & $m_{\rmssalpha}/m_{\rm e}$ & $ 7294.299\,541\,71(17)$ & & $ 2.4\times 10^{-11}$ \\
alpha particle to proton mass ratio & $m_{\rmssalpha}/m_{\rm p}$ & $ 3.972\,599\,690\,252(70)$ & & $ 1.8\times 10^{-11}$ \\
alpha particle molar mass $N_{\rm A}m_{\rmssalpha}$& $M({\rmalpha}),M_{\rmssalpha}$ & $ 4.001\,506\,1833(12)\times 10^{-3}$ & kg mol$^{-1}$ & $ 3.1\times 10^{-10}$ \\
alpha particle rms charge radius & $r_{\rmssalpha}$ & $ 1.6785(21)\times 10^{-15} $ & m
&  $ 1.2\times 10^{-3}$ \\ \\
\multicolumn {5} {c} {\vbb PHYSICOCHEMICAL} \\
Avogadro constant & $N_{\rm A}$ & $ 6.022\,140\,76\times 10^{23}$ & mol$^{-1}$ & exact \\
Boltzmann constant  & $k$ & $ 1.380\,649\times 10^{-23}$ & J~K$^{-1}$ & exact \\
& & $ 8.617\,333\,262\ldots\times 10^{-5}$ & eV~K$^{-1}$ & exact \\
\b & $k/h$ & $ 2.083\,661\,912\ldots\times 10^{10}$ & Hz~K$^{-1}$ & exact \\
\b & {$k/h c$} & $ 69.503\,480\,04\ldots$ & [m$^{-1}~$K$^{-1}$]$^{\rm a}$ & exact \\
atomic mass constant$^{\rm c}$ & & & & \\
\, {$m_{\rm u}=\frac {1}{12}m(^{12}{\rm C})=2 hc\,R_\infty/\alpha^2 c^2 A_{\rm r}({\rm e})$} & 
                             $m_{\rm u}$ & $ 1.660\,539\,068\,92(52)\times 10^{-27}$ & kg & $ 3.1\times 10^{-10}$ \\
\b energy equivalent & $m_{\rm u}c^2$ & $ 1.492\,418\,087\,68(46)\times 10^{-10}$ & J & $ 3.1\times 10^{-10}$ \\
& & $ 931.494\,103\,72(29)$ & MeV & $ 3.1\times 10^{-10}$ \\
molar mass constant$^{\rm e}$ & $M_{\rm u}$ & $ 1.000\,000\,001\,05(31)\times 10^{-3}$ & kg mol$^{-1}$ & $ 3.1\times 10^{-10}$ \\
molar mass$^{\rm e}$ of carbon-12 $A_{\rm r}(^{12}{\rm C})M_{\rm u}$ & $M(^{12}{\rm C})$ & $ 12.000\,000\,0126(37)\times 10^{-3}$ & kg mol$^{-1}$ & $ 3.1\times 10^{-10}$ \\
molar Planck constant & $N_{\rm A}h$ & $ 3.990\,312\,712\ldots\times 10^{-10}$ & {J~Hz$^{-1}$~mol$^{-1}$} & exact \\
molar gas constant $N_{\rm A}k$ & $R$ & $ 8.314\,462\,618\ldots$ & J~mol$^{-1}~$K$^{-1}$ & exact \\
Faraday constant $N_{\rm A}e$ & $F$ & $ 96\,485.332\,12\ldots$ & C~mol$^{-1}$ & exact \\
standard-state pressure  & &   100\,000 & Pa  & exact \\
standard atmosphere  & &   101\,325 & Pa  & exact \\
molar volume of ideal gas $RT/p$ & & & & \\
\b\b $T=273.15\ {\rm K},\, p=100\ {\rm kPa}$ & $V_{\rm m}$ & $ 22.710\,954\,64\ldots\times 10^{-3}$ & m$^{3}~$mol$^{-1}$ & exact \\
\b\b\b\b or standard-state pressure &  & & &\\
\b Loschmidt constant $N_{\rm A}/V_{\rm m}$ & $n_0$ & $ 2.651\,645\,804\ldots\times 10^{25}$ & m$^{-3}$ & exact \\
molar volume of ideal gas $RT/p$ & & & & \\
\b\b $T=273.15\ {\rm K},\,  p=101.325\ {\rm kPa}$ & $V_{\rm m}$ & $ 22.413\,969\,54\ldots\times 10^{-3}$ & m$^{3}~$mol$^{-1}$ & exact \\
\b\b\b\b or standard atmosphere &  & & &\\
\b Loschmidt constant $N_{\rm A}/V_{\rm m}$ & $n_0$ & $ 2.686\,780\,111\ldots\times 10^{25}$ & m$^{-3}$ & exact \\
Sackur-Tetrode (absolute entropy) constant$^{\rm f}$   & & & & \\
\, {$\frac {5} {2}+\ln[(m_{\rm u}kT_1/2\rmpi \hbar^2)^{3/2}kT_1/p_0]$} & & & & \\
\b $T_1=1\ {\rm K},\, p_0=100\ {\rm kPa}$ & $S_0/R$ & $ -1.151\,707\,534\,96(47)$ & & $ 4.1\times 10^{-10}$ \\
\b\b\b\b or standard-state pressure &  & & &\\
\b $T_1=1\ {\rm K},\, p_0=101.325\ {\rm kPa}$ & & $ -1.164\,870\,521\,49(47)$ & & $ 4.0\times 10^{-10}$ \\
\b\b\b\b or standard atmosphere &  & & &\\
Stefan-Boltzmann constant & & & & \\
\, ($\rmpi^2/60)k^4\!/\hbar^3c^2$ & $\sigma$ & $ 5.670\,374\,419\ldots\times 10^{-8}$ & W~m$^{-2}~$K$^{-4}$ & exact \\
first radiation constant for spectral &     &   &   & \\
\b radiance {$2 h c^2\,{\rm sr}^{-1}$} & $c_{\rm 1L}$ & $ 1.191\,042\,972\ldots\times 10^{-16}$ & [W~m$^{2}$~sr$^{-1}$]$^{\rm g}$ & exact \vbox to 10 pt{}\\
first radiation constant {$2\rmpi h c^2=\rmpi\, {\rm sr}\,c_{1\rm L}$} & $c_1$ & $ 3.741\,771\,852\ldots\times 10^{-16}$ &  [W~m$^{2}$]$^{\rm g}$ & exact \\
second radiation constant {$h c/k$} & $c_2$ & $ 1.438\,776\,877\ldots\times 10^{-2}$ & [m~K]$^{\rm a}$ & exact \\
Wien displacement law constants & & & & \\
\, $b=\lambda_{\rm max}T=c_2/4.965\,114\,231...$ & $b$ & $ 2.897\,771\,955\ldots\times 10^{-3}$ &  [m~K]$^{\rm a}$ & exact \\
\, $b^\prime=\nu_{\rm max}/T=2.821\,439\,372...\,c/c_2 $ & $b^\prime$ & $ 5.878\,925\,757\ldots\times 10^{10}$ & Hz~K$^{-1}$ & exact \\ \\
\hline
\hline
\end{longtable*}

\begin{widetext}
{\small 
$^{\rm a}${The full description of [m]$^{-1}$ is cycles or periods
per meter and that of [m] is meter per cycle (m/cycle).  The scientific
community is aware of the implied use of these units.  It traces back to
the conventions for phase and angle and the use of unit Hz versus
cycles/s. No solution has been agreed upon.}

$^{\rm b}$Value recommended by the PDG \numcite{2022048}.

$^{\rm c}$Based on the ratio of the masses of the W and Z bosons $m_{\rm
W}/m_{\rm Z}$ recommended by the PDG \numcite{2022048}. The value for
${\rm sin}^2{\theta}_{\rm W}$ they recommend, which is based on a
variant of the modified minimal subtraction $({\scriptstyle {\rm
\overline{MS}}})$ scheme, is ${\rm sin}^2\hat{\theta}_{\rm W}(M_{\rm Z})
= 0.223\,05(25)$.

$^{\rm d}$This and other constants involving $m_{\rmsstau}$ are based on
$m_{\rmsstau}c^2$ in MeV recommended by the PDG \numcite{2022048}.

$^{\rm e}$The relative atomic mass $A_{\rm r}(X)$ of particle $X$
with mass $m(X)$ is defined by $A_{\rm r}(X) = m(X)/m_{\rm u}$, where
$m_{\rm u} = m(^{12}{\rm C})/12 = 1~{\rm u}$ is the atomic mass constant
and u is the unified atomic mass unit.  Moreover, the mass of particle
$X$ is $m(X) = A_{\rm r}(X)$~u and the molar mass of $X$ is $M(X) =
A_{\rm r}(X)M_{\rm u}$, where $M_{\rm u}=N_{\rm A}\ {\rm u}$ is the
molar mass constant and $N_{\rm A}$ is the Avogadro constant.

$^{\rm f}$The entropy of an ideal monoatomic gas of relative atomic mass
$A_{\rm r}$ is given by $S = S_0 +\frac{3}{2} R\, \ln A_{\rm r} -R\,
\ln(p/p_0) +\frac{5}{2}R\,\ln(T/{\rm K}).$

$^{\rm g}${The full description of [m]$^2$ is ${\rm m}^{-2}\times
({\rm m/cycle})^4$. See also the first footnote.}}
\end{widetext}

\clearpage

\begin{table}[h!]
\caption[Values of some x-ray-related quantities]{Values of some
x-ray-related quantities based on the 2022 CODATA adjustment of the
constants.}
\label{tab:xrayvalues}

\tablehead{
\hline\hline
 &&&& \multicolumn{1}{c}{Relative std.}    \\
 \multicolumn{1}{c}{Quantity}& \multicolumn{1}{c}{Symbol} & \multicolumn{1}{c}{Value}  &  
 \multicolumn{1}{c}{Unit}   & \multicolumn{1}{c}{uncert. $u_{\rm r}$ } \\
 \hline
 }

\noindent

\renewcommand{\thefootnote}{\fnsymbol{footnote}}
\begin{center}
\begin{supertabular}{lllll}
Cu x unit: $\lambda({\rm CuK}{\rm \alpha}_{\rm 1}) / 1\,537.400 $ & ${\rm xu}({\rm CuK}{\rm \alpha}_{\rm 1})$ & $ 1.002\,076\,97(28)\times 10^{-13}$ & m & $ 2.8\times 10^{-7}$ 
\vbox to 12 pt {} \\ Mo x unit: $\lambda({\rm MoK}{\rm \alpha}_{\rm 1}) / 707.831 $ & ${\rm xu}({\rm MoK}{\rm \alpha}_{\rm 1})$ & $ 1.002\,099\,52(53)\times 10^{-13}$ & m & $ 5.3\times 10^{-7}$ \\
{\AA}ngstr\"om star$: \lambda({\rm WK}{\rm \alpha}_{\rm 1}) / 0.209\,010\,0 $ & \AA$^{\ast}$ & $ 1.000\,014\,95(90)\times 10^{-10}$ & m & $ 9.0\times 10^{-7}$ \\
lattice parameter\footnotemark[1]   of Si \ (in vacuum, 22.5 $^\circ$C)~~ & $a$ & $ 5.431\,020\,511(89)\times 10^{-10}$ & m & $ 1.6\times 10^{-8}$ \vbox to 9 pt {}\\
\{220\} lattice spacing of Si $a/\sqrt{8}$ & $d_{\rm 220}$ & $ 1.920\,155\,716(32)\times 10^{-10}$ & m & $ 1.6\times 10^{-8}$ \\
\,\, (in vacuum, 22.5 $^\circ$C)&&&&\\
molar volume of Si \ $M({\rm Si})/\rho({\rm Si})=N_{\rm A}a^{3}\!/8$ & $V_{\rm m}$(Si) & $ 1.205\,883\,199(60)\times 10^{-5}$ & m$^{3}$ mol$^{-1}$ & $ 4.9\times 10^{-8}$ \\
\,\, (in vacuum, 22.5 $^\circ$C)&&&&\\ \\
\hline
\hline
\end{supertabular}
\end{center}   %

{\small
\footnotemark[1] This is the lattice parameter (unit cell edge
length) of an ideal single crystal of naturally occurring Si 
with natural isotopic Si abundances, free of impurities and
imperfections. \\ }
\end{table}
\begin{table*}
\caption[Values for non-SI units]{Non-SI units based on the 2022 CODATA
adjustment of the constants, although eV and u are accepted for use with
the SI.} 
\label{tab:nonsi}
\begin{tabular}{lllll}
\hline\hline
 &&&& \multicolumn{1}{c}{Relative std.}    \\
\multicolumn{1}{c}{Quantity}& \multicolumn{1}{c}{Symbol} & \multicolumn{1}{c}{Value}  &  
\multicolumn{1}{c}{Unit}   & \multicolumn{1}{c}{uncert. $u_{\rm r}$ } \\
\hline
electron volt: ($e/{\rm C}$) {\rm J} & eV & $ 1.602\,176\,634\times 10^{-19}$ & J & exact \\
(unified) atomic mass unit: ${1\over12}m(^{12}$C)~~ & u & $ 1.660\,539\,068\,92(52)\times 10^{-27}$ & kg & $ 3.1\times 10^{-10}$ \\
\multicolumn {5} {c} {} \\ \multicolumn {5} {c} {Natural units (n.u.)} \\ n.u. of velocity & $c$ & $ 299\,792\,458$ & m s$^{-1}$ & exact \\
n.u. of action  & $\hbar$ & $ 1.054\,571\,817\ldots\times 10^{-34}$ & J s & exact \\
& & $ 6.582\,119\,569\ldots\times 10^{-16}$ & eV s & exact \\
& $\hbar c$ & $ 197.326\,980\,4\ldots$ & MeV fm & exact \\
n.u. of mass & $m_{\rm e}$ & $ 9.109\,383\,7139(28)\times 10^{-31}$ & kg & $ 3.1\times 10^{-10}$ \\
n.u. of energy & $m_{\rm e}c^2$ & $ 8.187\,105\,7880(26)\times 10^{-14}$ & J & $ 3.1\times 10^{-10}$ \\
& & $ 0.510\,998\,950\,69(16)$ & MeV & $ 3.1\times 10^{-10}$ \\
n.u. of momentum & $m_{\rm e}c$ & $ 2.730\,924\,534\,46(85)\times 10^{-22}$ & kg m s$^{-1}$ & $ 3.1\times 10^{-10}$ \\
& & $ 0.510\,998\,950\,69(16)$ & MeV/$c$ & $ 3.1\times 10^{-10}$ \\
n.u. of length: $\hbar/m_{\rm e}c$& $\lbar_{\rm C}$ & $ 3.861\,592\,6744(12)\times 10^{-13}$ & m & $ 3.1\times 10^{-10}$ \\
n.u. of time & $\hbar/m_{\rm e}c^2$ & $ 1.288\,088\,666\,44(40)\times 10^{-21}$ & s & $ 3.1\times 10^{-10}$ \\
\multicolumn {5} {c} {} \\ \multicolumn {5} {c} {Atomic units (a.u.)} \\ a.u. of charge & $e$ & $ 1.602\,176\,634\times 10^{-19}$ & C & exact \\
a.u. of mass & $m_{\rm e}$ & $ 9.109\,383\,7139(28)\times 10^{-31}$ & kg & $ 3.1\times 10^{-10}$ \\
a.u. of action & $\hbar$ & $ 1.054\,571\,817\ldots\times 10^{-34}$ & J s & exact \\
a.u. of length: Bohr radius (bohr)~~& & & & \\
\, $\hbar/\alpha m_{\rm e}c$& $a_0$ & $ 5.291\,772\,105\,44(82)\times 10^{-11}$ & m & $ 1.6\times 10^{-10}$ \\
a.u. of energy: Hartree energy (hartree)~~ & & & & \\
\, $\alpha^2m_{\rm e}c^2=e^2\!/4\rmpi\epsilon_0a_0=2hc\,R_\infty$ & $E_{\rm h}$ & $ 4.359\,744\,722\,2060(48)\times 10^{-18}$ & J & $ 1.1\times 10^{-12}$ \\
a.u. of time & $\hbar/E_{\rm h}$ & $ 2.418\,884\,326\,5864(26)\times 10^{-17}$ & s & $ 1.1\times 10^{-12}$ \\
a.u. of force & $E_{\rm h}/a_0$ & $ 8.238\,723\,5038(13)\times 10^{-8}$ & N & $ 1.6\times 10^{-10}$ \\
a.u. of velocity: $\alpha c$ & $a_0E_{\rm h}/\hbar$ & $ 2.187\,691\,262\,16(34)\times 10^{6}$ & m s$^{-1}$ & $ 1.6\times 10^{-10}$ \\
a.u. of momentum & $\hbar/a_0$ & $ 1.992\,851\,915\,45(31)\times 10^{-24}$ & kg m s$^{-1}$ & $ 1.6\times 10^{-10}$ \\
a.u. of current & $eE_{\rm h}/\hbar$ & $ 6.623\,618\,237\,5082(72)\times 10^{-3}$ & A & $ 1.1\times 10^{-12}$ \\
a.u. of charge density & $e/a_0^3$ & $ 1.081\,202\,386\,77(51)\times 10^{12}$ & C m$^{-3}$ & $ 4.7\times 10^{-10}$ \\
a.u. of electric potential & $E_{\rm h}/e$ & $ 27.211\,386\,245\,981(30)$ & V & $ 1.1\times 10^{-12}$ \\
a.u. of electric field & $E_{\rm h}/ea_0$ & $ 5.142\,206\,751\,12(80)\times 10^{11}$ & V m$^{-1}$ & $ 1.6\times 10^{-10}$ \\
a.u. of electric field gradient & $E_{\rm h}/ea_0^2$ & $ 9.717\,362\,4424(30)\times 10^{21}$ & V m$^{-2}$ & $ 3.1\times 10^{-10}$ \\
a.u. of electric dipole moment & $ea_0$ & $ 8.478\,353\,6198(13)\times 10^{-30}$ & C m & $ 1.6\times 10^{-10}$ \\
a.u. of electric quadrupole moment & $ea_0^2$ & $ 4.486\,551\,5185(14)\times 10^{-40}$ & C m$^2$ & $ 3.1\times 10^{-10}$ \\
a.u. of electric polarizability & $e^2a_0^2/E_{\rm h}$ & $ 1.648\,777\,272\,12(51)\times 10^{-41}$ & C$^2$ m$^2$ J$^{-1}$ & $ 3.1\times 10^{-10}$ \\
a.u. of 1$^{\rm st}$ hyperpolarizability & $e^3a_0^3/E_{\rm h}^2$ & $ 3.206\,361\,2996(15)\times 10^{-53}$ & C$^3$ m$^3$ J$^{-2}$ & $ 4.7\times 10^{-10}$ \\
a.u. of 2$^{\rm nd}$ hyperpolarizability & $e^4a_0^4/E_{\rm h}^3$ & $ 6.235\,379\,9735(39)\times 10^{-65}$ & C$^4$ m$^4$ J$^{-3}$ & $ 6.2\times 10^{-10}$ \\
a.u. of magnetic flux density & $\hbar/ea_0^2$ & $ 2.350\,517\,570\,77(73)\times 10^{5}$ & T & $ 3.1\times 10^{-10}$ \\
a.u. of magnetic dipole moment: $2\mu_{\rm B}$ & $\hbar e/m_{\rm e}$ & $ 1.854\,802\,013\,15(58)\times 10^{-23}$ & J T$^{-1}$ & $ 3.1\times 10^{-10}$ \\
a.u. of magnetizability & $e^2a_0^2/m_{\rm e}$ & $ 7.891\,036\,5794(49)\times 10^{-29}$ & J T$^{-2}$ & $ 6.2\times 10^{-10}$ \\
a.u. of permittivity    & $e^2/a_0E_{\rm h}$ & $ 1.112\,650\,056\,20(17)\times 10^{-10}$ & F m$^{-1}$ & $ 1.6\times 10^{-10}$ \\ \\
\hline
\hline
\end{tabular}
\end{table*}

\def\hsp{\hbox to 7 pt {}}
\begin{turnpage}
\begin{table*}[!]
\def\vb{\vbox to 8 pt{}}
\caption[Values of some energy equivalents I.]{The values of some energy
equivalents derived from the relations $E=mc^2 = h c/\lambda = h\nu =
kT$ and based on the 2022 CODATA adjustment of the values of the
constants; 1~eV~$=(e/{\rm C})$~J, 1~u $= m_{\rm u} =
\textstyle{1\over12}m(^{12}{\rm C})$, and $E_{\rm h} = 2 hc\, R_{\rm
\infty} = \alpha^2m_{\rm e}c^2$ is the Hartree energy (hartree).  }
\label{tab:enconv1}
\begin{tabular} {l@{\hsp}l@{\hsp}l@{\hsp}l@{\hsp}l}
\hline
\hline
\multicolumn{5} {c} {Relevant unit} \\
\hline

&   \s{30}J &  \s{30}kg &  \s{30}[m]$^{-1}$\footnote{The full
description of [m]$^{-1}$ is cycles or periods per meter.}&   \s{30}Hz
\vb   \\
\hline
&&&&\\
1~J    & $(1\ {\rm J})=$    &  (1 J)/$c^2=$     &        (1 J)/$h c=$   &        (1 J)/$h=$              \\
 & 1 J   & $ 1.112\,650\,056\ldots\times 10^{-17}\,$kg  & $ 5.034\,116\,567\ldots\times 10^{24}\,$m$^{-1}$ & $ 1.509\,190\,179\ldots\times 10^{33}\,$Hz \\
        &        &             &           &  \\
1~kg    &        (1 kg)$c^2=$   &   $(1 \ {\rm kg})=$       &     (1 kg)$c/h =$ &        (1 kg)$c^2/h=$          \\
 &  $ 8.987\,551\,787\ldots\times 10^{16}\,$J & 1 kg   & $ 4.524\,438\,335\ldots\times 10^{41}\,$m$^{-1}$ & $ 1.356\,392\,489\ldots\times 10^{50}\,$Hz \\
 &           &            &         &   \\
 1~[m$^{-1}$]\footnotemark[1] &   (1 m$^{-1})h c=$   &        (1 m$^{-1})h/c=$    &  $(1$ m$^{-1})=$    &    (1 m$^{-1}) c =$  \\
 &  $ 1.986\,445\,857\ldots\times 10^{-25}\,$J  & $ 2.210\,219\,094\ldots\times 10^{-42}\,$kg  & 1 m$^{-1}$  & $ 299\,792\,458\,$Hz \\
 &           &            &         &   \\
1~Hz   &  (1 Hz)$h=$  &  (1 Hz)$h/c^{2}=$   &        (1 Hz)$/c=$  &  $(1$ Hz$)=$  \\
 &  $ 6.626\,070\,15\times 10^{-34}\,$J  & $ 7.372\,497\,323\ldots\times 10^{-51}\,$kg & $ 3.335\,640\,951\ldots\times 10^{-9}\,$m$^{-1}$ & 1 Hz \\
 &           &            &         &   \\
1~K  &  (1 K)$k=$  &   (1 K)$k/c^{2}=$  &     (1 K)$k/h c=$ &  (1 K)$k/h=$   \\
 &  $ 1.380\,649\times 10^{-23}\,$J  & $ 1.536\,179\,187\ldots\times 10^{-40}\,$kg & $ 69.503\,480\,04\ldots\,$m$^{-1}$ & $ 2.083\,661\,912\ldots\times 10^{10}\,$Hz \\
         &            &            &           &        \\
1~eV    &  (1 eV) =  &  $(1~{\rm eV})/c^{2}=$  &    $(1~{\rm eV})/h c=$   &     $(1~{\rm eV})/h=$   \\
 &  $ 1.602\,176\,634\times 10^{-19}\,$J  & $ 1.782\,661\,921\ldots\times 10^{-36}\,$kg & $ 8.065\,543\,937\ldots\times 10^{5}\,$m$^{-1}$ & $ 2.417\,989\,242\ldots\times 10^{14}\,$Hz \\
         &          &             &         &     \\
1~u   &   $(1~{\rm u})c^{2}=$   &  (1 u) =   &   $(1~{\rm u})c/h =$  &  $(1~{\rm u})c^{2}/h=$    \\
 &  $ 1.492\,418\,087\,68(46)\times 10^{-10}\,$J  & $ 1.660\,539\,068\,92(52)\times 10^{-27}\,$kg & $ 7.513\,006\,6209(23)\times 10^{14}\,$m$^{-1}$ & $ 2.252\,342\,721\,85(70)\times 10^{23}\,$Hz \\
         &          &             &        &       \\
1~$E_{\rm h}$   &  $(1~E_{\rm h})=$  & $(1~E_{\rm h})/c^2=$ &   $(1~E_{\rm h})/h c=$  &  $(1~E_{\rm h})/h=$   \\
 &  $ 4.359\,744\,722\,2060(48)\times 10^{-18}\,$J  & $ 4.850\,870\,209\,5419(53)\times 10^{-35}\,$kg & $ 2.194\,746\,313\,6314(24)\times 10^{7}\,$m$^{-1}$ & $ 6.579\,683\,920\,4999(72)\times 10^{15}\,$Hz \\ \\
\hline
\hline
\end{tabular}
\end{table*}
\end{turnpage}

\def\hsp{\hbox to 13 pt {}}
\begin{turnpage}
\def\vb{\vbox to 8 pt {}}
\begin{table*}

\caption[Values of some energy equivalents II]{The values of some energy
equivalents derived from the relations $E=mc^2 =  \hbar c/\lambda = h\nu
= kT$ and based on the 2022 CODATA adjustment of the values of the
constants; 1~eV~$=(e/{\rm C})$~J, 1~u $= m_{\rm u} =
\textstyle{1\over12}m(^{12}{\rm C})$, and $E_{\rm h} =  2 hc\, R_{\rm
\infty} = \alpha^2m_{\rm e}c^2$ is the Hartree energy (hartree).  }
\label{tab:enconv2}
\begin{tabular} {l@{\hsp}l@{\hsp}l@{\hsp}l@{\hsp}l}
\hline
\hline
\multicolumn{5} {c} {Relevant unit} \\
\hline
 &   \s{30}K &  \s{30}eV &  \s{30}u  &   \s{30}$E_{\rm h}$  \vb   \\
\hline
 &&&&\\
1~J     & (1 J)/$k=$   &  (1 J) =   &       (1 J)/$c^2$ = &   (1 J) =              \\
 &  $ 7.242\,970\,516\ldots\times 10^{22}$ K  & $ 6.241\,509\,074\ldots\times 10^{18}$ eV  & $ 6.700\,535\,2471(21)\times 10^{9}$ u & $ 2.293\,712\,278\,3969(25)\times 10^{17} \, E_{\rm h}$ \\
        &        &             &           &  \\
1~kg    &        (1 kg)$c^2/k=$    &   (1 kg)$c^2$ =    &    (1 kg) =  &  (1 kg)$c^2=$   \\
 &  $ 6.509\,657\,260\ldots\times 10^{39}$ K  & $ 5.609\,588\,603\ldots\times 10^{35}$ eV  & $ 6.022\,140\,7537(19)\times 10^{26}$ u  & $ 2.061\,485\,788\,7415(22)\times 10^{34} \, E_{\rm h}$ \\
 &           &            &         &   \\
 1~[m]$^{-1}$\footnote{The full description of [m]$^{-1}$ is cycles or periods per
 meter.}   &  (1 [m]$^{-1})h c/k=$  &    (1 [m]$^{-1})h c=$  &  (1
 [m]$^{-1})h/c$ =   &  (1 [m]$^{-1})h c=$ \\
 &  $ 1.438\,776\,877\ldots\times 10^{-2}$ K  & $ 1.239\,841\,984\ldots\times 10^{-6}$ eV  & $ 1.331\,025\,048\,24(41)\times 10^{-15}$ u & $ 4.556\,335\,252\,9132(50)\times 10^{-8} \, E_{\rm h}$ \\
         &           &            &            &      \\
1~Hz   &  (1 Hz)$h/k=$  &  (1 Hz)$h=$  &   (1 Hz)$h/c^2$ = & (1 Hz)$h=$  \\
 &  $ 4.799\,243\,073\ldots\times 10^{-11}$ K  & $ 4.135\,667\,696\ldots\times 10^{-15}$ eV  & $ 4.439\,821\,6590(14)\times 10^{-24}$ u & $ 1.519\,829\,846\,0574(17)\times 10^{-16} \, E_{\rm h}$ \\
           &             &          &           &        \\ 
1~K  & $(1$ K$)=$  &   (1 K)$k=$ &    (1 K)$k/c^2=$ &  (1 K)$k=$   \\
 & 1 K   &   $ 8.617\,333\,262\ldots\times 10^{-5}$ eV & $ 9.251\,087\,2884(29)\times 10^{-14}$ u & $ 3.166\,811\,563\,4564(35)\times 10^{-6} \, E_{\rm h}$ \\
         &            &            &           &        \\
1~eV    &  (1 eV)/$k=$  &  $(1$ eV$)=$  &   $(1~{\rm eV})/c^2=$  &     $(1~{\rm eV})=$   \\
 &  $ 1.160\,451\,812\ldots\times 10^{4}$ K  & 1 eV  & $ 1.073\,544\,100\,83(33)\times 10^{-9}$ u & $ 3.674\,932\,217\,5665(40)\times 10^{-2} \, E_{\rm h}$ \\
         &          &             &         &     \\
1~u   &   $(1~{\rm u})c^{2}/k=$   &  $(1~{\rm u})c^2=$ &  $(1$ u$)=$  &  $(1~{\rm u})c^2=$   \\
 &  $ 1.080\,954\,020\,67(34)\times 10^{13}$ K  & $ 9.314\,941\,0372(29)\times 10^{8}$ eV  & 1 u & $ 3.423\,177\,6922(11)\times 10^{7} \, E_{\rm h}$ \\
         &          &             &        &       \\
1~$E_{\rm h}$   &  $(1~E_{\rm h})/k=$  & $(1~E_{\rm h})=$ &  $(1~E_{\rm h})/c^2=$ & $(1~E_{\rm h})=$    \\
 &  $ 3.157\,750\,248\,0398(34)\times 10^{5}$ K  & $ 27.211\,386\,245\,981(30)$ eV  & $ 2.921\,262\,317\,97(91)\times 10^{-8}$ u  & $1~E_{\rm h}$  \\ \\
\hline
\hline
\end{tabular}
\end{table*}
\end{turnpage}

\clearpage

\section{Summary and Conclusion}
\label{sec:sumconcl}

Here we (i) compare the 2022 to the 2018 recommended values of a
representative group of  constants with a focus on the changes in their
values since the 2018 adjustment; (ii) discuss some notable features of
the 2022 adjustment; and (iii) identify work that might possibly
eliminate the need for applying expansion factors to the uncertainties
of those constants for which it was necessary in the 2022 adjustment.

\begin{table}[hbt!]
\caption[Comparison of the 2022 and 2018 CODATA recommended values of a
representative group of constants]{Comparison of the 2022 and 2018
CODATA recommended values of a representative group of constants $C_i$.
The column labeled $D_{\rm r}$ is the 2022 absolute value $|C_i(2022)|$
minus the 2018 absolute value $|C_i(2018)|$ divided by the uncertainty
$u_i$ of $C_i(2018)$.  Calculations were performed with extra digits to
eliminate rounding inconsistencies.  A minus sign for a $D_{\rm r}$
value indicates $|C_i(2018)| > |C_i(2022)|$; and an uncertainty ratio
less than 1.0 means the 2018 uncertainty is smaller than the 2022
uncertainty.}
\label{tab:compare}
\begin{tabular}{c@{\qquad}c@{\qquad}r@{\qquad}c@{\qquad}l}
\hline\hline
Item & Constant & $D_{\rm r}$ & $\tfrac{u(2018)}{u(2022)}$ & $u_{\rm
r}(2022)$ \\
\hline \\[-7 pt]
{\it 1 } &$ \alpha                    $&$  -4.5 $&$  0.97 $&$ 1.6\times10^{-10}$ \\
{\it 2 } &$ \mu_0                     $&$  -4.5 $&$  0.97 $&$ 1.6\times10^{-10}$ \\
{\it 3 } &$ \epsilon_0                $&$   4.5 $&$  0.97 $&$ 1.6\times10^{-10}$ \\
{\it 4 } &$ Z_0                       $&$  -4.5 $&$  0.97 $&$ 1.6\times10^{-10}$ \\
{\it 5 } &$ a_0                       $&$  -4.5 $&$  0.97 $&$ 1.6\times10^{-10}$ \\
{\it 6 } &$ \lambda_{\rm C}           $&$  -4.5 $&$  0.97 $&$ 3.1\times10^{-10}$ \\
{\it 7 } &$ r_{\rm e}                 $&$  -4.5 $&$  0.97 $&$ 4.7\times10^{-10}$ \\
{\it 8 } &$ \sigma_{\rm e}            $&$  -4.5 $&$  0.97 $&$ 9.3\times10^{-10}$ \\
{\it 9 } &$ m_{\rm u}                 $&$   4.6 $&$  0.97 $&$ 3.1\times10^{-10}$ \\
{\it 10} &$ M_{\rm u}                 $&$   4.6 $&$  0.97 $&$ 3.1\times10^{-10}$ \\
{\it 11} &$ R_\infty                  $&$  -0.3 $&$  1.74 $&$ 1.1\times10^{-12}$ \\
{\it 12} &$ E_{\rm h}                 $&$  -0.3 $&$  1.74 $&$ 1.1\times10^{-12}$ \\
{\it 13} &$ r_{\rm p}                 $&$  -0.4 $&$  2.91 $&$ 7.7\times10^{ -4}$ \\
{\it 14} &$ r_{\rm d}                 $&$  -0.4 $&$  2.73 $&$ 1.3\times10^{ -4}$ \\
{\it 15} &$ A_{\rm r}({\rm e})        $&$  -1.3 $&$  1.65 $&$ 1.8\times10^{-11}$ \\
{\it 16} &$ A_{\rm r}(\rmmu)          $&$   0.1 $&$  1.00 $&$ 2.2\times10^{ -8}$ \\
{\it 17} &$ A_{\rm r}(\rmtau)         $&$   0.0 $&$  1.00 $&$ 6.8\times10^{ -5}$ \\
{\it 18} &$ A_{\rm r}({\rm p})        $&$  -0.8 $&$  6.36 $&$ 8.3\times10^{-12}$ \\
{\it 19} &$ A_{\rm r}({\rm n})        $&$   0.2 $&$  1.21 $&$ 4.0\times10^{-10}$ \\
{\it 20} &$ A_{\rm r}({\rm d})        $&$  -5.0 $&$  2.70 $&$ 7.4\times10^{-12}$ \\
{\it 21} &$ A_{\rm r}({\rm t})        $&$  -2.0 $&$  1.17 $&$ 3.4\times10^{-11}$ \\
{\it 22} &$ A_{\rm r}({\rm h})        $&$  -2.5 $&$  1.31 $&$ 2.5\times10^{-11}$ \\
{\it 23} &$ A_{\rm r}(\rmalpha)       $&$   0.0 $&$  1.02 $&$ 1.6\times10^{-11}$ \\
{\it 24} &$ m_{\rm e}                 $&$   4.5 $&$  0.97 $&$ 3.1\times10^{-10}$ \\
{\it 25} &$ m_{\rm p}/m_{\rm e}       $&$   0.0 $&$  3.47 $&$ 1.7\times10^{-11}$ \\
{\it 26} &$ m_{\rmssmu}/m_{\rm e}     $&$   0.1 $&$  1.00 $&$ 2.2\times10^{ -8}$ \\
{\it 27} &$ a_{\rm e}                 $&$  -4.7 $&$  0.96 $&$ 1.6\times10^{-10}$ \\
{\it 28} &$ a_{\rmssmu}               $&$  -0.4 $&$  1.54 $&$ 3.5\times10^{ -7}$ \\
{\it 29} &$ g_{\rm e}                 $&$  -4.7 $&$  0.96 $&$ 1.8\times10^{-13}$ \\
{\it 30} &$ g_{\rmssmu}               $&$  -0.4 $&$  1.54 $&$ 4.1\times10^{-10}$ \\
{\it 31} &$ \mu_{\rm B}               $&$  -4.5 $&$  0.97 $&$ 3.1\times10^{-10}$ \\
{\it 32} &$ \mu_{\rm N}               $&$  -4.4 $&$  0.98 $&$ 3.1\times10^{-10}$ \\
{\it 33} &$ \mu_{\rm e}/\mu_{\rm B}   $&$  -4.7 $&$  0.96 $&$ 1.8\times10^{-13}$ \\
{\it 34} &$ \mu_{\rm e}/\mu_{\rmssmu} $&$   0.1 $&$  1.00 $&$ 2.2\times10^{ -8}$ \\
{\it 35} &$ \mu_{\rm e}/\mu_{\rm p}   $&$   0.0 $&$  1.02 $&$ 3.0\times10^{-10}$ \\
{\it 36} &$ \mu_{\rm e}/\mu_{\rm n}   $&$   0.1 $&$  1.00 $&$ 2.4\times10^{ -7}$ \\
{\it 37} &$ \mu_{\rm e}/\mu_{\rm d}   $&$   0.1 $&$  1.00 $&$ 2.6\times10^{ -9}$ \\
{\it 38} &$ \mu_{\rm e}/\mu_{\rm B}   $&$   0.1 $&$  1.00 $&$ 2.2\times10^{ -8}$ \\
{\it 39} &$ \mu_{\rm p}/\mu_{\rm N}   $&$   0.0 $&$  1.00 $&$ 2.9\times10^{-10}$ \\
{\it 40} &$ \mu_{\rm n}/\mu_{\rm N}   $&$   0.1 $&$  1.00 $&$ 2.4\times10^{ -7}$ \\
{\it 41} &$ \mu_{\rm d}/\mu_{\rm N}   $&$  -0.1 $&$  1.00 $&$ 2.6\times10^{ -9}$ \\
{\it 42} &$ \gamma_{\rm p}^\prime     $&$   1.5 $&$  2.66 $&$ 4.1\times10^{ -9}$ \\
{\it 43} &$ \sigma_{\rm p}^\prime     $&$  -1.6 $&$  2.66 $&$ 1.6\times10^{ -4}$ \\
{\it 44} &$ \mu_{\rm h}(^3\mbox{He})  $&$   1.6 $&$ 13.52 $&$ 8.7\times10^{-10}$ \\
\hline\hline
\end{tabular}

\end{table}

\subsection{Comparison of 2022 and 2018 CODATA recommended values}
\label{ssec:comparison}

Table~\ref{tab:compare} compares the 2022 and 2018 recommended values of
a representative group of constants.  However, the constants $c$, $h$,
$e$, $k$, and $N_{\rm A}$ and those that are combinations of them, for
example, the Josephson constant $K_{\rm J} = 2e/h$, molar gas constant
$R = N_{\rm A}k$, and Stefan-Boltzmann constant $\sigma =
(2\rmpi^5/15)k^4/h^3c^2$, are not included in the table.  This is
because for the CODATA 2018 adjustment, these four constants were already
exactly known as a result of the 1983 redefinition of the meter in terms
of an exact value of $c$ and the 2019 revision of the SI which redefined
the kilogram, ampere, kelvin, and mole by assigning exact values to $h$,
$e$, $k$, and $N_{\rm A}$.  Another consequence of the exactness of
these five constants is that the energy equivalency factors in Tables
\ref{tab:enconv1} and \ref{tab:enconv2} for J, kg, m$^{-1}$, Hz, K, and
eV are exactly known.  Also not included in Table \ref{tab:compare} are
the Newtonian gravitational constant $G$ and any of the x-ray-related
quantities in Table \ref{tab:xrayvalues} because their 2022 values are
identical to their 2018 values.

As Table~\ref{tab:compare} shows, the decrease in the recommended value
of the fine-structure constant $\alpha$ by 4.5 times its 2018
uncertainty significantly influences the recommended value of many other
constants.  Indeed, 15 of the other 43 constants listed are so affected.
These and their dependence on $\alpha$ are as follows:\\
\begin{center}
\begin{tabular}{c@{\qquad}l}
{\it 2 } & $\mu_0 = (2h/e^2c)\,\alpha$ \\[2 pt]
{\it 3 } & $\epsilon_0 = (e^2/2hc)\,\alpha^{-1}$ \\[2 pt]
{\it 4 } & $Z_0 = (2h/e^2)\,\alpha$ \\[2 pt]
{\it 5 } & $a_0 = (1/4\rmpi R_\infty)\,\alpha$ \\[2 pt]
{\it 6 } & $\lambda_{\rm C} = (1/2R_\infty)\,\alpha^2$ \\[2 pt]
{\it 7 } & $r_{\rm e} = (1/4\rmpi R_\infty)\,\alpha^3$ \\[2 pt]
{\it 8 } & $\sigma_{\rm e} = (1/6\rmpi R_\infty^2)\,\alpha^6$ \\[2 pt]
{\it 9 } & $m_{\rm u} 
= [2h R_\infty/cA_{\rm r}({\rm e})]\,\alpha^{-2}$ \\[2 pt]
{\it 10} & $M_{\rm u} 
= [2N_{\rm A}h R_\infty/cA_{\rm r}({\rm e})]\,\alpha^{-2}$ \\[2 pt]
{\it 24} & $m_{\rm e} = (2h R_\infty/c)\,\alpha^{-2}$ \\[2 pt]
{\it 27} & $a_{\rm e} \approx (1/2\rmpi)\,\alpha$ \\[2 pt]
{\it 29} & $g_{\rm e} \approx -2(1+\alpha/2\rmpi)$ \\[2 pt]
{\it 31} & $\mu_{\rm B} = (ec/8\rmpi R_\infty)\,\alpha^2$ \\[2 pt]
{\it 32} & $\mu_{\rm N} = [ec/(m_{\rm p}/m_{\rm e})8\rmpi R_\infty]\,\alpha^2$ \\[2 pt]
{\it 33} & $\mu_{\rm e}/\mu_{\rm B} \approx -(1+\alpha/2\rmpi)$
\end{tabular}
\end{center}
The constants $\mu_0$, $\epsilon_0$, and $Z_0$, {\it 2, 3}, and {\it 4},
depend only on exactly known constants and $\alpha^n$, where $n$ is 1,
$-1$, and 1, respectively, hence their relative uncertainties are
identical and equal to that of $\alpha$.  Constants $a_0$, $\lambda_{\rm
C}$, $r_{\rm e}$, and $\sigma_{\rm e}$, {\it 5, 6, 7,} and {\it 8},
depend on $\alpha^n$, where n is 1, 2 , 3, and 6, respectively, hence
their relative uncertainties are 1, 2, 3, and 6 times that of $\alpha$.
(The fact that the relative uncertainty of $\alpha$ is actually
$1.558\times10^{-10}$ explains the uncertainties of $\lambda_{\rm C}$,
$r_{\rm e}$, and $\sigma_{\rm e}$ in the table.) Although  the
expressions for these four constants also contain the Rydberg constant
$R_\infty$, which is an adjusted constant, it contributes negligibly to
the change in their values and uncertainties compared to $\alpha$.

The atomic mass constant $m_{\rm u}$ and molar mass constant $M_{\rm
u}$, {\it 9} and {\it 10}, are proportional to $\alpha^{-2}$ and thus
their relative uncertainties are twice that of $\alpha$.  They too
depend on $R_\infty$, and on the relative atomic mass of the electron
$A_{\rm r}({\rm e})$, which like $R_\infty$ is an adjusted constant.
However, like $R_\infty$, $A_{\rm r}({\rm e})$ contributes negligibly to
the change in the values and uncertainties of $m_{\rm u}$ and $M_{\rm
u}$ compared to $\alpha$.  (The Avogadro constant in the expression for
$M_{\rm u}$, {\it 10}, is of no consequence since it is exactly known.)
These constants are important because $m(X) = A_{\rm r}(X)m_{\rm u}$ and
$M(X) = A_{\rm r}(X)M_{\rm u}$.  That is, $m_{\rm u}$ ``converts'' the
relative atomic mass of an atomic particle $X$ to its mass $m(X)$ in kg,
and $M_{\rm u}$ converts $A_{\rm r}(X)$ to its molar mass $M(X)$ in
kg/mol, and vice versa.

Only a few comments are necessary for the remaining six constants in the
above list.  Since the relative uncertainty of $R_\infty$ is
$1.1\times10^{-12}$, the equation for the electron mass $m_{\rm e}$,
{\it 24}, suggests that in the foreseeable future, the relative
uncertainty of $m_{\rm e}$ will be twice that of $\alpha$.  The
``approximately equal'' sign, $\approx$, in the expressions for $a_{\rm
e}$, $g_{\rm e}$, and $\mu_{\rm e}/\mu_{\rm B}$, {\it 27, 29,} and {\it
33}, is a reminder that there are higher-order terms involving $\alpha$
and other contributions to the theoretical expression for $a_{\rm e}$,
$g_{\rm e}$, and $\mu_{\rm e}/\mu_{\rm B}$.  Also, in the equation for
$\mu_{\rm N}$, {\it 32}, like $R_\infty$ and $A_{\rm r}({\rm e})$,
$m_{\rm p}/m_{\rm e}$ contributes negligibly to the change in the value
and uncertainty of $\mu_{\rm N}$ compared to $\alpha$. 

The revised International System of Units based on exact values of $h$,
$e$, $k$, and $N_{\rm A}$ went into effect on 20 May 2019 and the values
chosen for them are based on the special CODATA adjustment carried out
in the summer of 2017 \cite{2018015,2018018}.  It was recognized that in
the revised SI the magnetic constant and molar mass of $^{12}$C would no
longer have their previously exact values, $\mu_0 = 4\rmpi\times10^{-7}$
N A$^{-2}$ and $M(^{12}{\rm C}) = 0.012$ kg mol$^{-1}$, but would become
experimentally determined constants.  However, since $M(^{12}{\rm C}) =
12M_{\rm u}$, the above expression {\it 2} for $\mu_0$ and {\it 10} for
$M_{\rm u}$ imply that the consistency of the recommended values of
$\mu_0$ and $M(^{12}{\rm C})$ with their previous exact values will
likely change from one adjustment to the next because of their
dependence on $\alpha$.  The following are the ratios:
\begin{center}
\begin{tabular}{c@{\qquad}c@{\qquad}c}
 & $\tfrac{\mu_0}{4\rmpi\times10^{-7} \mbox{ N A}^{-2}}$ 
 & $\tfrac{M(^{12}{\rm C})}{0.012 \mbox{ kg mol}^{-1}}$ \\[15 pt]
2017 & $1+20(23)\times10^{-11}$ & $1+37(45)\times10^{-11}$ \\[10 pt]
2018 & $1+55(15)\times10^{-11}$ & $1-35(30)\times10^{-11}$ \\[10 pt]
2022 & $1-13(16)\times10^{-11}$ & $1+105(31)\times10^{-11}$ \\[10 pt]
\end{tabular}
\end{center}
The 2017 values met the requirement for both $\mu_0$ and $M(^{12}{\rm
C})$, and $\mu_0$ still meets it in 2022 since the deviation is smaller
than its uncertainty, but the deviation for $M(^{12}{\rm C})$ in 2022
exceeds its uncertainty by over a factor of 3.

The experimental and theoretical input data discussed in
Secs.~\ref{sec:elmagmom} and \ref{sec:atomrecoil} determine the 2022
recommended value of $\alpha$.  These data, the uncertainties of which
were multiplied by the expansion factor 2.5 in the 2022 adjustment to
reduce their inconsistencies to an acceptable level, are the six items
D1 through D6 in Table~\ref{tab:pdata}.  Of these, the key results are
$h/m(^{133}{\rm Cs})$, $h/m(^{87}{\rm Rb})$, and $a_{\rm e}$(exp)
reported respectively by \citet{2018033,2020064,2023002}. Each of these
papers gives a value of $\alpha^{-1}$ derived by their authors based on
their data and are respectively 137.035\,999\,046(27) [Berkeley-18],
137.035\,999\,206(11) [LKB-20], and 137.035\,999\,166(15) [NW-23].  The
weighted mean of these three values after applying an expansion factor
of 2.5 is 137.035\,999\,178(21), which is essentially the same as the
2022 recommended value 137.035\,999\,177(21).  In the same vein, though
an expansion factor of 2.5 is applied to each of the six items D1
through D6, the $1.6\times 10^{-10}$ $u_{\rm r}$ of the 2022 recommended
value of $\alpha$ does not differ significantly from the $1.5\times
10^{-10}$ uncertainty of the 2018 value.

Returning to Table~\ref{tab:compare}, we consider those several
constants with a $|D_{\rm r}|$ or a 2018-to-2022 uncertainty ratio
greater than 2.0 that is not due to the change in $\alpha$.  The first
of these, $r_{\rm p}$ and $r_{\rm d}$, {\it 13, 14,} have uncertainty
ratios of 2.9 and 2.7, respectively.  This improvement arises from the
use of the recently available improved theory of the Lamb shift in
muonic hydrogen and deuterium \cite{Pachucki2022} together with the
previously available Lamb-shift measurement in $\rmmu$H \cite{2013011}
and in $\rmmu$D \cite{2016037} as discussed in Sec.~\ref{sec:mulsrprd}.

The relative atomic masses $A_{\rm r}$(p), $A_{\rm r}$(d), and $A_{\rm
r}$(h), {\it 18, 20, and 22,} also meet our criteria for discussion.  In
the 2022 adjustment it was decided to use cyclotron frequency ratio
measurements to determine the relative atomic masses of the light-mass
nuclei, because of the availability of new and highly accurate frequency
ratio data with parts in $10^{11}$ uncertainties.  The seven such input
data used are D15 through D21 in Table~\ref{tab:pdata}, and of these
only D19 and D20, $\omega_{\rm c}({\rm HD}^+)/\omega_{\rm c}(^3{\rm
He}^+)$ \cite{2017099} and $\omega_{\rm c}({\rm t})/\omega_{\rm
c}(^3{\rm He}+)$ \cite{2015002}, are employed in CODATA 2022 with no
change in value from that used in CODATA 2018.  The important ratio
$\omega_{\rm c}(^{12}{\rm C}^{6+})/\omega_{\rm c}({\rm p})$ reported by
\citet{2017056} was included in CODATA 2018 but its updated value
\cite{2019074} is the value included in CODATA 2022.  The other four
frequency ratios are $\omega_{\rm c}(^{12}{\rm C}^{6+})/\omega_{\rm
c}({\rm d})$ \cite{2020060}, $\omega_{\rm c}({\rm H}_2^+)/\omega_{\rm
c}({\rm d})$ \cite{2021042}, $\omega_{\rm c}(^{12}{\rm
C}^{4+})/\omega_{\rm c}({\rm HD}^+)$ \cite{2020060}, and $\omega_{\rm
c}(^{12}{\rm C}^{4+})/\omega_{\rm c}({\rm HD}^+)$ \cite{2006036}.  The
one non-cyclotron frequency input datum that contributes to the
determination of $A_{\rm r}$(e), $A_{\rm r}$(p), and $A_{\rm r}$(d) is
the newly available experimental and theoretical determination of the
rovibrational transition frequencies in HD$^+$ discussed in
Sec.~\ref{sec:HDplus}; based on the work of \citet{2023021}, these are
items D27 through D32 in Tables~\ref{tab:pdata} and \ref{tab:pobseqsb}.

We conclude our discussion of Table~\ref{tab:compare} with its last
three constants, $\gamma_{\rm p}^\prime$, $\sigma_{\rm p}^\prime$, and
$\mu_{\rm h}(^3{\rm He})$, {\it 42, 43, 44.} The reduction in their
uncertainties by the factors 2.7, 2.7, and 13.5, respectively, arise
from the new input datum $\mu_{\rm h}(^3{\rm He}^+)/\mu_{\rm N}$ with
$u_{\rm r} = 8.1\times 10^{-10}$; discussed in Sec.~\ref{ssec:dmhe3p},
it is item D45 in Table~\ref{tab:pdata} and is due to \citet{2022029}.
As can be seen from its observational equation in
Table~\ref{tab:pobseqsb}, it contributes to the determination of the
adjusted constant $\mu_{\rm e}/\mu_{\rm p}^\prime$ and thus also to the
determination of the recommended value of $\sigma_{\rm p}^\prime$ since
$(\mu_{\rm e}/\mu_{\rm p})/(\mu_{\rm e}/\mu_{\rm p}^\prime) = \mu_{\rm
p}^\prime/\mu_{\rm p} = 1-\sigma_{\rm p}^\prime$.  Further, it
contributes to the determination of the recommended value of
$\gamma_{\rm p}^\prime$ since $\gamma_{\rm p}^\prime = 4\rmpi\mu_{\rm
p}(1-\sigma_{\rm p}^\prime)/h$.  Finally, it is the dominant contributor
to the determination of the adjusted constant $\mu_{\rm h}(^3{\rm
He})/\mu_{\rm p}^\prime$, which yields the recommended value of
$\mu_{\rm h}(^3{\rm He})$ with $u_{\rm r} = 8.7\times10^{-10}$ when
multiplied by the recommended value of $\mu_{\rm p}^\prime$ and taking
into account the correlation of the two constants.

\subsection{Notable features of the CODATA 2022 adjustment}
\label{ssec:nfda}

\subsubsection{Impact on metrology and chemistry} 

The CODATA 2018 adjustment was the first to reflect the revision of the
SI based on exact values of $h$, $e$, $k$, and $N_{\rm A}$. This
revision significantly impacted electrical metrology and physical
chemistry because it eliminated the conventional electrical units
$V_{90}$, ${\it \Omega}_{90}$, $A_{90}$, {\it etc.} introduced in 1990
and because it made important physicochemical constants such as the
molar gas constant $R$, Faraday constant $F$, and Stefan-Boltzmann
constant $\sigma$ exactly known. By comparison, the impact of the 2022
adjustment on these fields is much less significant, but it does provide
the current values of the ratios $\mu_0/(4\rmpi\times10^{-7} \mbox{ N
A}^{-2})$ and $M(^{12}{\rm C})/(0.012 \mbox{ kg mol}^{-1})$.  These
ratios are given in the previous section and show that currently $\mu_0$
is fractionally {\it smaller} than its previous exact value by 13(16)
parts in $10^{11}$ and $M(^{12}C)$ is fractionally {\it larger} than its
previous exact value by 105(35) parts in $10^{11}$.

\subsubsection{Importance of theory}

Theory plays a significant role in the 2022 adjustment, perhaps even to
a greater extent than in previous adjustments; here is a summary of its
current role.
\begin{itemize}

\item

The improved theoretical values for the ionization energies $E_{\rm
I}({\rm H}_2^+)/hc$ and $E_{\rm I}({\rm HD}^+)/hc$ discussed in
Sec.~\ref{ssec:frmd}, items D25 and D26 in Tables~\ref{tab:pdata} and
\ref{tab:pobseqsb}, contribute to the determination of $A_{\rm r}$(e),
$A_{\rm r}$(p) $A_{\rm r}$(d), and $A_{\rm r}$(h).

\item

The theory of rovibrational transition frequencies in the molecular ion
HD$^+$ together with their experimentally determined values, items D27
through D32 in Tables~\ref{tab:pdata} and \ref{tab:pobseqsb} and
discussed in Sec.~\ref{sec:HDplus}, has allowed the experimental data to
be included as input data in the adjustment and contribute to the
determination of $A_{\rm r}$(e), $A_{\rm r}$(p), and $A_{\rm r}$(d).

\item

The theory of hydrogen and deuterium energy levels as discussed in
Secs.~\ref{ssec:hdel} and \ref{par:teu} combined with the experimentally
measured H and D transition frequencies contribute to the determination
of $R_\infty$.

\item

The improved theory of the Lamb shift in the muonic atoms $\rmmu$H and
$\rmmu$D discussed in Sec.~\ref{sec:mulsrprd} provide, in combination
with the experimentally measured Lamb shifts, accurate recommended
values of the rms charge radius of the proton and deuteron, $r_{\rm p}$
and $r_{\rm d}$. These in turn contribute to the determination of
$R_\infty$ with a reduced uncertainty. The theory of the Lamb shift in
$\rmmu^4{\rm He}^+$ discussed in the same section together with its
experimentally measured value provide for the first time a recommended
value of the rms charge radius of the alpha particle $r_{\rmssalpha}$;
its relative uncertainty is $1.2\times 10^{-3}$.

\item

The updated theory of the electron magnetic-moment anomaly $a_{\rm e}$
discussed in Sec.~\ref{sec:elmagmom} in combination with its
experimentally determined value, item D1 in Table~\ref{tab:pdata}, also
discussed in Sec.~\ref{sec:elmagmom}, provides one of the three most
accurate values of $\alpha$ available for inclusion in the 2022
adjustment.

\item

The theory of the muon magnetic-moment anomaly $a_{\rmssmu}$ is updated
and reviewed in Sec.~\ref{ssec:amuth}.  However, because of the
$4.2\,\sigma$ difference between the theoretical and experimental values
and questions about the hadronic contributions to the theory, especially
those from lattice QCD, the TGFC decided that it should not be included
in the CODATA 2022 adjustment. Consequently, the recommended value of
$a_{\rmssmu}$ is the experimentally determined value. Continued work on
$a_{\rmssmu}$  is of critical importance, because if the difference
between theory and experiment were to become unquestionable, it would be
a challenge for the Standard Model.

\item

The theory of the electron $g$-factors $g_{\rm e}(^{12}{\rm C}^{5+})$
and $g_{\rm e}(^{28}{\rm Si}^{13+})$ in the hydrogenic ions $^{12}$C$^{5+}$
and $^{28}$Si$^{13+}$ is updated in Sec.~\ref{ssec:thbegf}. They are
used in the respective observational equations for input data
$\omega_{\rm s}(^{12}{\rm C}^{5+})/\omega_{\rm c}(^{12}{\rm C}^{5+})$
and $\omega_{\rm s}(^{28}{\rm C}^{13+})/\omega_{\rm c}(^{28}{\rm
C}^{13+})$, D7 and D10 in Tables~\ref{tab:pdata} and \ref{tab:pobseqsb},
and contribute to the determination of $A_{\rm r}({\rm e})$.

\item

The theory of the hyperfine splitting in muonium is reviewed in
Sec.~\ref{ssec:muhfs} and found not to require any change from that used
in the 2018 adjustment. Similarly, the theory of various bound particle
to free-particle $g$-factor ratios is reviewed in
Secs.~\ref{ssec:thbfrats} and \ref{ssec:thbfratsmol} and evaluated with
2022 recommended values.  However, the values of these ratios, given in
Table~\ref{tab:gfactrat}, are unchanged from their 2018 values. The
ratios are taken as exact and are used in observational equations D41,
D42, and D43 in Table~\ref{tab:pobseqsb} and contribute to the
determination of $\mu_{\rm e}/\mu_{\rm p}$ and $\mu_{\rm d}/\mu_{\rm e}$

\end{itemize}

\subsubsection{Lack of data}

For the first time in recent CODATA adjustments, there is not a new
x-ray related datum to include in the 2022 adjustment nor is there a new
value of the Newtonian constant of gravitation, $G$.

\subsubsection{Decreased and increased uncertainties}

Table~\ref{tab:compare} indicates that the new data that became
available for the 2022 adjustment have led to significant reductions in
the uncertainties of the 2022 recommended values of many constants.
Nevertheless, it also shows that the uncertainties of a number of
constants have increased slightly, mainly because of the increase in the
uncertainty of $\alpha$.

\subsubsection{Changes in recommended values of constants}

The 2022 recommended values of many constants, of which
Table~\ref{tab:compare} gives only a small sample, have changed from the
2018 values. We recognize that using the one standard deviation
uncertainty of the 2018 value as the reference for calculating $D_{\rm
r}$ in Table ~\ref{tab:compare}, our chosen measure of these changes,
indicates larger changes than would be the case if the reference
uncertainty was the square root of the sum of the squares of the
uncertainties of the 2018 and 2022 values. Nevertheless, we believe that
emphasizing the changes in this way is useful, and of course, ideally
most changes should be smaller than the one standard deviation
uncertainty assigned to the value from the previous adjustment.

\subsubsection{Post closing date results}

Inevitably, useful new data became available after the 31 December
adjustment deasline, and the 2022 adjustment with its closing date of 31
December 2022 is no exception. Three such data are $A_{\rm
r}(\rmalpha)$ \cite{2023022}, $R^\prime_{\rmssmu}$ \cite{2023025}, and
both $\sigma_{\rm h}(^3{\rm He}^+)$ and $\sigma_{\rm h}(^3{\rm He})$
\cite{2023039}. The new value of $A_{\rm r}(\rmalpha)$ is briefly
discussed at the end of Sec.~\ref{ssec:frmd} and that of
$R^\prime_{\rmssmu}$ at the end of Sec.~\ref{ssub:fnal}.
\citet{volkov24} recently posted a preprint giving his result for the
total $10^{\rm th}$-order QED contribution to the lepton magnetic-moment
anomalies.

\subsection{Suggestions for future work}

We focus here on those constants for which the uncertainties of the data
that determined their recommended values in the 2022 adjustment required
an expansion factor to reduce their inconsistencies to an acceptable
level. Of course, innovative new experiments and theoretical
calculations that will lead to reduced fundamental constant
uncertainties should always be encouraged because it can never be known
in advance what new physics might be uncovered in the next decimal
place.

\subsubsection{Fine-structure constant $\alpha$}

Table~\ref{tab:compare} and its accompanying discussion of the
importance of $\alpha$ for the determination of the recommended values
of other constants, and the fact that the uncertainties of the six input
data that determine its 2022 recommended value have to be increased by a
factor 2.5, make clear why work to improve our knowledge of the value of
$\alpha$ should be given high priority.  Moreover, obtaining a more
accurate value of $\alpha$ that depends only weakly on QED theory and
comparing it with a more accurate value from $a_{\rm e}$(exp) and
$a_{\rm e}$(th) can provide an improved test of the Standard Model.

The large change in the 2018 recommended value of $\alpha$ is primarily
due to the change in the value of $h/m(^{87}{\rm Rb})$ resulting from
the new LKB atom-recoil experiment reported by \citet{2020064}; it is
input D3 in Table~\ref{tab:pdata}. As discussed in
Sec.~\ref{sec:atomrecoil}, this new experiment with much improved
apparatus and methodology uncovered significant systematic effects in
the earlier LKB experiment \cite{2011014} that could not be corrected
retroactively.  Presumably, the most significant of these effects
would have been reduced in the similar experiment to measure
$h/m(^{133}{\rm Cs})$ at Berkeley reported by \citet{2018033}; the
latter is input datum D4 in Table~\ref{tab:pdata}.  Nevertheless,
Fig.~\ref{fig:alpha} of Sec.~\ref{sec:atomrecoil} shows that the $a_{\rm
e}$ and $h/m(^{87}{\rm Rb})$ recoil values of $\alpha^{-1}$ are in
better agreement with each other than the $h/m(^{133}{\rm Cs})$ recoil
value is with either of them. This inconsistency is apparent from the
4.7 normalized residual of $h/m(^{133}{\rm Cs})$ in the 2022 adjustment
before any expansion factor is applied. Continued work on experiments to
determine $a_{\rm e}$ and $h/m$ of atoms including their relative atomic
masses, and, as already mentioned, on the theory of $a_{\rm e}$, are
encouraged.

\subsubsection{Newtonian constant of gravitation $G$}

The value of $G$ has been the least well known of the major fundamental
constants for decades. The uncertainties of the 16 values in
Table~\ref{tab:bg} on which the CODATA 2022 recommended value is based
required an expansion factor of 3.9 to reduce their inconsistencies to
an acceptable level and no new values have become available for the 2022
adjustment. Recently, \citet{2023018} experimentally investigated if the
inconsistency of the two values of $G$ identified as BIPM-01 and BIPM-14
in Table~\ref{tab:bg} with other values arises from stray alternating
current (AC) magnetic fields in the vicinity of the BIPM $G$ experiments
but concluded this possibility was unlikely. Before the uncertainties of
the 16 values of $G$ in Table~\ref{tab:bg} were multiplied by 3.9, the
largest three normalized residuals were 7.7, 6.8, and 4.8 for BIPM-14,
JILA-18, and BIPM-01, respectively. Two review articles make clear why
$G$ is such a difficult constant to determine accurately
\cite{2019041,2017105}. The detailed review of the two BIPM
determinations of $G$, BIPM-01 and BIPM-14, by \citet{2014120} is also
insightful. We note that the balance used in the BIPM-14 experiment was
transferred to NIST for use there to measure $G$ \cite{schlametal}.  Two
new approaches that may eventually provide a reliable value of $G$ and
the pursuit of which should be encouraged have also been proposed, one
by using atom interferometry \cite{2021056, 2018005} and the other using
precision displacement sensors \cite{2020015}.

\subsubsection{Transition frequencies of H and D}

In the initial 2022 adjustment, six of the 29 experimentally determined H
and D transition frequencies in Table~\ref{tab:rydfreq} had a normalized
residual $r_i$ greater than 2. These are items A12 through A15 and A22
and A23 in that table with initial $r_i$ values 3.1, 2.5, 2.5, 3.1. 2.7,
and 3.4, respectively. An expansion factor of 1.7 applied to the
uncertainties of all 29 transition frequencies in the final adjustment
reduced all six of these to 2 or less and at the same time maintained the
relative weights of these data in the adjustment.  For this reason, but
also for the other reasons given in the next-to-last paragraph of
Sec~\ref{ssec:cd} above, the same expansion factor is applied to the 25
additive corrections in Table~\ref{tab:deltaRyd} and to the six muonic
atom Lamb-shift data in Table~\ref{tab:muhdata}.

The four 2S$-$8D transition results A12 through A15 are from an LKB and
SYRTE collaborative effort in Paris and are reported by \citet{1997001}.
The A12 measurement used hydrogen and the A13, A14, and A15 measurements
used deuterium. The self-sensitivity coefficients $S_{\rm c}$ of these
four data are only 0.08\,\%, 0.04\,\%, 0.05\,\%, and 0.07\,\%, which means that
they play an insignificant role in determining the adjusted value of
their transition and hence $R_\infty$. The 1S$_{1/2}-$3S$_{1/2}$
transition result A22 was obtained using hydrogen in a more recent LKB
effort and is reported by \citet{2018042}; its $S_{\rm c}$ is 0.28\,\%.
The 2S$_{1/2}-$8D$_{5/2}$ transition result A23, the most recent and the
last of the six, was also obtained using hydrogen.  It was determined at
Colorado State University in the USA (CSU) and reported by
\citet{2022002}; its 3.4 initial residual is the largest of the six but
its $S_{\rm c}$ of 1.8\,\% is also the largest.

Unfortunately, the current H and D transition frequency situation is
problematic. There are six data that lead to an expansion factor of 1.7
for some 60 input data, but $S_{\rm c}$ of four of those six data is
less than 0.1\,\%, for another it is less than 0.3\,\%, and for the last
one it is less than 2\,\%. Research that could eliminate the need for an
expansion factor would have significant benefits; for example, it would
reduce the uncertainty of the recommended values of $R_\infty$, $r_{\rm
p}$, and $r_{\rm d}$. Yet another re-examination of the four LKB/SYRTE
results reported some 35 years ago and new measurements of the six
transitions in question would be valuable.  The fact that the most
recently reported transition result, A23 with $u_{\rm r} =
2.6\times10^{-12}$, has an initial residual of 3.4 may indicate the
existence of unrecognized systematic effects in such experiments.

As discussed in Sec.~\ref{ssec:vrads}, the values of $r_{\rm p}$ and
$r_{\rm d}$ obtained from the measurements and theory of electronic H
and D transition frequencies exceed by $2.8\,\sigma$ the values obtained
from the measurements and theory of the Lamb shift in muonic hydrogen
and deuterium. This disagreement provides further evidence of the need
for additional experimental and theoretical research in both of these
areas.

We conclude our report on the 2022 CODATA adjustment with a thought
expressed at the conclusion of the 2018 CODATA report. The key idea is
this: It would be useful if researchers kept in mind the limited
robustness of the data set on which CODATA adjustments are based in
planning their research. All too often there is only one input datum for
a quantity and too many important input data are too many years old, as
an inspection of the tables listing the input data show. Unknown
systematic errors can often be identified if the same quantity is
measured by a different method in a different laboratory, and similarly
for theoretical calculations. Repeating a previous experiment or
calculation may not be as glamorous as doing it for the first time even
if done by a different method, but doing so may be the only way to
ensure that a result is correct and that the magnitude of the changes in
the recommended values of the constants from one CODATA adjustment to
the next will not be unduly large.

\section{List of symbols and abbreviations} \label{sec:nom}

\begin{longtable}{lp{6.75cm}}
\endfirsthead
\endhead
\endlastfoot{}
\endfoot{}
\\

ASD & NIST Atomic Spectra Database (online) \\

AMDC & Atomic Mass Data Center, Institute of Modern Physics, Chinese
Academy of Sciences, Lanzhou, People's Republic of China.\\

$A_{\rm r}(X)$ & Relative atomic mass of $X$: $A_{\rm r}(X) =
m(X)/m_{\rm u}$ \\

$a_0$ & Bohr radius: $a_0=\hbar/\alpha m_{\rm e} c$ \\

$a_{\rm e}$ & Electron magnetic-moment anomaly: $a_{\rm e} = (|g_{\rm e}|-2)/2$ \\

$a_{\rmssmu}$ & Muon magnetic-moment anomaly: $a_{\rmssmu} =
(|g_{\rmssmu}| -2$)/2  \\

Berkeley  &  University of California at Berkeley, Berkeley, California, USA\\ 

BIPM & International Bureau of
Weights and Measures, S\`evres, France \\

BNL & Brookhaven National Laboratory, Upton, New York, USA \\

CGPM &  General Conference on Weights and Measures \\

CIPM &  International Committee for Weights and Measures \\

CODATA & Committee on Data of the International Science Council  \\

CREMA &  The international collaboration {\it Charge Radius Experiment with Muonic Atoms} at the Paul Scherrer Institute, Villigen, Switzerland \\

CSU & Colorado State University, Fort Collins, Colorado, USA \\

$c$ & Speed of light in vacuum and one of the seven defining constants of the SI \\

d & Deuteron (nucleus of deuterium D, or $^2$H) \\

$d_{220}$ & $\{220\}$ lattice spacing of an ideal silicon crystal with natural
isotopic Si abundances \\

$d_{220}({\scriptstyle X })$ & $\{220\}$ lattice spacing of crystal
$X$ of silicon with natural isotopic Si abundances \\

$E_{\rm h}$ & Hartree energy: $E_{\rm h} = 2R_\infty hc = \alpha^2m_{\rm e}c^2$\\
	
e   & Symbol for either member of the electron-positron 
pair; when necessary, e$^-$ or e$^+$ is used to indicate
the electron or positron  \\

$e$ & Elementary charge: absolute value of the charge of the electron and one of the seven defining constants of the SI\\

FNAL & Fermi National Accelerator Laboratory, Batavia, Illinois, USA\\

FSU & Florida State University, Tallahassee, Florida, USA \\

FSUJ & Friedrich-Schiller University, Jena, Germany \\

$G$ & Newtonian constant of gravitation \\

$G_{\rm F}$ & Fermi coupling constant\\

$g_{\rm d}$ & Deuteron $g$-factor: $g_{\rm d} = \mu_{\rm d}/\mu_{\rm N}$ \\

$g_{\rm e}$ & Electron $g$-factor: $g_{\rm e} = 2\mu_{\rm e}/\mu_{\rm B}$ \\
 
$g_{\rm p}$ & Proton $g$-factor: $g_{\rm p} = 2\mu_{\rm p}/\mu_{\rm N}$ \\

$g^\prime_{\rm p}$ & Shielded proton $g$-factor: 
$g_{\rm p}^\prime = 2\mu_{\rm p}^\prime/\mu_{\rm N}$ \\

$g_{\rm t}$ & Triton $g$-factor: $g_{\rm t} = 2\mu_{\rm t}/\mu_{\rm N}$ \\

$g_{X}(Y)$ & $g$-factor of particle $X$ in the ground (1S) state of \
hydrogenic atom $Y$ \\

$g_{\rmssmu}$ & Muon $g$-factor: $g_{\rmssmu} = 2\mu_{\rmssmu}/(e\hbar/2m_{\rmssmu})$ \\

Harvard & HarvU also.  Harvard University, Cambridge, Massachusetts, USA \\

HD & A hydrogen-deuterium molecule  \\

HHU & Heinrich-Heine-Universit\"at, D\"usseldorf, Germany\\

HT & A hydrogen-tritium molecule  \\

HUST & Huazhong University of Science and Technology, Wuhan, People's Republic of China \\

h & Helion (nucleus of $^{3}$He) \\

$h$ & Planck constant and one of the seven defining constants of the SI  \\

$\hbar$ & Reduced Planck constant \\

ILL & Institut Max von Laue-Paul Langevin, Grenoble, France \\

INRIM  & Istituto Nazionale di Ricerca Metrologica, Torino, Italy \\

JILA & JILA, University of Colorado and NIST, Boulder, Colorado, USA \\

J-PARC & Japan Proton Accelerator Research Complex \\

$k$ & Boltzmann constant and one of the seven defining constants of the SI\\

KEK & High Energy Accelerator Research Organization, Tsukuba, Japan \\

LAMPF & Clinton P. Anderson Meson Physics Facility at Los Alamos National Laboratory, Los Alamos, New Mexico, USA \\

LANL & Los Alamos National Laboratory, Los Alamos, New Mexico, USA \\

LENS & European Laboratory for Non-Linear Spectroscopy, University of Florence, Italy \\

LKB & Laboratoire Kastler-Brossel, Paris, France \\

LSA & Least-squares adjustment \\

MIT & Massachusetts Institute of Technology, Cambridge, Massachusetts, USA \\

MPIK & Max-Planck-Institut f\"ur Kernphysik, Heidelberg, Germany \\

MPQ & Max-Planck-Institut f\"ur Quantenoptik, Garching, Germany \\

MSL & Measurement Standards Laboratory, Lower Hutt, New Zealand \\

$M(X)$ & Molar mass of $X$: $M(X) = A_{\rm r}(X) M_{\rm u}$ \\

$M(^{12}{\rm C})$ & Molar mass of carbon-12. $M(^{12}{\rm C})= 12M_{\rm u}=12 N_{\rm A}m_{\rm u} \approx 0.012~{\rm kg}/{\rm mol}$\\

$M_{\rm u}$ & Molar mass constant: $M_{\rm u} = N_{\rm A}m_{\rm u}$ \\

Mu & Muonium (${\rmmu}^{+} {\rm e}^{-}$ atom) \\

$m_{\rm u}$ & Unified atomic mass constant: $m_{\rm u} = m(^{12}{\rm C})/12 = 2hcR_\infty/\alpha^2c^2A_{\rm r}({\rm e})$\\

$m_{X}$, $m(X)$ & Mass of $X$ (for the electron e, proton p, and
other elementary particles, the first symbol is used, i.e.,
$m_{\rm e}$, $m_{\rm p}$, etc.) \\

$N_{\rm A}$ & Avogadro constant and one of the seven defining constants of the SI\\

NIST & National Institute of Standards and Technology, Gaithersburg, 
Maryland and Boulder, Colorado, USA  \\

NMR & Nuclear magnetic resonance \\

NPL  &  National Physical Laboratory, Teddington, UK\\

n & Neutron \\

$p(\chi^2|\nu)$ & Probability that an observed value of chi square for
$\nu$ degrees of freedom would exceed $\chi^2$ \\

p & Proton \\

PDG & Particle Data Group \\

PTB & Physikalisch-Technische Bundesanstalt, Braunschweig and Berlin,
Germany \\

QCD & Quantum chromodynamics \\

QED & Quantum electrodynamics \\

$R$ &  Molar gas constant; $R=N_{\rm A}k$\\

$R_{\rm B}$ & Birge ratio: $R_{\rm B} = (\chi^{2}/\nu) ^\frac {1}{2}$ \\ 

$R_\infty$ & Rydberg constant: $R_\infty = m_{\rm e}c\alpha^{2}
/2h$ \\

$r_i$  &    Normalized residual of an input datum $X_i$ in a
least-squares adjustment: \\ 
 &  $r_i = (X_i -\langle X_i\rangle) /u(X_i)$  \\

$r_{\rm d}$ & Bound-state rms charge radius of the deuteron \\

$r_{\rm p}$ & Bound-state rms charge radius of the proton \\

$r(X,Y)$ & Correlation coefficient of quantity or constant  $X$ and $Y$:
$r(X,Y) = u(X,Y)/[u(X) u(Y)]$ \\

rms & square root of the mean of the squares \\

SI & Syst\`eme international d'unit\'es (International System of Units) \\

StPtrsb &  D. I. Mendeleyev All-Russian Research Institute for Metrology (VNIIM), St. Petersburg, Russian Federation \\

Sussex  &  University of Sussex, Brighton, UK\\

SYRTE & Syst\`emes de r\'ef\'erence Temps Espace, Paris, France \\

TGFC & Task Group on Fundamental Constants of the Committee
on Data of the International Science Council (CODATA)\\

TR\&D & Tribotech Research and Development Company, Moscow, Russian Federation \\

t & Triton (nucleus of tritium T, or $^3$H) \\

UBarc &  Universitat Aut\`onoma de Barcelona, Barcelona, Spain\\

UCB  &  University of California at Berkeley, Berkeley, California, USA \\ 

UCI & University of California at Irvine, Irvine, California, USA \\

UMZ &  Institut f\"ur Physik,
Johannes Gutenberg-Universit\"at Mainz, Mainz, Germany \\

UWash & University of Washington, Seattle, Washington, USA \\

UWup & University of Wuppertal, Wuppertal, Germany \\

UZur & University of Zurich, Zurich, Switzerland \\

u & Unified atomic mass unit (also called the dalton, Da): 1 u = $m_{\rm u}$ = $m(^{12}$C)/12 \\

$u(X)$ & Standard uncertainty (i.e., estimated standard deviation)
of quantity or constant $X$ \\

$u_{\rm r}(X)$ & Relative standard uncertainty of a quantity or constant $X$: 
$u_{\rm r}(X) = u(X)/|X|, \ X \ne 0$ (also simply $u_{\rm r}$) \\

$u(X,Y)$ & Covariance of quantities or constants $X$ and $Y$ \\
 
$u_{\rm r}(X,Y)$ & Relative covariance of quantities or constants $X$ and $Y$:
$u_{\rm r}(X,Y) = u(X,Y)/(X Y)$ \\ 

$u_0$ &  Type of uncertainty in the theory of the energy levels of
hydrogen and deuterium:  The contribution to the energy has correlated
uncertainties for states with the same $\ell$ and $j$. See also
entry $u_{\rm n}$.\\

$u_{\rm n}$ & Type of uncertainty in the theory of the energy levels
of hydrogen and deuterium: The contribution has uncorrelated
uncertainties. See also entry $u_0$.  \\

VUA & Vrije Universiteit Amsterdam, Amsterdam, The Netherlands \\

WarsU & University of Warsaw, Warszawa, Poland \\

Yale & Yale University, New Haven, Connecticut, USA \\

York & York University, Toronto, Canada\\

$\alpha$ & Fine-structure constant: \\
& $\alpha = e^2/4\rmpi\epsilon_0\hbar c \approx 1/137 $ \\

${\rmalpha}$ & Alpha particle (nucleus of $^{4}$He) \\

$\Delta{E}_{\rm B}(^AX^{n+})$ & Energy required to remove $n$ electrons from a neutral atom \\

$\Delta{E}_{\rm I}(^AX^{i+})$ & Electron ionization energies, $i = 0$ to $n-1$ \\

$\Delta{E}_{\rm Mu}$ & Ground-state muonium hyperfine splitting energy\\

$\Delta {\cal E}_{\rm LS}({\rmmu}{\rm H},{\rmmu}{\rm D})$  &  Transition energy of Lamb shift in muonic hydrogen or muonic deuterium \\	

$\delta_{\rm H,\rm D}{(X)}$& Additive correction to the theoretical expression for the energy of a specified level in hydrogen or deuterium \\

$\delta_{\rm th}{(X)}$ & Additive correction to a specified theoretical expression \\	

$\doteq$ & Symbol used to relate an input datum to its observational equation \\

$\theta_{\rm W}$ & Weak mixing angle \\

$\lbar_{\rm C}$ & Reduced Compton wavelength: $\lbar_{\rm C}=\hbar/m_{\rm e}c$\\

${\rmmu}$ & Symbol for either member of the muon-antimuon pair; when necessary, ${\rmmu}^{-}$ or ${\rmmu}^{+}$
is used to indicate the negative muon or positive antimuon \\

${\rmmu}{\rm D}$ & Muonic deuterium (an atom comprising a deuteron and a muon)\\

${\rmmu}{\rm H}$ & Muonic hydrogen (an atom comprising a proton and a muon)\\

$\mu_{\rm B}$ & Bohr magneton: $\mu_{\rm B} = e \hbar/2m_{\rm e}$ \\

$\mu_{\rm N}$ & Nuclear magneton: $\mu_{\rm N} = e\hbar/2m_{\rm p}$ \\

$\mu_{X}(Y)$ & Magnetic moment of particle $X$ in atom or molecule $Y$. \\

$\mu_{X}$,~$\mu^\prime_{X}$ & 
Magnetic moment, or shielded magnetic moment, of particle $X$ \\

$\nu$ & Degrees of freedom of a particular least-squares adjustment: $\nu = N - M$, $N$ number of input data, $M$ number of variables or adjusted constants \\

$\sigma$ & Stefan-Boltzmann constant: \\
& $\sigma = ({\rmpi}^{2}/60)k^{4}/\hbar^{3}c^{2}$ \\

${\rmtau}$ & Symbol for either member of the tau-antitau pair; 
when necessary, ${\rmtau}^{-}$ or
${\rmtau}^{+}$ is used to indicate the negative or positive tau lepton\\

${\rm \chi}^{2}$ & The statistic ``chi square'' 

\end{longtable}

\section*{Acknowledgments}

We thank our fellow CODATA Task Group members for their invaluable
guidance, suggestions, and contributions to the 2022 adjustment.  We
gratefully acknowledge our many colleagues throughout the world who
provided results prior to formal publication and promptly answered many
questions about their work.

\end{document}